\documentclass[phd,bottom,nosig]{usbthesis}
\usepackage{graphicx}
\usepackage{color}
\usepackage{bm}
\usepackage{amsmath}
\usepackage{amssymb}
\usepackage{setspace}
\usepackage{array}
\usepackage{tabularx}
\usepackage{multirow}
\usepackage{afterpage}
\usepackage{hhline}
\usepackage{url}
\usepackage{verbatim}
\usepackage{hyperref} 
\usepackage{footmisc} 
\usepackage{xspace}
\usepackage[numbers,sort&compress]{natbib}
\usepackage{pdfpages}

\def\simge{%
    \mathrel{\rlap{\raise 0.511ex
    \hbox{$>$}}{\lower 0.511ex \hbox{$\sim$}}}}
\def\simle{%
    \mathrel{\rlap{\raise 0.511ex
    \hbox{$<$}}{\lower 0.511ex \hbox{$\sim$}}}}

\renewcommand{\vec}[1]{\boldsymbol{#1}}

\DeclareMathAccent{\ring}{\mathalpha}{operators}{"17}

\providecommand*{\eu}{\ensuremath{\mathrm{e}}}

\providecommand*{\iu}{\ensuremath{\mathrm{i}}}

\providecommand{\renewoperator}[3]{\renewcommand*{#1}{\mathop{#2}#3}}
\renewoperator{\Re}{\mathrm{Re}}{\nolimits}
\renewoperator{\Im}{\mathrm{Im}}{\nolimits}

\makeatletter
\providecommand*{\diff}{\@ifnextchar^{\DIfF}{\DIfF^{}}}
\def\DIfF^#1{\mathop{\mathrm{\mathstrut d}}\nolimits^{#1}\gobblespace}
\def\gobblespace{\futurelet\diffarg\opspace}
\def\opspace{%
    \let\DiffSpace\!%
    \ifx\diffarg(%
        \let\DiffSpace\relax
    \else
        \ifx\diffarg[%
            \let\DiffSpace\relax
        \else
            \ifx\diffarg\{%
                \let\DiffSpace\relax
            \fi\fi\fi\DiffSpace}

\newcommand{\ket}[1]{\ensuremath{|{#1}\rangle}}

\newcommand\blfootnote[1]{%
  \begingroup
  \renewcommand\thefootnote{}\footnote{#1}%
  \addtocounter{footnote}{-1}%
  \endgroup
}

\author{Sahal Kaushik}%
\title{Magnetic and Optical Response of Chiral Fermions}%

\month{August}
\year{2021}%
\program{Physics}%
\director{Dmitri E. Kharzeev}{Distinguished Professor, Department of Physics and Astronomy}
\directorr{Jennifer Cano}{Assistant Professor, Department of Physics and Astronomy}%
\chairman{Edward Shuryak}{Distinguished Professor, Department of Physics and Astronomy}%
\fstmember{Xu Du}{Associate Professor, Department of Physics and Astronomy}%
\outmember{Deyu Lu}{Physicist with Continuing Appointment}{Brookhaven National Laboratory}%
\dean{Eric Wertheimer}%

\begin{document}

\singlespacing %
\pagenumbering{roman} %
\maketitle %
\makeapproval 

\begin{abstract}
Dirac and Weyl materials possess chiral fermions, which are characterized by nontrivial topology and large Berry curvature. Chiral fermions have nontrivial interactions with magnetic fields and light. In this work, we propose three different mechanisms for magnetic photocurrents caused by chiral fermions, with different requirements on the symmetry group of the crystal. We also study quantum oscillations of the anomalous current in the presence of a magnetic field, showing it has a different phase from the Ohmic current. We formulate a mechanism for THz emission observed in TaAs in response to ultrafast pulses. We propose a strain-induced anomalous current. We also show that an external magnetic field can create a difference between left and right handed fermions, which can be controlled by changing the field.
\end{abstract}

\begin{dedication}
\textit{To the late Dr. N. Ratnasree}
\end{dedication}

\tableofcontents %
\listoffigures %
\listoftables %
\begin{acknowledgements}
    I would first like to thank my advisor Dmitri Kharzeev for guiding me in my development as an independent researcher. The work described in Chapters \ref{chOscillations}, \ref{chCMP}, \ref{chTHz}, \ref{chTHME}, and \ref{chStrain} was supervised by him.

I would also like to thank Jennifer Cano, my co-advisor for the last one and a half years, for her guidance, and especially for helping me to apply for future positions. The work described in Chapters \ref{chTunable} and \ref{chMHME} was supervised by her.

I next acknowledge my colleague and collaborator Evan John Philip. Working with him was a very enjoyable experience. He contributed to the work described in Chapters \ref{chCMP}, \ref{chTHz}, \ref{chTHME}, \ref{chStrain}, and \ref{chTunable}.

I thank Mengkun Liu and Jingbo Qi for giving us the opportunity to contribute to the theoretical explanation of their observation. The work described in Chapter \ref{chTHz} is a result of this collaboration.   

I also thank Lanlan Gao, who conceived the project on which Chapter \ref{chStrain} is based.

I am also grateful to Qiang Li for hosting me at BNL for a summer, and Xu Du and Alexander Abanov for insightful group meetings and Yuta Kikuchi for useful discussions.

I have learnt a lot from coursework at Stony Brook. In particular I acknowledge Peter van Nieuwenhuizen whose group theory course was extremely illuminating.

I thank my friends, in particular Kaushik Roy, Mukul Sholapurkar, and Sharmila Duppala for being part of the memorable time I had at Stony Brook.

Finally, I thank my family for their moral support over these last 6 years.

\vspace{6 mm}

This work was supported in part by the US Department of Energy under Awards DE-SC-0017662 and DE-FG- 88ER40388. I also acknowledge the support of an OPVR Seed Grant from Stony Brook University.

\end{acknowledgements}

\begin{paperslist}
This dissertation is based on the following published works and preprints under review:

\begin{enumerate}
    \item Kaushik, S. and Cano, J. Magnetic Photocurrents in Multifold Weyl Fermions. \textit{arXiv preprint arXiv:2107.05106}. July 2021.
    \item Kaushik, S., Philip, E. J., and Cano, J. Tunable chiral symmetry breaking in symmetric Weyl materials. \textit{Phys Rev B}, 103(8), 085106.\\  \textit{arXiv:2011.00970}. Feb 2021.
    \item Gao, L. L., Kaushik, S., Kharzeev, D. E., and Philip, E. J. Chiral kinetic theory of anomalous transport induced by torsion. \textit{arXiv preprint arXiv:2010.07123}. Oct 2020
    \item Kaushik, S., Kharzeev, D. E., and Philip, E. J. Transverse chiral magnetic photocurrent induced by linearly polarized light in mirror-symmetric Weyl semimetals. \textit{Phys Rev Research}, 2(4), 042011. \textit{arXiv:2006.04857}. Oct 2020.
    \item Gao, Y., Kaushik, S., Philip, E.J., Li, Z., Qin, Y., Liu, Y.P., Zhang, W.L., Su, Y.L., Chen, X., Weng, H., Kharzeev, D.E., and Qi, J. Chiral terahertz wave emission from the Weyl semimetal TaAs. \textit{Nature Communications}, 11(1), 1-10. \textit{arXiv:1901.00986}. Feb 2020.
    \item Kaushik, S., Kharzeev, D. E., and Philip, E. J. Chiral magnetic photocurrent in Dirac and Weyl materials. \textit{Phys Rev B}, 99(7), 075150. \textit{arXiv:1810.02399}. Feb 2019.
    \item Kaushik, S. and Kharzeev, D. E. Quantum oscillations in the chiral magnetic conductivity. \textit{Phys Rev B}, 95(23), 235136. \textit{arXiv:1703.05865}. June 2017.
\end{enumerate}
\end{paperslist}

\pagestyle{thesis}
\newpage
\pagenumbering{arabic}
\chapter{Introduction}
\section{Outline}
This dissertation is organized as follows: In this Chapter, we first give a brief introduction to Dirac and Weyl fermions and photocurrents in Weyl materials. 

In Chapter \ref{chOscillations} we describe quantum oscillations in the anomalous conductivity of Dirac and Weyl materials. This chapter is adapted from \cite{SahalOsc}.

In Chapter \ref{chCMP} we discribe a proposed mechanism of magnetic photocurrent in response to \textit{circularly} polarized light in both Dirac and Weyl materials, due to anomalous pumping of chiral from light to matter. This chapter is adapted from \cite{kaushik2019chiral}.

In Chapter \ref{chTHz} we discuss an observation of elliptically polarized terahertz emission from TaAs in response to circularly polarized ultrafast optical pulses, due to the chirality of Weyl cones. This chapter is adapted from \cite{gao2020chiral}.

In Chapter \ref{chTHME} we discuss a photocurrent transverse to applied magnetic field in response to \textit{linearly} polarized light in TaAs and related Weyl materials, due to the Berry curvature of Weyl cones. This chapter is adapted from \cite{SahalTHME}.

In Chapter \ref{chStrain}, we discuss chiral charge pumping by nonuniform and time-varying strain, which has similar effects on Weyl fermions as electromagnetic field. This chapter is adapted from \cite{LanLanStrain}.

In Chapter \ref{chTunable}, we describe how an external magnetic field can break all symmetries relating left and right handed Weyl fermions, in a material that has mirror or inversion symmetry when unperturbed. This chapter is adapted from \cite{SahalAsymm}.

In Chapter \ref{chMHME}, we show that multifold Weyl fermions can have a photocurrent parallel to external magnetic field in response to \textit{linearly} polarized light, even in the ``ideal" situation when the Hamiltonian is perfectly spherically symmetric and linear. This chapter is adapted from \cite{SahalMHME}.

Finally, we summarize these results and mention possible future directions.

\section{Chirality}
Chirality is a concept of ``handedness". This word was introduced by Lord Kelvin, in the context of crystal structures \cite{kelvin}. An object is said to be chiral if it is not equivalent to its mirror image, or in other words, if it is not related to its mirror image by any rotation, translation, or combination of rotation and translation. More precisely, it does not have any reflection symmetry, or a symmetry that is a combination of a mirror symmetry and translations and/or rotations. However, it is possible for chiral objects to be invariant under rotations, translations, and/or their combinations. As an example, helices are chiral: they can be left or right handed, and helices of opposite handedness cannot be superimposed on each other.

In 3 dimensions, or any odd number of dimensions, chiral objects necessarily lack inversion symmetry $P$, which flips all coordinates ($\vec{x}\to-\vec{x}$). However, not all objects that lack inversion symmetry are chiral. For example, it is possible that an object has a plane of reflection and has the symmetry $(x,y,z)\to(-x,y,z)$, but not inversion symmetry. In even dimensions, inversion is a rotation or combination of rotations, and it is possible for chiral objects to be invariant under inversion.

Example of chiral objects include biological molecules such as DNA, proteins, and sugars. Chiral molecules often form crystals with macroscopically chiral shapes, e.g. tartarate salts. The chirality of subcellular structures ultimately results in chiral gross anatomy in many organisms - for example, in the human body, certain organs always form on one side.

\begin{figure}
    \centering
    \includegraphics[scale=0.15]{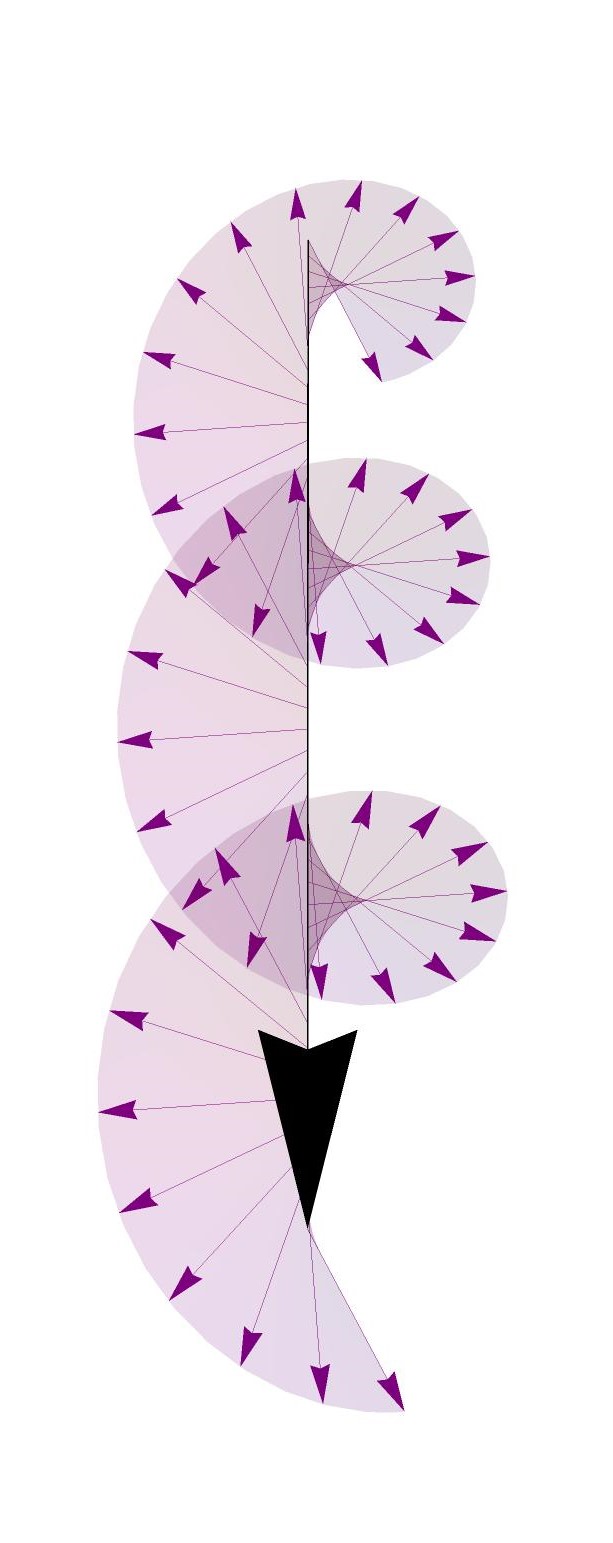}\hspace{0.15\linewidth}\includegraphics[scale=0.15]{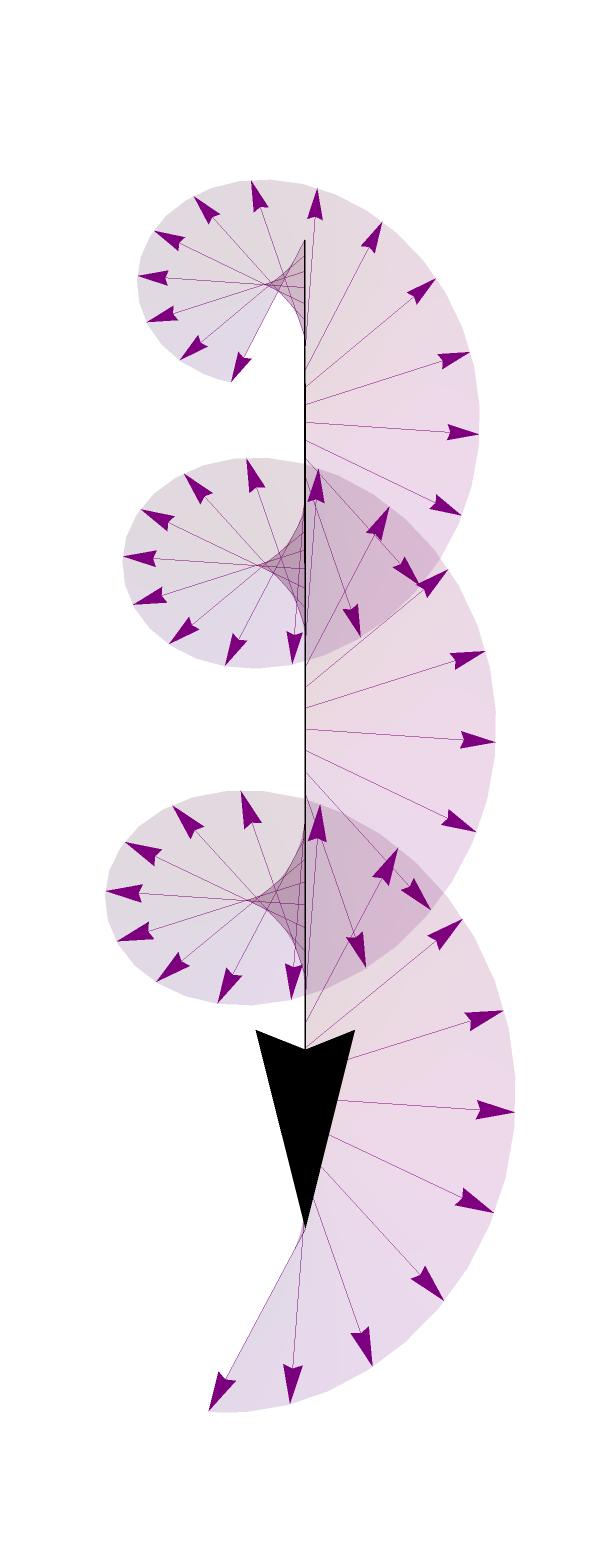}
    \caption{Circularly polarized light is not equivalent to its mirror image and is chiral}
    \label{fig:xlight}
\end{figure}

Circularly polarized light is also chiral, as depicted in Fig~\ref{fig:xlight}. A classic effect involving chirality is optical rotation: when linearly polarized light passes through a solution of chiral molecules, such as glucose, its plane of polarization gets rotated. This is because linearly polarized light is a superposition of left and right handed circularly polarized light. In a chiral medium, the two circularly polarized components have slightly different refractive indices, and acquire a phase difference. This phase difference is equivalent to the angle of rotation. 

\begin{figure}
    \centering
    \includegraphics[scale=0.25]{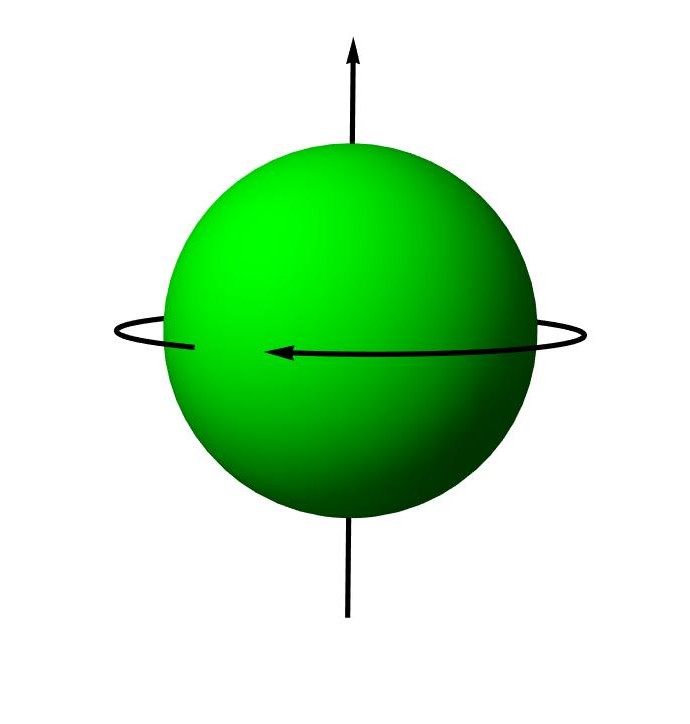}\hspace{0.1\linewidth}\includegraphics[scale=0.25]{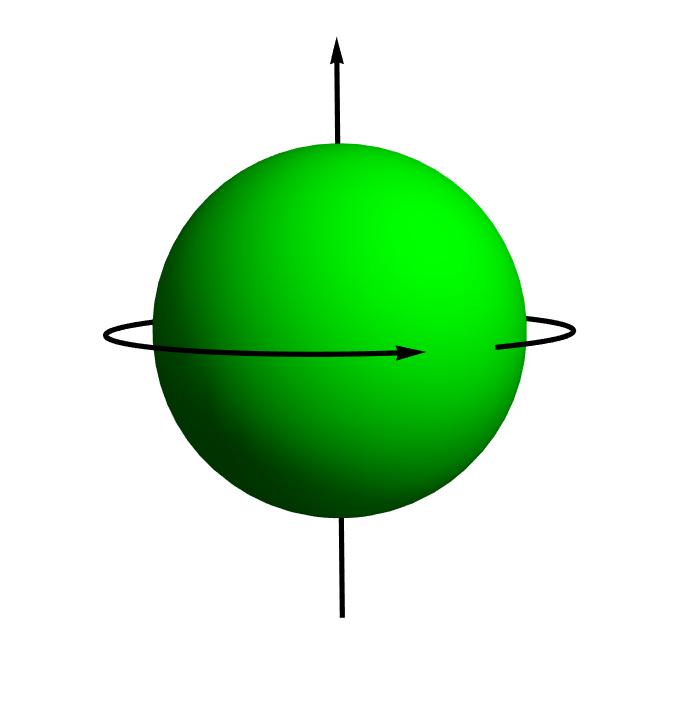}
    \caption{Particles with angular momentum parallel or antiparallel to linear momentum are chiral}
    \label{fig:xmatter}
\end{figure}

Matter particles with both linear and angular momentum also have chirality, as illustrated in Fig~\ref{fig:xmatter}. The chirality is defined as $\chi = \mathrm{sgn}(\vec{p}\cdot\vec{s})$ where $\vec{p}$ is the linear momentum and $\vec{s}$ is the angular momentum. It is $-1$ for left handed and $+1$ for right handed particles. Beta decay, which is mediated by the weak interaction is a parity-violating (i.e. chiral) process\cite{PViolationTh,PViolationExpt}, producing an excess of left handed electrons. This is because the electroweak gauge fields couple differently with left and right handed leptons and quarks.

The physics of non-radioactive ordinary matter at low energies is almost exclusively governed by electromagnetism, and is parity-invariant to a very good approximation. Therefore, to create a chiral effect, there must be an underlying source of chirality. For example, producing only one chirality of a chiral molecule requires chiral precursors or chiral catalysts (which are often obtained from biological sources).

\section{Chiral Fermions}
Dirac and Weyl fermions are characterized by gapless dispersion relations and well-defined chirality. The prototypical Dirac Hamiltonian \cite{DiracEqn}, which describes a spin-1/2 relativistic particle, such as an electron, in free space, is
\begin{equation}
    H = c p_i \alpha_i + mc^2 \beta
\end{equation}

where the Dirac matrices are $\alpha_i = \begin{pmatrix}-\sigma_i & 0\\0 & \sigma_i \end{pmatrix}$ and $\beta = \begin{pmatrix} 0 & I_2\\I_2 & 0 \end{pmatrix}$. The chirality is $\gamma_5 = \begin{pmatrix} -I_2 & 0 \\ 0 & I_2\end{pmatrix}$. For a massive Dirac fermion, the chirality is not conserved. However, for a massless Dirac fermion, the Hamiltonian is of the form
\begin{equation}
    H = cp_i\alpha_i
\end{equation}
the Hamiltonian commutes with $\gamma_5$ and chirality is conserved. In the presence of electromagnetic field, $p_i$ can be replaced by $p_i - e A_i$.
\begin{equation}
    H = v(p_i-eA_i)\alpha_i
\end{equation}
The Hamiltonian still commutes with $\gamma_5$, and in the first quantized framework, chirality is still conserved. 

The Weyl Hamiltonian \cite{WeylEqn} was proposed for massless fermions which travel at the speed of light. It has the form
\begin{equation}
    H = \chi c p_i \sigma_i
\end{equation}
It is essentially one half of a Dirac Hamiltonian. It describes a fermion with definite chirality $\chi$, which is $-1$ for left and $+1$ for right handed fermions. Like the massless Dirac Hamiltonian, the Weyl Hamiltonian also does not admit terms that violate chirality conservation in the first quantized framework. For a long time, neutrinos were believed to be Weyl particles, but they are now known to have a very small mass.

\section{Dirac and Weyl Materials}
In condensed matter systems, fermions with definite chirality include massless (gapless) Dirac fermions \cite{Young12,Wang12,Liu14,Liu14a,Steinberg14,nagaosa2014,bradlyn2016beyond,cano2019multifold,klemenz2020systematic,wieder2020strong} and Weyl fermions \cite{Wan11,Weng15,Huang15,xu2015discovery,lv2015Nat,xu2015,lv2015PRX,2015Xiong} (see \cite{BalatskyRev, armitage2018} for reviews). Instead of travelling with the speed of light, they travel with a substantially lower (by a factor of $\sim 1000$) Fermi velocity, which can also be highly anisotropic. 

Gapless Dirac fermions can be symmetry-protected if they lie on high-symmetry axes \cite{nagaosa2014}.

The degeneracy in Weyl fermions does not need any symmetry protection; it is topologically protected. In the isotropic case, the Hamiltonian is
\begin{equation}
    H = \chi v k_i \sigma_i
\end{equation}
where $v$ is the velocity of the fermions. The general linear Weyl Hamiltonian, taking into account anisotropy and tilt is,
\begin{equation}
    H = v^i_a k_i \sigma_a + v^i_t k_i + E_0
\end{equation}
where $v^i_a$ is responsible for the chirality of the fermion, $v^i_t$ is the tilt, and $E_0$ is the energy of the node. The velocity is different in different directions. The chirality is $\chi = \mathrm{sgn}\,\mathrm{det}(v^i_a)$. A small perturbation cannot lift the degeneracy, only shift it in momentum and energy.

A material with both inversion $P$ and time-reversal $T$ cannot have have Weyl fermions, only Dirac fermions. This is because the momentum is odd under both $P$ and $T$, and therefore even under $PT$, while the chirality is odd under $P$ and even under $T$, therefore odd under $PT$. If a material has $PT$, a chiral fermion at a certain momentum corresponds to a fermion of opposite chirality, also at the same momentum. These fermions would form a Dirac cone, not a Weyl cone.

\begin{figure}
    \centering
    \includegraphics[scale=0.5]{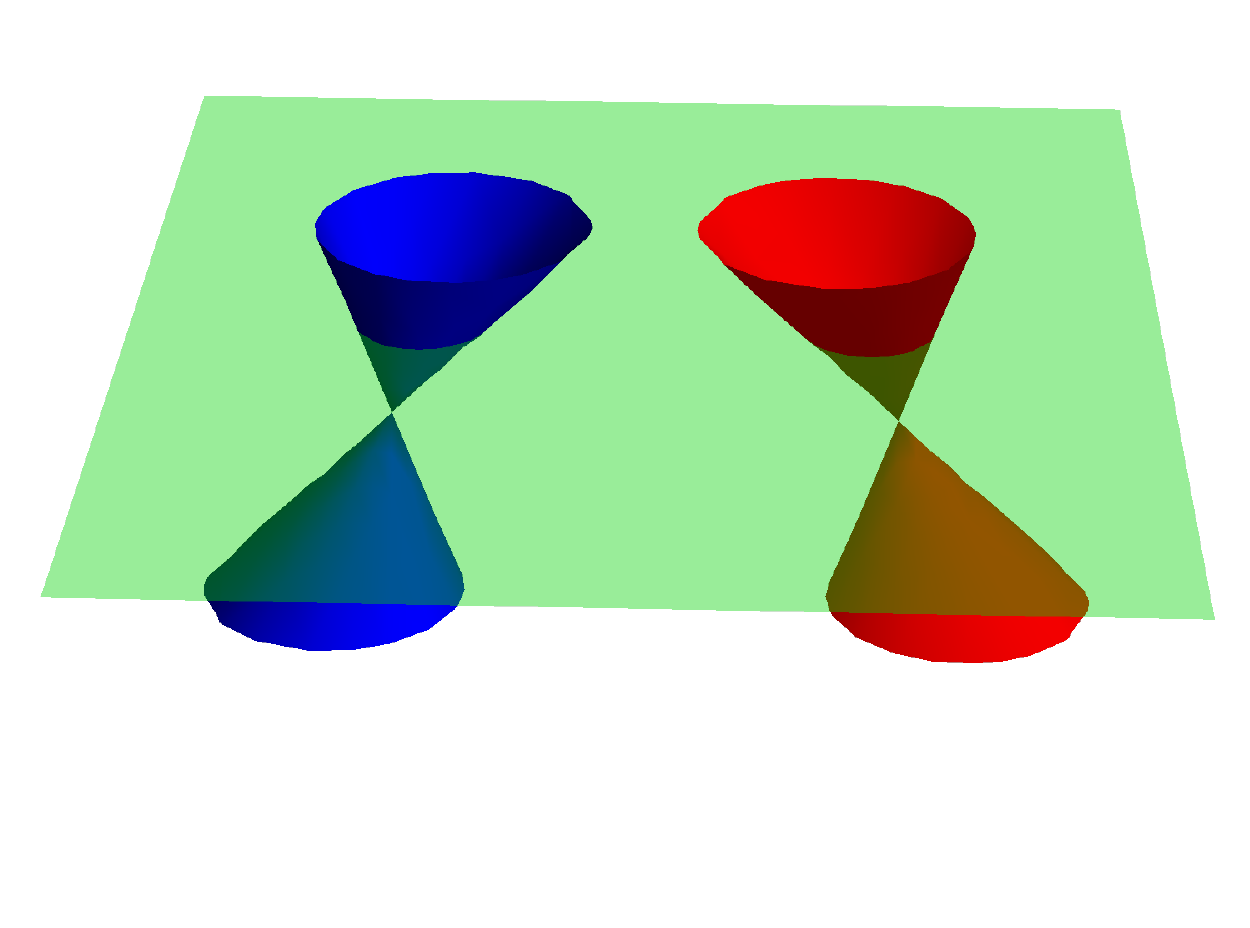}\includegraphics[scale=0.5]{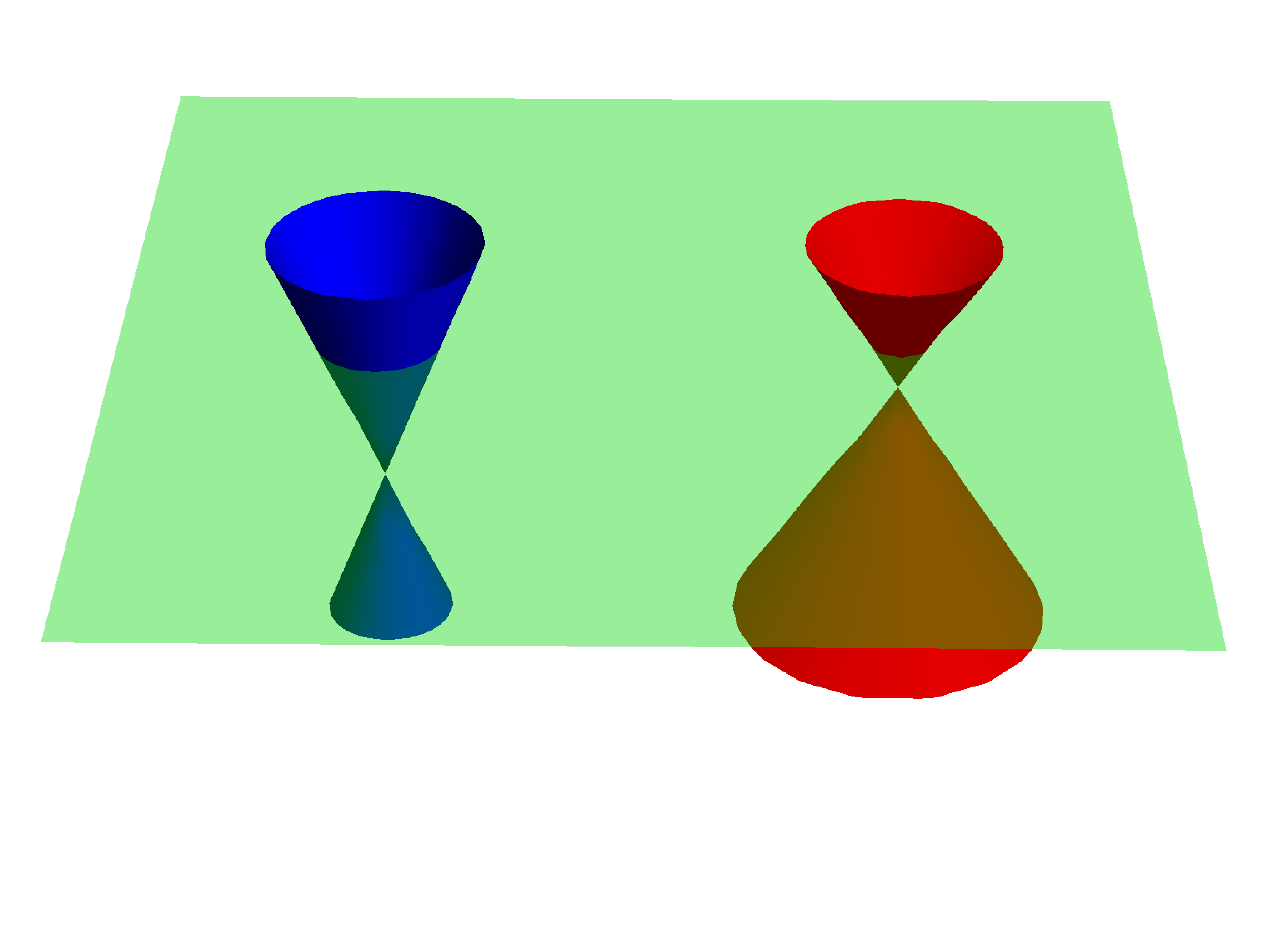}
    \caption{The dispersion relations of cones of a symmetric Weyl material with tilted cones (left) and an asymmetric Weyl material (right). The color indicates chirality and the green plane indicates the Fermi level.}
    \label{fig:cones1}
\end{figure}

In some Weyl materials, there is an inversion or mirror symmetry relating left and right handed cones; we refer to them as symmetric Weyl materials. In asymmetric Weyl materials, there is no symmetry relating cones of opposite chirality, and they can have different velocities, energies, or both. Asymmetric Weyl materials necessarily have chiral crystal structures. Possible cones of symmetric and asymmetric Weyl materials are illustrated in Fig \ref{fig:cones1}.

Weyl fermions are monopoles of Berry curvature. The Berry monopole charge is $2\pi C$ where $C$ is the Chern number; it is quantized in units of $2\pi$ because the phase can be uniquely defined up to multiples of $2\pi$. The Berry curvature in momentum space of the Hamiltonian $H = \chi v\Vec{k}\cdot\Vec{\sigma}$ is $\pm \chi \hat{k}/2k^2$ where the $+$ and $-$ are for the upper and lower band, respectively. The Berry monopole charge, i.e. the integral of Berry curvature on a surface surrounding the cone, is $\pm \chi 2\pi$. For a more general linear Weyl cone, the form of the Berry curvature is different, but the Berry charge is still the same, because it is quantized. Even in the general case, the chirality is equivalent to the Chern number. This explains the topological protection of Weyl cones: a small perturbation cannot change the quantized Berry charge, it can only move the Weyl node in momentum space.

Dirac cones have $SU(2)$ Berry curvature because of their twofold degeneracy, but they have vanishing $U(1)$ curvature, and vanishing Chern number, and are therefore not topologically protected. They can be gapped by perturbations, which can introduce a ``mass" term. Depending on the space group of the crystal and the little group of the Dirac node momentum, a symmetry-breaking perturbation might be required to create a gap.

It is possible that the tilt velocity is larger than the untilted velocity. Such cones are called type II Weyl cones and do not have ellipsoidal Fermi surfaces with well-defined non-zero Berry flux \cite{soluyanov2015}. Weyl cones with tilts smaller than their untilted velocities are called type I. In this work, we focus primarily on type-I Weyl materials.

In a Weyl material, the total chirality of all left and right handed cones cancels; the total Berry charge in the Brillouin zone vanishes. This is the Nielsen-Ninomiya theorm \cite{1981Nielsen}. Weyl materials also have Fermi arcs, which are gapless states connecting the projections of Weyl points at surfaces \cite{Wan11}.

In certain space groups, there are symmetry-protected higher order Weyl points\cite{huang2016, FourWeyl} which have higher Chern number than $\pm 1$, and/or multifold fermions \cite{bradlyn2016beyond}, which involve more than 2 bands, and usually also have Chern numbers greater than $\pm 1$.   

\section{Chiral Anomaly}

The chiral anomaly results in the nonconservation of chirality, which is expected to be conserved semiclassically \cite{Adler1969,Bell1969, nielsen1983adler}. It is a strictly second quantized effect. Parallel electric and magnetic fields create a chiral charge: the rate of charge pumping at \textit{each} Weyl cone (or each chirality of each Dirac cone) is
\begin{equation}
    \Dot{\rho} = \frac{e^2}{4\pi^2}\Vec{E}\cdot\Vec{B}
\end{equation}
In a real crystal, there is chirality flipping scattering, however, its timescale $\tau_V$ is often much longer than that of ordinary scattering $\tau$. The rate of chiral charge generation for each pair of Weyl cones, or a single Dirac cone is
\begin{equation}
    \dot{\rho_5} = \frac{e^2}{2\pi^2}\Vec{E}\cdot\Vec{B} - \frac{1}{\tau_V}\rho_5
\end{equation}
Here $\rho_5 = \rho_R - \rho_L$ is the difference in the density of the left and right handed fermions.

If there there is an imbalance in the chemical potentials of left and right handed fermions, a magnetic field results in a current of \textit{real} charge. This is the chiral magnetic effect (CME) \cite{CME} (see \cite{Kharzeev:2013ffa,Kharzeev:2012ph} for reviews)
\begin{equation}
    \Vec{j} = \frac{e^2}{2\pi^2}\mu_5\Vec{B}
    \label{eq1}
\end{equation}
where the effective chemical potential corresponding to chiral charge $\mu_5 = (\mu_R - \mu_L)/2$ is half the difference between the chemical potentials of the right and left handed fermions.

Apart from the anomaly, manipulation of chiral charge has been proposed using light \cite{zhong2016gyrotropic, ma2015chiral}, strain \cite{cortijo2016}, pseudoscalar phonons \cite{song2016detecting}, and AC voltage in asymmetric Weyl materials\cite{meyer2018}. 

One consequence of the chiral anomaly is a negative quadratic magnetoresistance \cite{CME,Son2012Berry,2013Son,Zyuzin:2012tv,Basar:2013iaa,vazifeh2013electromagnetic,goswami2013axionic,2014Burkov, aji2012adler}: When there are parallel electric and magnetic fields, chiral charge is pumped, which results in an additional contribution to the current along the magnetic field. The anomalous part of the conductivity is:

\begin{equation}
    \sigma_{CME} = \frac{e^4}{4\pi^4}\tau_V \left(\frac{\partial \rho_5}{\partial \mu_5}\right)^{-1} B^2
\end{equation}

This negative magnetoresistance has been observed in a number of Dirac and Weyl materials \cite{2016Li,kim2013dirac,2015Xiong,li2015giant,Huang2015,Wang2016,Zhang2016,shekhar2015large,yang2015chiral, Arnold2016}

If we consider only the anomalous contributions, the work on done the material by the external electric field acting on the chiral magnetic current is 

\begin{equation}
   \vec{E}\cdot\vec{j} =  \frac{e^2}{2\pi^2}\mu_5\vec{E}\cdot\Vec{B}
\end{equation}

The change in energy due to anomalous pumping of chiral charge is:

\begin{equation}
    \mu_5 \dot{\rho_5} = \frac{e^2}{2\pi^2}\mu_5\vec{E}\cdot\Vec{B}
\end{equation}

The two are equal; the anomaly itself is a strictly non-dissipative process. However, chirality flipping is a dissipative process in general, and therefore the anomalous conductivity is a dissipative phenomenon. 

\section{Landau Levels}

The Landau levels of Dirac and Weyl fermions are different from those of topologically trivial fermions. The Landau levels of a band with parabolic dispersion relation with effective mass $m$ are:

\begin{equation}
    E = \left(n+\frac{1}{2}\right) \frac{eB}{m} + \frac{1}{2m} k_z^2
\end{equation}
with $n = 0,1,2...$

The Landau levels for a gapless Dirac fermion with velocity $v$ are:

\begin{equation}
    E = \pm \sqrt{2neBv^2 + v^2 k_z^2}
\end{equation}

\begin{figure}
    \centering
    \includegraphics[scale=0.15]{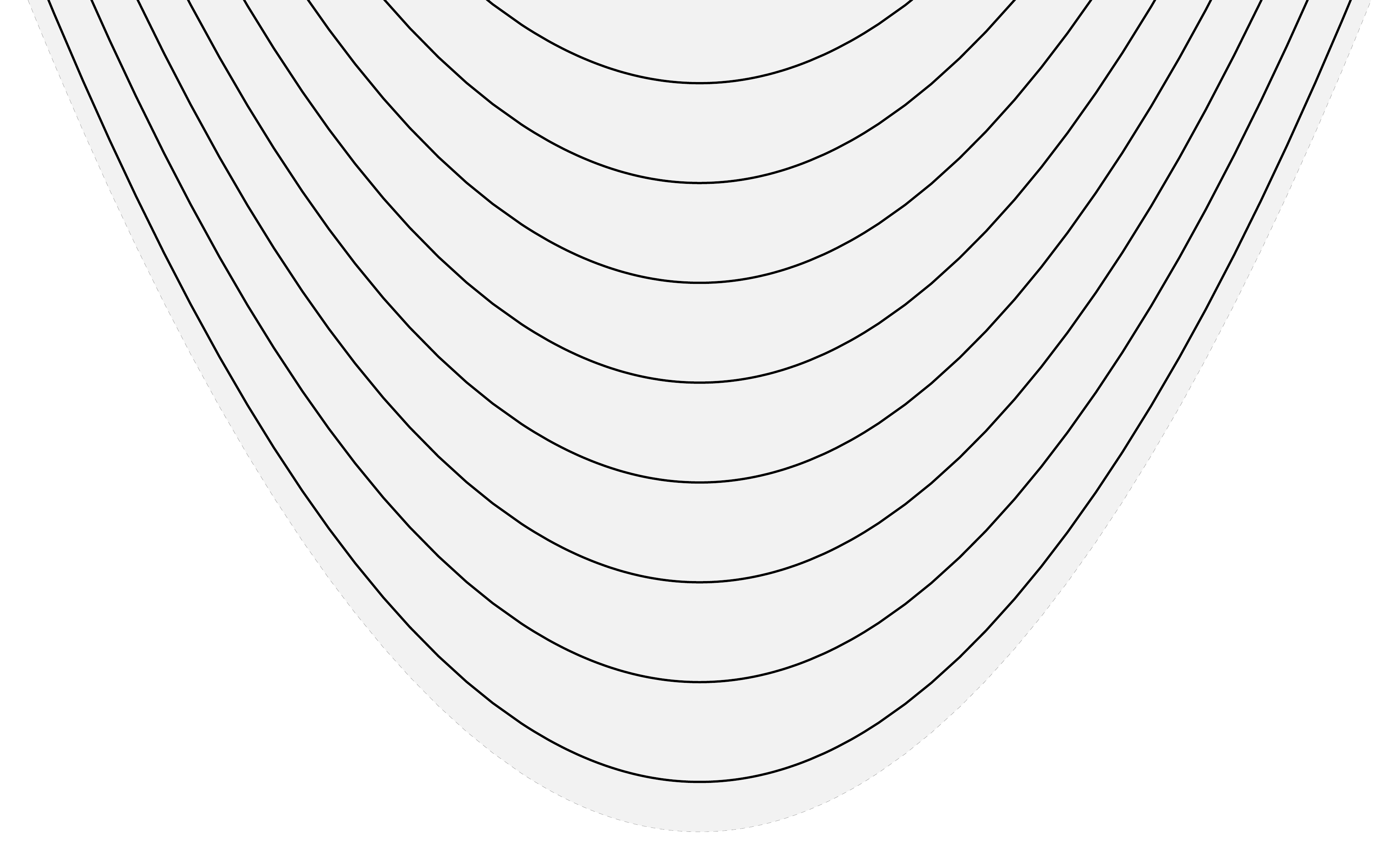}\hspace{0.1\linewidth}\includegraphics[scale=0.15]{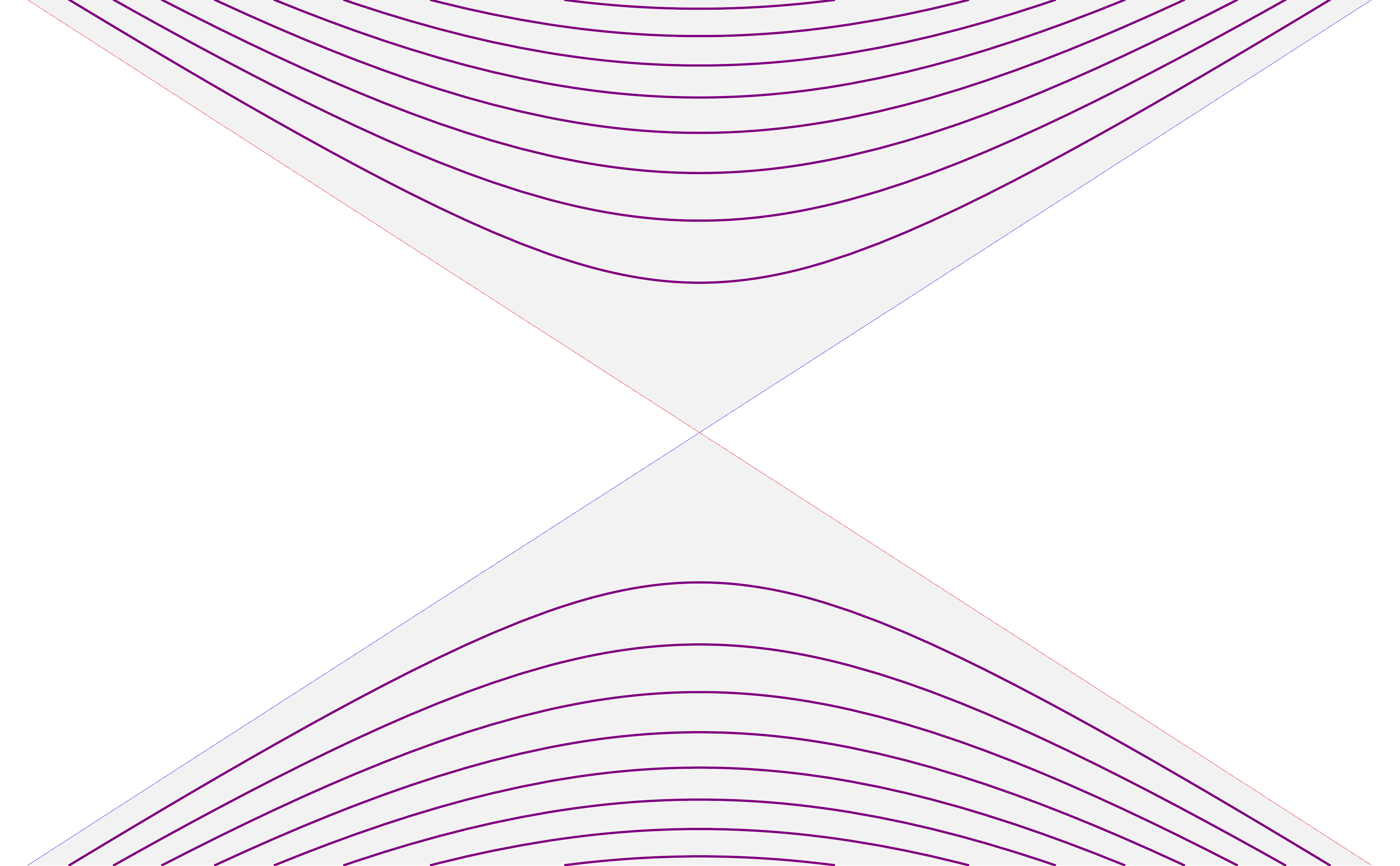}
    \caption{The Landau levels of a massive fermion (left) and Dirac fermion (right). Shading represents energy levels without magnetic field. Color indicates chirality: purple represents degenerate modes with both chiralities, red and blue represent left and right handed chiralities.}
    \label{fig:LLC}
\end{figure}

Note the lack of the $1/2$ term. This means quantum oscillations in chiral fermions have a phase of $\pi$ compared to massive fermions \cite{murakawa2013detection}. The Landau levels of massive and Dirac fermions are depicted in Fig~\ref{fig:LLC}

The Landau levels still have well-defined chirality. The modes with $n>0$ are doubly degenerate, having both chiralities while those with $n = 0$ are singly degenerate. The mode travelling in the direction of $e\vec{B}$ is right handed while the one travelling in the opposite direction is left handed. In Weyl fermions, the higher Landau levels not degenerate, while the lowest Landau level propagates only in one direction.

The chiral anomaly can be understood heuristically through these energy spectra. If there are parallel electric and magnetic fields, the electric field causes $k_z$ of all fermions to increase or decrease with time, causing the number of fermions of one chirality to increase and those of the other chirality to decrease. The total number of fermions remains unchanged, of course, because the Landau levels are ultimately connected deep below the Fermi surface. In a sense, an electric field pulls fermions from one cone to another through the Dirac sea. The transverse density of each Landau level is $eB/2\pi$. The rate of change of longitudinal momentum is $eE_z$, which corresponds to a change in longitudinal density $eE_z/2\pi$. Therefore, the pumping of chiral charge at each cone is $e^2\vec{B}\cdot\vec{E}/4\pi^2$. 

The chiral magnetic effect can also be understood in a similar way: The density of states of the chiral mode is $(eB/2\pi)\times(1/2\pi v)$. If we increase the chemical potential of the right handed cones by $\mu_5$ and decrease that of the left handed cones by the same amount, the current contributed by each pair of cones is $2\times ev \times (eB/2\pi)\times(1/2\pi v)\times \mu_5 = e^2 \mu_5 \vec{B}/2\pi^2$.

When the magnetic field is very large, so that $\sqrt{2eBv^2} > |\mu|$, only the lowest Landau level contributes. In this extreme quantum limit, we essentially have a 1+1-dimensional chiral fermion. The chiral anomaly in 1+1 dimensions is:
\begin{equation}
    \dot{\rho_5} = \frac{1}{\pi} e E
\end{equation}

Multiplying by the transverse density of states $eB/2\pi$, we get the rate of chiral charge generation by the 3-dimensional anomaly.

\section{Chiral Circular Photocurrents}

Chirality also plays a role in the optical response of Weyl materials. One contribution to the photocurrent is the injection photocurrent: the current produced when electrons are excited to states with different velocity. When circularly polarized light interacts with chiral fermions, selection rules involving angular momentum are involved in the excitations. For example, if we consider light with angular momentum along the $\hat{z}$ direction interacting with a right handed cone, the electromagnetic field can cause transitions from angular momentum $-\hbar\hat{z}/2$ to $+\hbar\hat{z}/2$ but not vice-versa. This is a transition of a fermion with momentum along $+\hat{z}$, with the direction of velocity changing from $-\hat{z}$ to $+\hat{z}$. So there is an average change in velocity in the $+\hat{z}$ direction. Similarly, in left handed cones, the same electromagnetic field will induce transitions of fermions with momentum along $-\hat{z}$, with the velocity direction changing from $+\hat{z}$ to $-\hat{z}$. The left and right cones each contribute to the photocurrent, but their contributions will cancel, unless they respond differently to light.

The Pauli blockade can play a role: if the cones are tilted, or, if the left and right handed cones have different energies, certain transitions might be forbidden because both states are completely filled or completely empty. 

In symmetric Weyl materials, a photocurrent in response to circularly polarized light has been predicted \cite{lee2017}, and observed in TaAs \cite{MaTaAs}. This is a consequence of tilted cones, which are blockaded only on one side. While the left and right handed cones have tilts of the same magnitude, they are tilted in different directions, and in general, depending on the polarization of the photons, cones of one chirality are more strongly blockaded than those of the other chirality (for a detailed analysis see the supplement of \cite{MaTaAs}).

\begin{figure}[hb]
    \centering
    \includegraphics[scale=0.18]{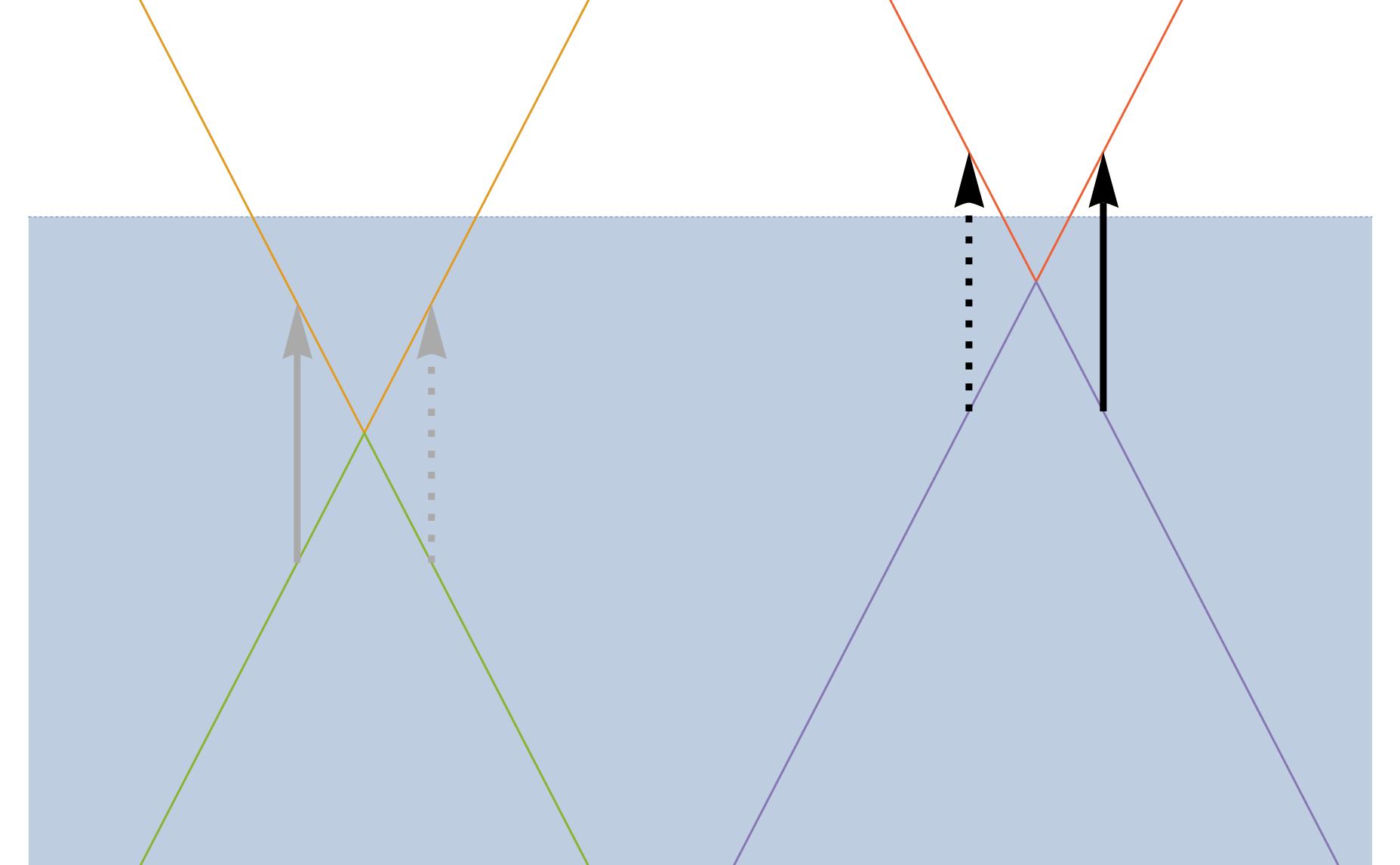}
    \caption{The effects of selection rules and Pauli blockade on an asymmetric Weyl material. Blockaded transitions are greyed out, and transitions forbidden by selection rules are dashed. The only allowed transitions contribute a change in velocity towards the right.}
    \label{fig:blockade}
\end{figure}

In asymmetric Weyl materials, it is possible that cones of one chirality are blockaded, while those of the other chirality are not; this results in a quantized photocurrent in the direction of the angular momentum of light \cite{dejuan17}. The quantization holds even if the cones are not linear. This effect is also possible in materials with multifold fermions, and is still quantized as long as the bands are perfectly linear \cite{FlickerMultifold}. This has been observed in RhSi\cite{RhSiCurrent} and CoSi \cite{CoSiCurrent}.

The effects of selection rules and Pauli blockade are illustrated in Fig \ref{fig:blockade}. The photocurrent in a symmetric Weyl material with tilted cones is a strictly 3 dimensional effect and cannot be effectively depicted in a 1+1 dimensional sketch.

\section{Chiral Kinetic Theory}

The dynamics of chiral fermions can be investigated \textit{semiclassically} by Chiral Kinetic Theory \cite{xiao2005berry,Son2012Berry, stephanov2012chiral, 2013Son}. In chiral kinetic theory, position and momentum are treated on equal footing, and Berry curvature and gauge field are also treated on equal footing. The phase space volume and equations of motion are modified as follows:

\begin{align}
    d^3 k &\to (1+e\Vec{\Omega}\cdot\Vec{B}) d^3 k\nonumber\\
    \partial_t \Vec{x} &= \nabla_{\vec{p}} \epsilon + \partial_t \Vec{p}\times \Vec{\Omega}\\
    \partial_t \Vec{p} &= e\Vec{E} + \partial_t \Vec{x}\times \Vec{B}\nonumber
\end{align}

where the momentum derivative of energy $\nabla_{\vec{p}} \epsilon$ is the velocity in the absence of electromagnetic field. The deformation of phase space corresponds to the chiral Landau level, which exists only on one side of the Weyl cone.

In the absence of electric field, the deformation of phase space and velocity is:

\begin{align}
    d^3 k &\to (1+e\Vec{\Omega}\cdot\Vec{B}) d^3 k\nonumber\\
   \vec{v} &\to \frac{\vec{v} + e(\vec{v}\cdot\vec{\Omega})\vec{B}}{1+e\Vec{\Omega}\cdot\Vec{B}}
\end{align}

These equations give the correct expression of the chiral anomaly and chiral magnetic effect. However, this framework does not consider Landau quantization and therefore cannot be used to analyse effects such as quantum oscillations.
\chapter{Quantum Oscillations in the Chiral Magnetic Conductivity}\label{chOscillations}

\blfootnote{This chapter is based on \cite{SahalOsc}.}

In a condensed matter system in a magnetic field, the Shubnikov - de Haas (SdH) and De Haas - van Alphen oscillations appear 
due to the quantized Landau levels and the presence of Fermi surface\cite{lifshits_kosevich, shoenberg_2009}; SdH oscillations  have been observed in the transverse magnetoconductivity of some Dirac and Weyl (semi)metals \cite{liang2015ultrahigh, Huang2015, hu2016pi}. The phase of the SdH oscillations depends on the Berry curvature; this fact can be used to distinguish materials with massive carriers from Dirac and Weyl materials \cite{Huang2015, hu2016pi, murakawa2013detection, luk2006dirac}. The oscillating part of the transverse conductivity has the form
\begin{equation}
\sigma_{xx} = A(B) \cos\left[2\pi\left(\frac{B_0}{B} - \gamma + \delta\right)\right] ,
\end{equation}
where $\gamma$ is $0$ for Dirac and Weyl carriers and $1/2$ for massive carriers, and $\delta$ varies between $-1/8$ and $1/8$ for 3D materials.

In this work, we point out the existence of a new type of quantum oscillations that emerge in strong magnetic fields due to a non-linear relation between the chiral chemical potential $\mu_5$ and the density of chiral charge $\rho_5$.

\section{Chiral Susceptibility}
\begin{figure}[ht]\label{landau}
\begin{center}
\includegraphics[scale=0.4]{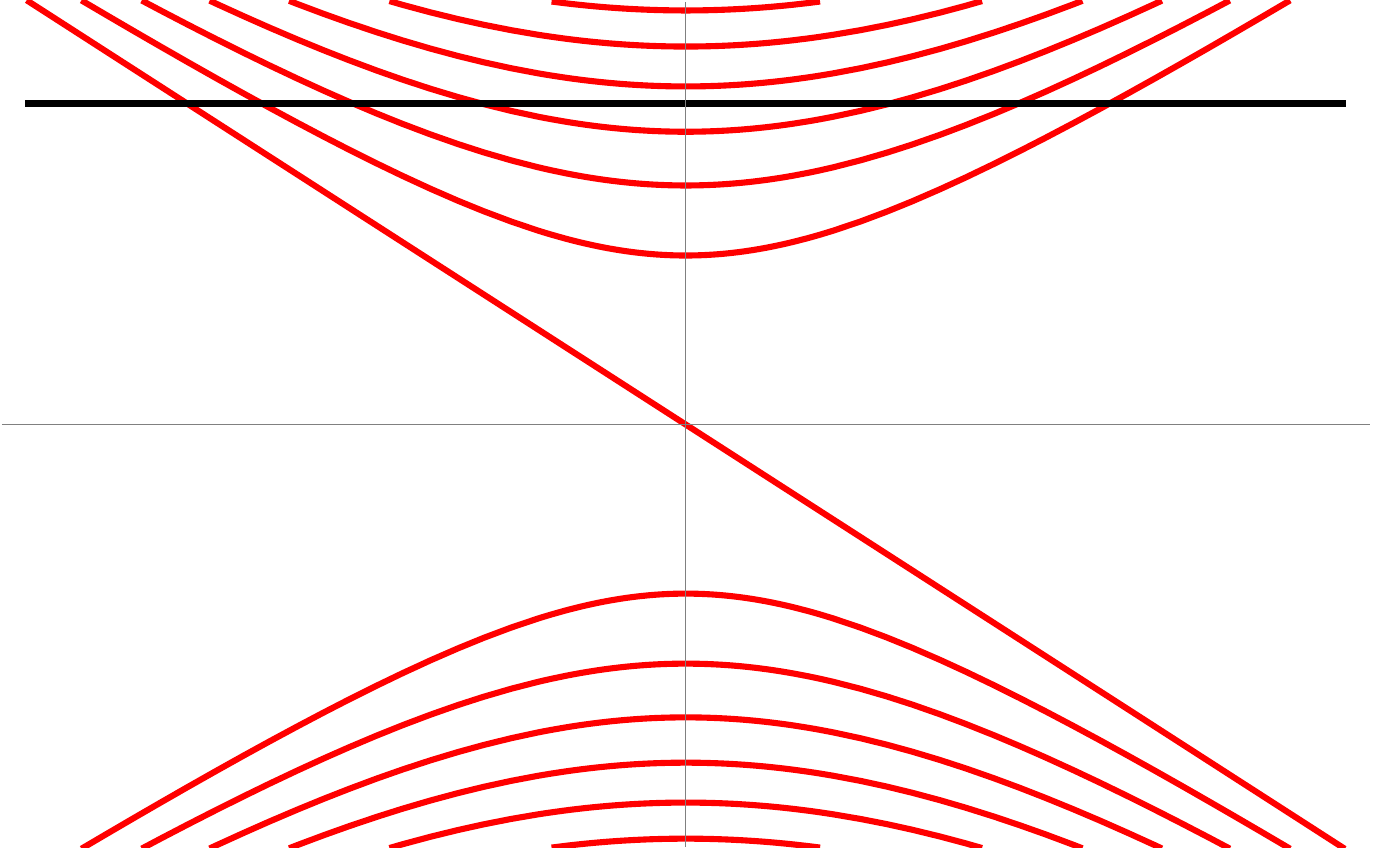} \hspace{0.05\linewidth} \includegraphics[scale=0.4]{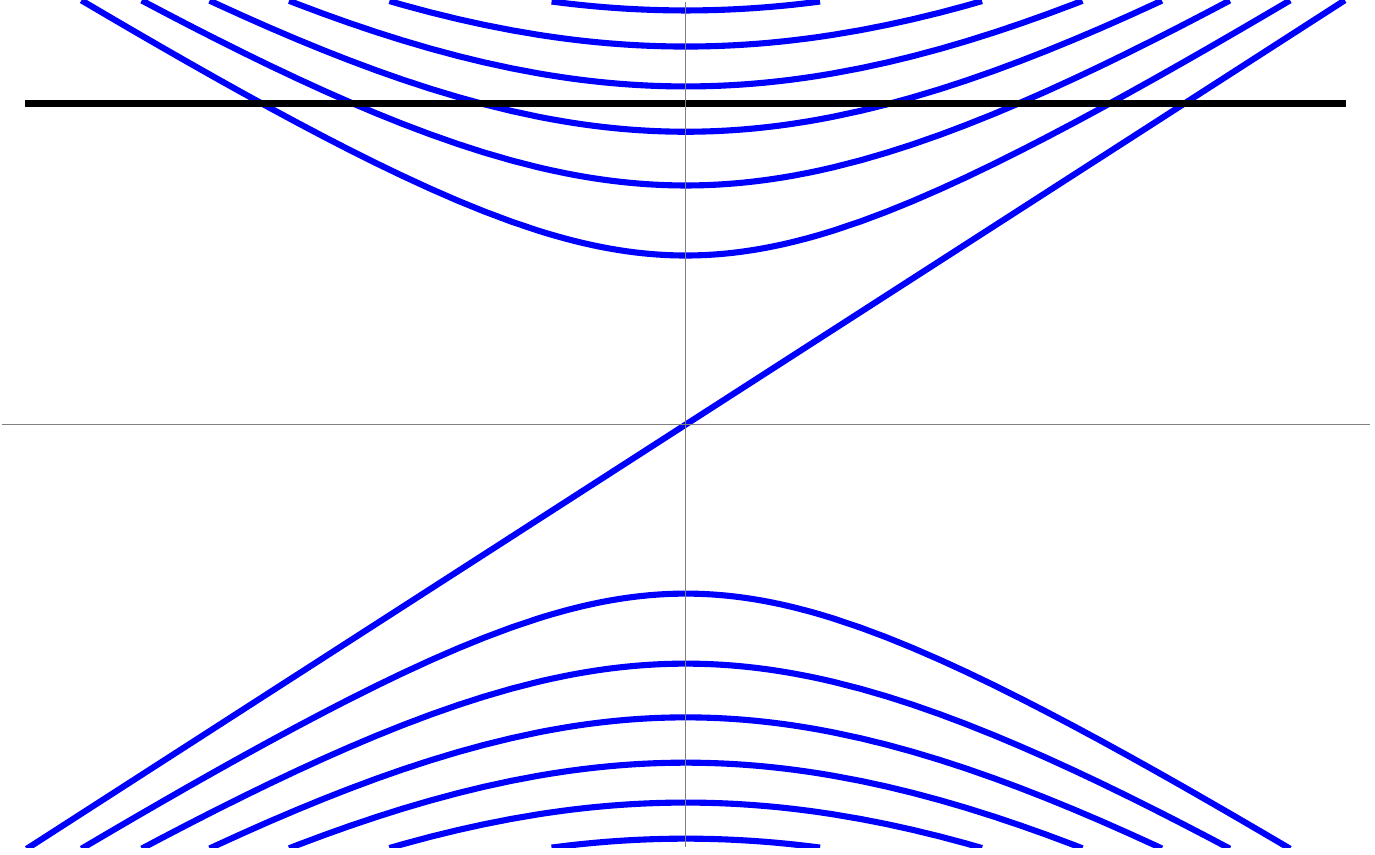}
\end{center}
\caption{The energy structure of left-handed and right-handed fermions in a magnetic field. The black line indicates $\mu$.}
\end{figure}

In a uniform and constant magnetic field, the energies of the lowest Landau levels are $\epsilon = -vp_z$ for left-handed and $\epsilon = + vp_z$ for right-handed fermions, see Fig. \ref{landau} ($v$ is the Fermi velocity; we assume that magnetic field $B$ is along the $z$-axis). The energies of excited Landau levels are $E = \pm v\sqrt{p_z^2 + 2eBn}$ for $n\geq 1$, for both chiralities.  The density of Landau levels in the $xy$ plane is $eB/2\pi$, whereas the density of states in the $z$-direction is $p_z/2\pi$.
Because the lowest Landau level is not degenerate in spin, it has right-handed fermions of positive charge traveling along $B$ and left-handed ones of negative charge traveling in the opposite direction. This induces the CME current (\ref{eq1}).

The density of the chiral charge $\rho_5$ is related to the chiral chemical potential $\mu_5$ through the chiral susceptibility $\chi \equiv \partial \rho_5  / \partial \mu_5$ -- at small $\mu_5$, $\rho_5 = \chi \mu_5 + ...$ so that $\mu_5 \simeq  \chi^{-1} \rho_5$. Note that in the absence of chirality loss corresponding to $\tau_V \to \infty$ the CME current would grow linearly in time -- in other words, it would behave as a superconducting current, see \cite{kharzeev2016chiral} for a discussion. 

At finite $\tau_V$, the density of the chiral charge saturates at the value $\rho_5 = e^2/2\pi^2\ {\bf{E}}\cdot{\bf{B}} \ \tau_V$, and the longitudinal CME conductivity for parallel ${\bf E}$ and ${\bf B}$ is given by 
\begin{equation}\label{CMEsigma}
\sigma_{\rm CME} = \frac{e^4 B^2}{4\pi^4 \chi (i\omega + 1/\tau_V)} ,
\end{equation}
where $\omega$ is the frequency of an external field.

In the presence of a Fermi surface with a chemical potential $\mu = (\mu_R + \mu_L)/2$, in weak magnetic fields with $2eBv^2 \ll \mu^2$ the Landau quantization can be ignored, and the chiral susceptibility is given by
\begin{equation}\label{susc_weak}
\chi = \frac{\mu^2}{\pi^2 v^3} + \frac{T^2}{3v^3}.
\end{equation}
The DC CME conductivity \cite{CME,2013Son, Zyuzin:2012tv, 2016Li} is then
\begin{equation}
\sigma_{\rm CME} = \frac{e^4 v^3 \tau_V B^2}{4\pi^2 (\mu^2 + \pi^2 T^2/3)} .
\end{equation}
In strong magnetic fields with $2eBv^2 \gg \mu^2$, only the lowest Landau level contributes, and $\mu = eB/2v\pi^2$, so  the DC CME conductivity has a linear dependence on $B$: 
\begin{equation}
\sigma_{\rm CME} = \frac{e^2 v\ \tau_V B}{2\pi^2}.
\end{equation}
 In a real material, the conductivity is the sum of the CME conductivity and the Ohmic conductivity,  $\sigma_{zz} = \sigma_{\rm CME} + \sigma_{\rm Ohm}$. 

\section{Quantum Oscillations}
 
 We now consider a system with $\tau_V \gg \tau$ (i.e. the chirality flips are relatively rare), $T\ll\mu$, $\mu_5\ll T, \mu$ and 
$2eBv^2 \ll \mu^2$. We focus on Dirac materials and Weyl materials in which the Weyl points have the same energy and Fermi velocity.


The density of states in energy $E$ for each chirality is
\begin{equation}\label{DoS}
g(E) = \frac{E_L^2}{8\pi^2 v^3}\left[1+2\sum_{n=1}^\infty \Theta(E^2 - nE_L^2) \sqrt{\frac{E^2}{E^2 - n E_L^2}}\right]
\end{equation}
where $E_L^2 = 2eBv^2$ is the difference in the squares of the Landau level energies. The factor of $2$ is because we have particles traveling in both directions for higher levels.

The total number density of particles, for each chirality is given by
\begin{equation}
\rho_{R,L} (\mu, T) = \int_{E_-}^{E_+} g(E) f(\mu - E, T) dE ,
\end{equation}
where $f(V, T) = \frac{e^{V/T}}{e^{V/T}+1}$ is the Fermi distribution function, and $E_-, E_+$ are the cutoff energies, which can be taken to be the lowest and highest energies in the bands containing the Dirac or Weyl points. Therefore,
\begin{equation}
\chi = \frac{\partial \rho_5}{\partial \mu_5} = 2 \frac{\partial \rho_{L,R}}{\partial \mu_{R,L}} = 2 \int_{E_-}^{E_+} g(E) f'(\mu - E, T) dE
\end{equation}
yielding
\begin{align}
\chi = \frac{E_L^2}{2\pi^2 v^3}  \int_{-\infty}^\infty & \left[\frac{1}{2}+\sum_{n=1}^\infty \Theta(E^2 - nE_L^2) \sqrt{\frac{E^2}{E^2 - n E_L^2}}\right] \nonumber\\ &\times f'(\mu-E,T) dE
\end{align}
For small fields (when many Landau levels contribute), the sum can be approximated by an integral, and we recover (\ref{susc_weak}):
\begin{equation}
\chi(B=0) = \frac{\mu^2}{\pi^2 v^3} + \frac{T^2}{3v^3} .
\end{equation}
Let us now evaluate the quantum corrections to this expression that will be responsible for the quantum oscillations in CME conductivity. We are concerned only with energies close to $\mu$, so 
\begin{align}
\sqrt{\frac{E^2}{E^2 - n E_L^2}} &\approx \sqrt{\frac{\mu^2}{\mu^2 + 2\mu V - n E_L^2}} 
\end{align}
where $x = \mu^2/E_L^2$ and $V = E-\mu$.

We can define the contribution of the $n$th level to the susceptibility $\chi_n$ as
\begin{align}
\chi_n = \frac{E_L^2}{2\pi^2 v^3}\int_{-\infty}^\infty &\Theta(E^2 - nE_L^2) \sqrt{\frac{E^2}{E^2 - n E_L^2}}f'(\mu-E,T) dE 
\end{align}
Now, in the expression
\begin{equation}
\chi = \sum_{n=1}^\infty \chi_n + \frac{1}{2} \chi_0
\end{equation}
we can extend the sum to $-\infty$ and approximate the contribution of the fictitious negative levels by an integral:
\begin{align}
\chi &\approx \sum_{n=-\infty}^\infty \chi_n - \int^0_{-\infty} \chi_n dn \\ &= \sum_{n=-\infty}^\infty \chi_n - \int^\infty_{-\infty} \chi_n dn + \int^\infty_0 \chi_n dn \\ &= \sum_{n=-\infty}^\infty \chi_n - \int^\infty_{-\infty} \chi_n dn + \frac{\mu^2}{\pi^2 v^3} + \frac{T^2}{3v^3}
\end{align}
We can then use the Poisson summation in 
\begin{equation}
\chi = \frac{\mu^2}{\pi^2 v^3} + \frac{T^2}{3v^3} + \sum_{l=1}^\infty 2\Re (\chi_l)
\end{equation}
to evaluate the Fourier transform of $\chi_n$,
\begin{equation}
\chi_n = \alpha\left[(x-n)\frac{E_L^2}{2\mu}\right]
\end{equation}
where 
\begin{align}
\alpha(z) &= \frac{E_L^2}{2\pi^2 v^3}\int_{-\infty}^\infty \Theta(V+z)\sqrt{\frac{\mu}{2(V + z))}} f'(V,T) dV \\ &\equiv \frac{E_L^2}{2\pi^2 v^3}(\beta\star\gamma)(z) ,
\end{align}
with $\beta(z) = \Theta(z)\sqrt{\frac{\mu}{2z}}$ and $\gamma(z) = f'(z,T)$. Here we have used the fact that $f'(V,T)$ is even in $V$. So according to the Poisson summation,
\begin{equation}
\chi_l = \frac{2\mu}{E_L^2}\sqrt{2\pi}\exp(-2\pi i l x) \tilde{\alpha}\left(2\pi l \frac{2\mu}{E_L^2}\right) .
\end{equation}
From the convolution theorem, 
\begin{equation}
\tilde{\alpha}(k) = \frac{E_L^2}{2\pi^2 v^3} \sqrt{2\pi} \tilde{\beta}(k)\tilde{\gamma}(k) .
\end{equation}
The Fourier transforms are
\begin{equation}
\tilde{\beta}(k) = \sqrt{\frac{\mu}{2}}\frac{1}{2}\frac{|k| + ik}{|k|^{\frac{3}{2}}} ,
\end{equation}
\begin{equation}
\tilde{\gamma}(k) = \sqrt{\frac{\pi}{2}}\frac{kT}{\sinh(\pi kT)} ,
\end{equation}
and
\begin{equation}
\chi_l = \frac{\mu E_L}{4\pi^2 v^3} \frac{(1+i)}{\sqrt{l}}\frac{\left(l\frac{4\pi^2 \mu T}{E_L^2}\right)}{\sinh\left(l\frac{4\pi^2 \mu T}{E_L^2}\right)} \exp(-2\pi i l \mu^2/E_L^2) .
\end{equation}
In a real material, the scattering caused by impurities smears the Landau levels. The density of states is thus the convolution of (\ref{DoS}) with a Lorentzian distribution $\frac{1}{\Gamma\pi}\frac{\Gamma^2}{\Gamma^2 + (E-E_0)^2}$, where $\Gamma$ is the Dingle factor. Therefore, we must multiply each harmonic in the oscillating term by factor of $\exp(-4\pi l\Gamma\mu/E_L^2)$ (the Fourier transform of the Lorentzian):
\begin{align}\label{chi_sus}
\chi \approx &\frac{\mu^2}{\pi^2 v^3} + \frac{T^2}{3v^3} + \frac{\mu E_L}{2\pi^2 v^3} \sum_{l=1}^\infty \frac{1}{\sqrt{l}}\frac{\left(l\frac{4\pi^2 \mu T}{E_L^2}\right)}{\sinh\left(l\frac{4\pi^2 \mu T}{E_L^2}\right)}\nonumber \\ &\times \exp(-4\pi l\Gamma\mu/E_L^2) [\cos(2\pi l \mu^2/E_L^2) + \sin(2 \pi l \mu^2/E_L^2)]
\end{align}

\begin{figure}[ht]\label{figosc}
\begin{center}
\includegraphics[scale=0.65]{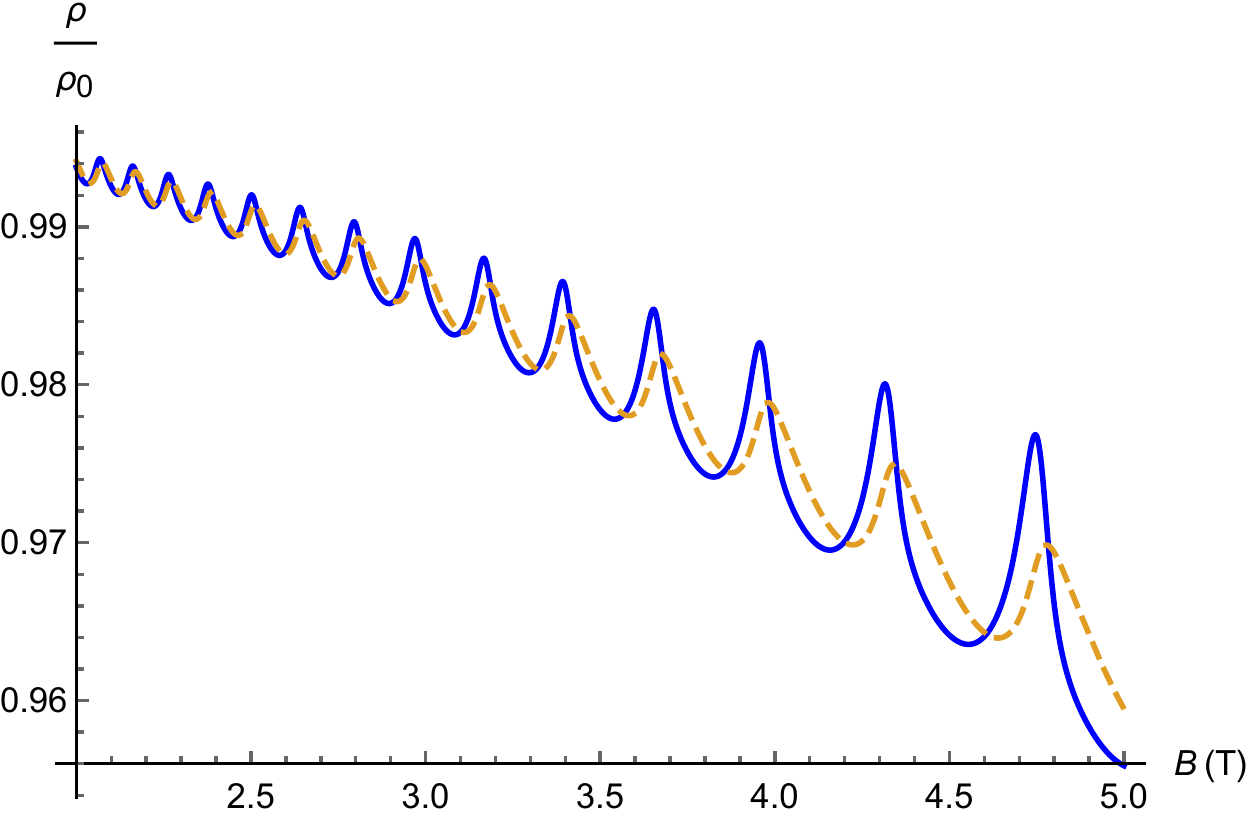}
\end{center}
\caption{$\rho_{zz}/\rho_0$ vs $B$ for $\mu = 150\ \mathrm{meV}$, $v = c/600$,  $T = 1.74\ \mathrm{K}$, $\Gamma = 0.3\ \mathrm{meV}$, and $\tau_V/\tau = 20$.  The solid line represents the full prediction taking account of the quantum CME oscillations, see (\ref{Total}); the dashed line represents only the SdH oscillations given by (\ref{SdH}). The quantum CME oscillations become larger than the SdH oscillations at $B \simeq 3$ T.}
\end{figure}

\begin{figure}[ht]\label{resi}
\begin{center}
\includegraphics[scale=0.65]{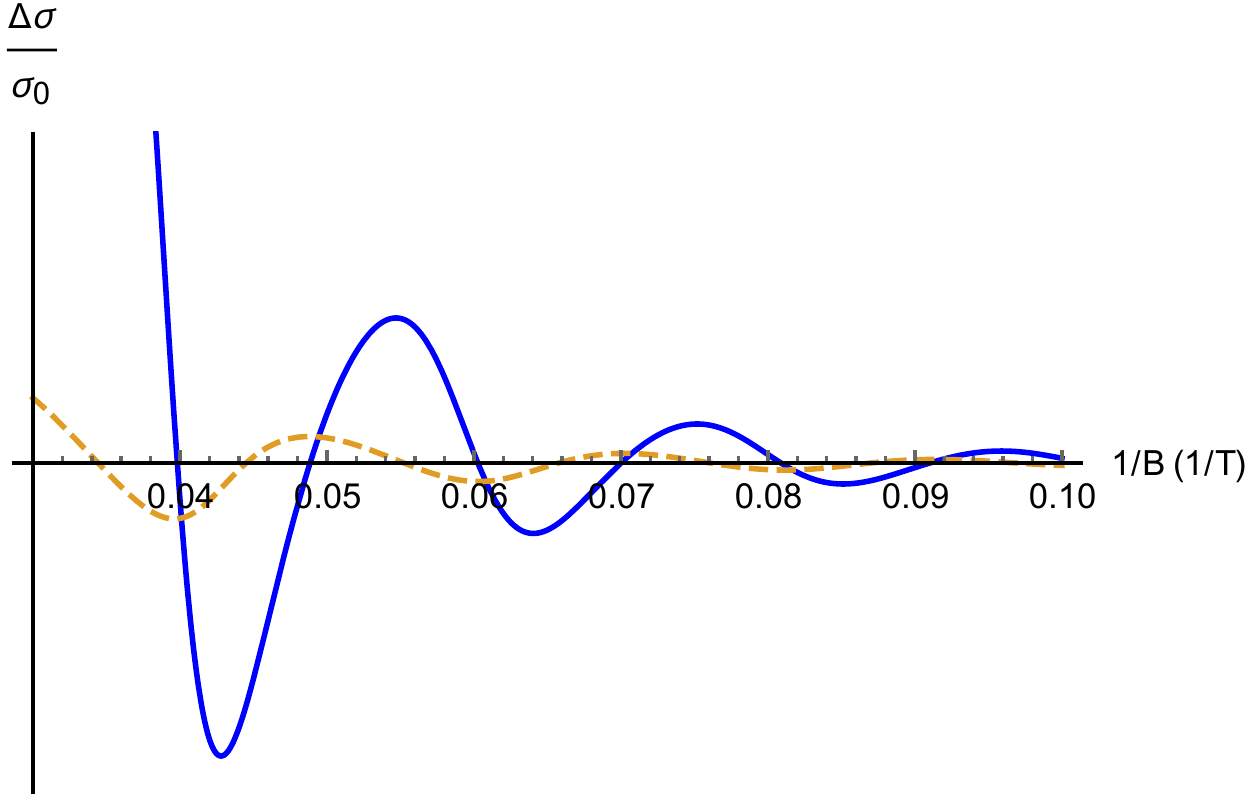}
\end{center}
\caption{The residue of $\sigma_{zz}$ after subtracting the constant and quadratic in $B$ contributions, plotted as a function of $1/B$ for $\mu = 150\ \mathrm{meV}$, $v = c/600$,  $T = 34.8\ \mathrm{K}$, $\Gamma = 0.3\ \mathrm{meV}$, and $\tau_V/\tau = 20$. 
The solid line represents the prediction of (\ref{Total}) while the dashed line represents the predictions of (\ref{SdH}) that ignore the quantum CME oscillations.}
\end{figure}

Since $\chi$ oscillates as a function of magnetic field, the CME conductivity also acquires these quantum oscillations. 
The Ohmic conductivity also oscillates with $B$; these oscillations for 3D chiral materials have been evaluated using the chiral kinetic theory in \cite{2015gustavo}. In our notations,
\begin{align}\label{SdH}
\frac{\sigma_{zz} (B)}{\sigma_0} \approx & 1 + \frac{3}{16}\frac{\tau_V}{\tau} \frac{E_L^4}{\mu^4} - \frac{3}{20} \frac{E_L^4}{\mu^4} - \frac{3}{8\pi} \frac{E_L^3}{\mu^3} \sum_{l=1}^\infty \frac{1}{l^{3/2}}  \nonumber\\ &\times \frac{\left(l\frac{4\pi^2 \mu T}{E_L^2}\right)}{\sinh\left(l\frac{4\pi^2 \mu T}{E_L^2}\right)}\exp(-4\pi l\Gamma\mu/E_L^2)\nonumber\\ &\times[\cos(2\pi l \mu^2/E_L^2) - \sin(2\pi l \mu^2/E_L^2)] ,
\end{align}
where $\tau$ is the (chirality-preserving) scattering time and $\sigma_0 \equiv \sigma(B=0) = \frac{\mu^2 e^2 \tau}{3\pi^2 v^2}$. The $\frac{3}{16}\frac{\tau_V}{\tau} \frac{E_L^4}{\mu^4}$ term, which is quadratic in $B$, comes from the CME conductivity;  to account for the variation of $\chi$ with $B$, we should now include a factor of $\frac{\mu^2}{\pi^2 v^3 \chi}$ in this term. Note that in weak magnetic fields, according to (\ref{susc_weak}), this factor is equal to unity, $\frac{\mu^2}{\pi^2 v^3 \chi} = 1$, but quantum corrections to $\chi$ given by (\ref{chi_sus}) will now induce additional oscillations in longitudinal magnetoconductivity. All other terms in (\ref{SdH}) represent the Ohmic conductivity. Therefore, the total longitudinal conductivity as a function of $E_L = \sqrt{2eB} v$ is
\begin{align}\label{Total}
\frac{\sigma_{zz} (B)}{\sigma_0} \approx & 1 + \frac{3}{16}\frac{\mu^2}{\pi^2 v^3 \chi}\frac{\tau_V}{\tau} \frac{E_L^4}{\mu^4} - \frac{3}{20} \frac{E_L^4}{\mu^4}  \nonumber \\ &- \frac{3}{8\pi} \frac{E_L^3}{\mu^3} \sum_{l=1}^\infty \frac{1}{l^{3/2}}  \frac{\left(l\frac{4\pi^2 \mu T}{E_L^2}\right)}{\sinh\left(l\frac{4\pi^2 \mu T}{E_L^2}\right)} \exp(-4\pi l\Gamma\mu/E_L^2)\nonumber \\ &\times [\cos(2\pi l \mu^2/E_L^2) - \sin(2\pi l \mu^2/E_L^2)] ,
\end{align}
where the chiral susceptibility $\chi$ that enters the second term oscillates with $B$ according to (\ref{chi_sus}).
When the temperature or the Dingle factor are large enough so that the first term in the Fourier series dominates, the longitudinal conductivity is given by
\begin{align}\label{approx}
\frac{\sigma_{zz} (B)}{\sigma_0} \approx  1 + \left(\frac{3}{16}\frac{\tau_V}{\tau} - \frac{3}{20}\right) \left(\frac{B}{B_0}\right)^2\nonumber\\
- A(B) \left[\cos\left(\frac{B_0}{B}+\frac{\pi}{4}\right) + \frac{\pi}{4}\frac{\tau_V}{\tau}\frac{B}{B_0}\cos\left(\frac{B_0}{B}-\frac{\pi}{4}\right)\right] ,
\end{align} 
where $B_0 = \mu^2/2ev^2$ and $A(B)$ is a positive non-oscillating factor which represents the effects of the temperature and the Dingle factor. For a material with $\mu = 150\ \mathrm{meV}$ and Fermi velocity $v = c/500$, the value of $B_0$ is $B_0 \approx 48\ \mathrm{T}$. We have plotted the oscillations in the resistivity, considering oscillations in both parts of the conductivity, and also considering oscillations in the Ohmic conductivity, in Fig~\ref{figosc}. We have also plotted the oscillatory part of the conductivity vs $1/B$ in Fig~\ref{resi}

When chirality flipping time is much longer than the scattering time $\tau_V/\tau \gg 1$, and in strong magnetic field, the quantum CME oscillations dominate over the SdH ones; these CME oscillations have a phase of $-\pi/4$. On the other hand, in weak fields the SdH oscillations are dominant, with the phase of $\pi/4$. 

In a material with multiple pairs of Weyl points with different energies and Fermi velocities, $\chi$ would be the sum of the contribution to $\chi$ of each Weyl point. $\chi$ and therefore the CME conductivity would have oscillations with multiple frequencies. The SdH oscillations would also have multiple frequencies.
\begin{align}
\frac{\sigma_{zz}}{\sigma_0} = 1 + \alpha B^2 + &A_1 (B) \cos\left(\frac{B}{B_1}+\phi_1 (B)\right)\nonumber\\ + &A_2 (B) \cos\left(\frac{B}{B_2}+\phi_2 (B)\right)
\end{align}
\vskip0.3cm

\section{Conclusion}
To summarize, we have demonstrated that the non-linear relation between the density of chiral charge and the chiral chemical potential induces a new type of quantum oscillations in longitudinal magnetoconductivity of Dirac and Weyl (semi)metals. In strong magnetic fields and in materials that approximately preserve chirality (when the chirality flipping time is much longer than the scattering time), these new quantum oscillations dominate over the SdH ones. The  phase of these quantum CME oscillations differs from the SdH oscillations by $\pi/2$ which makes it possible to 
isolate them in experiment. 

\chapter{Chiral Magnetic Photocurrent}\label{chCMP}

\blfootnote{This chapter is based on \cite{kaushik2019chiral}.}

Circularly polarized light (CPL) breaks the symmetry between left and right and thus possesses a non-zero chirality. In the interactions of CPL with matter, the chirality of the electromagnetic field can couple to the chirality of matter. Various quantities have been used to describe the chirality of the electromagnetic field. A notable example is the ``zilch" introduced by Lipkin~\cite{Lipkin1964}: 
\begin{align}
Z^0 &= \boldsymbol{E} \cdot (\nabla\times\boldsymbol{E}) + \boldsymbol{B} \cdot (\nabla\times\boldsymbol{B})  \\
\boldsymbol{Z} &= \boldsymbol{E} \times \dot{\boldsymbol{E}} + \boldsymbol{B} \times \dot{\boldsymbol{B}}.
\end{align} Lipkin's zilch is gauge-invariant and obeys the continuity equation in free space as a consequence of Maxwell equations: $\partial_\mu Z^\mu =0$, with $Z^\mu = (Z^0, \boldsymbol{Z})$. However, interactions can transfer chirality from the electromagnetic field to matter. Chirality conservation for electromagnetic field and its role in light-matter interactions in chiral materials are the subjects of active current interest. In particular, 
Lipkin's zilch has been used to describe the interaction of CPL with chiral molecules~\cite{Tang2010,Bliokh2011,Coles2012}. CPL has also been proposed to cause a photovoltaic Hall effect~\cite{Oka2009,Yudin2015} in graphene, which is a material with two-dimensional  relativistic Dirac fermion quasiparticles. 

In this article, we discuss the interaction of CPL with the three-dimensional chiral quasiparticles in recently discovered Dirac and Weyl semimetals~\cite{Taguchi2016,Ebihara2016,Chan2016,dejuan17,lee2017}. We will show that the transfer of chirality from the electromagnetic field to chiral fermions can be described in a model-independent way by using the chiral anomaly~\cite{Adler1969,Bell1969}.
Because of the focus on the effect of the chiral anomaly, 
our treatment will be based on a measure of chirality that is different from Lipkin's zilch. Namely, we will use the Chern-Simons current~\cite{Chern1974}
\begin{align}
h^0 &= \boldsymbol{A}\cdot \boldsymbol{B}\\
\boldsymbol{h} &= A^0\, \boldsymbol{B} - \boldsymbol{A}\times\boldsymbol{E},
\end{align} to describe the chirality of light and its transfer to the chirality of matter. 

The Chern-Simons current is proportional to the helicity of the free electromagnetic field~\cite{Afanasiev1996}, which describes the difference between left and right circularly polarized photons. The chirality density $h^0$ is  well-known in magneto-hydrodynamics, where magnetic helicity~\cite{Woltjer1958,Moffatt1969,Arnold1998,Taylor1974} is defined as~$\int \boldsymbol{A}\cdot\boldsymbol{B}\, \mathrm{d^3}r$. Note that the helicity of the electromagnetic field is not conserved, even in free space. 

The gauge dependence of helicity is essential in describing the interactions mediated by the chiral anomaly. Indeed, the chiral anomaly results in the absence of invariance of the chiral charge $\int h^0 d^3r$ under ``large" gauge transformations that change the global topology of the gauge field. The chirality conservation law that we propose below is a consequence of the change of chirality under large gauge transformations that results from the transfer of chirality from electromagnetic field to the chiral fermion zero modes.

The chiral anomaly is known to result in the transport of charge in parallel electric and magnetic fields through the chiral magnetic effect ~\cite{CME,Kharzeev2009} by creating a chirality imbalance. The resulting longitudinal negative magnetoresistance~\cite{2013Son, burkov2015negative} has been observed in Dirac semi-metals such as $\mathrm{ZrTe_5}$~\cite{2016Li} and $\mathrm{Na_3 Bi}$~\cite{2015Xiong} and Weyl semi-metals such as TaAs~\cite{Huang2015}. It has been proposed that chiral pumping in three-dimensional Dirac materials by a rotating electromagnetic field can produce a separation between left and right handed cones in momentum space, resulting in the generation of axial current and polarization of electric charge density~\cite{Ebihara2016}.

The interactions of light with Dirac and Weyl materials have recently attracted significant attention. A photocurrent in response to CPL, proposed for Weyl materials with tilted cones~\cite{Chan2016}, has been recently observed in TaAs~\cite{MaTaAs}. A photocurrent in the presence of magnetic field has been proposed for asymmetric Weyl materials with tilted cones~\cite{yuta2018} and a quantised topological photocurrent in response to CPL has been proposed in asymmetric Weyl materials in which the left and right handed cones have different energies~\cite{dejuan17}. All of the above effects would be absent in Dirac materials such as $\mathrm{ZrTe_5}$ which possess both inversion and time reversal symmetries. 

In this paper, we propose an effect that does not rely on the breaking of inversion or time reversal symmetries of the crystal and arises solely from the helicity transfer from light to the chiral fermions in an external magnetic field. The proposed effect thus provides a clean way to probe the chiral anomaly and the chiral magnetic effect. 

The effect can be briefly described as follows. If the charged chiral fermion quasiparticles are massless (the corresponding linear band has no gap), and no chirality-changing interactions are present in the Hamiltonian, the helicity of external gauge fields can be transferred to the chirality carried by the charged quasiparticles, and vice versa. The only conserved quantity is the total chirality of the fermion quasiparticles and the gauge field. 
Once the light is absorbed by the material, the helicity of the light gets fully transferred to the material. The resulting chiral imbalance between the left- and right-handed chiral fermions, as we will see, is completely fixed by the chiral anomaly. This chiral imbalance in an external magnetic field is known to induce a current due to the chiral magnetic effect ~\cite{CME,Kharzeev2009}. Therefore, in an external magnetic field, CPL will induce an electric current.

\section{Conservation of Total Chirality}

In the interaction of CPL with an optically thick Dirac or Weyl semimetal (for a mid-infrared laser, the light penetration length for these materials is of the order of a few hundred nanometers), the helicity of the absorbed light gets fully transferred to the chirality of the fermions. 
The chiral fermions in 
Dirac and Weyl semimetals are described by the Hamiltonian which in the simplest isotropic case is given by 
\begin{equation}
\hat{H} = \pm v_\text{F} k_i \sigma_i,
\end{equation} 
where $v_\text{F}$ is the Fermi velocity and $k_i$ is the crystal momentum; the $\pm$ signs refer to the left and right handed fermions; the matrices $\sigma_i$ act over pseudospin degrees
of freedom. In a Dirac material, the left and right handed cones are located at the same positions in the Brillouin zone, whereas in Weyl materials they are separated. Each Weyl cone is a monopole of the Berry curvature, and since the total Berry charge inside a Brillouin zone is zero, the left- and right-handed Weyl cones always appear in pairs.

The chirality carried by the chiral fermion quasiparticles is described by the axial current  $j_5^\mu$; the temporal component of this current is the density of chiral charge, $j_5^0 \equiv \rho_5 = \rho_\text{R} - \rho_\text{L}$. The chiral anomaly causes non-conservation of $j_5^\mu$ in the presence of an electromagnetic field $F^{\mu \nu}$, as given by~\cite{Adler1969,Bell1969,Wilczek1987,Carroll1990,Sikivie1983} 
\begin{equation}\label{chir_an}
\partial_\mu j_5^\mu = \frac{e^2}{16\pi^2} \epsilon^{\mu\nu\rho\sigma}F_{\mu\nu}F_{\rho\sigma};
\end{equation}
note that the quantity on the right hand side is odd under parity and thus vanishes for linearly polarized light.
The quantity on the right is given by the full derivative of the  Chern-Simons current~\cite{Chern1974}
\begin{equation}\label{CS}
h^\mu = -\frac{e^2}{8\pi^2}\epsilon^{\mu\nu\rho\sigma}A_\nu F_{\rho\sigma}
\end{equation}
that describes the helicity density and flux carried by the electromagnetic field. 

We deduce from~\eqref{chir_an} and~\eqref{CS} that the total chirality, which is the sum of the helicity of the electromagnetic field and the chirality of the fermions, is conserved:
\begin{equation}\label{an_cons}
\partial_\mu (j_5^\mu + h^\mu) = 0.
\end{equation}
It is this conservation law that causes transfer of chirality from light to chiral fermions. This is valid separately for each Dirac cone or pair of Weyl cones.

The helicity density and helicity flux are given (in SI units) by
\begin{equation}\label{chir_dens}
h^0 = \frac{e^2}{4\pi^2\hbar^2} \boldsymbol{A}\cdot \boldsymbol{B}
\end{equation}
and
\begin{equation}\label{chi_flux}
\boldsymbol{h} = \frac{e^2}{4\pi^2\hbar^2} (A^0\, \boldsymbol{B} - \boldsymbol{A}\times\boldsymbol{E}).
\end{equation}
In our case, the prefactor of~$e^2/(4\pi^2\hbar^2)$ appears in equation~\eqref{chir_dens} because, as it can be seen from equation~\eqref{an_cons}, it is the chirality density available for the transfer to the chiral fermions.

In a real Dirac or Weyl material, there is also chirality-flipping scattering, with a characteristic relaxation time $\tau_\mathrm{V}$, so equation~\eqref{chir_an} (in SI units) becomes
\begin{equation}\label{chir_rel}
\dot{\rho_5} + \nabla\cdot\boldsymbol{j}_5 = \frac{e^2}{2\pi^2 \hbar^2} \boldsymbol{E}\cdot\boldsymbol{B} - \frac{\rho_5}{\tau_\mathrm{V}}.
\end{equation}
The last term on the right hand side of equation~\eqref{chir_rel} does not affect the balance of chirality transfer if the frequency of light $\omega$ is large compared to ${\tau_\mathrm{V}}^{-1}$.

For an oscillating electric field  with $\boldsymbol{E}(t,\boldsymbol{r}) = \Re(\eu^{-\iu\omega t}\boldsymbol{\mathcal{E}}(\boldsymbol{r}))$, in the Coulomb gauge with $A^0 = 0$, the time-averaged helicity flux~\eqref{chi_flux} is
\begin{equation}
\langle\boldsymbol{h}\rangle = \frac{e^2}{8\pi^2\hbar^2\omega}\Re(\iu\,\boldsymbol{\mathcal{E}}\times\boldsymbol{\mathcal{E}}^*).
\end{equation}
For light traveling in the $z$~direction, this becomes
\begin{equation}
\langle h^z \rangle = \frac{e^2}{4\pi^2\hbar^2\omega}\Re(\iu\,\mathcal{E}_x \mathcal{E}_y^*);
\end{equation}
note that this quantity vanishes for linearly polarized light.

\section{Chirality Transfer from Light to Fermions}

Let us now describe CPL using the Chern-Simons helicity. 
For CPL with $\boldsymbol{\mathcal{E}} = E_0\,(\boldsymbol{\hat{x}} \pm \iu\,\boldsymbol{\hat{y}})$,
\begin{equation}
\langle h^z \rangle = \pm \frac{e^2}{4\pi^2\hbar^2\omega} E_0^2.
\end{equation}
The ratio of this helicity flux to the energy flux $\langle S^z \rangle$ (the Poynting vector) is given by
\begin{equation}\label{per_phot}
\frac{\langle h^z \rangle}{\langle S^z \rangle} = \pm\frac{1}{\pi}\frac{e^2}{4\pi \epsilon_0 \hbar c}\frac{1}{\hbar\omega}.
\end{equation}
Hence, the helicity per photon available for transfer to each Dirac cone or pair of Weyl cones of charged fermions is~$\pm\alpha/\pi$, where~$\alpha$ is the fine structure constant; the signs refer to the two circular polarizations of light. Of course, the helicity per photon in the beam is~$\pm 1$, since photons are massless vector particles. The factor of~$\alpha/\pi$ in equation~\eqref{per_phot}, according to equation~\eqref{chir_rel}, describes the coupling of photons to the charged fermions through the chiral anomaly.  

Note that $\boldsymbol{E}\cdot\boldsymbol{B}$ is zero for CPL in vacuum, but it is non-zero if it is being attenuated in a material. This is what allows chiral charge to be generated according to~\eqref{chir_rel}.

If CPL is incident on a 3D chiral material, as the light is absorbed by the material, its helicity flux is converted into the chirality of fermions. The total chirality generated per unit area is equal to the helicity flux of light transmitted at the interface. In a real material, if $\omega \tau_\mathrm{V} \gg 1$, the chirality will saturate at a constant value proportional to $\tau_\mathrm{V}$ due to chirality relaxation, as dictated by equation~\eqref{chir_rel}. Using equation~\eqref{an_cons}, for light incident perpendicular to the interface at $z=0$,
\begin{equation}
\int_0^\infty \rho_5\, dz = \tau_\mathrm{V} \langle h^z\rangle |_{z=0} = \pm\tau_\mathrm{V}\frac{\alpha}{\pi} \frac{I_\mathrm{in}}{\hbar \omega} \Re(a_x a_y^*),
\end{equation}
where $I_\mathrm{in}$ is the intensity of the incident light and $a_{x,y}$ are the transmission amplitudes of the two linear polarizations. The chiral charge is distributed in the material over a length scale determined by the diffusion length of fermion quasiparticles and the attenuation length of light, but the total chiral charge integrated over the depth is unaffected by this.

\begin{figure}
 \begin{center}
 \includegraphics[scale=0.14]{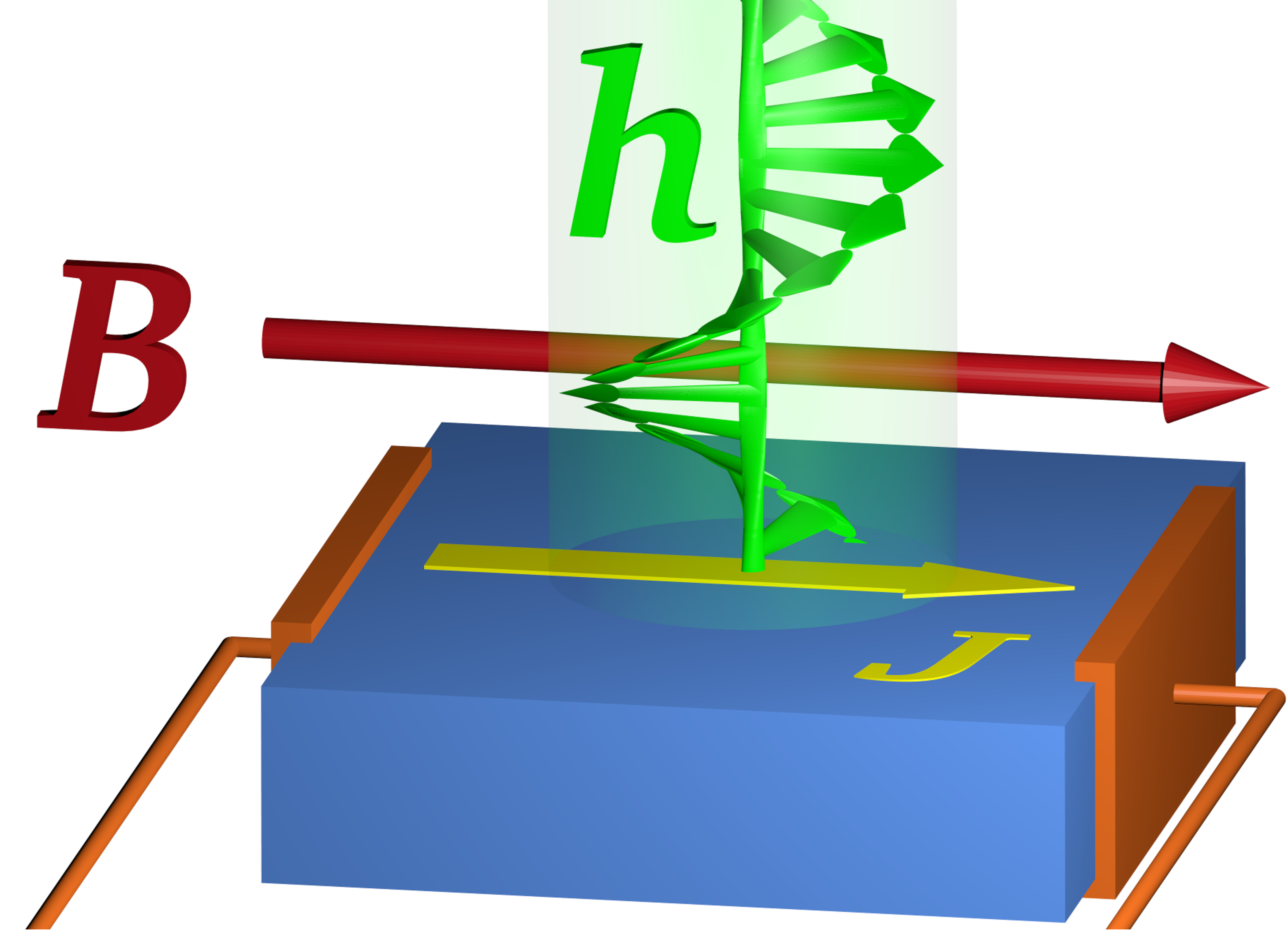}
 \caption{The helicity of circularly polarized light (CPL) is characterized by the Chern-Simons current $\boldsymbol{h}$. When incident on a Dirac or Weyl semimetal, as a consequence of the chiral anomaly, CPL induces an asymmetry between the number of left- and right-handed chiral quasiparticles. In an external magnetic field $\boldsymbol{B}$, this chiral asymmetry induces a chiral magnetic photocurrent $\boldsymbol{J}$ along the direction of the externally applied magnetic field. }
 \label{fig:expt}
 \end{center}
\end{figure}

The chiral charge density of fermions $\rho_5$ translates into a chiral chemical potential $\mu_5 \simeq \chi^{-1} \rho_5$, where $\chi=\partial \rho_5/\partial \mu_5$ is the chiral susceptibility. If the whole system is placed in a constant magnetic field $\boldsymbol{B}_\mathrm{ext}$ perpendicular to the incident light, a current 
\begin{equation}
\boldsymbol{J}_\text{CME} = \frac{e^2}{2\pi^2\hbar^2}\, \boldsymbol{B}_\mathrm{ext}\, \mu_5
\end{equation}
due to the chiral magnetic effect (CME)~\cite{CME,Kharzeev2009} is generated along the direction of the magnetic field, as shown in Fig.~\ref{fig:expt}. The linear density $\boldsymbol{\kappa}_\text{CMP}$ of the resulting chiral magnetic photocurrent is given by the integral over the depth:
\begin{align}
\label{cur_dens}
\boldsymbol{\kappa}_\text{CMP} &= \int_0^\infty \frac{e^2}{2\pi^2\hbar^2}\, \boldsymbol{B}_\text{ext}\, \mu_5\, dz\nonumber\\& = 
\pm \frac{e^2}{2\pi^2\hbar^2}\,\boldsymbol{B}_\mathrm{ext}\,\frac{\tau_\mathrm{V}}{\chi}\,\frac{\alpha}{\pi}\, \frac{I_\mathrm{in}}{\hbar \omega}\, \Re(a_x a_y^*).
\end{align}
The formula for the CME conductivity is given by~\cite{2013Son,2015Xiong,2016Li}
\begin{align}
\left(\frac{e^2}{2\pi^2\hbar^2}\right)^2 \frac{\tau_\text{V}}{\chi} B_\text{ext}^2.
\end{align}
If the only contribution to the magnetic field dependence of the longitudinal conductivity is from the CME, then the quadratic coefficient of the longitudinal conductivity is
\begin{align}
\sigma^{(2)}_{zz} = \left(\frac{e^2}{2\pi^2\hbar^2}\right)^2 \frac{\tau_\text{V}}{\chi}
\end{align}
for each Dirac cone or pair of Weyl cones. In terms of $\sigma^{(2)}_{zz}$
\begin{equation}\label{expt_formula}
\boldsymbol{\kappa}_\text{CMP}= 
\pm \frac{2 \pi^2 \hbar^2}{e^2}\,\boldsymbol{B}_\mathrm{ext}\, \sigma^{(2)}_{zz} \,\frac{\alpha}{\pi}\, \frac{I_\mathrm{in}}{\hbar \omega}\, \Re(a_x a_y^*).
\end{equation}
This is the main result of this chapter; note that this formula does not depend on the number of cones.

\section{Numerical Estimates}

The linear photocurrent density in~\eqref{expt_formula} can be integrated over the diameter of the spot of light to estimate the magnitude of the observed photocurrent. We assume a mid-infrared laser of power $10\,\mathrm{mW}$, intensity $I_\mathrm{in}=5 \times 10^{6}\, \mathrm{W/m}^2$, spot diameter $50\,\mu\mathrm{m}$,  and wavelength $10\,\mathrm{\mu m}$. We also assume that the square of the optical transmission amplitude is $\Re(a_x a_y^*)=0.1$, and use magnetic field $B_\text{ext} = 2\, \mathrm{T}$ and temperature $5\, \mathrm{K}$. We need the coefficient $\sigma^{(2)}_{zz}$ and the ratio of the resistance of the sample to the load (suppression factor) to estimate the photocurrent.

For the Weyl material TaAs used in~\cite{Zhang2016}, using the same suppression factor as the  setup in~\cite{MaTaAs}, we get a current of approximately $25\,\mathrm{nA}$. For the Dirac material $\mathrm{ZrTe}_5$ used in~\cite{2016Li}, assuming the same resistance for the external load as in~\cite{MaTaAs}, we get a current of approximately $50\,\mathrm{nA}$. This current scales linearly with the magnetic field and wavelength, and should be much stronger if a THz source is used.

This estimate can be compared to the photocurrent of about $40\,\mathrm{nA}$ observed recently in TaAs~\cite{MaTaAs} that is said to exceed the photocurrent observed in other materials currently used for detecting the mid-infrared radiation by a factor of 10-100. 

\section{Discussion}

In contrast to the photocurrents that have been proposed~\cite{Chan2016} and observed~\cite{MaTaAs} recently and the photocurrents proposed in~\cite{dejuan17} and~\cite{yuta2018} which utilize asymmetries of the crystals, chiral magnetic photocurrent solely depends on the imbalance between the densities of right- and left-handed chiral fermions.  The direction of the photocurrent depends only on the magnetic field and the circular polarization of light, but not on the crystal axes. The observation of the chiral magnetic photocurrent would provide a strong independent evidence for the importance of chiral anomaly in condensed matter systems.

Since chiral magnetic photocurrent depends on transitions from one Weyl cone to another, instead of interband transitions, there is no lower cut-off frequency -- as a result, the photocurrent is predicted to be strong in the THz frequency range. This could potentially be used to detect circularly polarized THz radiation. In large magnetic fields and at low frequency of light, the strength of the chiral magnetic photocurrent may exceed the currently observed chiral photocurrents. This opens possibilities for applications in photonics and optoelectronics, especially in the THz frequency range.

\def\eu{\mathrm{e}}
\def\iu{\mathrm{i}}
\def\du{\mathrm{d}}
\def\Tr{\mathrm{Tr}}
\def\Det{\mathrm{det}}

\chapter{Chiral Terahertz Emmission in TaAs}\label{chTHz}

\blfootnote{This chapter is based on \cite{gao2020chiral}.}

\begin{figure}
    \centering
	\includegraphics[width=12cm]{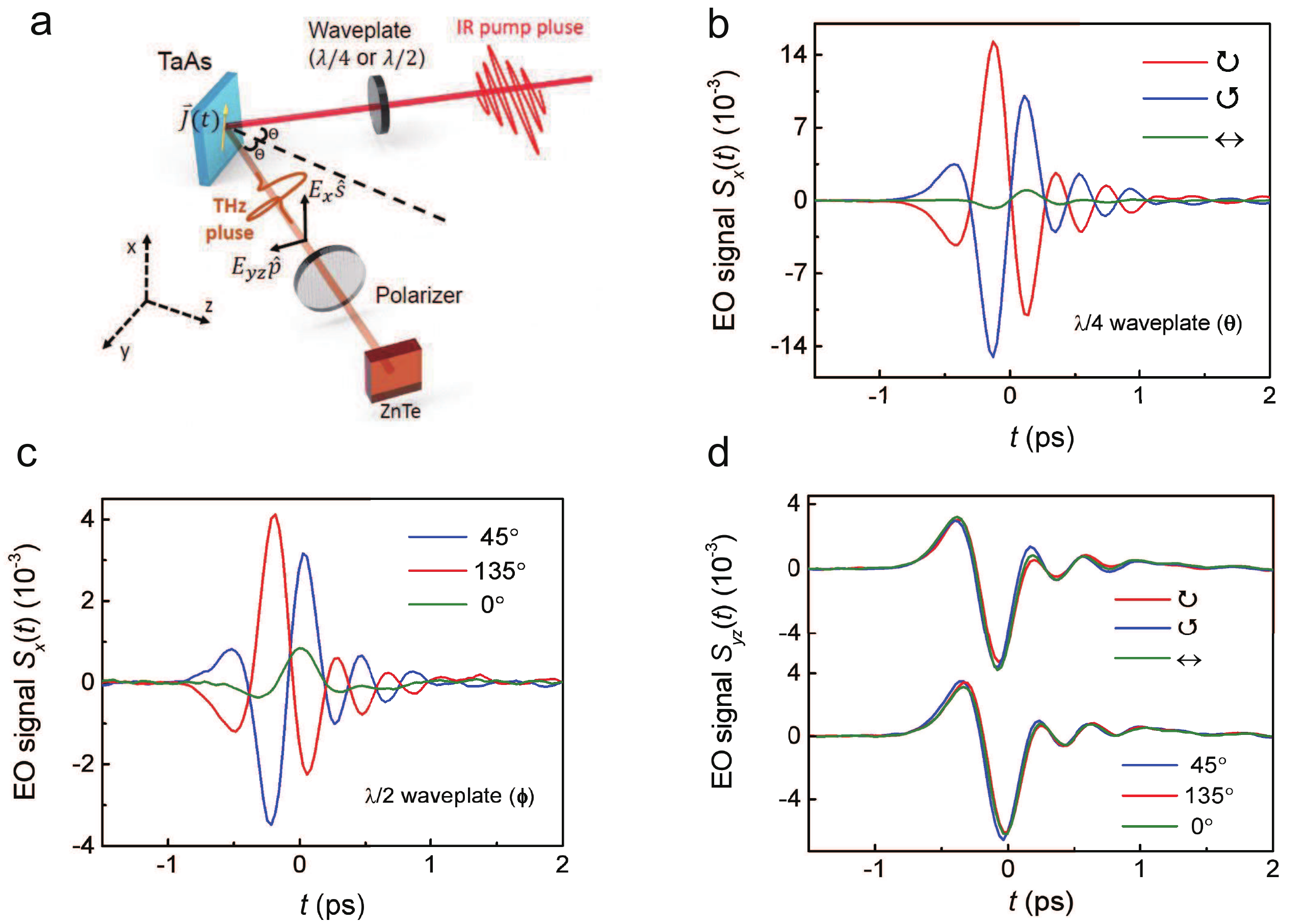}
	\caption{\label{fig:mainTHzresults} \textbf{a}. Schematic of the
		THz emission spectroscopy. Excitation of a fs laser pulse with an incident angle $\Theta$ onto a TaAs single crystal initiates a photocurrent burst and, consequently, emission of a THz pulse $\vec{E}(t)$[$=E_x(t)\hat{s}+E_{yz}(t)\hat{p}$]. Measurement of the components $E_x(t)$ and $E_{yz}(t)$ by the EO sampling provides access to the sheet current density $\vec{J}(t)$ flowing inside the sample. \textbf{b-d}. Typical THz EO signal components $S_x(t)$ and $S_{yz}(t)$ along the $\hat{s}$ and $\hat{p}$ directions were measured at various settings for pump polarization via rotating the $\lambda/4$ or $\lambda/2$ waveplate, characterized by the angle $\theta$ or $\phi$. Here, $\leftrightarrow$ ($\theta$=0$^\circ$), $\circlearrowright$ ($\theta$=45$^\circ$), and $\circlearrowleft$ ($\theta$=135$^\circ$) represent the $p$, right-handed, and left-handed circularly polarized light, respectively. The angle $\phi$ stands for the linear polarization state with respect to the $p$ polarized light ($\phi$=0$^\circ$).}
\end{figure}


In this chapter, we describe an experimental generation of ellipitcally polarized terahertz radiation from TaAs in response to ultrafast near-infrared and optical pulses. We quantitatively elucidate the colossal ultrafast photocurrents in TaAs for the first time. We find that the polarization of the elliptically polarized THz wave can be easily manipulated on a fs timescale, which is unprecedented. Such control arises from the colossal chiral ultrafast photocurrents whose direction and magnitude can be manipulated in an ultrafast manner using the circularly and/or linearly polarized fs optical pulses. The excitation pulse can have a broad spectral range from visible to mid-infrared light, and generate maximum photocurrent around 1.5 eV. We unravel that the Weyl fermions play the key role in generating the giant chiral ultrafast photocurrents. 

A schematic of the experiment is shown in Fig.~\ref{fig:mainTHzresults}\textbf{a}. Femtosecond laser pulses are used to induce ultrafast photocurrents. According to the Maxwell equations, a change in the current density $\vec{j}(z,t)$ on the picosecond (ps) timescale will result in electromagnetic radiation in the THz spectral range (1 THz = 1 ps$^{-1}$) \cite{Kampfrath_NNano_2013}. The transient electric field $\vec{E}(t)$ is generated with a polarization parallel to the direction of the current. Therefore, one can use the time-domain spectra $\vec{E}(t)$ of the THz radiation as a probe for the ultrafast sheet current density given by $\vec{J}(t)=\int dz\vec{j}(z,t)$. The orthogonal components $J_x$ and $J_{yz}$ via the generalized Ohm's law determine the THz near-field $\vec{E}(t)$ on top of the sample surface, i.e., the $s$-polarized $E_x$ along $\hat{x}$ and the $p$-polarized $E_{yz}$ in the $yz$ plane. Experimentally, the THz far-field electro-optic (EO) signal $\vec{S}(t)$ was collected, and the THz near-field $\vec{E}(t)$ was derived via inversion procedures based on a linear relationship between these two quantities \cite{Kampfrath_NNano_2013}.  Therefore, $\vec{S}(t)$ is a qualitative indicator of the ultrafast photocurrent, whose genuine properties shall be quantitatively obtained by analyzing $\vec{J}(t)$.    

\section{Results and discussions}
\begin{figure}
    \centering
	\includegraphics[width=12cm]{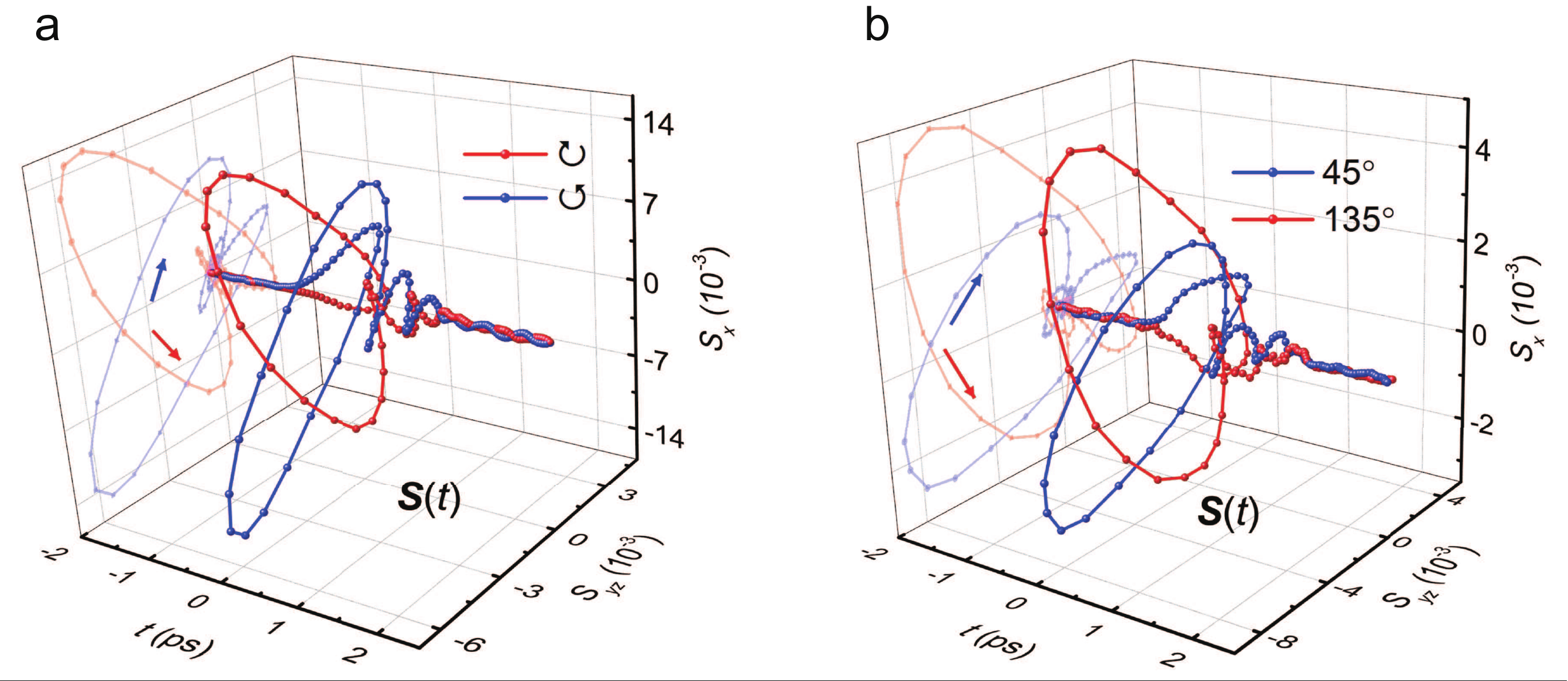}
	\caption{\label{fig:3D} \textbf{a} and \textbf{b} are the far-field EO signals $\vec{S}(t)$ $[=S_x(t)\hat{s}+S_{yz}(t)\hat{p}]$ for circularly and linearly polarized pump light, respectively. The coloured arrows indicate different optical chiralities: left-handed (blue) and right-handed (red).}
\end{figure}

\textbf{THz emission from TaAs.} In the present experiments, unless noted in the text, we mainly focus on the results obtained for the TaAs(112) single crystal with an incident angle $\Theta\simeq3^\circ$ using the excitation light with a wavelength of 800 nm. $\hat{x}$ is along the [$\overline{1}$10] direction. Figs.~\ref{fig:mainTHzresults}\textbf{b-d} show strong time-domain THz far-field EO signals $\vec{S}(t)$ detected from the sample. Clearly, both the magnitude and temporal shape of the THz waveform $S_x(t)$ depend strongly on the light polarization. The key observation is that signals $S_x(t)$ taken with right- ($\circlearrowright$) and left-handed ($\circlearrowleft$) circularly polarized light are completely out of phase (Fig.~\ref{fig:mainTHzresults}\textbf{b}). A similar observation was found for the 45$^\circ$ and 135$^\circ$ linearly polarized light (Fig.~\ref{fig:mainTHzresults}\textbf{c}). In terms of the peak values, $S_x$ induced by the linearly polarized light is approximately three times smaller than that due to excitation by circularly polarized light. By contrast, $S_{yz}(t)$ is almost polarization-independent and differs substantially from $S_x(t)$ (see Fig.~\ref{fig:mainTHzresults}\textbf{d}). Such distinct $S_x$ and $S_{yz}$ components lead to a elliptically polarized transient THz field $\vec{S}(t)$ (or $\vec{E}(t)$), which exhibits opposite chirality for different circularly or linearly polarized pump light (see Figs.~\ref{fig:3D}\textbf{a} and \textbf{b}), as we will discuss in detail below.

\begin{figure}
    \centering
	\includegraphics[width=12cm]{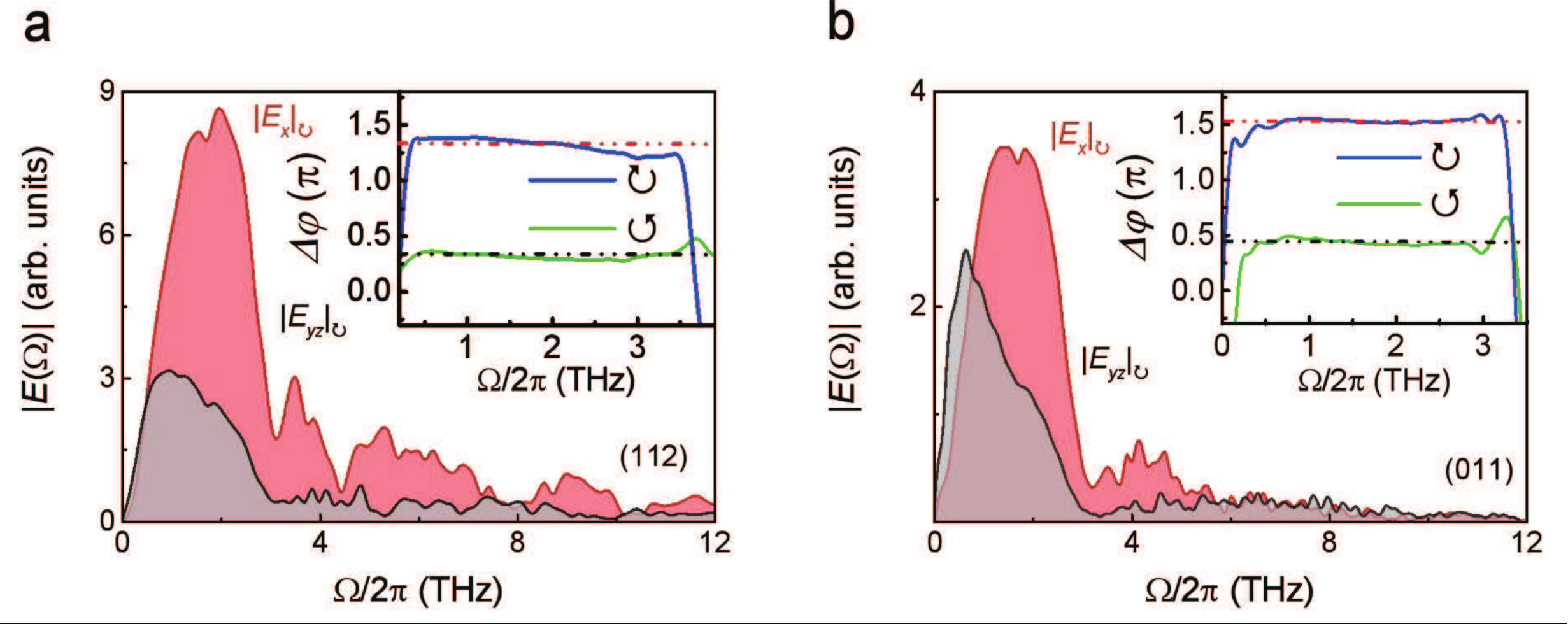}
	\caption{\label{fig:Fourier} \textbf{a} and \textbf{b} are Fourier transform spectra for the THz near-field $\vec{E}(t)$ from (112) and (011) faces for circularly polarized pump light, respectively. The insets show the phase difference ($\Delta\varphi$) between $E_x(\Omega)$ and $E_{yz}(\Omega)$ for each circularly polarized pump light. The dashed lines represent the average value of $\Delta\varphi$.}
\end{figure}

In the frequency domain of the THz near-field $\vec{E}(t)$ (see Fig.~\ref{fig:Fourier}\textbf{a}-\textbf{b}), the dominant spectra for both $E_x$ and $E_{yz}$ sit below $\sim$3 THz. At approximately 1.7 and 3.1 THz ($\sim$57 and 104 cm$^{-1}$), there exist two obvious dips. The former might be due to the infrared active phonon mode in TaAs. The latter can be attributed to the absorption of the Raman active $E(1)$ mode \cite{Liu_PRB_2015}. The high-frequency tails extend almost to 12 THz, consistent with a time resolution of $\sim$80 fs. Strikingly, for 0.2$\lesssim\Omega\lesssim$3 THz, we discovered that the phase difference, $\Delta\varphi$, between $E_x$ and $E_{yz}$ is nearly constant. This phase difference is independent of the incident angle $\Theta$ for a given pump polarization. However, its value differs between different faces (insets of Figs.~\ref{fig:Fourier}\textbf{a} and \textbf{b}), i.e. for (112), $\Delta\varphi_\circlearrowleft\simeq\pi/3$ and $\Delta\varphi_\circlearrowright\simeq4\pi/3$; for (011), $\Delta\varphi_\circlearrowleft\simeq\pi/2$ and $\Delta\varphi_\circlearrowright\simeq3\pi/2$.  
The origin of $\Delta\varphi$ will be discussed later. Nevertheless, such extraordinary findings suggest that $\vec{E}(t)$ can be well regarded as a broadband elliptically polarized THz pulse with its detailed characteristics depending on the pump polarization and the sample faces and, hence, has a defined chirality. According to the polarization trajectory ($S_{yz}(t)$,$S_x(t)$) in Fig.~\ref{fig:3D}\textbf{a} and \textbf{b}, chirality of the THz pulse can be instantaneously switched by varying the circular or linear polarization of pump light. 

\textbf{Polarization dependence of the THz signals.} To understand the peculiar THz wave emission from TaAs, it is necessary to elucidate the mechanism(s) generating the underlying time-resolved photocurrents. We measured the dependence of $S_x(t)$ and $S_{yz}(t)$ on the degree of circular polarization of the incident light, which can be controlled by rotating the quarter-wave plate by an angle $\theta$ (Fig.~\ref{fig:mainTHzresults}\textbf{a}). The experimental results were found to be well fitted by the following equation \cite{Osterhoudt_arXiv_2018,Ganichev_JPhys_2003,McIver_NNano_2011} 
\begin{eqnarray}
S_\lambda(t,\theta)&=&C_\lambda(t)sin2\theta+L_{1\lambda}(t)sin4\theta+L_{2\lambda}(t)cos4\theta \nonumber\\
& & +D_\lambda(t), 
\label{Eq:angle1}
\end{eqnarray}
where $\lambda$=$x$ or $yz$. $C_\lambda$ represents the contribution from helicity-dependent photocurrents. $L_{1\lambda}$ depends on the linear polarization and is phenomenologically associated with a quadratic nonlinear optical effect. $L_{2\lambda}$ and $D_\lambda$ arise from a thermal effect related to the light absorption. All four terms on the right side of Eq.~(\ref{Eq:angle1}) depend monotonically on the optical pump power, which agrees with our experimental observation

\begin{figure}
    \centering
	\includegraphics[width=12cm]{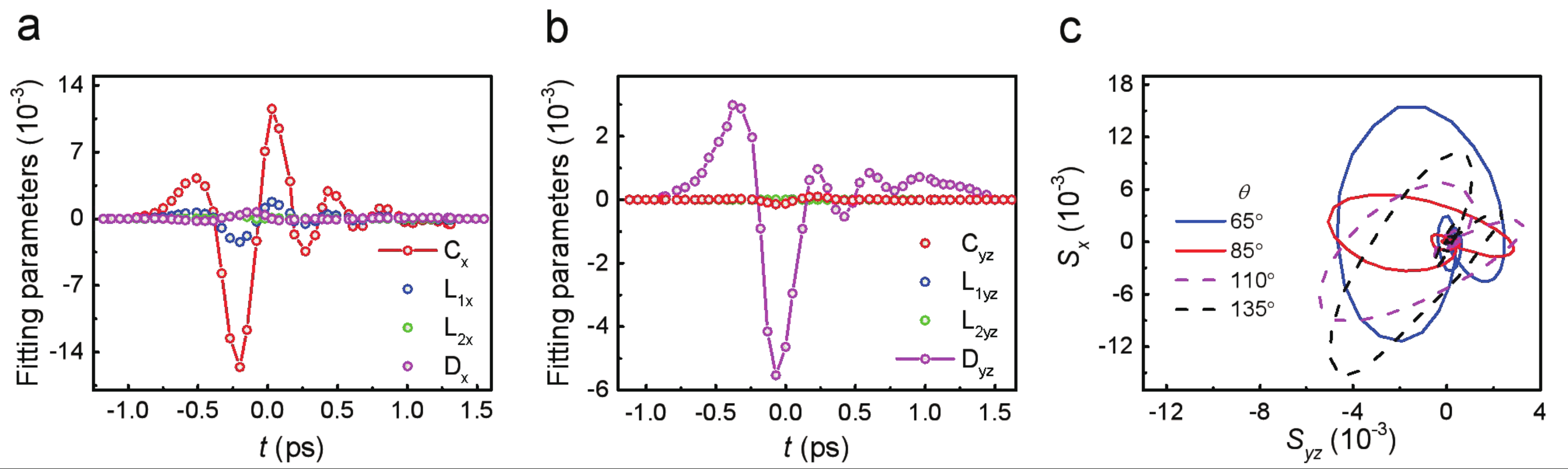}
	\caption{\label{fig:parameters} \textbf{a} and \textbf{b}. Time-dependent fitting parameters $C_\lambda$, $L_{1\lambda}$, $L_{2\lambda}$ and $D_\lambda$ ($\lambda=x,yz$) in Eq.~(\ref{Eq:angle1}). \textbf{c} shows the polarization trajectory ($S_{yz}(t)$,$S_x(t)$) under elliptically polarized pump light with different $\theta$. The solid and dashed curves represent opposite chiralities.} 
\end{figure}

\begin{figure}
    \centering
	\includegraphics[width=12cm]{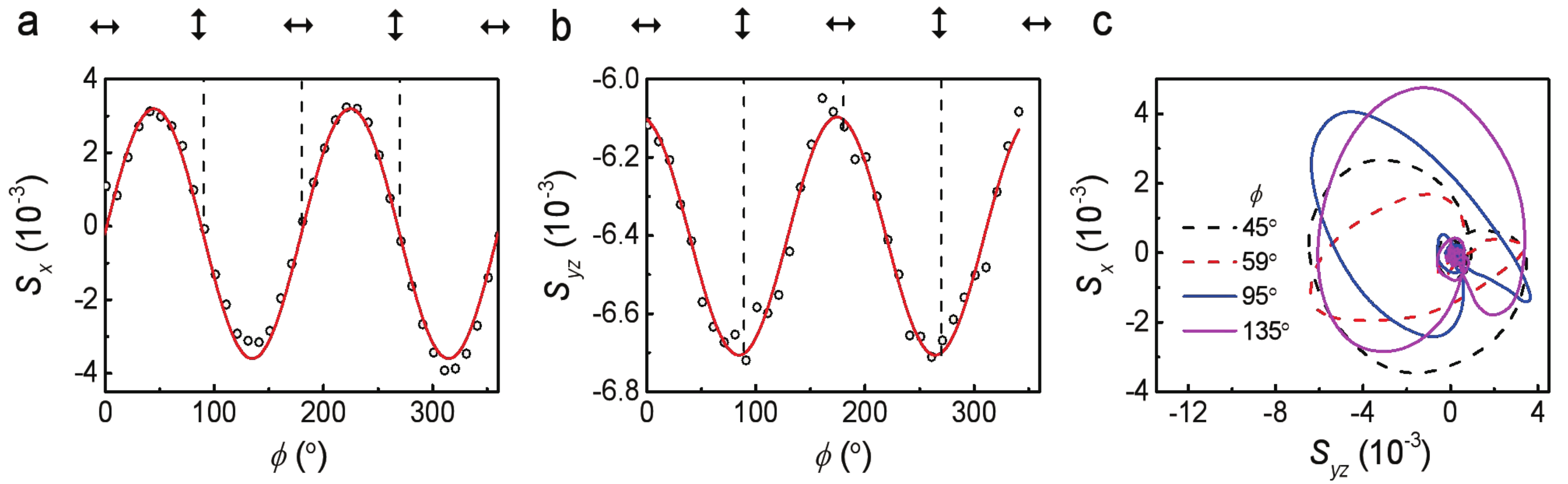}
	\caption{\label{fig:angle} \textbf{a} and \textbf{b} display the EO signals for $S_x(t=0.03$ ps) and $S_{yz}(t=-0.02$ ps) (near the peak values) as a function of the linear-polarization angle, $\phi$. The red solid lines show the fitted results. \textbf{c} shows the polarization trajectory ($S_{yz}(t)$,$S_x(t)$) for different linearly polarized pump light with several typical $\phi$. The solid and dashed curves represent opposite chiralities.} 
\end{figure}

Figs.~\ref{fig:parameters}\textbf{a} and \textbf{b} display the time-dependent parameters $C_\lambda$, $L_{1\lambda}$, $L_{2\lambda}$ and $D_\lambda$, which were obtained by fitting the experimental $S(t,\theta)$ using the Eq.~\ref{Eq:angle1}. Based on our results, $S_x$ is unambiguously dominated by $C_x$, and has a non-negligible contribution from $L_{1x}$. Both the amplitude and phase of $S_x(t)$ change with $\theta$, while $L_{2x}$ and $D_{x}$ can be omitted. On the other hand, $S_{yz}(t)$ is dominated by a polarization-independent $D_{yz}(t)$. $C_{yz}$ plays a very small role, while $L_{1yz}$ and $L_{2yz}$ can be neglected. These results suggest that the ultrafast photocurrents leading to the THz signal $E_x$ (or $S_x$) is polarization-dependent (or $\theta$-dependent), in contrast to the polarization-independent thermally related photocurrent inducing $E_{yz}$ (or $S_{yz}$). Therefore, as demonstrated in Fig.~\ref{fig:angle}\textbf{c}, one can control the ellipticity and chirality of the elliptically polarized THz pulse by changing the quarter-wave plate angle $\theta$ (the elliptical polarization of the pump light). Realization of the broadband circularly polarized THz pulses also becomes possible, e.g., THz emission from the (011) face with $\Delta\varphi\simeq\pi/2$.

Photocurrents arising from the linearly polarized light can be uncovered by measuring the dependence of $S_x$ and $S_{yz}$ on the linear polarization angle $\phi$, using a half-wave plate. The angle dependence of $S_x(t,\phi)$ near the peak values are shown in Figs.~\ref{fig:angle}\textbf{a} and \textbf{b}. We found that $S_x(\phi)$ can be well described by a second-order nonlinear optical process after considering the crystal symmetry, as demonstrated by the fitted curves in Figs.~\ref{fig:angle}\textbf{a} and \textbf{b}. On the other hand, $S_{yz}$ is only slightly modulated by the linear-polarization dependent signal and is dominated by a polarization-independent background. Similarly, we can manipulate the elliptically polarized THz pulse by changing the linear polarization state of the pump light, as illustrated in Fig.~\ref{fig:angle}\textbf{c}.  

\begin{figure}
    \centering
	\includegraphics[width=12cm]{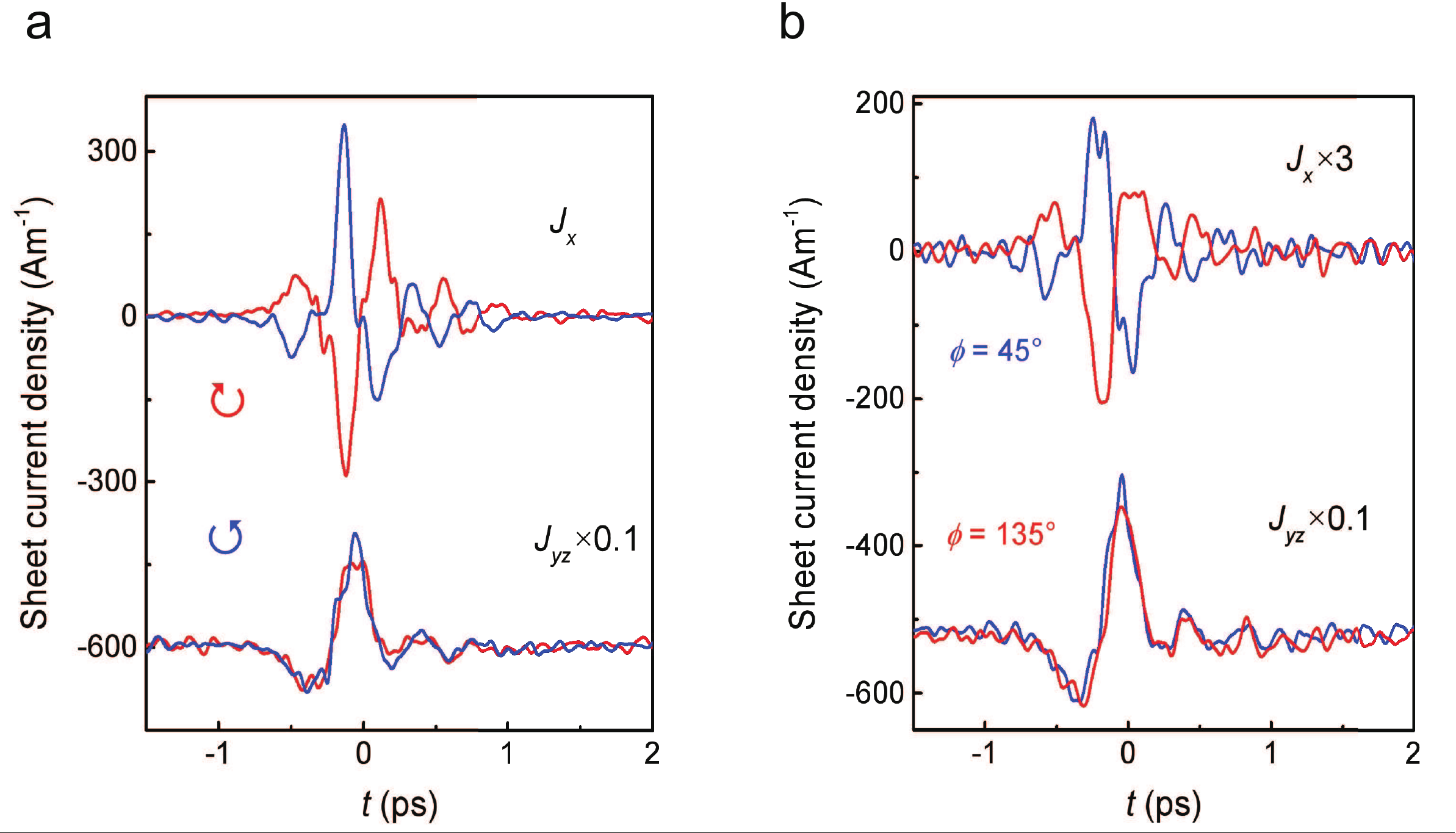}
	\caption{\label{fig:current} \textbf{a} and \textbf{b}. Extracted sheet photocurrent densities $J_x(t)$ and $J_{yz}(t)$ for different circular or linear pump polarization. Curves are offset for clarity.}
\end{figure}

\textbf{Ultrafast photocurrents in TaAs.} Observations of chiral broadband THz pulses indicate that the amplitude and phase of the ultrafast photocurrents can be fully controlled by polarized fs optical pulses. One can use the measured THz signals to quantitatively extract the ultrafast photocurrents, which are displayed in Figs.~\ref{fig:current}\textbf{a} and \textbf{b}. The data unambiguously demonstrates that switching of the current direction of $J_x(t)$ occurs instantaneously on a fs timescale using circularly or linearly polarized light, while $J_{yz}(t)$ is nearly unchanged for different polarized light. Along the time axis, $\vec{J}(t)$ shows spiral behavior. As a result, $\vec{J}(t)$ has chirality, which can be manipulated by the polarized pump light. Such results are rarely seen in conventional materials and, hence, lead to peculiar elliptically polarized ultrafast THz pulses. With regard to the dynamics of $J_x(t)$ and $J_{yz}(t)$, after an initial onset, $J_x(t)$ generally proceeds much faster than $J_{yz}(t)$. The former notably shows a strong oscillatory behaviour, which might be attributed to plasma oscillation (or plasmon) of the charge carriers. In fact, both previous FTIR \cite{Xu_PRB_2015} and our ultrafast optical transient reflectivity studies show that the Drude scattering time in TaAs has a timescale of $\sim$400 fs, which is consistent with the current relaxation time observed in Figs.~\ref{fig:current}\textbf{a} and \textbf{b}.

To determine the origin of $J_x$ and $J_{yz}$, we need to consider the mechanisms for the photocurrent generation. Microscopically, the ultrafast photocurrents can be generated during processes such as optical transitions, phonon- or impurity-scatterings, and electron-hole recombinations \cite{Braun_NC_2016}. Photocurrents induced by the optical transitions, occurring within the pulse duration, can in principle be controlled non-thermally in an ultrafast way \cite{Ganichev_JPhys_2003,McIver_NNano_2011,lee2017,MaTaAs}. Of particular interest are the photocurrents due to the circular photogalvanic effect (CPGE) and linear photogalvanic effect (LPGE) \cite{Ganichev_JPhys_2003,Osterhoudt_arXiv_2018}, which are often respectively referred to as injection and shift currents \cite{Sipe_PRB_2000,Nastos_PRB_2006}. The former depends on the helicity of the pump light, while the latter is dependent on the crystal symmetry or the linear polarization state of the light. Based on the obtained sheet current densities, the injection currents due to CPGE play the main role in $J_x(t)$. We thus put the main focus on the injection currents.
    
\textbf{CPGE - Injection currents.} The scenario for CPGE is displayed in Fig.~\ref{fig:mechanisms}\textbf{a} \cite{McIver_NNano_2011}, where circularly polarized light introduces asymmetric population (depopulation) of the excited (initial) states complying with the angular momentum selection rules ($\Delta m_J=\pm1$). Due to band velocity differences among these states, an instantaneous charge current emerges, which is proportional to the average band velocity ($\Delta\vec{v}_w$). This scenario together with the cone tilting was employed to explain the helicity-dependent DC photocurrent in TaAs \cite{lee2017,MaTaAs}. However, in contrast to those studies, where direct optical transitions only occurred within the Weyl cones due to usage of the long-wavelength infrared light with a photon energy of $\sim$120 meV, our experiments directly access the interband transitions between the Weyl cones and high-lying excited states above $E_f$ using excitation energies greater than $\sim$470 meV \cite{Weng15,Buckeridge_PRB_2016}. 

We thus carried out detailed theoretical calculations to clarify our observations. Note that the electromagnetic radiation is driven by the acceleration of the charge and is therefore proportional to the difference between the initial and final velocities of the charge resulting from its interaction with an external field. In our case, when the quasiparticle is excited by the laser light from/to a linear Weyl band with a large momentum-independent velocity $v = \partial E(q) / \partial q$ to/from a band with a much smaller velocity, the velocity difference is very large -- this makes Weyl semimetals ideal sources of induced radiation. The current then relaxes in the material over a typical time scale of $\sim 1$ ps, with this rapid deceleration of electric charges accompanied by electromagnetic radiation in the THz frequency range.	

\section{Calculations of the chiral photocurrent}

Let us derive the expression for the CPGE in the case when the transition is from a linear band to a massive band. The calculation would be analogous for transitions from a massive band to linear band, with the same sign for the photocurrent. This is because the holes have opposite helicity and opposite charge as electrons, so the final expression would have the same sign.

We assume the Hamiltonian in which a single Weyl cone makes a contribution is described by
\begin{equation}\label{hamilt}
\hat{H} =  \hbar v^i_a\, \sigma_a\, q_i + \hbar v_t^i\, \sigma_0\, q_i \equiv \hat{H}_{\text W} + \hat{H}_{\text t},
\end{equation}
where $\sigma_a$ are the Pauli matrices, $\sigma_0$ the identity matrix, $a$ is the pseudospin index and $i$ is the spatial index; the vector $q_i$ is the momentum measured from the Weyl point. The first term $\hat{H}_\text{W}$ contains information about the chirality and the velocity of the Weyl fermion, and the second term $\hat{H}_\text{t}$ describes  the tilt in the direction determined by the constant vector $\vec v_{\text{t}}$; $q_i$ is the quasiparticle's momentum measured from the position of the Weyl node; $v_{ia}$ is the velocity matrix. The interaction with the electromagnetic field $\vec{A}$ is obtained from (\ref{hamilt}) through the Peierls substitution $\vec{q} \to \vec{q} - \frac{e}{\hbar}\vec{A}$; this leads to the electromagnetic interaction Hamiltonian $\hat{H}_{\text EM}$
\begin{equation}\label{hamilt_em}
\hat{H}_\text{EM} =  -e v^i_a\, \sigma_a\, A_i - e\, v^i_t\, \sigma_0\, A_i \equiv \hat{H}_{\text W EM} + \hat{H}_{\text t EM}. 
\end{equation}
Here, the second term (which we denote by $\hat{H}_{\text t EM}$) is diagonal in spin space and does not contribute if one considers transitions between the Weyl bands; however, it will in general contribute once other nonlinear bands are excited.

Using the Hamiltonian (\ref{hamilt_em}), the induced electric current density can now be readily computed basing on two physical assumptions: i) we can neglect the band velocity of an excited band compared to the velocity on the Weyl band \cite{Weng15,Buckeridge_PRB_2016}, so the energy of the excited band can be assumed to be approximately independent of momentum; and ii) once the photons enter the material, they will induce an excitation with unit probability. Assumption ii) may not be realistic due to other excitations induced by the photons, e.g., the shift photocurrents discussed in later sections. Based on these assumptions and using Fermi's golden rule, we can write the current density integrated over the penetration depth (the DC sheet current density) as
\begin{align}
&\vec{J}(\omega,\vec{k}_p, \vec{\varepsilon}) = \int \vec{j}(\omega,\vec{k}_p, \vec{\varepsilon})\, d z =  \nonumber \\
& \frac{-eI}{\hbar \omega}\,  \frac{\sum_{l} \tau_l a_l \int \frac{\du^3q}{2\pi^3}\, \delta(E_{l-}(q)-E_{l0})\, (0-\vec{v}_{l-}(q))\, \sum_i \langle {s_{li}| \hat{H}_{\text EM}|q_{l-}}\rangle^2}{\sum_{l} a_l \int \frac{\du^3q}{2\pi^3}\, \delta(E_{l-}(q)-E_{l0})\, \sum_i \langle{s_{li}| \hat{H}_{\text EM}|q_{l-}}\rangle^2}\label{jdef},
\end{align}
where the summation over $l$ is the summation over the 24 Weyl cones of TaAs. $\omega, \vec{k}_p, \vec{\varepsilon}$ and $I$ are the frequency, momentum, polarization and intensity of the pump light entering the material. $v_-(q)$ is the band velocity; $E_-$ is the energy on the valence Weyl band; $E_{0}$ is the energy of the excited band minus $\hbar\omega$; $\ket{s_i}$ is the spin state of the excited band, and $\tau_l$ is the current relaxation time. The relaxation time appears in Eq.\eqref{jdef} because Fermi's golden rule yields the number of transitions per unit time and we have to integrate it over the lifetime of the current. The prefactor of $\frac{I}{\hbar\omega}$ is the flux of photons entering the material. The factor $a_l$ is the spatial overlap between the wave functions of the Weyl band and the excited band. We assume that the difference between $E_{-}(q)$ and the Fermi energy is much greater than the temperature. We neglect the momentum transfer from light to the quasiparticle due to the small incident angle $\Theta$.

After performing the integrals, Eq.\eqref{jdef} will be of the form 
\begin{align}
J^i(\omega,\vec{k}_p, \vec{\varepsilon})  
&= \frac{-eI}{\hbar \omega}\,  \frac{\sum_{l} \tau_l a_l \chi_l N^i_{(l)j} L^j}{\sum_{l} a_l D^{ij}_{(l)} \varepsilon_i \varepsilon^*_j},
\end{align}
where $N^i_{(l)j}$ and $D^{ij}_{(l)}$ are tensors that depend on the dispersion relations of the cones, and are independent of the frequency of light as long as the Weyl bands are linear; $\chi_l = \pm 1$ is the chirality of each Weyl cone (the $+$ and $-$ signs correspond to right- and left-handed cones, respectively); $\vec{L} = \iu\vec{\varepsilon}\times\vec{\varepsilon}^*$ is the angular momentum per photon. For circularly polarized light, $\vec{L}=\pm\hbar\hat{k}_p$. Explicit expressions for the tensors $N^i_{(l)j}$ and $D^{ij}_{(l)}$ are given in the next section

\begin{figure}[ht]
    \centering
	\includegraphics[width=12cm]{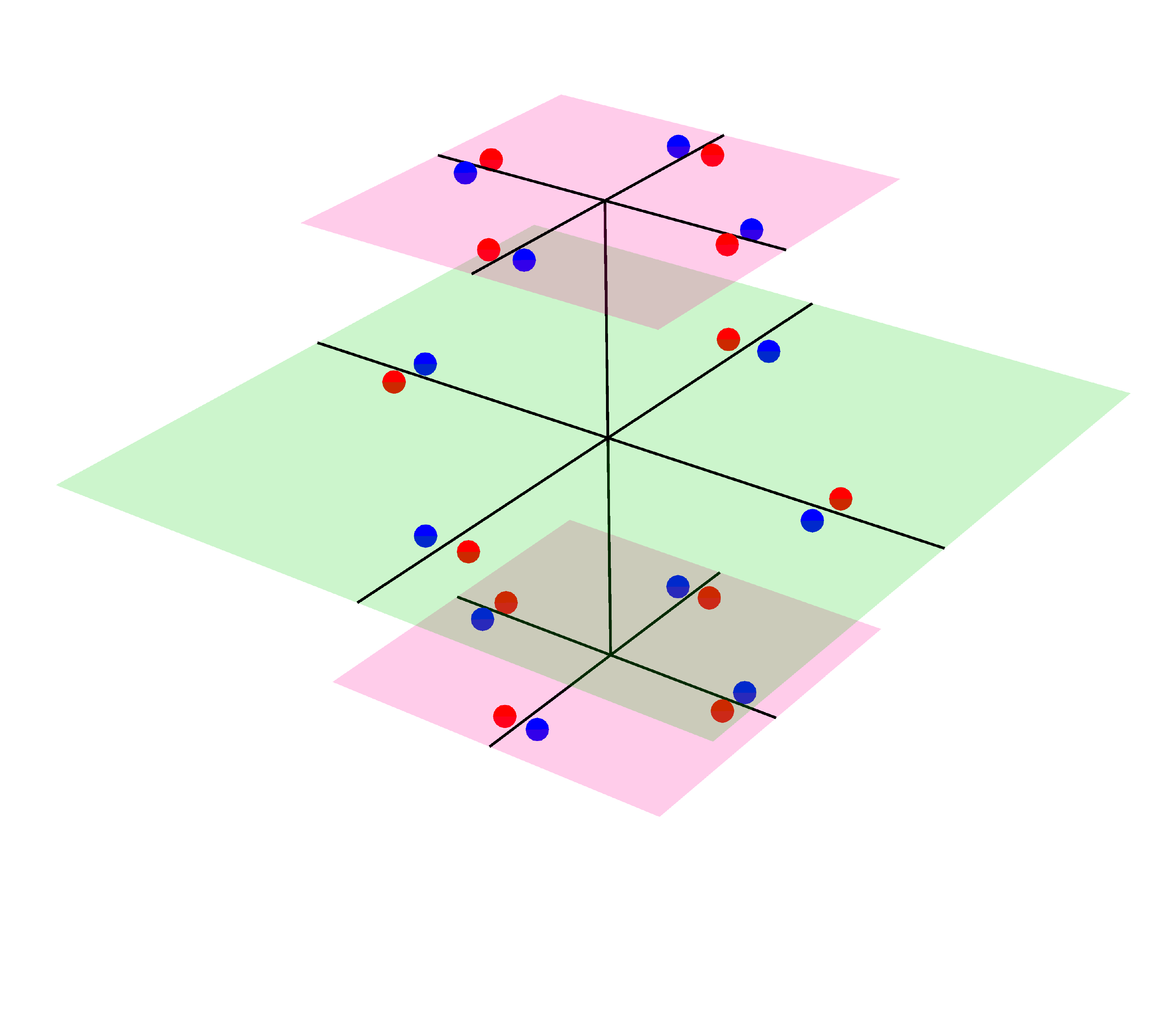}
	\caption{\label{sfig:NvsOmega1} Illustration of the 24 Weyl cones of TaAs in momentum space. The colors represent chirality.}
\end{figure}

TaAs has tetragonal symmetry, i.e. 4-fold rotational symmetry about an axis and reflection symmetry about 4 planes containing that axis. It also has time reversal symmetry. This means the 24 Weyl points exist as a set of 8 ($W_1$) and a set of 16 ($W_2$), with the cones in each set related by the crystal symmetries. Chirality is invariant under rotations and time reversal, and flips sign under reflections. This means each set has an equal number of left and right handed cones. See Fig~\ref{sfig:NvsOmega1} for a sketch of the 24 cones. If we take the sum over a set of cones, the symmetric components of $N^i_{(l)j}$ cancel due to the tetragonal symmetry of the crystal and the only possible non-canceling contribution is from $N^x_{(l)y} - N^y_{(l)x}$. Therefore, the chiral photocurrent in Eq.\eqref{jdef} is $\vec{J}\propto \pm\hat{k}_p \times \hat{c}$, where $+$ and $-$ signs refer to the right- and left-handed polarizations of light. $N^x_{(l)y} - N^y_{(l)x}$ is non-zero only if the tilt Hamiltonian $\hat{H}_t$ is non-zero, the untilted part of the Hamiltonian $\hat{H}_W$ is anisotropic, and the tilt is not aligned with principal axes of the untilted part (the vector $v^i_t$ is not along any of the principal axes of the tensor $v^i_a v^j_a$).

Based on the assumption that once the photons enter the material they will induce excitations with unit probability, for a single Weyl cone, we can write
\begin{equation}\label{trace}
\Tr\sum_i\langle{s_i|\hat{H}_{\text EM}|q_-}\rangle^2 = \mathrm{Tr}(\hat{H}_{\text EM}|q_-\rangle\langle q_-|\hat{H}_{\text EM}^*).
\end{equation}

If we define the dispersion relation
\begin{equation}\label{disper}
\sum a^{\mu \nu}\, p_\mu p_\nu =0
\end{equation}
and compare it to Eq.(2), we find that 
\begin{align}
a^{00} &= 1 \\
a^{0i} &= - 2 v^i_t \\
a^{ij} &= v^i_t\,v^j_t - v^i_a\, v^j_a
\end{align}  

To calculate \eqref{trace}, we first find
\begin{equation}
|q_-\rangle\langle q_-| = \frac{\sigma_0 \sqrt{-\Det(\hat{H}_\text{W})}-\hat{H}_\text{W}}{2\sqrt{-\Det(\hat{H}_\text{W})}}
\end{equation} 

\begin{equation}
-\Det(\hat{H}_\text{W}) = (v^i_t\,v^j_t - a^{ij}) q_i q_j
\end{equation} 

\begin{align}
\Tr(\hat{H}_{\text EM}|q_-\rangle\langle q_-\hat{H}_{\text EM}^*) = &  (v^i_t\,v^j_t - a^{ij} ) \varepsilon_i \varepsilon_j^* + v^i_t\,v^j_t\varepsilon_i \varepsilon_j^*\\ - & \frac{\Tr(\varepsilon_i\, v^i_a\, \sigma_a\, v^j_b\, q_j\, \sigma_b\, \varepsilon_k^*\, v^k_c\, \sigma_c)}{2\sqrt{(v^i_t\,v^j_t - a^{ij}) q_i q_j}}\nonumber\\
- & 2 \frac{ (v^i_t\,v^k_t - a^{ik}) \varepsilon_i q_k v^j_t \varepsilon_j }{\sqrt{(v^i_t\,v^j_t - a^{ij}) q_i q_j}}\nonumber
\end{align} 
where $\varepsilon_{i}$ is the polarization of the photon.

\begin{align}
\Tr(\varepsilon_i\, v^i_a\, \sigma_a\, v^j_b\, q_j\, \sigma_b\, \varepsilon_k^*\, v^k_c\, \sigma_c) &= 2 \iu \epsilon_{abc}\, \varepsilon_i\, v^i_a\, v^j_b\, q_j\, \varepsilon_k^*\, v^k_c\\
& = 2 \iu \Det\left[(\varepsilon q \varepsilon^*) (v^i_a)\right]\\
&= 2 \vec{q}.\vec{L} \Det(v^i_a).
\end{align}

Now, since $v^i_a v^i_b = v^i_t\,v^j_t - a^{ij}$, $\Det(v^i_a) = \chi\sqrt{\Det(v^i_t\,v^j_t - a^{ij})}$, where $\chi$ is the chirality of the Weyl fermion. \eqref{trace} becomes
\begin{align}\label{tracesimp}
\sum_i\langle{s_i|\hat{H}_{\text EM}|q}\rangle^2 = (2 v^i_t\,v^j_t - a^{ij}) \varepsilon_i \varepsilon_j^* + & \frac{\chi\vec{q}.\vec{L}\sqrt{\Det(v^i_t\,v^j_t - a^{ij})}}{\sqrt{(v^i_t\,v^j_t - a^{ij}) q_i q_j}}\nonumber\\
- & 2 \frac{ (v^i_t\,v^k_t - a^{ik}) \varepsilon_i q_k v^j_t \varepsilon_j }{\sqrt{(v^i_t\,v^j_t - a^{ij}) q_i q_j}}
\end{align}

TaAs possesses invariance with respect to time reversal $t$. Time reversal takes momentum $\vec{k}$ to $-\vec{k}$ and $\sigma_a$ to $-\sigma_a$, so the chirality is preserved under $t$-reversal. The velocity $\vec{v}$ that enters equation (4) is odd under time reversal. In \eqref{tracesimp}, the second term is parity odd and the other two terms are parity even. When \eqref{tracesimp} is integrated and summed over the Weyl cones, only the first and third term survive. Eliminating the terms that cancel, Eq.(4) becomes
\begin{align}\label{integ}
\vec{J} = \int \vec{j}\, \du z = \\ \frac{eI}{\hbar \omega}\, \frac{\sum \tau a \int \du^3q\, \delta(E_-(q)-E_0)\, \vec{v}_-(q)\, \frac{\chi\vec{q}.\vec{L}\sqrt{\Det((E_\text{W}^2)^{ij})}}{\sqrt{(E_\text{W}^2)^{ij} q_i q_j}}}{\sum a \int \du^3q\, \delta(E_-(q)-E_0)\, \left(\left(v^i_t v^j_t + (E^2_\text{W})^{ij}\right) \varepsilon_i \varepsilon_j^*
	- 2 \frac{ (E^2_\text{W})^{ik} \varepsilon_i q_k v^j_t \varepsilon_j }{\sqrt{(E_\text{W}^2)^{ij} q_i q_j}}\right)}\nonumber
\end{align}
where $(E_\text{W}^2)^{ij} = v^i_t v^j_t - a^{ij} = v^i_a v^j_a $. The sum is over the Weyl cones.

When we integrate the numerator, we get the tensor:
\begin{equation}
N_{(l) j}^i = E_{l0}^2 \left[\frac{\frac{\alpha}{1-\alpha^2} -\tanh^{-1} \alpha}{\alpha^3} \delta_j^i  + \frac{3 \tanh^{-1} \alpha - \frac{3\alpha-2\alpha^3}{1-\alpha^2}}{\alpha^5}(E_W^2)^{-1}_{jk} v^k_t v^i_t \right]
\end{equation}
and integrating the denominator, we get the tensor
\begin{equation}
D^{ij}_{(l)} = E_{l0}^2 \frac{2}{1-\alpha^2} \frac{(E_W^2)^{ij} - v^i_t v^j_t}{\sqrt{\det (E_W^2)^{ij}}}  
\end{equation}
where $E_{l0}$ is measured from the Weyl point (not the Fermi surface), and the parameter $\alpha$, which quantifies the ``tiltedness" of the cone, is defined as:
\begin{equation}
\alpha^2 = (E_W^2)^{-1}_{ij} v^i_t v^j_t
\end{equation}

Substituting these tensors into \ref{integ}, we get

\begin{equation}
J_i(\omega,\vec{k}_p, \vec{\varepsilon})  
= \frac{-eI}{\hbar \omega}\,  \frac{\sum_{l} \tau_l a_l \chi_l N^i_{(l)j} L^j}{\sum_{l} a_l D^{ij}_{(l)} \varepsilon_i \varepsilon^*_j},
\end{equation}

Because of the tetragonal symmetry of TaAs, the only surviving component of the tensor $\sum_{l} \tau_l a_l \chi_l N^i_{(l)j}$ would come from the antisymmetric $x-y$ component $N^x_{(l)y} - N^y_{(l)x}$. The diagonal components cancel for left and right handed cones, and all other off-diagonal components cancel due to the 4-fold rotational symmetry. 

\begin{figure}[ht]
    \centering
	\includegraphics[width=12cm]{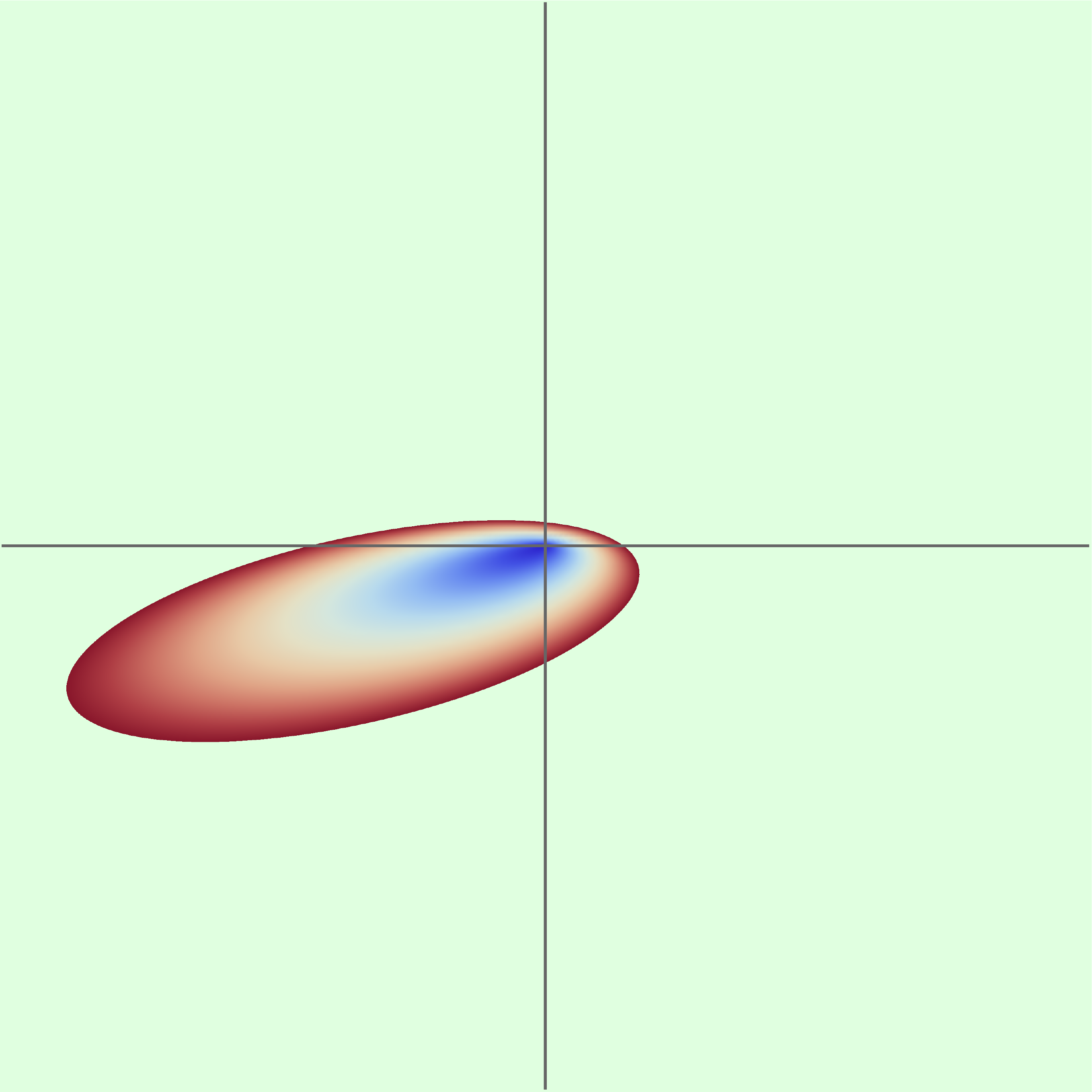}
	\caption{\label{sfig:NvsOmega} Sketch of a tilted and anisotropic Weyl cone. The shading represents energy. The axes are the principal axes of the untilted Hamiltonian; the origin is the Weyl point. The tilt parameter is $\alpha = 5/6$.}
\end{figure}

The only contribution to the antisymmetric part of $N^i_{(l)j}$ is from the second term $(E_W^2)^{-1}_{jk} v^k_t v^i_t$; it is non-zero only if $v^i_t$ is not along any of the principal axes of $(E_W^2)^{ij} = v^i_a v^j_a$, i.e. the tilt of the Hamiltonian is not along any principal axis of the untilted part. Such a cone is depicted in Fig~\ref{sfig:NvsOmega}.

The chiral photocurrent is perpendicular to both the [001] crystallographic axis and the momentum of light $\vec{k}_p$, and it reverses sign for different circular polarization or opposite direction of $\hat{c}$-axis. The chiral nature of Weyl cones cannot contribute to the LPGE and hence produce a current along the [001] axis. We consider the light incident approximately normal to the (112) face of the crystal. In this case, the chiral photocurrent is along the $[\bar{1}10]$ direction.

For excitation light with a wavelength of 800 nm, the numerical evaluation of Eq.\eqref{jdef} yields a value of $J_1 = + 940$ nA/m for the contribution of the 8 Weyl cones $W_1$ to the sheet current density. For the classification of the Weyl cones with different chiralities in TaAs, we follow the supplementary materials of Ref.~\cite{MaTaAs} and use an optical penetration depth of 25 nm. For the 16 Weyl cones $W_2$, the sheet current density $J_2 = - 1340$ nA/m. Due to the lack of detailed information about the probabilities of excitation for the two sets of cones $W_1$ and $W_2$, we are only able to reliably obtain the range of the sheet current density from $-1340$ nA/m to $940$ nA/m. If we assume that 8 Weyl cones $W_1$ and 16 Weyl cones $W_2$ are excited with equal probabilities, we obtain -580 nA/m (for right circular polarization) or +580 nA/m (for left circular polarization). Experimentally, the peak value of the helicity-dependent $J_x(t)$ reaches almost 350 A/m. Considering the photocurrent flows over a time $<\tau_l>\simeq$400 fs at a repetition rate of $f_{rep}$=1 KHz, we evaluated an equivalent DC sheet current density via $\overline{J_x}\simeq$max($J_x$)$f_{rep}<\tau_l>$, which gives $\overline{J_x}\sim$140 nA/m. This value is consistent with the our theoretical result, although there exists some discrepancy between their numbers, which could partly be due to i) our rough estimation of the experimental $\overline{J_x}$ and ii) our assumption that the Weyl bands absorb all of the photons and contribute solely to the helicity-dependent photocurrents. In fact, observation of the shift photocurrents indicates that this approximation is not very accurate. Extraordinarily, we found that the sheet current density is inversely proportional to the frequency $\omega$ of the incident light. This dependence holds for frequencies above the threshold ($\sim$360 THz) for excitation of the high lying non-Weyl bands and below the energy ($\sim$400 THz) at which non-linearity of the Weyl bands sets in, as indicated by the Regions I and II in Fig.~\ref{fig:mechanisms}\textbf{a}.

\section{Comparison of CPGE and LPGE}
\begin{figure}
    \centering
	\includegraphics[width=12cm]{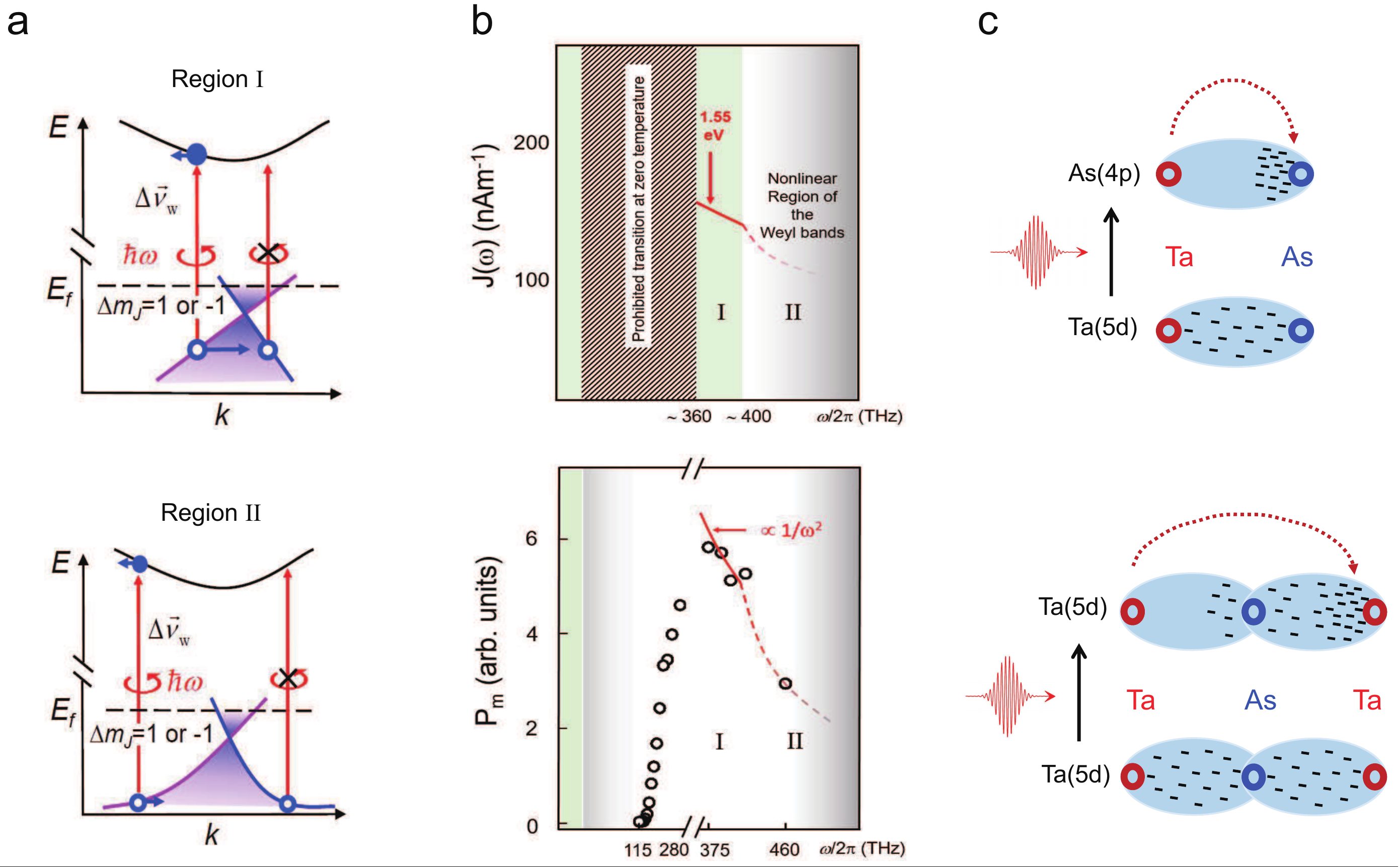}
	\caption{\label{fig:mechanisms} \textbf{a}. Schematic of CPGE. The asymmetric population of high-lying bands and depopulation of the Weyl cones leading to an average band velocity of $\Delta\vec{v}_w$ results in a nonzero charge current using circularly polarized light. During the optical transitions, the total angular momentum should be conserved, and its quantum number $m_J$ satisfies the relation: $\Delta m_J$=1 or -1, depending on the helicity of the pump pulse. The top is for the Weyl bands in the linear region (Region I). The bottom is for the Weyl bands in the nonlinear region (Region II), where the pump light has much higher photon energies. \textbf{b}. Theoretical and experimental results for the sheet current density and THz emission. The top displays our calculated sheet current density as a function of pump-light frequency (the red solid line). The calculated photocurrent is proportional to 1/$\omega$ (Region I). The bottom represents the measured THz peak power as a function of pump-light frequency (open dots). The red solid line in Region I is a fit using $1/\omega^2$ derived from our theory. The red dashed lines in Region II inside both figures are only for guideline. \textbf{c}. Schematic of the shift photocurrents (LPGE) is shown on the bottom. Due to the initial states being dominated by Ta 5$d$ orbitals and the final states contributed by both Ta 5$d$ and As 4$p$ orbitals the shift current is attributed to the ultrafast transfer of electron density along both the Ta-(As)-Ta and Ta-As bonds.}
\end{figure}

According to the above description, the peak power of THz radiation for incident light with the photon energy $\hbar\omega$, defined by $P_m\propto|E_{max}|^2$, is also proportional to the square of DC current density  $\vec{J}(\omega,\vec{k}_p, \vec{\varepsilon})$ and inversely proportional to the duration of the current:
\begin{equation}
P_m\propto \frac{1}{\tau^2} \lvert J \rvert^2\propto \frac{1}{\omega^2}.
\end{equation} 
Because the sheet current density is inversely proportional to the frequency $\omega$ of the incident light, the radiation power is proportional to $1/\omega^2$ within the frequency range described above. As shown in the Region I of Fig.~\ref{fig:mechanisms}\textbf{b}, our experimental data consistent well with the calculations. We note, however, that the CPGE-induced THz signal or photocurrent in experiments does not suddenly drop to zero in the frequency range, where our theory predicts that the optical transitions are prohibited at zero temperature (the shaded region), and instead is manifested by gradual decrease. Such behavior may arise from the band tailing effect due to the non-zero temperature and defects in the sample \cite{Stern_PR_1966,Mieghem_RMP_1992}.     

Therefore, we demonstrate both experimentally and theoretically that for excitation light with high photon energies Weyl physics is the key determinant for the observed giant helicity-dependent photocurrents and the resultant coherent THz emission. This is because the excitation from the Weyl band to the high-lying band is accompanied by a large and rapid change in the effective velocity of the charged quasiparticles. The microscopic mechanism is quite different from previous findings in the WSMs upon light excitations with small photon energies \cite{lee2017,Zhang_PRB_2018,MaTaAs}, where the Weyl band tilting plays the pivotal role.

One would argue that helicity-dependent photocurrents may arise from other mechanisms, such as the circular photon drag effect (CPDE) \cite{Shalygin_PRB_2016} or the spin-galvanic effect (SGE) \cite{Ganichev_JPhys_2003}. The CPDE current requires additional transfer of the light momentum along the charge current direction. This effect will be irrelevant for the case here with a small incidence angle. For the SGE current, its decay is determined by the spin relaxation time, $\tau_s$ \cite{Ganichev_JPhys_2003}. Based on our time-resolved Kerr rotation measurement, the observed $J_x(t)$ cannot be explained by SGE since $\tau_s\simeq$60 fs is too small 

\textbf{Photo-thermal currents.} $J_{yz}$ is largely independent of any pump polarization. Therefore, discussing its mechanism can basically exclude the CPGE and LPGE since they only show a little contribution to the signal. Based on Eq.~(\ref{Eq:angle1}), the dominance of $D_{yz}$ in the $S_{yz}$ (or $E_{yz}$) signal indicates $J_{yz}$ has a thermal origin. Its relatively slow response in the time domain further suggests that scattering processes following the initial optical transitions are involved during the generation of $J_{yz}$. Possible candidates are the photo-Dember effect \cite{Dekorsy_PRB_1996} and carrier drift, both of which depend strongly on the phonon or impurity scatterings. An estimation of $J_y(t)$ and $J_z(t)$ can be obtained using the data by varying the incident angle $\Theta$. Their magnitudes are close to that of the shift current.    

Therefore, the distinct polarization dependent $J_x(t)$ and $J_{yz}(t)$ arise from different physical mechanisms, non-thermal and thermal, respectively. A phase difference between these two components is expected. As a result, the chiral ultrafast photocurrent $\vec{J}(t)$ emerges, as evidenced by our experiment. Such phase difference directly determines the observed $\Delta\varphi$ between $E_x(t)$ and $E_{yz}(t)$. An estimation of $\Delta\varphi$ can be roughly made by $\Delta\varphi\simeq\bar{\Omega}<\tau_{ep}>$, where $\bar{\Omega}$ and $\tau_{ep}$ are the angular THz frequency and the electron-phonon scattering time, respectively. Using the frequency ($\sim$1.8 THz) at the peak magnitude of the Fourier transform and the average $<\tau_{ep}>$ value of $\sim$400 fs, we obtain $\Delta\varphi\simeq1.4\pi$, which is close to the values measured values, i.e. $4\pi/3$ for (112) and $3\pi/2$ for (011). Potential anisotropy of the detail electron-phonon scatterings corresponding to different faces may cause such various phases. 

The mechanism for both currents and the frequency dependence are shown in Fig~\ref{fig:mechanisms}

\section{Conclusions} 
The demonstrated ultrafast generation and control of chiral charge currents in TaAs offer unique opportunities for novel THz emission and THz spintronics. The mechanism of generating controllable elliptically/circularly polarized broadband THz pulses using WSMs is fundamentally different from previous methods. The intrinsic optical chiralities, i.e. the optical/chirality selection rules in WSM, which do not apply for previous THz emitters, such as ZnTe and LiNbO$_3$. The simplicity of polarization control is extremely powerful and useful for a wide range of applications. Other advantages include the low cost in sample preparation and the high THz emission efficiency. We further believe that our observation will benefit the study of other novel phenomena led by the Weyl physics, such as the quantized CPGE \cite{dejuan17}, and the Weyl-orbit effect\cite{Zhang_NC_2017}.
\chapter{Transverse Helical Magnetic Effect}\label{chTHME}

\blfootnote{This chapter is based on \cite{SahalTHME}.}

Photocurrents induced by magnetic fields have been theoretically studied and experimentally observed in quantum well systems ~\cite{2005belkov} 
 The helical magnetic effect has been proposed~\cite{yuta2018} to cause colossal photocurrents parallel to an applied magnetic field in asymmetric Weyl semimetals. This effect is due to a combination of Pauli blockade and the effects of Berry curvature and magnetic field. Strong magnetic fields have been predicted to induce a magnetogyrotropic photogalvanic current in Weyl semimetals with tilted cones, due to the quantization of Landau levels~\cite{2018golub}. In mirror-symmetric Dirac and Weyl semimetals, in the presence of a magnetic field, a photocurrent has also been predicted due to chiral anomaly~\cite{kaushik2019chiral}. Circularly polarized light has also been predicted to cause a topologically protected photocurrent through the quantized circlular photogalvanic effect in asymmetric Weyl semimetals~\cite{dejuan17}, and a photocurrent parallel to the direction of incident light due to an induced effective magnetic field~\cite{Taguchi2016}.

In all these previously discussed cases, one relies either on circularly polarized light or on an asymmetric crystal structure to induce the photocurrent. In this chapter, we propose a new class of photocurrents that appear when the material has a broken inversion symmetry ($x,y,z \to -x, -y, -z$) but unbroken time-reversal symmetry ${\rm T}$, and a plane of reflection ($x,y,z \to -x,y,z$) in the presence of a background magnetic field.

\section{Symmetry and Photocurrents}
Broken inversion symmetry allows a magnetic photocurrent current of the following structure:
\begin{equation}
j^i = (\sigma_B)^i_j \ B^j,
\end{equation}
where $B^j$ is the magnetic field. 
Since under inversion transformation $B^j \to B^j$ and $j^i \to - j^i$, the current is allowed by parity only if the material lacks the inversion symmetry under which $(\sigma_B)^i_j \to -(\sigma_B)^i_j$. 

In a material with a  plane of reflection symmetry, the components of current and magnetic field parallel ($\parallel$) and perpendicular ($\perp$) to the plane transform under reflection as
\begin{align*}
   \vec{j}_\perp &\to  -\vec{j}_\perp\,, \\
   \vec{j}_\parallel &\to  \phantom{-}\vec{j}_\parallel\,, \\
   \vec{B}_\perp &\to  \phantom{-}\vec{B}_\perp\,, \\
   \vec{B}_\parallel &\to  -\vec{B}_\parallel\,.
\end{align*}
Therefore, in a material with broken inversion symmetry but reflection symmetry about at least one plane, $(\sigma_B)^i_j$ cannot have a component proportional to $\delta^i_j$ but it can still have transverse components of the form $(\sigma_B)^\perp_\parallel$ and $(\sigma_B)_\perp^\parallel$.  If we  have neither inversion nor reflection symmetry, $(\sigma_B)^i_j$ is allowed to have a component proportional to $\delta^i_j$; this component corresponds to the helical magnetic effect~\cite{yuta2018}. For a related symmetry analysis, see \citet{silva2020magneticconductivity}.

Transition metal monopnictides such as TaAs are examples of mirror-symmetric Weyl semimetals; even though they have broken inversion symmetry, they have reflection symmetry about multiple planes. They have tetragonal symmetry with the space group $I 4_1 md$ (No. 109): in particular, they have fourfold rotation symmetry about the $c$ axis, reflection symmetries about the $ac$ and $bc$ planes, but no reflection symmetry about the $ab$ plane.

The photocurrents in TaAs in response to circularly polarized light discussed and observed so far~\cite{MaTaAs, lee2017, gao2020chiral} are necessarily of the form $\vec{j} \sim \hat{c}\times\vec{J}$, where $\vec{J}$ is the angular momentum of the incident light. This is because of the combination of fourfold rotation and reflection symmetries, as we will now explain.

We can write the photocurrent as $j^i = (\sigma_P)^i_k J^k$. Since angular momentum and magnetic field have the same transformations under reflection,  reflection symmetry imposes the same constraints on the tensors $(\sigma_P)^i_j$ and $(\sigma_B)^i_j$. Since there is reflection symmetry about the $ac$ plane, the only non-vanishing components are $(\sigma_P)^a_b$, $(\sigma_P)^b_a$, $(\sigma_P)^c_b$, and $(\sigma_P)^b_c$. Similarly, due to reflection symmetry about the $bc$ plane, the components $(\sigma_P)^c_b$, and $(\sigma_P)^b_c$ also vanish. Due to the fourfold rotation symmetry about the $c$ axis, there is only one independent component $(\sigma_P)^b_a = - (\sigma_P)^a_b \equiv \sigma_P$, and the photocurrent is therefore 
\begin{equation}
    \vec{j} = \sigma_P\; \hat{c}\times\vec{J}.
\end{equation}

Since magnetic field $\vec{B}$ is even under parity and odd under time reversal, just like  angular momentum $\vec{J}$, we can use similar symmetry analyses for photocurrents in the presence of magnetic field and photocurrents induced by circularly polarized light. Such an analysis for the photocurrent due to unpolarized light in the presence of a magnetic field yields that the current should necessarily be of the form 
\begin{equation}\label{eq:current1}
    \vec{j} = \sigma_B\; \hat{c}\times\vec{B}
\end{equation}
Unlike the helical magnetic photocurrent~\cite{yuta2018} and chiral magnetic photocurrent~\cite{kaushik2019chiral} , this photocurrent is perpendicular to the magnetic field.

\begin{figure}[htp]
\centering
  \includegraphics[scale=1.25]{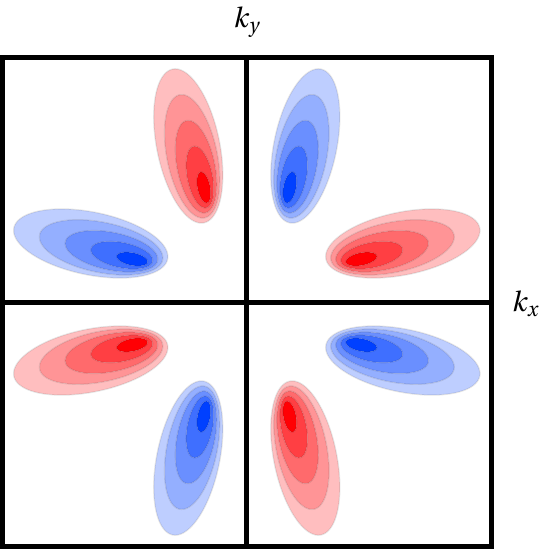}
  \caption{Weyl cones which possess both anisotropy and tilt. These Weyl cones are similar to the $W_1$ cones of the prototypical type-I Weyl semimetal used for numerical calculations in this chapter. The shading represents energy and the color represents chirality.}
  \label{fig:TiltAniso}
\end{figure}

The symmetry analysis above shows that this photocurrent is not forbidden. We will now demonstrate how the anisotropy and tilt of the Weyl cones (shown in Figure~\ref{fig:TiltAniso}), combined with the chirality of quasi-particles, enable this photocurrent. Recall that the resulting current in Eq.~\eqref{eq:current1} is transverse to the direction of the applied magnetic field. This is why we call it the \emph{transverse chiral magnetic photocurrent}. The transverse chiral magnetic photocurrent is due to the interplay between the modification of velocity of chiral quasi-particles by Berry curvature and the tilt and anisotropy of the Weyl cones that prevent the cancellation of the resulting current due to the symmetries of the Brillouin zone. 

\begin{figure}[htp]
     \centering
     
         \includegraphics[scale=1.25]{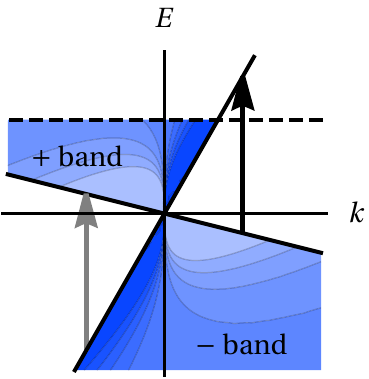}\hspace{0.1\linewidth}\includegraphics[scale=1.25]{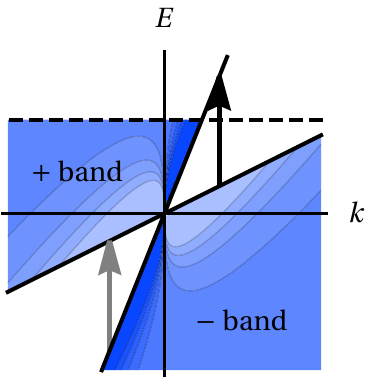}
    
        \caption{Transitions at zero temperature for type-I (left) and type-II (right) Weyl semimetals. The shading represents the effect of the Berry curvature on the density of states. The black arrow represents transitions that are allowed, and the grey arrow represents transitions forbidden by Pauli blockade at zero temperature.}
        \label{fig:WeylCones}
\end{figure}

We will now calculate the magnitude of the transverse chiral magnetic photocurrent for a prototypical Weyl semimetal with tetragonal symmetry, similar to transition metal monopnictides that include TaAs.
We assume that the crystalline structure has a plane of reflection, combination of rotation and reflection, or a combination of translation and reflection as a symmetry. The Hamiltonian of a linear Weyl cone has a general structure  
\begin{equation}\label{eq:Ham}
H = v_T^i q_i + v_a^i \sigma_a q_i,
\end{equation}
where $q_i$ is the crystal momentum of the quasi-particle and $\sigma_a$ are the Pauli matrices. For each cone, the vector $v_T^i$ characterizes its tilt. We can define the tensor $(v_W^2)^{ij} = v_a^i v_a^j$ and its inverse  $(v_W^{-2})_{ij}$ which characterize the anisotropy of the second term of the Hamiltonian in Eq.~\eqref{eq:Ham}. The chirality of the Weyl cone is defined as $\chi = \mathrm{sgn}\left( \mathrm{det}\left(v_a^i\right)\right)$. We can also define a dimensionless tilt parameter $\alpha = \sqrt{(v_W^{-2})_{ij}v_T^i v_T^j}$. In a type-I Weyl semimetal $\alpha < 1$ and in a type-II Weyl semimetal $\alpha>1$. Type-I Weyl semimetals have point-like (or in general compact) Fermi surfaces and type-II Weyl semimetals have open Fermi surfaces~\cite{soluyanov2015}. The Weyl cones for type-I and type-II Weyl materials are shown in Figure~\ref{fig:WeylCones}.

\section{Berry Curvature and Phase Space}

Weyl points are monopoles of Berry curvature and their chirality is the sign of the Berry monopole charge. Because the total Berry flux through the Brillioun zone is zero, there is always an equal number of left-handed and right-handed Weyl points (Nielsen-Ninomiya theorem~\cite{1981Nielsen}). The Berry curvature depends on the eigenstates, not on the corresponding energy eigenvalues of the Hamiltonian, and is therefore independent of the tilt velocity and depends only on the untilted part of the Hamiltonian. For our Hamiltonian, the Berry curvature is
\begin{equation}
\Omega_i^\pm = \pm\chi \frac{1}{2}\frac{q_i}{(q_j (v_W^2)^{jk}q_k)^{3/2}}\sqrt{\det(v_W^2)},
\end{equation}
where the $+$ and $-$ are for the upper and lower Weyl bands (shown in Figure~\ref{fig:WeylCones}).

In the semiclassical limit, i.e. when the cyclotron frequency $\omega_c = eBv/k$ is much smaller than the inverse of the scattering time $1/\tau$ and the temperature $T$, we can ignore the Landau quantization~\cite{2015gustavo}, and model the combined effects of a static magnetic field and momentum-space Berry curvature as a modification of the phase space ~\cite{xiao2005berry,stephanov2012chiral,son2013kinetic}
\begin{equation}
\frac{d^3 q_i}{(2\pi)^3} \to (1+e\Omega_j B^j)\frac{d^3 q_i}{(2\pi)^3}
\end{equation}
and velocity is modified as
\begin{equation}
v^i \to \frac{1}{1+e\Omega_j B^j} [v^i + e(v^j\Omega_j) B^i].
\end{equation}
Since the rate of transitions $\Gamma$ depends on the number of initial states and the number of final states, the effect of phase space cancels out:
\begin{equation}
\Gamma \to \Gamma (1+e\Omega_j^+ B^j)(1+e\Omega_j^- B^j) \approx \Gamma ,
\end{equation}
and we only have to consider the modifications to velocity:
\begin{multline}\label{eq:velocity}
v^i_+ - v^i_-  \to (v^i_+ - v^i_-)  -(v^i_+ + v^i_-)(e\Omega_j^+ B^j)\\ + e(v^j_+ + v^j_-)\Omega_j^+ B^i .
\end{multline}
There is no photocurrent in the absence of $B^i$. By reflection symmetry, there can also be no term along $B^i$, so we will focus on the second term in Eq.~\eqref{eq:velocity}. Since $v^i_+ + v^i_- = 2v_T^i$, the only relevant term in Eq.~\eqref{eq:velocity} is given by
\begin{equation}\label{eq:simpl_v}
v^i_+ - v^i_-  \sim   -2v_T^i(e\Omega_j^+ B^j).
\end{equation}

Making the assumption that one absorbed photon excites exactly one electron, the sheet current density, that is the current density integrated over the penetration depth $\int \vec{j} dz$, can be obtained by convoluting the change of velocity given by  Eq.~\eqref{eq:velocity} with the rate of transitions induced by photons:
\begin{equation}\label{eq:current}
j^i = \frac{I_{tr}}{\hbar \omega} e  \frac{\sum_{cones}\tau \int \frac{d^3 q_j}{(2\pi)^3} \Gamma_W\Delta f (-2v_T^ie\Omega_k^+ B^k)\delta (\Delta E - \hbar \omega)} {\Gamma_{abs}(\omega) + \sum_{cones} \int \frac{d^3 q_j}{(2\pi)^3} \Gamma_W\Delta f \delta (\Delta E - \hbar \omega)} ,
\end{equation}

where $\tau$ is the relaxation time, $\Delta f = f(E_-) - f(E_+)$ where $f(E)$ is the Fermi distribution function, $I_{tr}$ is the intensity of the transmitted light. $\Gamma_{abs}(\omega)$ is the rate of absorption from mechanisms not involving chiral fermions, $\Delta E = E_+ - E_-$ and $\Gamma_W$ is the rate of transitions for Weyl fermions, which is given by Fermi's golden rule $\Gamma_W \sim |\langle \psi_- | V | \psi_+ \rangle|^2$. The interaction term $V$ can be obtained by Peierls substitution, by replacing $q_i$ by $q_i + eA_i$ in the Hamiltonian as 
\begin{equation}
V  = H(q_i+eA_i) - H(q_i) \sim v_T^i \epsilon_i + v_a^i \sigma_a \epsilon_i,
\end{equation}
where $\epsilon_i$ is the polarization vector. For linearly polarized light, the transition rate $\Gamma$ is of the form
\begin{equation}
\Gamma_W \sim \epsilon_i (v_W^2)^{ij} \epsilon_j - \frac{(\epsilon_i (v_W^2)^{ij} q_j)^2}{q_i (v_W^2)^{kl} q_k}.
\end{equation}

In the numerical calculations, we consider absorption only due to excitation of chiral fermions. Under this assumption, we can ignore the constant factors in $\Gamma_W$ because it appears in both the numerator and the denominator of Eq.~\eqref{eq:current}.

In order to simplify the numerical calculations, we utilised the symmetries of the system to decompose the numerator of Eq.~\eqref{eq:current} into component tensors, with corresponding numerical coefficients $N_1, N_2, N_3$. The contribution of one cone to the photocurrent is of the form
\begin{multline}
\chi\,v_T^i(N_1 [B^j (v_W^{-2})_{jk} v_T^k] [\epsilon_l (v_W^2)^{lm} \epsilon_m] + \\
N_2 [B^i\epsilon_i] [v_T^j\epsilon_j] +
N_3 [B^j (v_W^{-2})_{jk} v_T^k] [v_T^l\epsilon_l]^2)
\end{multline}
where the scalar coefficients $N_1, N_2, N_3$ depend on $\mu$, $T$, and $\omega$. In general, these coefficients are obtained by numerical integration. 

In a material with tetragonal symmetry, such as the transition metal monopnictides, for light of arbitrary polarization, the magnetic photocurrent has the form
\begin{align}
    \vec{j} =\quad &  A_1[(\hat{\epsilon}\cdot\hat{a})^2 + (\hat{\epsilon}\cdot\hat{b})^2] (\hat{c}\times\vec{B})\nonumber\\ + &A_2 (\hat{\epsilon}\cdot\hat{c})^2 (\hat{c}\times\vec{B})\nonumber\\ + &A_3 (\vec{B}\cdot \hat{c})(\hat{\epsilon}\cdot\hat{c})(\hat{\epsilon}\times\hat{c}) \nonumber\\ + &A_4 [\vec{B}\cdot(\hat{\epsilon}\times\hat{c})](\hat{\epsilon}\cdot\hat{c})\hat{c} \nonumber\\ + &A_5 (\hat{\epsilon}\cdot\hat{a})(\hat{\epsilon}\cdot\hat{b})[(\vec{B}\cdot\hat{a})\hat{a} - (\vec{B}\cdot\hat{b})\hat{b}] \nonumber\\ + &A_6 [(\hat{\epsilon}\cdot\hat{a})^2 - (\hat{\epsilon}\cdot\hat{b})^2][(\vec{B}\cdot\hat{a})\hat{b} + (\vec{B}\cdot\hat{b})\hat{a}]. 
\end{align}

\section{Sample Calculation}

We numerically calculated this photocurrent for a prototypical material with tetragonal symmetry. We have assumed 8 tilted Weyl cones ($W_1$) related to each other by reflection and rotation symmetries, and 16 untilted cones ($W_2$), also related to each other by reflection and rotation symmetries, similar to the cone distribution in transition metal monopnictide class of materials which includes TaAs. The energies of $W_1$ cones are 10 meV below the Fermi surface, while the energies of $W_2$ cones are exactly at the Fermi surface (so they contribute to absorption but not to the photocurrent). Both sets of cones have the same untilted Hamiltonians. 

For numerical computations, we have chosen a set of parameters representative of the materials in the TaAs class~\cite{MaTaAs}. 
The matrix $(v_W^2)^{ij}$ is taken as 
\begin{equation*}\begin{pmatrix} 16 & 0 & 0\\ 0 & 4 & 0\\ 0 & 0 & 1
\end{pmatrix}\times 10^{10}\; \mathrm{m^2/s^2}. 
\end{equation*}
The tilt velocity for the type-I Weyl semimetal is assumed to be $v_T^i = (1.8, 1.2, 0) \times 10^5\; \mathrm{m/s}$ and for the type-II Weyl semimetal it is assumed to be $v_T^i = (3.6, 2.4, 0) \times 10^5\; \mathrm{m/s}$. These parameters are similar to those in \citet{MaTaAs}.

\begin{figure}[htp]
\centering
  \includegraphics[scale=0.8]{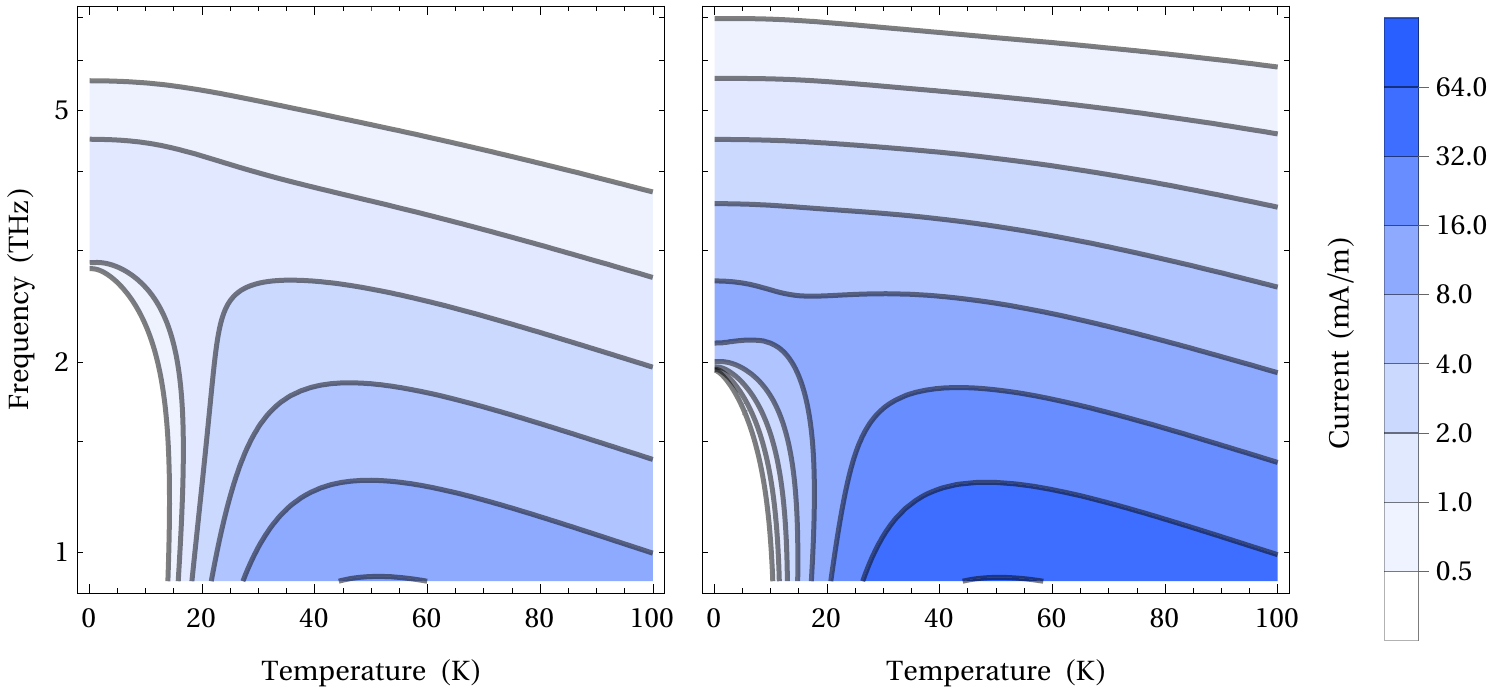}
  \caption{Photocurrent density vs temperature (horizontal axis) and frequency (vertical axis) of the prototypical type-I (left) and type-II (right) Weyl semimetal.}
  \label{fig:ResultPlot}
\end{figure}

We considered light polarized along the $c$ axis, incident on the $ac$ plane of the material, with magnetic field directed along the $b$ axis, so that the observed current would be along the $a$ axis, as predicted by Eq.~\eqref{eq:current1}. The other parameters used in the numerical calculations are given in Table~\ref{table:param} and the results are shown in Figure~\ref{fig:ResultPlot}.
\begin{table}[htb]
\centering
\begin{tabular*}{\linewidth}{p{0.2cm} l @{\extracolsep{\fill}} l}
\hline\hline
& Parameter &Value\\ [0.5ex]
\hline
& $\tau$ (relaxation time) &$4 \times 10^{-11}\; \mathrm{s}$ \\
& $B$ (magnetic field) &$0.5 \; \mathrm{T}$ \\
& $R$ (reflectivity) &0.95 \\
& $I$ (light intensity) &$10^{6}\; \mathrm{W/m^2}$\hspace{0.2cm} \\ 
[1ex]
\hline\hline
\end{tabular*}
\caption{Parameters used for numerical computations}
\label{table:param}
\end{table}

Since the magnitude of the current depends on the tilt $\alpha$, maximal current is predicted for the type-II Weyl semimetals, as shown in Figure~\ref{fig:ResultPlot}. For a beam of frequency $\omega/2\pi = 3\; \mathrm{THz}$ and spot size $0.5\;\mathrm{mm}$ at temperature $T=50\; \mathrm{K}$, the photocurrent is $0.75\; \mathrm{\mu A}$ for the type-I Weyl semimetal and $2.5\; \mathrm{\mu A}$ for the type-II  Weyl semimetal. The photocurrent is obtained by multiplying Eq.~\eqref{eq:current} with the spot size. Note that the resistance of the sample is usually smaller than the load resistance in experimental setups, resulting in a suppression of the detected current.

As seen in Figure~\ref{fig:ResultPlot}, the photocurrent is maximised at around $50\; \mathrm{K}$. The role of temperature is important since transitions cannot happen to completely occupied states; nor can they happen from completely empty states. Temperature allows to avoid the blockade due to Pauli exclusion principle by smearing the occupation fraction. On the other hand, if the temperature is too high, the magnitude of the photocurrent decreases since thermal smearing reduces the difference in occupation fraction between the two states between which transitions can occur.  Note that it is also possible to observe a photocurrent even at zero temperature if the frequency is high enough, since this allows transitions to access the region beyond the sharp Pauli blockade, which can also be seen in Figure~\ref{fig:ResultPlot}.

\section{Summary}
To summarize, we predict a new type of photocurrent to occur in Weyl semimetals with broken inversion symmetry, time reversal symmetry, and a symmetry with respect to a reflection plane. The class of materials that satisfy these conditions includes the monopnictides such as TaAs. In contrast to all previous proposals and observations of photocurrents in Weyl semimetals, the predicted \emph{transverse chiral magnetic photocurrent} can be induced even by a linearly polarized light and does not require a breaking of reflection symmetry of the crystal; the current is transverse to the direction of an applied magnetic field.  The magnitude of the resulting photocurrent is predicted to be  significant in the THz frequency range, about $0.75\; \mathrm{\mu A}$ for type-I and $2.5\; \mathrm{\mu A}$ for type-II Weyl semimetals. Therefore, the transverse chiral magnetic photocurrent, especially in type-II Weyl semimetals, can enable a significant increase in sensitivity of unpolarized THz radiation detectors.

\newcommand{\bx}{{\boldsymbol x}}
\newcommand{\bp}{{\boldsymbol p}}
\newcommand{\bq}{{\boldsymbol q}}
\newcommand{\bv}{{\boldsymbol v}}
\newcommand{\bu}{{\boldsymbol u}}
\newcommand{\bB}{{\boldsymbol B}}
\newcommand{\bJ}{{\boldsymbol J}}
\newcommand{\bSigma}{{\boldsymbol \Sigma}}
\newcommand{\bOmega}{{\boldsymbol \Omega}}
\newcommand{\bomega}{{\boldsymbol \omega}}
\newcommand{\tbv}{\tilde{\boldsymbol v}}
\newcommand{\tepsilon}{\tilde{\epsilon}}
\newcommand{\tF}{\tilde{F}}
\newcommand{\hp}{{\hat p}}
\newcommand{\hbp}{{\hat{\boldsymbol p}}}
\newcommand{\hbq}{{\hat{\boldsymbol q}}}
\newcommand{\hmu}{{\hat{\mu}}}
\newcommand{\hnu}{{\hat{\nu}}}
\newcommand{\hrho}{{\hat{\rho}}}
\newcommand{\hsigma}{{\hat{\sigma}}}
\newcommand{\hlambda}{{\hat{\lambda}}}
\newcommand{\halpha}{{\hat{\alpha}}}
\newcommand{\hbeta}{{\hat{\beta}}}
\newcommand{\hgamma}{{\hat{\gamma}}}
\newcommand{\hdelta}{{\hat{\delta}}}
\newcommand{\hzero}{{\hat{0}}}
\newcommand{\hone}{{\hat{1}}}
\newcommand{\htwo}{{\hat{2}}}
\newcommand{\hthree}{{\hat{3}}}
\newcommand{\lcalG}{\overleftarrow{\mathcal{G}}}
\newcommand{\lcalH}{\overleftarrow{\mathcal{H}}}
\newcommand{\lD}{\overleftarrow{D}}
\newcommand{\calC}{\mathcal{C}}
\newcommand{\calG}{\mathcal{G}}
\newcommand{\calH}{\mathcal{H}}
\newcommand{\calF}{\mathcal{F}}
\newcommand{\calP}{\mathcal{P}}
\newcommand{\calV}{\mathcal{V}}
\newcommand{\calA}{\mathcal{A}}
\newcommand{\calS}{\mathcal{S}}
\newcommand{\calR}{\mathcal{R}}
\newcommand{\calL}{\mathcal{L}}
\newcommand{\calUin}{\mathcal{U}_\text{in}}
\newcommand{\calUrot}{\mathcal{U}_\text{rot}}
\newcommand{\bbP}{\mathbb{P}}
\newcommand{\bbY}{\mathbb{Y}}
\newcommand{\bbT}{\mathbb{T}}
\newcommand{\bbM}{\mathbb{M}}

\newcommand{\intp}{\int_{\boldsymbol p}}
\newcommand{\feq}{f_\text{eq}}
\newcommand{\dis}{\displaystyle}
\renewcommand{\flat}{\text{flat}}
\newcommand{\red}[1]{{\color{red}{#1}}}
\newcommand{\blue}[1]{{\color{blue}{#1}}}

\renewcommand\a{\alpha}
\renewcommand\b{\beta}
\renewcommand\d{\delta}
\renewcommand\k{\kappa}
\renewcommand\l{\lambda}
\renewcommand\t{\tau}
\renewcommand\u{\upsilon}
\renewcommand\c{\chi}
\renewcommand\j{\psi}
\renewcommand\o{\omega}
\newcommand\e{\epsilon}
\newcommand\g{\gamma}
\newcommand\z{\zeta}
\newcommand\m{\mu}
\newcommand\n{\nu}
\newcommand\x{\xi}
\newcommand\p{\pi}
\newcommand\h{\theta}
\newcommand\s{\sigma}
\newcommand\f{\phi}
\newcommand\w{\eta}
\newcommand\ve{\varepsilon}
\newcommand\vh{\vartheta}
\newcommand\vf{\varphi}
\renewcommand\L{\Lambda}
\renewcommand\P{\Pi}
\renewcommand\S{\Sigma}
\renewcommand\O{\Omega}
\renewcommand\H{\Theta}
\newcommand\D{\Delta}
\newcommand\G{\Gamma}
\newcommand\F{\Phi}
\newcommand\J{\Psi}

\newcommand{\mb}[1]{\mathbf{#1}}
\newcommand\ra{\rightarrow}
\newcommand\la{\leftarrow}
\newcommand\pt{\partial}
\newcommand\mc{\mathcal}
\newcommand\ms{\mathscr}
\newcommand\na{\nabla}
\newcommand\ola{\overleftarrow}
\newcommand\lb{\left(}
\newcommand\rb{\right)}
\newcommand\ls{\left[}
\newcommand\rs{\right]}
\newcommand\lc{\left\{}
\newcommand\rc{\right\}}
\newcommand{\lan}{\langle}
\newcommand{\ran}{\rangle}
\newcommand{\com}[1]{{\sf\color[rgb]{0,0,1}[#1]}}
\newcommand{\non}{\nonumber\\}

\chapter{Chiral kinetic theory of anomalous transport induced by torsion
}\label{chStrain}

\blfootnote{This chapter is based on \cite{LanLanStrain}.}

 It is interesting to consider the effects of crystal deformation on anomalous transport, since deformations are known to lead to strong ``synthetic" gauge fields \cite{guinea2010energy,vozmediano2010gauge,
 cortijo2015elastic}; for example, in graphene nanobubbles, the synthetic magnetic field in excess of 300 T has been reported \cite{levy2010strain}. The geometrical torsional response of Weyl fermions has been addressed in \cite{Zhou:2012ix,PhysRevB.99.155152,PhysRevLett.124.117002,PhysRevResearch.2.033269,laurila2020torsional}, 
 including its relation to Neih-Yan anomaly \cite{nieh1982quantized}.
\medskip

 A natural question to ask is whether these synthetic gauge fields can be used to source the chiral anomaly, and thus drive the CME and other anomalous phenomena. At first glance, it may appear that deformations should be irrelevant for the CME -- indeed, the CME is captured \cite{Kharzeev:2009fn} by the topological Chern-Simons term $\sim \mu_5 \int \epsilon_{ijk} A^i F^{jk}$ in the effective action (where $\mu_5 = \mu_R - \mu_L$ is the chiral chemical potential, and $A^i$ and $F^{jk}$ are the gauge potential and field strength tensor, respectively) that does not depend on the space-time metric $g_{\mu\nu}$. Therefore, 
if one describes the deformation as a change in an effective metric, the chiral anomaly and thus the CME should seemingly not be affected. 
\medskip

However, this conclusion is premature since a time-dependent, inhomogeneous deformation can induce a change in the momentum space distribution of chiral quasiparticles, e.g. by deviating the chiral chemical potential $\mu_5$ away from its equilibrium value $\mu_5=0$ \cite{cortijo2016,Pikulin_2016}. In this case, the CME current will be induced by the deformation in the presence of a background magnetic field \cite{cortijo2016,Pikulin_2016}. The collective excitations mixing sound and chirality have been considered in \cite{song2019hear,Chernodub:2019lhw,PhysRevLett.124.126602}. 

\medskip

Kinetic theory provides a convenient framework for describing transport phenomena. For chiral fermions, the effect of chiral anomaly has been incorporated \cite{stephanov2012chiral,son2013kinetic} in this theory via the Berry curvature; see  \cite{xiao2010berry} for a review of earlier work on Berry curvature effects on transport. In the resulting chiral kinetic theory, Weyl cones are described as momentum-space monopoles of Berry curvature. In the presence of an external magnetic field, the combined effect of the momentum-space monopoles and magnetic field is the modification of the density matrix in the phase space, resulting in the anomalous Hall effect and the CME  \cite{stephanov2012chiral,son2013kinetic,Chen2013Berry,Dwivedi:2013dea,Basar:2013qia,Basar:2013iaa,Manuel:2013zaa,Chen:2014cla,Gorbar:2016ygi,Kharzeev:2016sut}.
\medskip

In this chapter, our goal is to construct a chiral kinetic theory for Weyl systems under torsion. We will first show that the effect of dynamical, time-dependent deformations on chiral fermions can be captured by Berry curvature in phase space (coordinate and momentum spaces combined).  This quantity has been introduced and used before in a variety of problems involving chirality and spatially inhomogeneous backgrounds  \cite{sundaram1999wave,volovik2003universe,xiao2005berry,freimuth2013phase}. In our problem, Berry curvature in phase space emerges because the spatially inhomogeneous, time-dependent deformations change the distribution of the fermions both in coordinate and momentum spaces.  We then derive the generalized chiral kinetic equations (\ref{eq:kineq4}) describing the effect of deformations on transport of chiral fermions, both with and without external electromagnetic fields. The corresponding anomaly equation (\ref{eq:kineq3}) has, as a source of chirality, an exterior derivative of the phase-space Berry curvature. Since it measures the charge of the Berry monopole in phase space, we call it the ``monopole charge function". 

The physical meaning of the anomaly equation is as follows: the twist creates a synthetic magnetic field $B_{eff}$, the time-dependent strain creates a synthetic electric field $E_{eff}$, and they combine to yield a source $E^i_{eff} B^i_{eff}$ for the chiral anomaly. We then use the generalized chiral kinetic theory to evaluate the magnitude of the CME current induced by torsion in the presence of magnetic field. The current has a linear dependence both on the chiral chemical potential and the magnetic field, similarly to the ``usual" CME. It is observable, and provides a way to discover the generation of chiral imbalance through synthetic gauge fields arising from the phase space Berry curvature. Throughout the chapter, we use as an example a concrete tight-binding Hamiltonian (\ref{tight}); however all of our derivations apply to any Hamiltonian of type $H = {\vec \sigma} {\vec p}(k, x, t) + \phi($k$, $x$, t)$.
\medskip

Let us consider a twisted Weyl semimetal with time-dependent compression along the axis of twist. The twist is chiral; the compression and twist together generate chirality and produce a chiral imbalance for the Weyl quasiparticles, as we will demonstrate. 
In order to solve this problem within the framework of chiral kinetic theory, we need to do the following. 
First, we need to understand how the elastic deformation in position space affects the Berry curvature, in both position and momentum spaces.
Second, we need to formulate and solve the kinetic equation, and calculate the anomalous current and chiral charge generation.

For the first problem, because the Hamiltonian in momentum space now depends on the spatial coordinates, it is impossible to separate the position space and  momentum space components of Berry curvature. 
Therefore, we will address this problem from the perspective of the phase space path integral; our starting point is a phase space Lagrangian, which we will obtain from the coordinate-dependent Hamiltonian.
To solve the second problem, we start from the definition of the Berry connection in phase space and  derive the corresponding  kinetic equation.


 \medskip
\section{Hamiltonian of deformed Weyl semimetal}
We will use a simple model Hamiltonian of an anisotropic Weyl semimetal proposed in \cite{mccormick2017minimal} as an example. This model has a simple tetragonal lattice with lattice constants $a$ and $b$. 
The tight-binding Hamiltonian of the model is given by
\begin{equation}\label{tight}
\begin{split}
H= &-\sum_i\,(c_{i-1x'}^+c_{i}+c_{i+1x'}^+c_{i})\, t_1\s_x\\&+2cos(k_{0} a)\cdot \sum_ic_i^+c_i\, t_1\s_x\\\ &+i\sum_i\,(c_{i+1\perp'}^+c_{i}-c_{i-1\perp'}^+c_{i})\,t_2\s_\perp\\ &+ \sum_i (c_{i+1\perp'}^+c_{i}+c_{i-1\perp'}^+c_{i})t_3
\end{split}
\end{equation}
where $a$ is the lattice spacing in x-direction, and $c^+$ and $c$ are creation and annihilation operators. The parameters $t_1$, $t_2$ and $t_3$ represent the strength of tight-binding interaction, and $k_{0}$ is the location of Weyl points in momentum space.

This model possesses a fourfold rotation symmetry around the $x$-axis, and inversion symmetry, but not the time-reversal symmetry. The deformation can be described by the following parameter change:
\begin{align}
    (c_{i-1x'}^+ c_i + c_{i+1x'}^+c_i) t_1\s_x \to &(c_{i-1x'}^+ c_i + c_{i+1x'}^+c_i) (t_1\s_x-\beta_1u_{xx}\s_x)\\& + i(c_{i-1x'}^+c_i - c_{i+1x'}^+c_i)\beta_2 u_{x\perp}\cdot \s_{\perp}\nonumber
\end{align}
\begin{equation}
t_2\s_\perp\rightarrow t_2\s_\perp -\beta_3 u_{\perp\perp}\cdot \s_{\perp},
\end{equation}
where $\b$ is the anisotropic Gruneisen parameter and $u_{ab}$ is the strain tensor.
We set the twist to be along the $x$-axis,
$u_{x\perp}=\gamma\varepsilon_{ij} r_j$ ; $\g$ describes the twist angle gradient, and $\varepsilon_{ij}$ is the rank two antisymmetric tensor in $y-z$ plane. We also apply a time-dependent compression $\lambda = - u_{xx}$.

With these substitutions, the strain-modified  Hamiltonian in momentum space becomes:

\begin{equation}
\label{eq:Ham1} 
\begin{split}
H= &-2t_1(\cos(k'_x a)-\cos(k_{0x} a))\s_x \\ &-2t_2\sin(k'_y b)\s_y
- 2t_2\sin(k'_z b)\s_z\\&-2\beta_2\gamma y \sin(k'_x a)\s_z+2\beta_2\gamma z \sin(k'_x a)\s_y\\ &+ 2 \b_1 \cos(k'_xa) \lambda\s_x + 2 [\cos(k'_y b) + \cos(k'_z b)] t_3 ,
\end{split}
\end{equation}
where the primed symbols denote the local lattice vector direction. The transformation of direction from global vierbein to the local one is given by
\begin{equation}
  \label{eq:metric1}
\begin{pmatrix}
   d x\\
d y\\
d z\\
    \end{pmatrix} =
    \begin{pmatrix}
    1-\lambda(t)& 0 & 0\\
0 &1&-\g x'\\
0& \g x'&1\\
    \end{pmatrix}\cdot \begin{pmatrix}
   d x'\\
dy'\\
d z'\\
    \end{pmatrix};
\end{equation}
 we will denote the transformation matrix as $M_i^j$.


Expanding the effective Hamiltonian in the vicinity of the Weyl point at momentum $K_i$,
we get a familiar form
\begin{equation}
\begin{split}
\label{eq:Ha}
&H_{eff}={e'}_a^i\s_a (k'_i-K'_i)+{W'}^i(k'_i-K'_i)+\mathcal E\\
&\qquad=e_a^i\s_a(k_i-K_i)+W^i(k_i-K_i)+\mathcal E,
\end{split}
\end{equation}
where the local Weyl points $K'$, dreibein $e'$ ,tilt vector $W'$ and energy of Weyl points $\mathcal E$ can be derived from equation~\eqref{eq:Ham1}. The corresponding dreibein and Weyl points in global coordinates, which are defined to be consistent with the untwisted lattice, are given by $K_i=(M^{-1})_i^j K'_j$, $e_a^i={e'}_a^j M^i_j$ and $W^i={W'}^{j}M_j^i$.

There are eight Weyl cones. In the lowest order:
\begin{equation}
\begin{split}
K_x &=K'_x= s_x [k_{0x} + \frac{\b_1}{t_1 a}\cot(k_{0x}a)\lambda]\\
K_y &=K'_y= s_y\frac{\beta_2 \gamma  z'}{t_2 b}s_x \sin(k_{0x} a) +\Theta(-s_y) \frac{\pi}{b}\\
K_z &=K'_z=-s_z\frac{\beta_2 \gamma y'}{t_2 b}s_x \sin(k_{0x} a) +\Theta(-s_z))\frac{\pi}{b}
 \end{split}
\end{equation}
where $s_x = sgn(\sin(K_x a))$ and $s_{y,z} = sgn(\cos(K_{y,z} a))$, the chirality $\chi$ is $s_x s_y s_z$.
The velocity of fermions in the $x$ direction is $v^x = 2at_1 cos(k_{0x}a)$ and in the $y,z$ direction is $v^{y,z} = 2t_2 b$.

\medskip

{\it Kinetic equations for the Hamiltonian $\sigma\cdot p(k,x,t)+\phi(k,x,t)$.---}
In the above analysis, we find that the elastic deformation affects the dreibein, tilt vector and position of Weyl points, but the Hamiltonian always maintains the form $\sigma\cdot p(k,x,t)+\phi(k,x,t)$. This Hamiltonian can be analyzed by chiral kinetic theory from a path integral perspective \cite{stephanov2012chiral}, where the action is given by
\begin{equation}
 \label{eq:I}
I=\int_{t_i}^{t_f}(k\cdot \dot{x}-\epsilon(p)-a_k\cdot \dot{k}-a_x\cdot \dot{x}-a_t )dt ;
\end{equation}
$\epsilon = \pm|p|+\phi $ is the eigenvalue of energy.  

Generalizing the Berry connection and curvature in momentum space \cite{stephanov2012chiral}, the Berry connection and Berry curvature in phase space can be defined as:
$a_\a=i\langle u|\frac{\pt p_i}{\pt \a}\pt_p|u\rangle =\frac{\pt p_i}{\pt \a}a^p_i $, $\Omega_{\a\b}=\pt_b a_\a-\pt_a a_\b=\frac{1}{2}\pt_\a p_m \pt_\b p_l\epsilon_{mln}\Omega_n^p$. Below, we use the Greek characters as the indices in phase space.

For an arbitrary Hamiltonian, the kinetic equations corresponding to ~\eqref{eq:I} are derived in \cite{Hayata:2016wgy}; see equations (C3,C8,C9) in that paper. For our Hamiltonian, the Berry curvatures satisfy the following identity:
\begin{equation}
 \label{eq:Weq-full}
 \begin{split}
  & \Omega_{\a\b}\Omega_{\g\s}+\Omega_{\a\g}\Omega_{\s\b}+\Omega_{\a\s}\Omega_{\g\b}=0.
 \end{split}
\end{equation}

Using this identity, the kinetic equations can be brought to the form
\begin{equation}
\label{eq:kineq4} 
\sqrt{G}\dot{x}^i=-\Omega_{k_i t}+(\delta^i_j(1+\Omega_{k_lx^l})-\Omega_{k_ix^j})\pt_{k_j}\epsilon+\Omega_{k_ik_j}\pt_{x^j}\epsilon,
\end{equation}
\begin{equation}
\label{eq:kineq2} 
\sqrt{G}\dot{k}_i=\Omega_{x^i t}-(\delta_i^j(1+\Omega_{k_lx^k})-\Omega_{k_ix^j})\pt_{x^j}\epsilon-\Omega_{x^ix^j}\pt_{k_j}\epsilon.
\end{equation}
where $\sqrt{G}=1+\Omega_{k_ix^i}$ describes the modification of the phase space density.

The corresponding anomaly equation can be written as:
\begin{equation}
\label{eq:kineq3} 
\frac{\pt\sqrt{G}}{\pt t}+\frac{\pt\sqrt{G}\dot{x}^i}{\pt x^i}+\frac{\pt\sqrt{G}\dot{k}_i}{\pt k_i}=\Theta_{k_i x^i t}+\Theta_{k_j x^jx^i}\pt_{k_i}\epsilon+\Theta_{x^jk_jk_i}\pt_{x^i}\epsilon.
\end{equation}
where 
\begin{equation}\label{mcf}
\Theta_{\a\b\g}=\pt_\a\Omega_{\b\g}+\pt_\b\Omega_{\g\a}+\pt_\g\Omega_{\a\b}
\end{equation}
 is the exterior derivative of the Berry curvature; we will call this quantity the monopole charge function, as it measures the charge of the Berry monopole in phase space. 
It is easy to prove that if the Berry connections are all continuous analytical functions, the equation~\eqref{eq:kineq3} becomes a classical Liouville equation, because $\Theta_{\a\b\g}=0$. However the Berry curvature possesses a singularity at the point $p=0$, where two degenerate bands cross. This is the reason for the chiral anomaly.
The chiral anomaly is linked to the topology of the system -- this is why it depends on the existence of Weyl cones, but not on their detailed shapes, as we will now demonstrate.

\medskip

{\it  Consistency check of kinetic equations.---}
Let us first check that our equations are consistent with \cite{stephanov2012chiral} in the absence of deformations. In this case, 
$p_i=k_i-A_i$; we will use the Coulomb gauge.
From (\ref{mcf}), we get for the monopole charge function
\begin{equation}
\Theta_{\a\b\g}=\varepsilon_{mnl}\frac{\pt p_m}{\pt \a}\frac{\pt p_n}{\pt \b}\frac{\pt p_l}{\pt \g}\Theta_p,
\end{equation}
where $\Theta_p=2\pi\d^3(p)$ is the well known Berry monopole in momentum space. Therefore,
\begin{eqnarray}
&\Theta_{k_ix^it}=2\pi(\vec{E}\cdot \vec{B})\d^3(p) ,\\
&\Theta_{k_jx^jx^i}\pt_{k_i}\epsilon+\Theta_{x^jk_jk^i}\pt_{x^i}\epsilon=0 .
\end{eqnarray}
 We thus obtain from (\ref{eq:kineq3}) the same expression for the ``Liouville anomaly" as in \cite{stephanov2012chiral}:
\begin{equation}
\frac{\pt\sqrt{G}}{\pt t}+\frac{\pt\sqrt{G}\dot{x}_i}{\pt x^i}+\frac{\pt\sqrt{G}\dot{k}_i}{\pt k_i}=2\pi(\vec{E}\cdot \vec{B})\d^3(p) .
\end{equation}
Let us now check the consistency of kinetic equation~\eqref{eq:kineq4}. For the Weyl Hamiltonian $H = \sigma_i k_i$ we get 
\begin{eqnarray}
 &\Omega_{k_it}= -\vec E\times  \frac{\hat{p}}{2|p|^2},\qquad\Omega_{k_lx^l}\pt_{k_i}\epsilon=(\vec B\cdot \vec \Omega)\hat{p},\\
 & -\Omega_{k_ix^j}\pt_{k_j}\epsilon+\Omega_{k_ik_j}\pt_{x^j}\epsilon =(\hat{p}\cdot\vec \Omega)\vec B-(\vec B\cdot \vec\Omega)\hat{p}.
 \end{eqnarray}
The kinetic equation is thus given by
\begin{equation}
\sqrt{G}\dot{x}=\hat{p}+\vec E\times  \frac{\hat{p}}{2|p|^2} +(\hat{p}\cdot\vec \Omega)\vec B ,
\end{equation}
which is exactly the equation derived by Stephanov and Yin \cite{stephanov2012chiral}. 
\medskip

\section{Topologically protected CME and anomaly for a general strained Weyl material}
Let us now apply the kinetic equations \eqref{eq:kineq3}-\eqref{eq:kineq4}
to the Weyl semimetal under torsion. 
We first use the anomaly equation \eqref{eq:kineq3} to identify the quantities responsible for the topological configuration of the system. In our case,  $p_a=e_a^i(x)(k_i-A_i-K_i)$, and $\phi=W^i(x)(k_i-A_i-K_i)+\mathcal E(x)$. 
Substituting these quantities into \eqref{eq:kineq3}, we get
\begin{equation}
\begin{split}
\label{eq:theta}
\Theta_{k_i x^i t}&=e_a^i e_b^j\frac{\pt (A_{j}+K_j)}{\pt x_i}e_c^l\frac{\pt (A_{l}+K_l)}{\pt t}\varepsilon_{abc}2\pi\d^3(p)\\
&-e_a^i\frac{\pt e_b^j}{\pt x^i}T_j e_c^l\frac{\pt (A_{l}+K_l)}{\pt t}\varepsilon_{abc}2\pi\d^3(p)\\
&=2\pi det(e)(\frac{d(\vec{A}+\vec{K})}{dt}\cdot[\nabla\times(\vec{A}+\vec{K})])\d^3(p)
\end{split}
\end{equation}
where $T_i =e^{-1}_{ij}p_j= k_i - A_i - K_i$. 
 
 The monopole charge functions $\Theta$ only depend on $p$, so \eqref{eq:theta} is tilt independent, and depends only on the existence of the Weyl points. 
One can further prove that the last two terms in \eqref{eq:kineq3} are also independent of tilt:
\begin{equation}
\Theta_{k_jx^jx^i}\pt_{k_i}\epsilon+\Theta_{x^jk_jk_i}\pt_{x^i}\epsilon
=2\pi det(e)(\nabla \mathcal E\cdot[\nabla\times(\vec{A}+\vec{K})])\d^3(p)
\end{equation}
Therefore, the anomaly equation for an isolated Weyl cone is:
\begin{align}
\label{eq:topochar}
\pt_\m j^\m &=\int \frac{d^3k}{(2\pi)^3}\left(\frac{\pt\sqrt{G}}{\pt t}+\frac{\pt\sqrt{G}\dot{x}^i}{\pt x^i}+\frac{\pt\sqrt{G}\dot{k}_i}{\pt k_i}\right)f\nonumber\\
&=\int sgn(J)\frac{J^{-1}d^3p}{(2\pi)^3}(\Theta_{k_i x^i t}+\Theta_{k_jx^jx^i}\pt_{k_i}\epsilon+\Theta_{x^jk_jk_i}\pt_{x^i}\epsilon)f\nonumber\\
&=\frac{sgn(J)}{4\pi^2}det(e)^{-1}det(e)\vec{B}_{eff}\cdot \vec{E}_{eff}\nonumber\\
&=\frac{\chi \vec{B}_{eff}\cdot \vec{E}_{eff} }{4\pi^2},
\end{align}
where the effective electric and magnetic fields are 
\begin{equation}\label{effective}
\vec{E}_{eff}=\frac{d(\vec{A}+\vec{K})}{dt}+\nabla\mathcal E,\\ \ \ 
 \vec{B}_{eff}=\nabla\times(\vec{A}+\vec{K}) ,
 \end{equation}
 and $J=det(e)$ is the Jacobian; its sign corresponds to the chirality of the Weyl cone.

It is clear from ~\eqref{eq:topochar} that the anomaly equation is independent of dreibein and tilt vector -- it 
depends only on the existence of the Weyl point, 
 due to its topological nature, but not on its shape.

Although the anomaly is not affected by dreibein and tilt vector, the general expression for the current still depends on them: 
\begin{align}\label{eq:cur}
 j^i&=\int \frac{d^3k}{(2\pi)^3} \sqrt{G}\dot{x^i}f=\int sgn(J)J^{-1} \frac{d^3p}{(2\pi)^3} \sqrt{G}\dot{x^i}f\\
&=\int sgn(J)J^{-1}  [-\Omega_{k_i t}+(\delta^i_j(1+\Omega_{k_lx^l})-\Omega_{k_ix^j})\pt_{k_j}\epsilon+\Omega_{k_ik_j}\pt_{x^j}\epsilon]f\frac{d^3p}{(2\pi)^3}\nonumber.
\end{align}
The first term $\Omega_{k_it}$ describes a transverse current analogous to the anomalous Hall effect. The second term $\pt_{k_j}\epsilon$  corresponds to the anomalous velocity induced by lattice motion. To study the CME, we focus on the last three terms:
\begin{eqnarray}
\label{eq:cur1}
&\Omega_{k_jx^j}\pt_{k_i}\epsilon=\varepsilon_{abc}(e_a^j\Delta_{bj}e_d^i\frac{p_cp_d}{2|p|^4}+e_a^j\Delta_{bj}\frac{W_ip_c}{2|p|^3}),\\
\label{eq:cur2}
&\Omega_{k_ix^j}\pt_{k_j}\epsilon= \varepsilon_{abc}(e_a^i\Delta_{bj}e_d^j\frac{p_cp_d}{2|p|^4}+e_a^i\Delta_{bj}\frac{W_jp_c}{2|p|^3}),\\
 \label{eq:cur3}
&\Omega_{k_ik_j}\pt_{x^j}\epsilon= \varepsilon_{abc}(e_a^ie_b^j\Delta_{dj}\frac{p_cp_d}{2|p|^4}+e_a^ie_b^j\frac{\Gamma_j p_c}{2|p|^3}), 
\end{eqnarray}
where $\Delta_{ai}=\frac{\pt p_a}{\pt x^i}=\omega^j_{ai}T_j-e^j_a\frac{\pt A_i}{\pt x^j}$.
$\Gamma_{i}=\frac{\pt W^j}{\pt x^i}T_j-W^j\frac{\pt A_j}{\pt x^i}+\frac{\pt\mathcal E}{\pt x^i}$, and we have defined $\omega^i_{aj}=\frac{\pt e^i_a}{\pt x^j}$.

Using the symmetry relation $\epsilon_{abc}p_d- \epsilon_{dbc}p_a-\epsilon_{adc}p_b-\epsilon_{abd}p_c=0$ one can simplify these equations, and get the following expression for the anomalous CME current:
\begin{align}
\label{eq:simp} 
j^i=&\int sgn(J)J^{-1} \frac{d^3p}{(2\pi)^3}
[\Omega_{k_jx^j}\pt_{k_i}\epsilon-\Omega_{k_ix^j}\pt_{k_j}\epsilon
+\Omega_{k_ik_j}\pt_{x^j}\epsilon]f\nonumber\\
=&\int sgn(J)J^{-1} \frac{d^3p}{(2\pi)^3}\left[e_c^ie_b^j\omega_{na}^j\frac{T_a\epsilon_{bcn}}{2|p|^2}+det(e)\frac{B_{eff}^i}{2|p|^2}\right.\nonumber\\
&+\varepsilon_{abc}(e_a^k\omega^j_{bk}T_j W^i-e_a^i\omega^j_{bk}T_jW^k +e_a^i e_b^j\Gamma_j)\frac{p_c}{2|p|^3}\\
&\left.+\frac{1}{2}\varepsilon_{abc}\varepsilon_{jlm}e_a^m e_b^l W^j\frac{p_c}{2|p|^3}B^i_{eff}\right]f\nonumber
\end{align}
Since the tilt parameter $W^i$ is of the order of $\gamma y e^j_a, \gamma z e^j_a$, and $\gamma y, \gamma z \ll 1$, we will ignore it in our calculations, and treat the cones as untilted. In such a cone, in a equilibrium distribution, $f(|p|)$ can be assumed to be an even function of $p_a$. So, most terms in eq~\eqref{eq:simp} should vanish except the second term. It is easy to show  that the second term is also independent of dreibein. Therefore the CME current is topologically protected. The contribution of each cone is
\begin{equation}
\label{eq:CME} 
j^i=\int sgn(J)J^{-1} \frac{d^3p}{(2\pi)^3}det(e)\frac{B^i_{eff}}{2|p|^2}f=\chi_n\frac{(\m-\mathcal E_n) B_{eff}^i}{4\pi^2}
\end{equation}
where $\chi_n, \mathcal E_n$ are the chirality and energy of the Weyl point. In this expression, as explained in \cite{Basar:2013iaa}, 
both $\m$ and $\mathcal E_n$ should be counted from the bottom of the filled band. 

The evolution of the chiral charge can be determined in the relaxation time approximation:
\begin{equation}
\begin{split}
&\pt_i j^i_n + \dot{\rho}_n = \chi_n\frac{e^3\vec{E}\cdot \vec{B}}{4\pi^2}-\frac{e}{\t}(\rho_n-(\rho_n)_{eq})
\label{eq:uncon}
\end{split}
\end{equation} 
where $(\rho_n)_{eq}$ denotes the equilibrium chiral charge density in position space, and $\tau$ is the chirality relaxation time. So the chemical potential $\mu_n$ associated with the chiral charge is \cite{CME,grushin2016inhomogeneous} 
\begin{equation}
\label{eq:ccp} 
\delta \m_n = \left(\frac{(\mu -\mathcal E_n)^2}{2\pi^2 v^x v^y v^z}\right)^{-1} \delta \rho_n \chi \frac{16 \pi^2 ab^2 \cos(k_{x0}a) t_1 t_2^2}{(\mu-\mathcal E_n)^2}\frac{\vec{E}_{eff}\cdot\vec{B}_{eff}}{4\pi^2}\t
\end{equation}
where $\mu$ is the chemical potential. The contribution of each cone to the chiral magnetic current is given by 
\begin{equation}
\vec{j}_n =\chi_n \frac{e^2}{4\pi^2}\delta \mu_n \vec{B};
\end{equation}
therefore, the total CME current is determined by the chiral imbalance \cite{CME}. 
The general form of the anomalous CME current in a deformed Weyl semimetal is given by ~\eqref{eq:simp}. 

\medskip
\section{The current in the model with torsion}
Following the general results derived above, we will now focus on the effective gauge field induced on the Weyl points by torsion. Since the effective electric field transforms as a 2-form under arbitrary diffeomorphisms, the effective electric and magnetic fields (\ref{effective}) are:
\begin{equation}
\vec{E}_{eff} = s_x \frac{\b_1}{t_1 a}\cot(k_{0x}a)\dot{\lambda} \hat{x} ;
\end{equation}
 \begin{equation}
\vec{B}_{eff} = -s_x (s_y + s_z) \frac{\beta_2 \gamma  }{t_2 b} \sin(k_{0x} a)\hat{x} .
\end{equation}


The chiral anomaly at each cone is
\begin{equation}
\begin{split}
\vec{E}_{eff}\cdot \vec{B}_{eff}= 
-(s_y + s_z)\frac{\b_1 \b_2\ \cos(ak_{0x})\g\dot{\lambda} }{abt_1 t_2},
\end{split}
\end{equation}



and the total charge is conserved, as expected:
\begin{equation}
\begin{split}
\sum_n (\pt_i j^i + \dot{\rho})_n=&\sum_n\frac{\chi_n}{4\pi^2}(E_{eff}\cdot B_{eff})_n\\=& \sum -s_y s_z (s_y + s_z)\frac{\b_1 \b_2\ \cos(ak_{0x})\g\dot{\lambda} }{abt_1 t_2}\\ = & 0 .
\end{split}
\end{equation}

\begin{figure}[hbtp]
\centering
\includegraphics[scale=0.6]{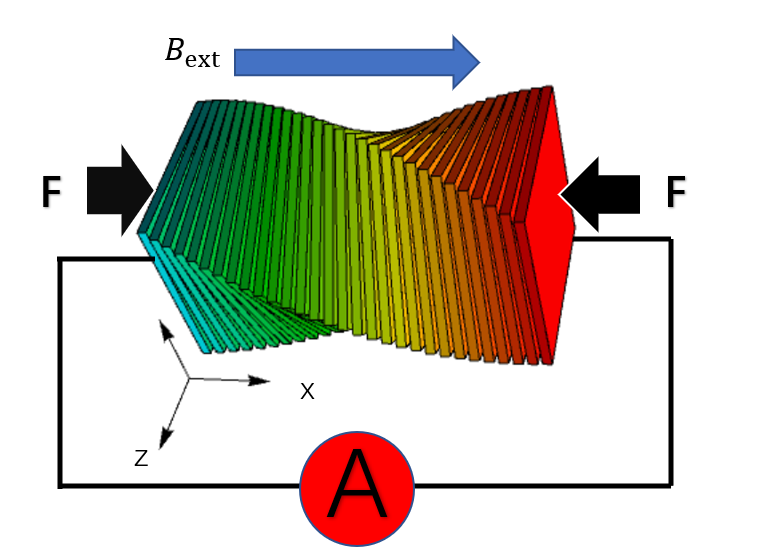}
\caption{The chiral magnetic current induced by torsion and compression in the presence of an external magnetic field.}
\label{fig:arch}
\end{figure}

The energies of the Weyl points are $\mathcal E_n = t_3 (s_y + s_z); $ therefore, 
combining equation~\eqref{eq:CME} and~\eqref{eq:ccp} we get
\begin{align}
j_{total}^i& = \sum_n(j^i)_n= \sum_n\frac{\chi_n \delta \m_n e^2 B_{total}^i}{4\pi^2}\\
&=-4e^2\frac{\b_1\b_2 b t_2  \cos(ak_{0x})\g \dot{\lambda}}{\pi^2\hbar^2}
\left(\frac{1}{(\m-4t_3)^2}-\frac{1}{(\m+4t_3)^2}\right)\t B_{ext}^i.\nonumber
\end{align}
This shows that the current is linear in the external magnetic field \cite{CME}, which is different from the CME induced in parallel (weak)  electric and magnetic fields, where $j\propto B^2$ \cite{2013Son,2014Burkov}. 
\medskip

For a numerical estimate, let us assume a statically twisted and dynamically compressed (along the axis of twist)  crystal, as depicted in Fig~\ref{fig:arch} with a square cross section 2 mm x 2 mm, and take parameters $t_1 = 2 eV,  t_2 = 0.5 eV,  t_3 = 0.05 eV,  \b_1 = 1.5 eV,  \b_2 = 1 eV,  a = 0.3 nm,  b = 0.4 nm,  \tau = 10^{-10} s,  \mu = 0.15 eV,  ak_{0x} = \pi/3$. Let us also assume an external magnetic field of 10 T, with a twist parameter of $\g = 10 m^{-1}$ and compression rate $\dot{\lambda} = 10 s^{-1}$. The anomalous current density would be $\sim 470\ \mu A/m^2$, and the total current about 2 nA. Note that the current has an inverse square dependence on $\mu - \mathcal E_n$; it will be maximized in materials with a Fermi surface very close to a Weyl point. For example, for our parameters, if we take $\mu - 4t_3$ to be 0.01 eV instead of 0.05 eV, our current will increase by a factor of 25 to approximately 50 nA. 

\medskip

\section{Summary}
Our generalized chiral kinetic equations (\ref{eq:kineq4}), (\ref{eq:kineq2}) and the anomaly equation (\ref{eq:topochar}) apply to any Weyl system with the Hamiltonian of the form $\sigma\cdot p(k,x,t)+\phi(k,x,t)$.  In the anomaly equation (\ref{eq:topochar}), torsion 
creates a synthetic magnetic field, while the time-dependent compressive strain creates a synthetic electric field -- so no external electric field is necessary to generate the chiral chemical potential. Detecting the resulting torsion-induced chiral magnetic current would thus allow to establish the anomalous coupling between the spatial and momentum-space chiralities,  without a background from the Ohmic current that exists in the longitudinal magnetoresistance measurements. 

The current measured in the corresponding experiment will also not get contributions from the piezoelectric effect (because the material is inversion symmetric and has no piezoelectricity), or from eddy currents (because the compression is along the magnetic field). 

\chapter{Tunable Chiral Symmetry Breaking in Weyl Materials}\label{chTunable}

\blfootnote{This chapter is based on \cite{SahalAsymm}.}

Certain phenomena in Weyl materials require the left- and right-handed fermions to have different energies or velocities. They include the quantized circular photogalvanic effect~\cite{dejuan17}, the helical magnetic effect~\cite{yuta2018} and the chiral magnetic effect without an external source of chirality (i.e., a chiral chemical potential is generated because of the chirality of the crystal structure, not because of external chiral fields)~\cite{meyer2018}. These effects are possible only in materials that have chiral crystal lattices which lack symmetries that reverse spatial orientation (these include inversion, mirrors, rotoinversions and their products with time reversal). We refer to such materials as asymmetric Weyl materials, and materials that do have orientation-reversing symmetries as symmetric Weyl materials. Asymmetric Weyl materials include RhSi~\cite{HasanRhSi} and CoSi~\cite{rao2019}.  The quantized circular photogalvanic effect has been observed in both these materials \cite{RhSiCurrent,CoSiCurrent}.

However, asymmetric Weyl materials are rare in nature compared to symmetric Weyl materials. One way to observe effects that depend on asymmetry is to break relevant symmetries by an external perturbation. 
It has recently been shown that a symmetric Weyl material can become asymmetric upon ordering magnetically if the magnetic moments break all symmetries that reverse spatial orientation \cite{ray2020}.

In this work, we investigate chiral symmetry breaking more generally.
Using the concept of true and false chirality introduced by Barron \cite{barron1986}, we derive a criterion 
to determine whether an external field or perturbation produces an asymmetric material, which depends on the symmetry of the perturbation and the space group of the crystal.
We then explicitly show that in the zincblende material InAs, applying a magnetic field in a low symmetry direction breaks all symmetries that reverse spatial orientation, and causes left- and right-handed fermions to have different velocities and energies.
This induced asymmetry allows for the observation of effects present only in asymmetric Weyl materials.
Furthermore, we show examples where the number of type I Weyl fermions \cite{soluyanov2015} of left- and right-chirality are not equal; the imbalance is compensated by the number of type II Weyl fermions of each chirality.

\section{Properties of Weyl Cones}
The chirality of a Weyl fermion is equivalent to its helicity: $\chi = \mathrm{sgn}(\vec{s}\cdot\vec{p})$ where $\vec{s}$ is the pseudospin and $\vec{p}$ is the momentum. The chirality is positive for right-handed fermions and negative for left- handed fermions.
Since Weyl cones are monopoles of Berry curvature and the total Berry charge in the Brillouin zone must be zero, there is always an equal number of left- and right-handed Weyl cones in a microscopic Hamiltonian. 

Under inversion symmetry ($P$), the pseudospin, momentum, and chirality transform as
\begin{align}
\vec{s} &\to \vec{s}\nonumber\\
\vec{p} &\to -\vec{p}\\
\chi &\to -\chi\nonumber
\end{align}

Under time-reversal symmetry ($T$), the pseudospin, momentum, and chirality transform as
\begin{align}
\vec{s} &\to -\vec{s}\nonumber\\
\vec{p} &\to -\vec{p}\\
\chi &\to \chi\nonumber
\end{align}

In a material that has both inversion and time-reversal symmetries, the chirality flips under $P$ and remains invariant under $T$. Crystal momentum $k_i$ flips sign under both $P$ and $T$. Therefore, under $PT$,
\begin{align}
k_i \to k_i\nonumber \\
\chi \to -\chi 
\end{align}
In such a material, the left- and right-handed fermions coincide in the Brillouin zone; thus, it has Dirac cones, not Weyl cones. Therfore, all Weyl materials lack time-reversal, inversion or both.

Symmetries that reverse spatial orientation transform left-handed fermions into right-handed fermion and vice versa.
Therefore, in materials with these symmetries, each left-handed cone has a right-handed partner cone at the same energy and with the same velocities.
Consequently, effects such as the quantized circular photogalvanic effect, which requires an asymmetry between left- and right-handed cones, are possible only in materials that have a chiral crystal lattice without orientation-reversing symmetries.

\begin{figure}[htp]
     \centering
         \includegraphics[width=0.75\linewidth]{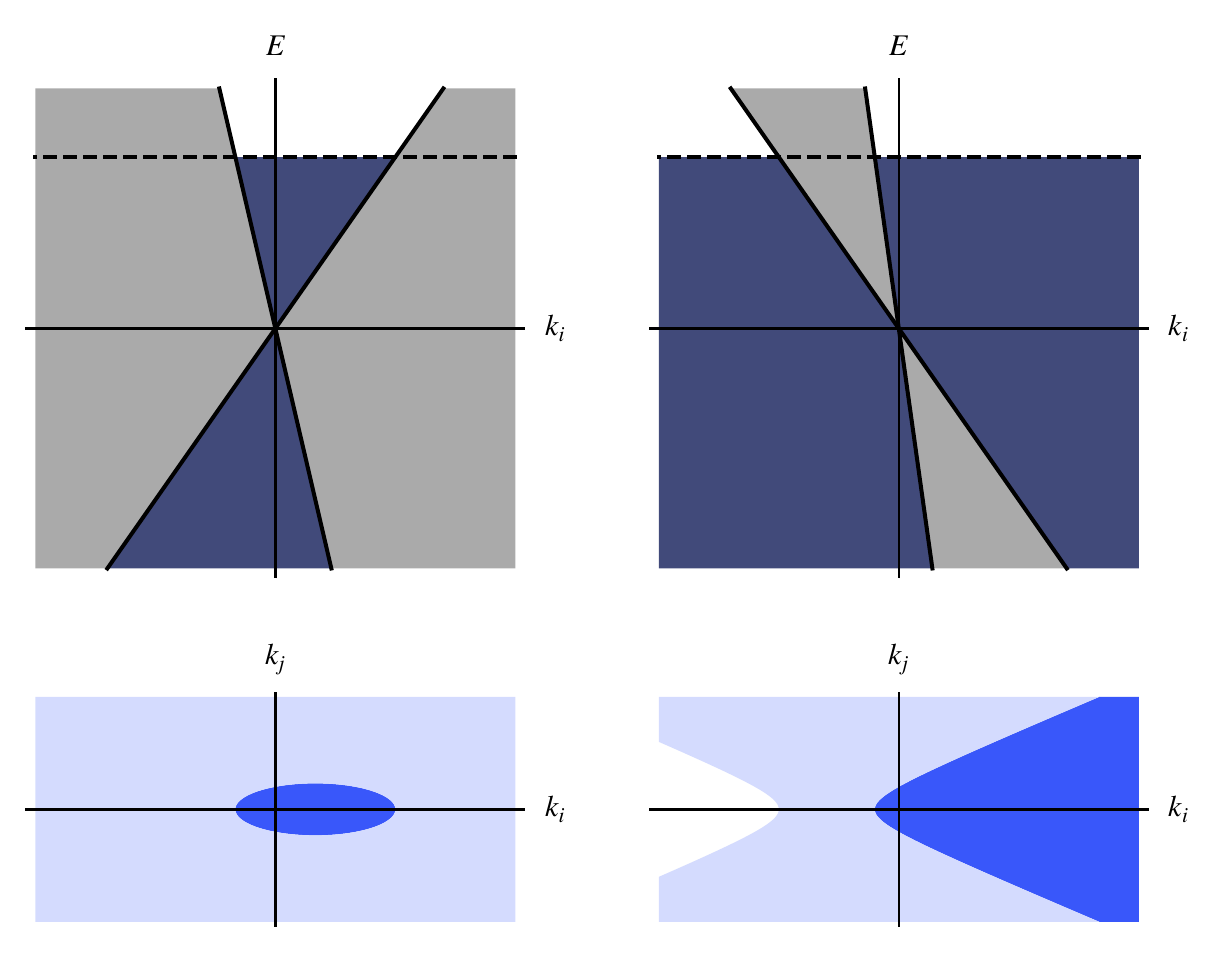}
        \caption{Dispersion relation (top) and Fermi surface (bottom) of type I (left) and type II (right) Weyl cones with tilt 0.5 and 1.5 respectively. In the upper figures, the dashed line shows the Fermi level, blue shows filled states, and white shows unfilled states; gray shows the region outside the cone. In the lower figures dark blue indicates both bands filled, light blue indicates one band filled, and white indicates both bands empty.}
        \label{types}
\end{figure}

The cones of Weyl fermions are generally tilted. Fermions with small tilts and elliptical Fermi surfaces are called type I Weyl fermions, while those with large tilts and hyperbolic Fermi surfaces are called type II Weyl fermions \cite{soluyanov2015}. The dispersion relations of type I and type II Weyl cones are sketched in Figure~\ref{types}. Many Weyl semimetals have exclusively type I fermions, such as TaAs~\cite{lv2015PRX,xu2015,yang2015,lv2015Nat}. Some materials have only type II fermions, such as $\mathrm{WTe_2}$~\cite{li2017} . The Weyl semimetal $\mathrm{OsC_2}$ is unusual in that it has both kinds of Weyl cones\textemdash24 of type I and 12 of type II \cite{zhang2018}. 


The linearized Hamiltonian for a general Weyl cone is
\begin{equation}
H = v_\text{t}^i q_i + v_a^i \sigma_a q_i + E,
\end{equation}
where $v_\text{t}^i$ is the ``tilt" velocity and $v^i_a$ represents the untilted part of the Hamiltonian. $E$ is the energy of the Weyl point. The matrices $\sigma_a$ act on spin or pseudospin. The chirality is
\begin{equation}\label{chi}
    \chi = \mathrm{sgn(det}(v^i_a)),
\end{equation} 
which is positive for a right-handed cone and negative for a left-handed cone. We define the product of velocities 
\begin{equation}\label{v1v2v3}
    v_1v_2v_3 = |\mathrm{det}(v^i_a)|,
\end{equation} 
the velocity tensor $(v_\text{W}^2)^{ij} = v_a^i v_a^j$ and its inverse $(v_\text{W}^{-2})_{ij}$. We also define a dimensionless measure of the tilt
\begin{equation}\label{tilt}
    \text{tilt parameter} = \sqrt{(v_\text{W}^{-2})_{ij}v_\text{t}^i v_\text{t}^j}.
\end{equation}The cone is type I if the tilt parameter is less than 1 and type II if it is greater than 1 \cite{soluyanov2015}.

 
\section{True and False chirality}

L. D. Barron introduced the idea of true and false chirality of a system \cite{barron1986}. 
A Weyl material is said to have false chirality if it possesses a symmetry $MT$, where $M$ is some symmetry that reverses spatial orientation, but does not have the symmetry $M$ itself.
Such a material transforms to its mirror image under time reversal. Systems with true chirality retain their chirality even under time reversal. Examples of systems with true chirality include glucose and DNA molecules and the electroweak part of the Standard Model. Systems with false chirality include a cone rotating about its axis and a crystal with inversion symmetry subjected to parallel uniform electric and magnetic fields.

Asymmetric Weyl materials, such as RhSi and CoSi, have crystal structures with true chirality. Materials with false chirality cannot be asymmetric Weyl materials, because the chirality of each cone is invariant under $T$ but flips under $M$ (since, by definition, $M$ reverses orientation), so left- and right-handed cones would be related by $MT$.

In symmetric Weyl materials, 
it is possible to break mirror symmetries by applying external perturbations. 
Such symmetry breaking may produce either true or false chirality, as we will demonstrate.
As a first example, consider the transition metal monopnitcide class of materials, which includes TaAs, TaP, NbAs, and NbP~\cite{lv2015PRX,xu2015,yang2015,lv2015Nat,xu2016, modic2019, xu2015discovery, yuan2020}.
These compounds have tetragonal symmetry and are in the space group $I4_1md$ (No. 109) with fourfold rotation about the [001] axis, reflection symmetry about the $[100], [010], [110], [\bar{1} 10]$ planes, and time-reversal symmetry, but no inversion symmetry or reflection symmetry about the $[001]$ plane.
If we apply a magnetic field along the $[001]$ direction, $B_z$ flips sign under reflections and therefore breaks all mirror symmetries. However, this perturbation does not introduce true chirality as $B_z$ also flips sign under time-reversal; therefore the perturbed system remains invariant under symmetries $MT$, where $M$ is a mirror reflection symmetry of the unperturbed system. Thus, TaAs with a magnetic field along the $c$ axis has false chirality and would therefore still be a symmetric Weyl material. 

Only systems with true chirality will display effects that depend on a difference in energy or velocity between left- and right-handed fermions.
Such effects will correspond to the expectation value of an operator, $Q$, such that $Q$ is invariant under all symmetries that preserve spatial orientation, but reverses sign under all symmetries that reverse spatial orientation (i.e. $Q$ has the same transformation as spatial orientation under the crystal symmetries).
For a perturbation $\lambda$ to produce true chirality in the system, there must exist a function $Q(\lambda)$ with this property.

For systems that possess time-reversal symmetry, a quantity that is of odd order in magnetic field does not qualify as $Q(\lambda)$ because it is odd under time-reversal, which preserves chirality. For example, in the transition metal monopnictides mentioned above, $B_z$ is invariant under rotation, and flips under reflections, but since it also flips under time-reversal, which preserves the chiralities of fermions, operators that are odd in $B_z$ will yield a vanishing expectation value.

\begin{figure}[htp]
\centering
  \includegraphics[width=0.9\linewidth]{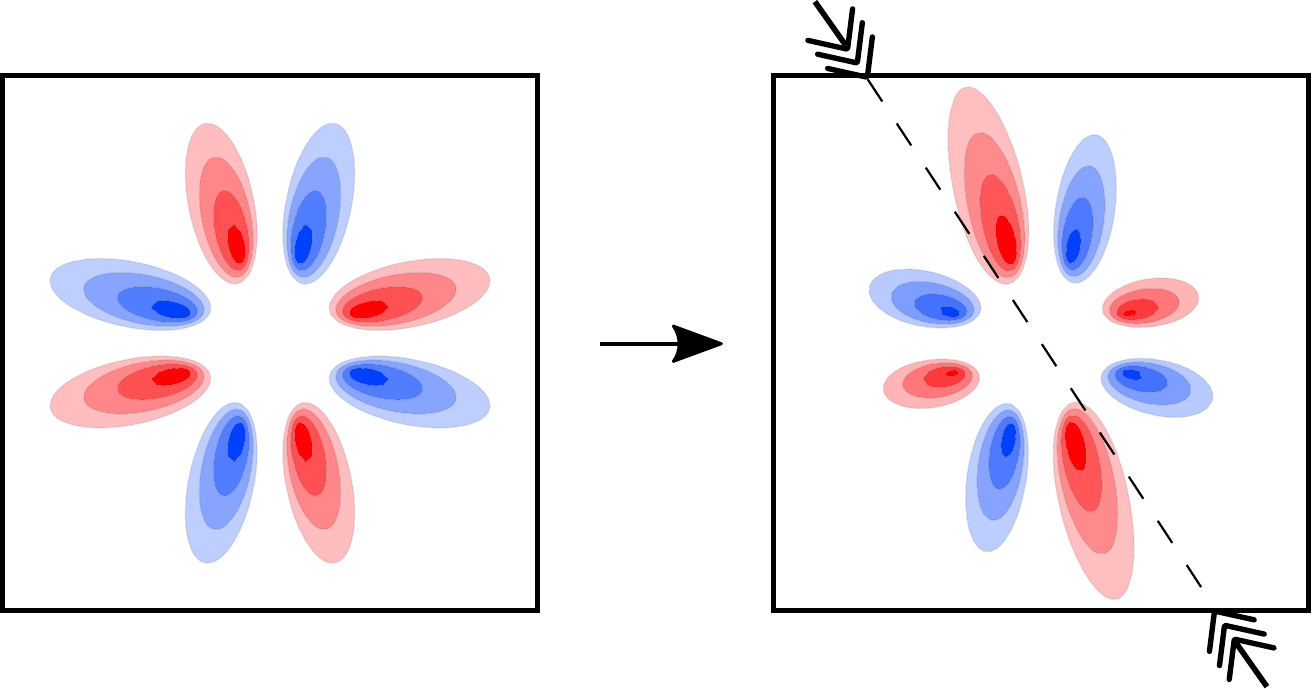}
  \caption{Slices of the Fermi surface for a Weyl material with fourfold rotation symmetry such as TaAs. The shading represents energy and the color represents chirality of the fermions. Each ellipse represents a Weyl cone. A perturbation along a low symmetry axis, such as a strain or a magnetic field, would affect left- and right-handed cones differently, resulting in a net chiral asymmetry.}
  \label{pert}
\end{figure}

In TaAs, the lowest order $Q$ as a function of magnetic field is $Q(\vec{B}) = B_x B_y (B_x^2 - B_y^2)$. Similarly, if we consider chirality induced by strain along a low symmetry axis, the lowest order $Q(S)$ is $S_{xy} (S_{xx} - S_{yy})$, where $S_{ij}$ is the strain tensor.
Therefore, in TaAs, we expect physical phenomena such as the helical magnetic effect to be of fourth order in $\vec{B}$ or second order in $S_{ij}$. 
The effects of a low symmetry perturbation are illustrated in Figure~\ref{pert}.
{As a second example, the rare earth carbides studied in \cite{ray2020} have the space group $Amm2$ (No. 38). In systems with this symmetry, the lowest order function of magnetization $\vec{\mu}$ that transforms in the same way as the chirality of fermions under all symmetries of the unperturbed system is $Q(\vec{\mu}) = \mu_x \mu_y$. }

In materials that possess inversion symmetry, external magnetic fields or uniform strain cannot induce chirality, because both magnetic field and strain are even under parity, and any functions of these quantities will also be even under parity. Since most Dirac materials such as $\mathrm{ZrTe_5}$ and $\mathrm{Na_3 Bi}$ have inversion symmetry, we cannot transform them into asymmetric Weyl materials through uniform strain or uniform external magnetic field. However, it is possible to create symmetric Weyl cones and other topological phases in materials such as $\mathrm{ZrTe_5}$ with Zeeman splitting \cite{sun2020topological,choi2020}.

We now specialize to the case of breaking the chiral symmetry in a material with no magnetic ordering by a uniform magnetic field. Since this magnetic field is translation invariant, the only symmetries that can change the magnetic field are the symmetries of the point group, and their products with time reversal. Therefore, the expressions $Q(\vec{B})$ depend only on the point group, rather than the whole space group.
Out of the 32 point groups, 11 are chiral and already asymmetric.
Another 11 are centrosymmetric; as discussed in the previous paragraph, magnetic field cannot induce chirality in these groups.
In the other 10 groups, we can use a magnetic field to break the chiral symmetry.
We list the function $Q(\vec{B})$ in Table~\ref{QB} for these 10 groups. 
The symmetry analysis in Table~\ref{QB} holds only for perturbations that can be described by uniform time-odd pseudovectors, such as a uniform magnetic field and ferromagnetism.

But a magnetic field is not the only perturbation that can create asymmetry in Weyl materials. For example, Weyl points can be created or manipulated by strain \cite{ruan2016, cortijo2016}, magnetic ordering \cite{ray2020,shekhar2018,ghimire2019}, incident light \cite{hubener2017}, ferroelectricity \cite{he2018,sharma2019}, or a superconducting condensate \cite{obrien2017}. The symmetry analysis in Table~\ref{QB} is not applicable to these perturbations. For example, an electric polarization along a low symmetry direction, which is described by a time-even vector, can break chiral symmetry even in materials with the $O_h$ point group. The relevant parameter that breaks chiral symmetry is $Q(\vec{P}) = P_x P_y P_z (P_x^2 - P_y^2)(P_y^2 - P_z^2)(P_z^2 - P_x^2)$.

\begin{table}[htp]
\centering
\begin{tabular}{c c}
\hline\hline
\textbf{Group} & $Q(\vec{B})$ \\
\hline
$m(C_s)$ & $B_x B_z$ \textbf{OR} $B_y B_z$ \\
$mm2(C_{2v})$ & $B_x B_y$\\
$\bar{4}(S_4)$ & $B_x B_y$ \textbf{OR} $B_x^2 - B_y^2$\\
$4mm(C_{4v})$ & $B_x B_y (B_x^2 - B_y^2)$\\
$\bar{4}2m(D_{2d})$ & $B_x^2 - B_y^2$\\
$3m(C_{3v})$ & $(B_1 - B_2)(B_2 - B_3)(B_3 - B_1)(B_1 + B_2 + B_3)$\\
$\bar{6}(C_{3h})$ & $B_z B_a B_b B_c$ \textbf{OR} $B_z (B_a - B_b)(B_b - B_c)(B_c - B_a)$\\
$6mm(C_{6v})$ & $B_a B_b B_c (B_a - B_b)(B_b - B_c)(B_c - B_a)$\\
$\bar{6}m2(D_{3h})$ & $B_z B_a B_b B_c$\\
$\bar{4}3m(T_d)$ & $(B_x^2 - B_y^2)(B_y^2 - B_z^2)(B_z^2 - B_x^2)$\\
\hline\hline
\end{tabular}
\caption{\label{QB} $Q(\vec{B})$ for the 10 point groups where $\vec{B}$ can break chirality. 
In the first five rows and the last row, $x, y, z$ indicate the orthorhombic crystal axes; 
in the sixth row, $1,2,3$ indicate the rhombohedral crystal axes;
in the remaining rows, $z, a, b, c$ indicate hexagonal lattice vectors, with $\hat{a} + \hat{b} + \hat{c} = \vec{0}$. 
When multiple functions $Q$ are listed, it is enough for one of these quantities to be non-zero to break chiral symmetry.}
\end{table}

In this work, we focus on a class of non-centrosymmetric materials that have the space group $F\bar{4}3m$ (No. 216) and the point group $\bar{4}3m (T_d)$. This class includes the half-Heusler compound GdPtBi, which is antiferromagnetic at low temperatures \cite{hirschberger2016} and the zincblende compounds HgTe and InSb. In a state without magnetic ordering, these materials have several axes of rotation and planes of reflection, as well as time-reversal symmetry, but they lack inversion symmetry. Their electronic structure is characterized by a fourfold degeneracy at the $\Gamma$ point and no Weyl cones. However, if we apply an external magnetic field~\cite{cano2017} or  strain~\cite{ruan2016}, or if there is magnetic ordering in the case of GdPtBi,~\cite{hirschberger2016,shekhar2018}, the degeneracy splits, and Weyl points appear. When the degeneracy is broken by a magnetic field along high symmetry axes, such as [111] and [100], all the emergent Weyl fermions come in pairs of opposite chirality related by symmetry \cite{cano2017}.
In the next section, we will show that this is not the case when the magnetic field is along a low-symmetry axis.


\section{Model}

We consider the Hamiltonian given in Ref.~\cite{cano2017}:
\begin{align}\label{Ham}
H = A k^2 I_4 + &C\left[(k_x^2 - k_y^2)\Gamma_1 + \frac{1}{\sqrt{3}}(2k_z^2 - k_x^2 - k_x^2)\Gamma_2\right]\nonumber\\  + &F(k_x k_y \Gamma_3 + k_x k_z \Gamma_4 + k_y k_z \Gamma_5)\nonumber\\ + &D(k_x U_x + k_y U_y + k_z U_z)\nonumber\\
+& \mu_B g (B_x J_x + B_y J_y + J_z B_z),
\end{align}
where $J_{x,y,z}$ are the spin-$\frac{3}{2}$ matrices, which can be expressed as
\begin{align}
J_x &= 
\begin{pmatrix}
 0 & \frac{\sqrt{3}}{2} & 0 & 0 \\
 \frac{\sqrt{3}}{2} & 0 & 1 & 0 \\
 0 & 1 & 0 & \frac{\sqrt{3}}{2} \\
 0 & 0 & \frac{\sqrt{3}}{2} & 0 \\
\end{pmatrix},  \\
J_y &= 
\begin{pmatrix}
 0 & -\frac{i \sqrt{3}}{2} & 0 & 0 \\
 \frac{i \sqrt{3}}{2} & 0 & -i & 0 \\
 0 & i & 0 & -\frac{i \sqrt{3}}{2} \\
 0 & 0 & \frac{i \sqrt{3}}{2} & 0 \\
\end{pmatrix},  \\
J_z &= 
\begin{pmatrix}
 \frac{3}{2} & 0 & 0 & 0 \\
 0 & \frac{1}{2} & 0 & 0 \\
 0 & 0 & -\frac{1}{2} & 0 \\
 0 & 0 & 0 & -\frac{3}{2} \\
\end{pmatrix}
\end{align}

The matrices $\Gamma_\mu$ are defined as
\begin{align}
\Gamma_1 &= \frac{1}{\sqrt{3}}(J_x^2 - J_y^2)\nonumber\\
\Gamma_2 &= \frac{1}{3}(2J_z^2 - J_x^2 - J_y^2)\nonumber\\
\Gamma_3 &= \frac{1}{\sqrt{3}}\{J_x,J_y\}\\
\Gamma_4 &= \frac{1}{\sqrt{3}}\{J_x,J_z\}\nonumber\\
\Gamma_5 &= \frac{1}{\sqrt{3}}\{J_y,J_z\}\nonumber
\end{align}
These matrices satisfy the anticommutation relations $\{\Gamma_\mu,\Gamma_\nu\} = 2\delta_{\mu\nu}$. The matrices $U_i$ are defined as
\begin{align}
U_x &= \frac{1}{2i}(\sqrt{3}[\Gamma_1,\Gamma_5] - [\Gamma_2,\Gamma_5])\nonumber\\
U_y &= \frac{-1}{2i}(\sqrt{3}[\Gamma_1,\Gamma_4] + [\Gamma_2,\Gamma_4])\\
U_z &= \frac{1}{i}[\Gamma_2,\Gamma_3]\nonumber
\end{align}
%
The unperturbed Hamiltonian (at zero magnetic field) has time-reversal and $T_d$ symmetry. The coefficient $D$ represents an inversion breaking term. 

We seek an expression $Q(\vec{B})$ that indicates true chirality. As discussed in the previous section, $Q(\vec{B})$ must be even(odd) under symmetries that maintain(reverse) the chirality of a Weyl cone.
The expression $B_x B_y B_z$ represents false chirality because it is odd under time-reversal and even under $S_4T$, where $S_4$ is the rotoinversion symmetry about the $z$ axis that maps $(x,y,z) \mapsto (y,-x,-z)$.
The lowest order polynomial of $B$ that breaks true chiral symmetry is $Q(\vec{B}) = (B_x^2 - B_y^2)(B_y^2 - B_z^2)(B_z^2 - B_x^2)$, as listed in Table~\ref{QB}. It is non-zero only when the magnetic field is away from high symmetry axes such as [100], [110], and [111]. 
Because the magnetic field itself lifts the degeneracy at the $\Gamma$ point and generates Weyl points, in what follows, we will use the dimensionless quantity
\begin{equation}\label{qprime}
    Q'(\hat{B}) = (B_x^2 - B_y^2)(B_y^2 - B_z^2)(B_z^2 - B_x^2)/B^6
\end{equation}
to measure how much chiral symmetry is broken, i.e. how much the left and right handed Weyl points might differ by each other.  $Q'(\hat{B})$ takes a maximum value of $ 0.0962$ when $\hat{B} = (0, 0.460, 0.888)$ or points in a symmetry-related direction.


The Hamiltonian defined by Eq.~\eqref{Ham} is a lowest order expansion in $k$ that is only valid for a finite range of crystal momentum and energy; at higher energies, mixing with other bands becomes important. 
Therefore, when we search for Weyl points in this Hamiltonian with an applied magnetic field, we choose a cut-off in crystal momentum and energy and only focus on Weyl points that are within this cutoff.
We ensure that the cutoff always contains the same number of left- and right-handed Weyl points.

We will now focus on the zincblende material InSb.
It has six bands closest to the Fermi level, consisting of a set of four valence bands that meet at $\Gamma$ very close to $E=0$ and a set of two conduction bands separated by a bandgap of 235 meV. 
The set of four is modeled by the Hamiltonian in Eq.~\eqref{Ham} with parameters as given in Table~\ref{parameters} \cite{cano2017,qu2016,nilsson2009,vurgaftman2001}.
\begin{table}[htp]
\centering
\begin{tabular}{lr}
\hline\hline
   $A$ & 8.85 eV \AA $^2$\\
   $C$ & 1.42 eV \AA $^2$\\
   $D$ & 0.01 eV \AA\\
   $F$ & -22 eV \AA $^2$\\
   $g$-factor & $51$\\
\hline\hline
\end{tabular}
\caption{\label{parameters} Hamiltonian parameters in Eq.~\eqref{Ham} corresponding to InSb.}
\end{table}

\section{Weyl fermions in indium antimonide in a magnetic field}
When a magnetic field is applied along a general low-symmetry direction, the induced Weyl points are not related to each other by any symmetry. There is no analytical expression for their positions and each one must be located numerically by diagonalizing Eq.~\eqref{Ham}. 
We will search for Weyl points within the cutoffs $k < 0.032$\AA$^{-1}$ and $E < 50\mathrm{meV}$. We focus on magnetic fields $\ge 0.5\mathrm{T}$, because for very small magnetic fields, the separation bewteen the Weyl points in momentum space and difference between the energies of the Weyl points and other bands are too small.  For all magnitudes and directions of magnetic field considered in our work, there is an equal number of left- and right-handed Weyl cones within these cutoffs. 

In addition to $Q'(\hat{B})$, we characterize the breaking of chiral symmetry by a dimensionless parameter 
\begin{equation}\label{deltav}
\delta_v = \sum\chi v_1v_2v_3/\sum v_1v_2v_3, 
\end{equation}
where the product of velocities $v_1 v_2 v_3$ is defined in Eq.~\eqref{v1v2v3} and the sum is over all the cones. We also characterize by how much chiral symmetry is broken by a dimensionful parameter 
\begin{equation}\label{deltaE}
\delta_E = \sum\chi E/n    
\end{equation}
where $n$ is the total number of right-handed Weyl cones (equal to the total number of left-handed Weyl cones). Physically, $\delta_v$ represents the average difference in velocities of the left- and right-handed Weyl cones normalized by the average velocity, while $\delta_E$ represents the average difference in energy between left- and right-handed Weyl cones.

Because $\delta_v$ is a dimensionless parameter, we expect it to vary strongly with $Q'(\hat{B})$ (which depends only on the direction of $\vec{B}$) and weakly, if at all, with the magnitude of $
\vec{B}$. However $\delta_E$, which is a dimensionful parameter, is expected to vary strongly with both $Q'(\hat{B})$ and the magnitude of $\vec{B}$.

While $Q'(\hat{B})$ serves as a quick check to determine which low-symmetry directions are likely to break chiral symmetry the most, $\delta_v$ and $\delta_E$ are directly related to known physical observables  \cite{dejuan17,yuta2018,meyer2018}. Note that $Q'(\hat{B})\neq 0$ is a necessary condition for the left- and right- handed cones to have different energies and velocities, and thus for $\delta_v$, $\delta_E$, and relevant physical observables to be non-zero.

In Table~\ref{detailedtable}, we have listed the energy, momentum, chirality, velocity and tilt of all ten Weyl cones that appear for a magnetic field of 0.75 T along the low symmetry direction $[147]$. The parameter $Q'(\hat{B})$ for this direction is 0.0826, close to the maximum possible value of 0.0962. 
For this field, we observe three type I Weyl cones and seven type II Weyl cones: unexpectedly, the type I (or type II) Weyl cones do not come in pairs of opposite chirality, which is only possible in asymmetric Weyl materials.

In Table~\ref{magtable}, we have listed the number of type I and type II cones of each chirality, and the parameters $\delta_v$ and $\delta_E$, for different magnitudes of magnetic field along the $[147]$ direction.
The values show that $\delta_v$, the dimensionless normalized average velocity difference between left- and right-handed Weyl cones, increases slightly with increasing magnitude of the magnetic field, even while $Q'(\hat{B})$ remains constant. The average energy difference, $\delta_E$, increases strongly with increasing field.
Thus, we expect physical observables that depend on a difference in energy or velocity between Weyl cones will increase as the magnetic field is increased along this direction.

In addition, Table~\ref{magtable} also shows that while the number of left and right handed cones of each chirality remains the same as we change the magnitude of the magnetic field, the number of type I (and type II) right handed cones changes. Type I and type II cones have very different Fermi surfaces as shown in Figure \ref{types}. When the tilt parameter is close to $1$, even a small change in the parameters of the Hamiltonian results in a drastic change in the Fermi surface.

The same quantities are recorded in Table~\ref{dirtable} for different directions of magnetic field and fixed magnitude 0.75~T.
Here the three quantities that characterize breaking of chirality symmetry, $Q'(\hat{B})$, $\delta_v$ and $\delta_E$ can be compared: $\delta_v$ and $\delta_E$ show similar trends as $Q'(\hat{B})$, but are not functions of $Q'(\hat{B})$.
Because of the complexities of the band structure and topology, they do not even necessarily vary monotonically with $Q(\vec{B})$ or $Q'(\hat{B})$: they can change sign or even be zero for some low-symmetry directions of $\vec{B}$.
Both Tables~\ref{magtable} and \ref{dirtable} show that it is not unusual to find different numbers of left- and right-handed Weyl cones of the same type in this model.


%
%

\begin{figure}[htp]
     \centering
         \includegraphics[width=0.3\textwidth]{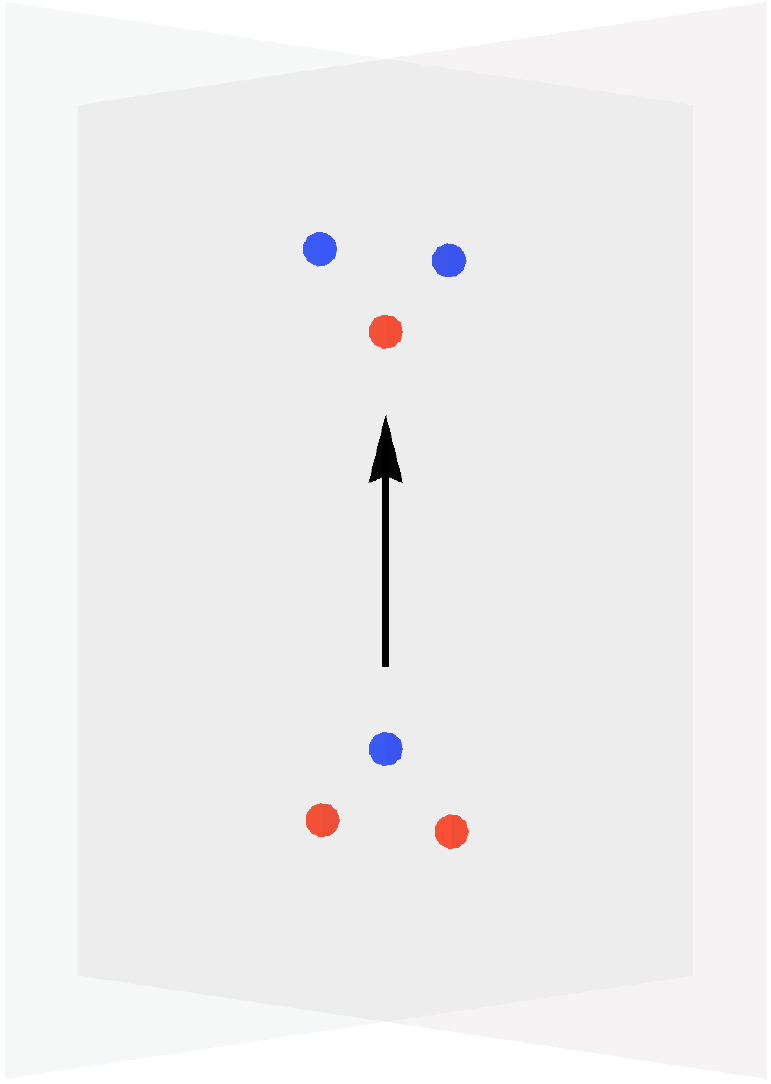}
         \includegraphics[width=0.3\textwidth]{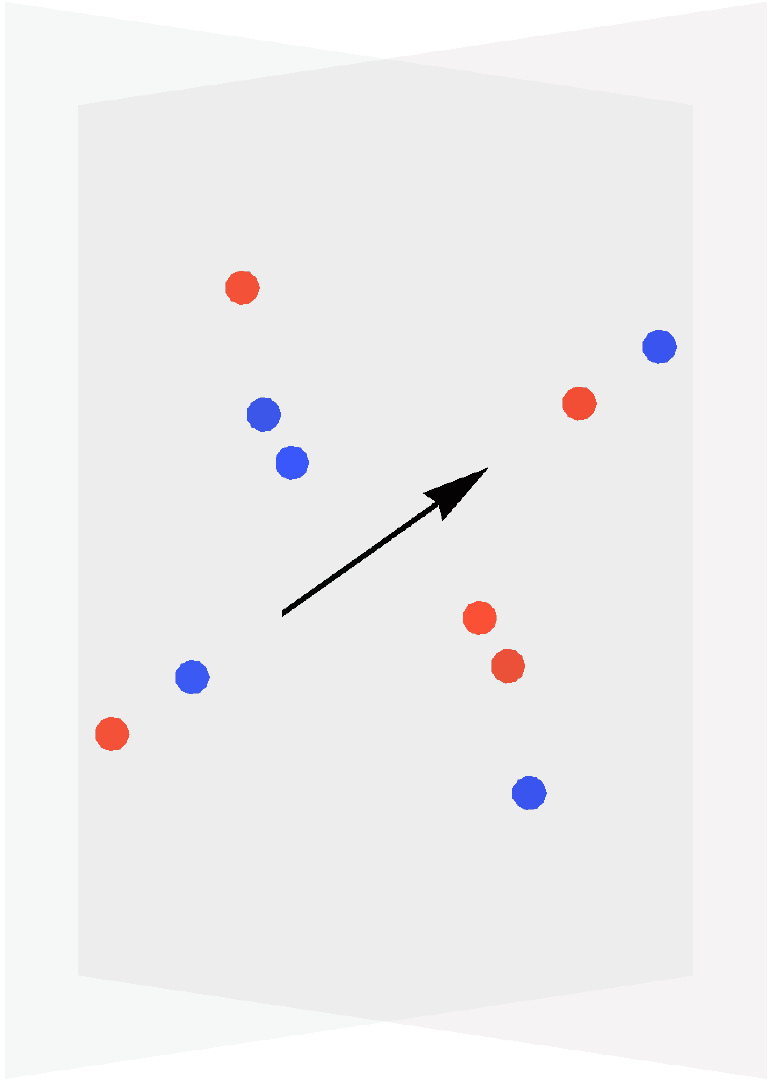}
         \includegraphics[width=0.3\textwidth]{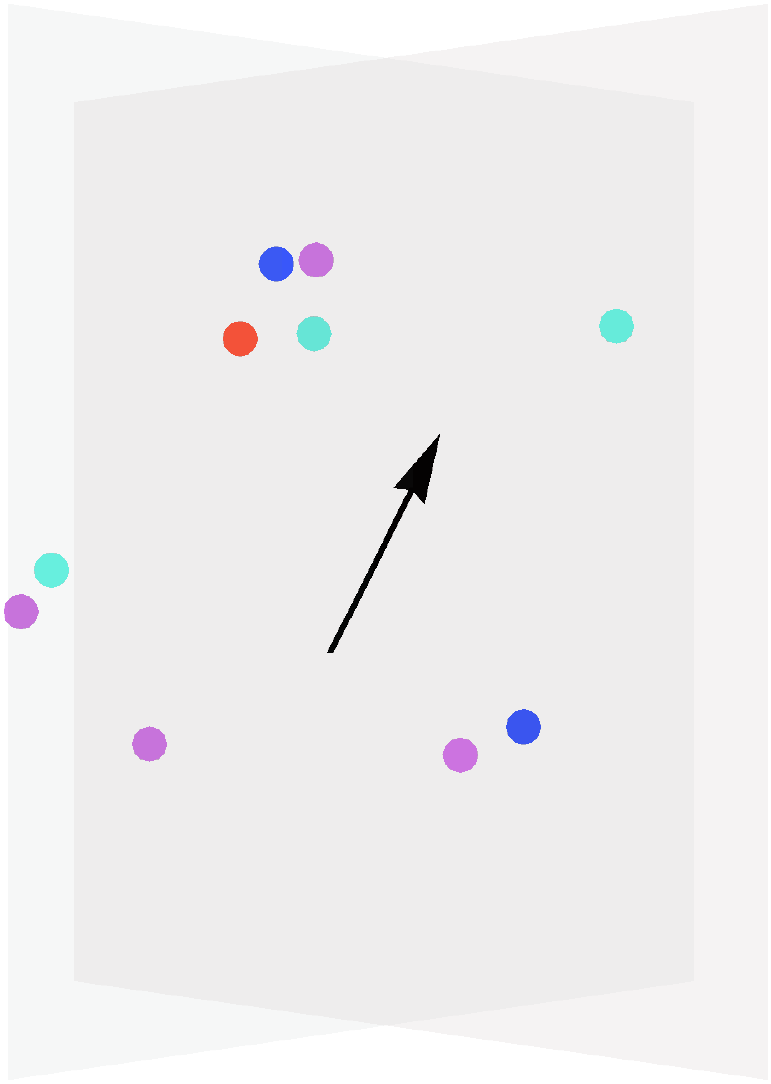}
        \caption{Distribution of Weyl points in momentum space for a magnetic field of 0.75 T along high symmetry directions $[001]$ (left) and $[111]$ (center) and the low symmetry direction $[147]$ (right). The arrow denotes the direction of magnetic field. The color represents the chirality and type of the cones: red indicates left-handed type I, magenta indicates left-handed type II, blue indicates right-handed type I, and cyan indicates right-handed type II. For the low symmetry direction $[147]$, there is no symmetry relating the cones.}
        \label{fig:cones}
\end{figure}

As an example, the positions of cones of different types and chiralities for a magnetic field of 0.75 T along the directions $[001]$, $[111]$, and $[147]$ are shown in Figure~\ref{fig:cones}.



\begin{table}[htp]
\centering
\begin{tabular}{rrrcrcrrc}
\hline\hline
\multicolumn{1}{c}{$k_x$} & \multicolumn{1}{c}{$k_y$} & \multicolumn{1}{c}{$k_z$} & \multicolumn{1}{c}{$\chi$} & \multicolumn{1}{c}{Energy} & \multicolumn{1}{c}{Bands} & \multicolumn{1}{c}{$v_1 v_2 v_3$} & \multicolumn{1}{c}{Tilt} & \multicolumn{1}{c}{Type} \\
\multicolumn{3}{c}{\multirow{2}{*}{($10^8\ \mathrm{m}^{-1}$)}} &  \hspace{0.5cm} & \multicolumn{1}{c}{\multirow{2}{*}{(meV)}} & & \multicolumn{1}{c}{$(\mathrm{eV^3}$} & \multicolumn{1}{c}{} & \multicolumn{1}{c}{} \\
\multicolumn{6}{c}{} & \multicolumn{1}{c}{$\mathrm{pm^3})$} & \multicolumn{2}{c}{} \\
\hline
-0.229 & -0.569 & 0.782  & L & -1.36 & 1-2 & 5830  & 0.706 & I \\
-0.755 & -0.572 & -0.802 & L & 1.82  & 2-3 & 13200 & 1.19  & II \\
0.812  & -0.432 & -0.782 & L & 1.32  & 2-3 & 9350  & 1.20  & II \\
-0.083 & -0.302 & 1.093  & L & 0.20  & 2-3 & 1780  & 1.87  & II \\
-2.114 & -0.120 & -0.304 & L & 11.72 & 3-4 & 12.6  & 333   & II \\
-0.016 & -0.579 & 1.058  & R & 0.10  & 2-3 & 2950  & 0.585 & I \\
0.310  & 0.462  & -0.733 & R & -1.59 & 1-2 & 4390  & 0.710 & I \\
0.735  & 0.533  & 0.837  & R & 1.81  & 2-3 & 11400 & 1.32  & II \\
-0.762 & 0.344  & 0.865  & R & 1.30  & 2-3 & 6430  & 1.57  & II \\
-1.648 & -0.335 & -0.119 & R & 7.45  & 3-4 & 44.7  & 40.1  & II\\     
\hline\hline
\end{tabular}
\caption{\label{detailedtable} Weyl cones for a magnetic field of $0.75\mathrm{T}$ along the low symmetry direction $[147]$. 
The first three columns specify the crystal momentum of the Weyl point; the next columns indicate its chirality; its energy; the two bands that comprise it, where band 1 has the lowest energy; its product of velocities; its tilt; and whether it is type 1 or type II.
The chirality $\chi$, product of velocities $v_1 v_2 v_3$ and tilt are defined in Eqs.~\eqref{chi},~\eqref{v1v2v3}, and~\eqref{tilt}, respectively.
}
\end{table}

\begin{table}[htp]
\centering
\begin{tabular}{rccccrr}
\hline\hline
\multicolumn{1}{c}{$B$} & \multicolumn{2}{c}{Type I} & \multicolumn{2}{c}{Type II} & \multicolumn{1}{c}{$\delta_v$} & \multicolumn{1}{c}{$\delta_E$} \\
\multicolumn{1}{c}{(T)} & \multicolumn{1}{c}{Left} & \multicolumn{1}{c}{Right} & \multicolumn{1}{c}{Left} & \multicolumn{1}{c}{Right} & \multicolumn{1}{c}{} & \multicolumn{1}{c}{(meV)} \\
\hline
0.500                   & 1                          & 2                           & 4                           & 3                            & -0.0820                          & -0.39                                \\
0.625                 & 1                          & 2                           & 4                           & 3                            & -0.0863                          & -0.64                                \\
0.750                  & 1                          & 2                           & 4                           & 3                            & -0.0907                          & -0.92                                \\
0.875                 & 1                          & 2                           & 4                           & 3                            & -0.0953                          & -1.25                                \\
1.000                     & 1                          & 1                           & 4                           & 4                            & -0.1012                          & -1.61     \\                          
\hline\hline
\end{tabular}
\caption{\label{magtable} Number of left- and right-handed cones of each type and parameters that characterize by how much chiral symmetry is broken for variable magnetic field $B$ along the low symmetry direction $[147]$. The parameters $\delta_v$ and $\delta_E$ are defined by Eq.~\eqref{deltav} and Eq.~\eqref{deltaE} respectively.}
\end{table}

\begin{table}[htp]
\centering
\begin{tabular}{lrccccrr}
\hline\hline
\multicolumn{1}{c}{Dir} & \multicolumn{1}{c}{$Q'(\hat{B})$} & \multicolumn{2}{c}{Type I} & \multicolumn{2}{c}{Type II} & \multicolumn{1}{c}{$\delta_v$} & \multicolumn{1}{c}{$\delta_E$} \\
& \multicolumn{1}{c}{} & \multicolumn{1}{c}{Left} & \multicolumn{1}{c}{Right} & \multicolumn{1}{c}{Left} & \multicolumn{1}{c}{Right} & \multicolumn{1}{c}{} & \multicolumn{1}{c}{(meV)} \\
\hline
{[}111{]} & 0                     & 5                          & 5                           & 0                           & 0                            & 0                               & 0                                    \\
{[}345{]} & 0.00806               & 2                          & 3                           & 2                           & 1                            & 0.00042                         & 0.0034                               \\
{[}123{]} & 0.0437                & 1                          & 2                           & 4                           & 3                            & -0.0175                          & -0.26                                \\
{[}147{]} & 0.0826                & 1                          & 2                           & 4                           & 3                            & -0.0907                          & -0.92                                \\
{[}001{]} & 0                     & 3                          & 3                           & 0                           & 0                            & 0                               & 0    \\
\hline\hline
\end{tabular}
\caption{\label{dirtable} Number of left- and right-handed cones of each type and parameters characterize by how much chiral symmetry is broken for magnetic field of magnitude 0.75 T along different directions (Dir). The parameters $Q'$, $\delta_v$ and $\delta_E$ are defined by Eq.~\eqref{qprime}, Eq.~\eqref{deltav} and Eq.~\eqref{deltaE} respectively. }
\end{table}

\section{Outlook}

Asymmetric Weyl materials are sought-after to observe effects that require breaking of chiral symmetry, such as the quantized circular photogalvanic effect~\cite{dejuan17}, the helical magnetic effect~\cite{yuta2018}, and the chiral magnetic effect without an external source of chirality~\cite{meyer2018}. Yet few of these materials have been shown to exist naturally. Therefore, in this work, we proposed inducing an asymmetry between the left- and right-handed Weyl cones in otherwise symmetric materials by an applied field, such as strain or magnetic field, along a low-symmetry direction. We have also provided a prescription for distinguishing true from false chirality, namely, the existence of an operator $Q$ that is even(odd) under all chirality-preserving(chirality-flipping) symmetries of the crystal.

We then studied how to induce true chirality in materials with $T_d$ symmetry. We introduced a parameter $Q'(\hat{B})$ which determines whether chiral symmetry is broken. We applied this analysis to the specific case of InSb. There we showed, by exact diagonalization of a low-energy model, that for a magnetic field along low symmetry directions, the energies, velocities, and tilts are different for left- and right-handed Weyl cones. We also introduced the parameters $\delta_v$ and $\delta_E$ to quantify the asymmetry between the left- and right-handed cones. The differences in energy will lead to the quantized circular photogalvanic effect, while differences in tilt and velocities will lead to the chiral magnetic effect without external source of chirality and the helical magnetic effect. 


For several directions and magnitudes of the magnetic field, the number of left- and right-handed type I cones (and number of left- and right-handed type II cones) are different. Of course, the total number of left- and right-handed cones remains equal, as required by topology. Since type I cones have a compact Fermi surface, and type II cones have a hyperbolic Fermi surface, this asymmetry between right- and left-handed Weyl cones of the same type will result in a highly nontrivial topology of Berry curvature.

Unlike intrinsically asymmetric Weyl materials, in the materials discussed in this chapter, effects that depend on breaking chiral symmetry can be turned on and off, and flipped in sign, which may be desirable for measuring certain effects. Our results can be generalized to other space groups and different types of symmetry-breaking perturbations.
\chapter{Magnetic Photocurrents in Multifold Weyl Fermions}\
\label{chMHME}

\blfootnote{This chapter is based on \cite{SahalMHME}.}

Asymmetric Weyl materials lack a symmetry that relates Weyl cones of opposite chiralities. 
Thus, left and right handed fermions can have different energies and velocities and, consequently, interact differently with electromagnetic fields, and exhibit effects not possible for symmetric Weyl materials.
For example, asymmetric Weyl materials are predicted to exhibit a quantized Circular Photogalvanic Effect (CPGE), i.e. a photocurrent in the direction of circularly polarized light, when Weyl cones of one chirality are fully Pauli blockaded \cite{dejuan17}. The Pauli blockade is only possible when the left and right handed cones are at different energies, which is why this effect is specific to asymmetric Weyl materials. 

In this chapter, we will study the Helical Magnetic Effect (HME), which predicts a photocurrent in the presence of a magnetic field in a tilted, asymmetric Weyl material \cite{yuta2018}. 
This effect can only occur in the absence of inversion and any mirror reflection (which is possible only in asymmetric Weyl materials) and in the absence of the product of inversion and particle-hole symmetry (which is possible only if the Weyl cones are tilted). 

A challenge in observing the quantized CPGE or the HME is the lack of asymmetric Weyl materials, which must also have tilted Weyl cones to exhibit the HME.
Asymmetric Weyl materials necessarily have a chiral crystal structure \cite{huang2016,HasanRhSi} or magnetic ordering \cite{ray2020} (although it is also possible to engineer an asymmetric Weyl material by applying an external field \cite{SahalAsymm})
and these materials are relatively rare.
However, recently, asymmetric chiral multifold fermions have been discovered in certain compounds with the B20 crystal structure \cite{tang2017multiple,HasanRhSi,sanchez2019topological,schroter2019chiral,RhSiCurrent,xu2020optical}.
Multifold fermions are generalizations of Weyl and Dirac fermions that exhibit either a higher degeneracy or a different topology \cite{bradlyn2016beyond,cano2019multifold}.
The chiral multifold fermions are also asymmetric and thus exhibit a quantized CPGE, which has been observed in experiment \cite{FlickerMultifold,RhSiCurrent,CoSiCurrent}. They have also been predicted to cause a quantized circular dichroism \cite{MandalDichroism}. But since the known multifold materials occur at high-symmetry momenta where a tilt is forbidden by crystal symmetry, naively they should not exhibit the HME.

The purpose of this work is to show that this naive expectation is incorrect: in fact, chiral multifold fermions are an ideal platform to exhibit the HME.
We show that chiral multifold fermions exhibit the HME even in the idealized limit where they have perfect spherical symmetry and a linear dispersion, as long as the Fermi level is not exactly at the degeneracy point. 
In this limit, the HME in multifold fermions takes a particularly simple form and is related to the quantized CPGE by a factor of the inverse of the number of Landau levels involved in the photoexcited transitions.
We plot the HME for both spin-1 and spin-3/2 fermions to explicitly demonstrate our results. 
Away from the idealized symmetric and linear limit, the HME is present, but symmetry breaking terms ruin its quantization.
We illustrate this for a double spin-1/2 fermion which splits into a spin-1 fermion and a trivial fermion by terms that break spherical symmetry; this example is relevant to the multifold fermions found in B20 compounds.
We end with a discussion of the relevance of these results to the experimentally characterized compounds CoSi and RhSi.


\section{Hamiltonians for Chiral Fermions}\label{secHam}

In this section, we review the Hamiltonian and Berry curvature of chiral symmetric and tilted Weyl and multifold fermions.
In the simplest incarnation, a Weyl cone is described by the continuum Hamiltonian
\begin{equation}
	H = \chi v_0\vec{k}\cdot\vec{\sigma},
	\label{eq:H0}
\end{equation}
where the chirality $\chi$ is $+1$ for a right handed cone and $-1$ for a left handed cone. This Hamiltonian yields a linear dispersion $E = \pm v_0k$ and a velocity $\vec{v} = \pm v_0\hat{k}$, where $\pm$ corresponds to the upper/lower band. The chirality of the Weyl cone in Eq.~(\ref{eq:H0}) can be also be defined as $\chi = \mathrm{sgn}(\vec{v}\cdot\vec{s})$ where $\vec{v}$ is the velocity and $\vec{s}$ the (pseudo)-spin of the fermion. This definition is valid for both bands of the Weyl cone: the upper/lower bands have opposite chirality corresponding to velocity aligned/anti-aligned with spin. 
An individual Weyl cone has electrons of fixed chirality; thus, an individual Weyl cone lacks inversion symmetry ($P$), which flips momentum but not spin.
Instead, in a crystal with inversion symmetry, the inversion symmetry operator will exchange Weyl cones of opposite chirality.

Eq.~(\ref{eq:H0}) has spherical symmetry, which is broken by the lattice.
More generally, a Weyl fermion can have tilt and anisotropy, and is described by the Hamiltonian:
\begin{equation}
	H = v^i_a k_i\sigma_a + v^i_t \sigma_0 k_i
	\label{tilted}
\end{equation}
where $v^i_a$ describes the untilted part of the Hamiltonian, which might be anisotropic, and $v^i_t$ describes the tilt. The chirality is $\chi = \mathrm{sgn}(\mathrm{det}\ v^i_a)$.

Weyl points are quantized monopole charges of Berry curvature. 
For a symmetric linear Weyl cone described by Eq.~(\ref{eq:H0}), the Berry curvature is of the form $\vec{\Omega} = \pm \chi \hat{k}/2k^2$, where $\pm$ correspond to the upper/lower band; more generally, the Berry curvature will be anisotropic. Whether isotropic or not, integrating the Berry curvature over a Fermi surface enclosing a single linear Weyl fermion of the form of Eq.~(\ref{eq:H0}) or (\ref{tilted}) yields $2\pi C$, where $C = \chi = \pm 1$ is the Chern number of the Fermi surface. By the Nielsen-Ninomiya theorem \cite{nielsen1983adler}, the total number of left and right handed cones in the Brillouin zone must be equal so that the total Berry flux vanishes.

Although Weyl fermions do not require any crystal symmetry, crystal symmetries can protect the following generalizations of Weyl fermions.
Rotation symmetries protect Weyl fermions with $|C|>1$, which have quadratic- or cubic-dispersions along certain directions \cite{fang2012multi,huang2016,FourWeyl}.

Chiral multifold fermions, which are higher-spin generalizations of Weyl fermions \cite{bradlyn2016beyond,cano2019multifold,FlickerMultifold}, can be protected by symmetry in chiral nonsymmorphic crystals.
A spin-$J$ chiral multifold fermion has $2J+1$ bands. The simplest (spherically symmetric) Hamiltonian for such a fermion is
\begin{equation}
H = \chi v_0 k_i S^i
\label{eq:HMF}
\end{equation}
where $S_i$ are the spin-$J$ matrices. 
Spin-1 (three-fold degeneracy) and spin-3/2 (four-fold degeneracy) flavors are possible in 3D crystals.
In addition, double spin-1/2 (four-fold degeneracy) and double spin-1 (six-fold degeneracy) fermions can also be symmetry-protected.
The Hamiltonian of a double spin-$J$ fermion is of the form
\begin{equation}
H = \chi \tau_0 v_0 k_i S^i
\label{eq:DHMF}
\end{equation}
where the Kronecker product is implied and $\tau_0$ is the $2\times 2$ identity matrix acting on some additional degree of freedom outside of the spin-$J$ multiplet; at least one crystal symmetry must be off-diagonal in the basis of $\tau$ matrices for the multifold fermion to be symmetry-protected.

Since the Hamiltonians in Eqs.~(\ref{eq:HMF}) and (\ref{eq:DHMF}) are spherically symmetric,
each band can be labelled by the projection of spin along momentum, $S_k$. The Chern number of a Fermi surface in a band with spin-projection $S_k$ is $2S_k$; the integral of the Berry curvature over that Fermi surface is $2\pi\times 2S_k$.
As discussed for Weyl fermions, the total Chern number in the Brillouin zone must always be zero. Thus, materials with multifold fermions can have multifold fermions of both chiralities or a multifold fermion of one chirality and the appropriate number of simple Weyl fermions of the other chirality.

In the following, we focus on spin-1 and spin-3/2 fermions.
The calculation of the HME for double spin-$J$ fermions is the same as single spin-$J$ fermions with an additional factor of two.

\section{Symmetry and Magnetic Photocurrent}\label{secSym}

As mentioned above, the HME requires a tilted Weyl cone.
To describe the tilt, we need to introduce charge-conjugation symmetry ($C$), which
maps one electron to another electron with opposite energy, momentum, and angular momentum, but same velocity. Since removal of an electron with velocity $\vec{v}$ and angular momentum $\vec{s}$ is equivalent to the creation of a hole with velocity $\vec{v}$ and angular momentum $-\vec{s}$,  charge-conjugation maps an electron to a hole with opposite chirality. 
This is analogous to the situation in high energy physics, where the antiparticle of a left-handed neutrino is a right-handed antineutrino. 

A \textit{single} untilted Weyl cone cannot have inversion symmetry (as discussed below Eq.~(\ref{eq:H0})) or charge-conjugation symmetry because these symmetries both flip chirality.
But a Weyl fermion whose low-energy bands are linear and whose dispersion relations of electrons and holes are similar, has \textit{approximate} $CP$ symmetry, which is broken by quadratic terms. 
This is analogous to the situation in the Standard Model, where the terms that break $P$ and $C$ are large, but the $CP$ violating terms are very small.

To reiterate, the energy, momentum, angular momentum, and velocity of \textit{electrons} transform under $P$, $C$, and $CP$ symmetries as follows:
\begin{align}
    P:\quad & E \to +E,\quad \vec{k} \to -\vec{k},\quad \vec{s} \to +\vec{s},\quad \vec{v} \to -\vec{v}\nonumber\\
    C:\quad & E \to -E,\quad \vec{k} \to -\vec{k},\quad \vec{s} \to -\vec{s},\quad \vec{v} \to +\vec{v}\\
    CP:\quad & E \to -E,\quad \vec{k} \to +\vec{k},\quad \vec{s} \to -\vec{s},\quad \vec{v} \to -\vec{v}\nonumber
\end{align}
The Hamiltonians of untilted simple Weyl cones (Eq.~(\ref{eq:H0})) and multifold fermions (Eqs.~(\ref{eq:HMF}) and (\ref{eq:DHMF})) satisfy $CP$, while the Hamiltonian of a tilted Weyl cone (Eq.~(\ref{tilted})) does not.

In the quantized CPGE \cite{dejuan17}, circularly polarized light, characterized by its angular momentum $\vec{J}$, produces a current $\vec{j}$, related by:
\begin{equation}
\vec{j} = \beta_l \vec{J}
\end{equation}
Since the current is odd under both $P$ and $C$, and the angular momentum is even under both, the coefficient $\beta_l$ is also odd under $P$ and $C$ but even under $CP$. This is also how a single Weyl cone transforms under $P$ and $C$. This is why the quantized CPGE occurs when there is an imbalance of occupied Weyl cones of each chirality.

\begin{figure}[t]
\centering
\includegraphics[scale=0.36]{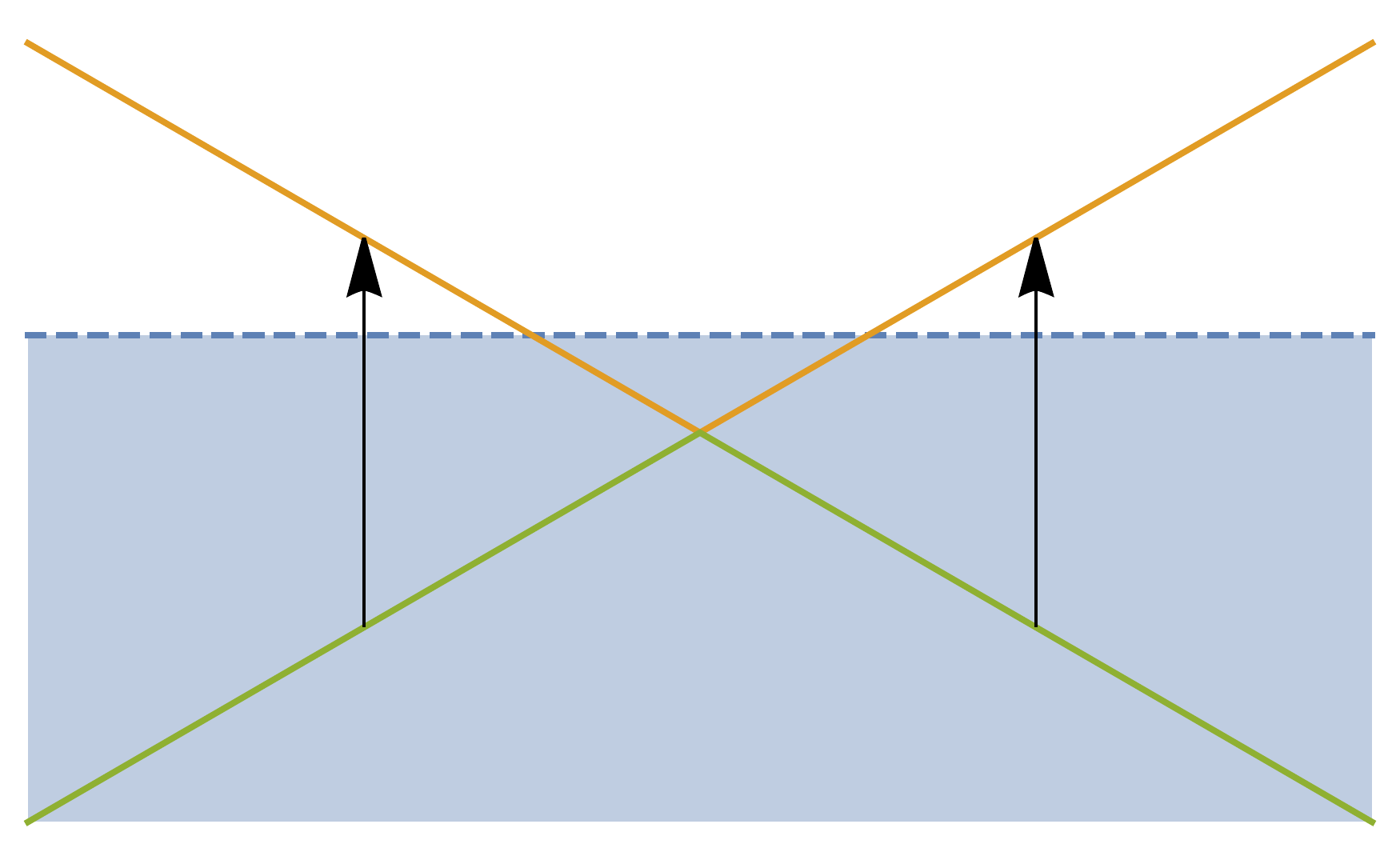}  \includegraphics[scale=0.36]{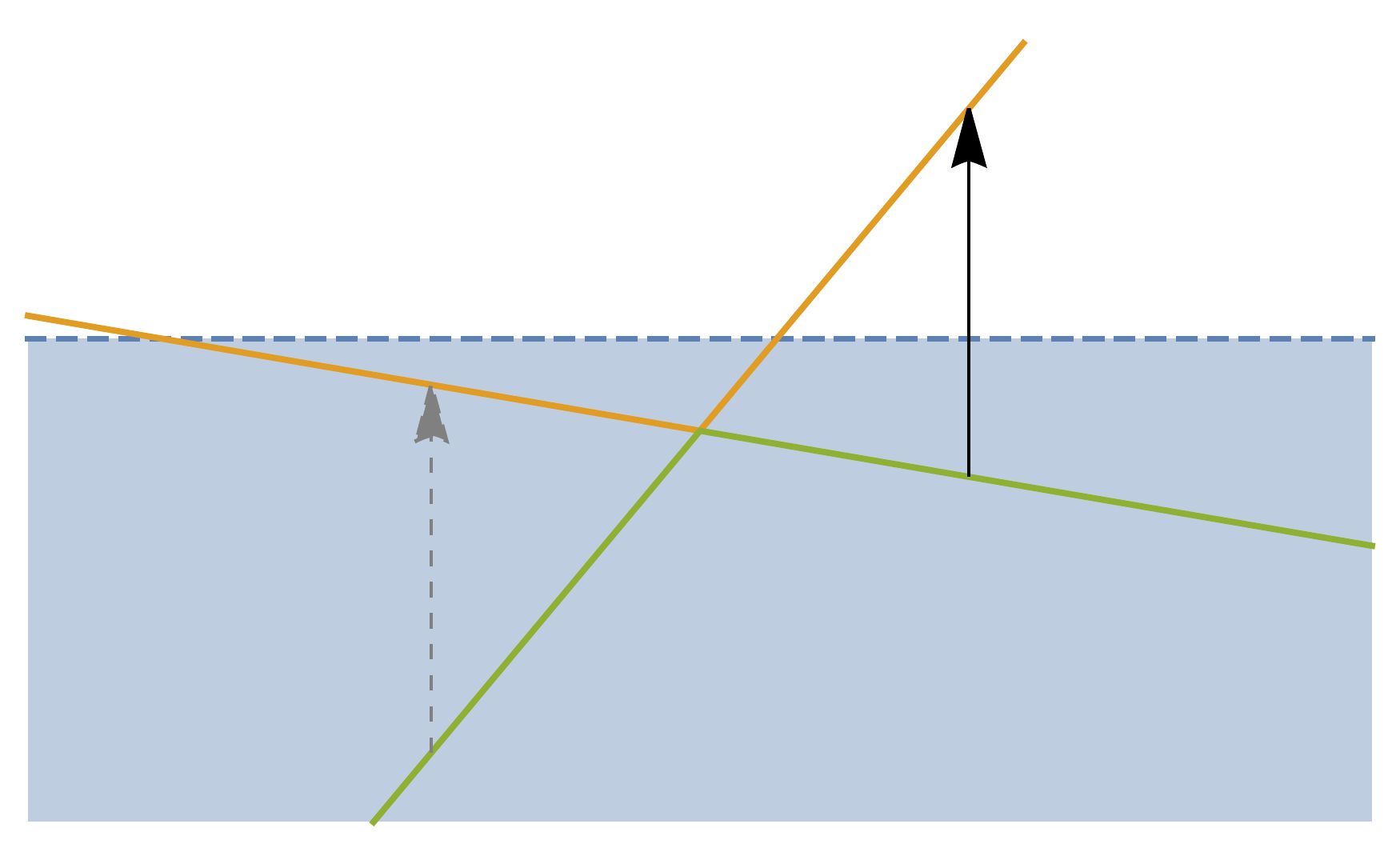}
\caption{Comparison of transitions in an untilted and tilted Weyl cone. In an untilted cone, the initial and final states are related by $CP$ and there is no HME. In the tilted cone, the initial and final states are not related by $CP$, and it is possible to partially blockade the cone. This allows the HME current to be non-zero.}
\label{NormalHME}
\end{figure}

In the HME, linearly polarized light is predicted to produce a current in the presence of a magnetic field \cite{yuta2018}:
\begin{equation}
\vec{j} = \beta_m \epsilon^2 \vec{B}
\end{equation}
where $\vec{\epsilon}$ is the polarization vector. In general, the coefficient $\beta_m$ is a rank-4 tensor. While current is odd under both $P$ and $C$, a magnetic field is even under $P$ and odd under $C$. Thus, the coefficient $\beta_m$ is odd under $P$, even under $C$, and odd under $CP$. 
Consequently, for a Weyl material to exhibit the HME, the (approximate) $CP$ symmetry must be broken by introducing a factor which affects electrons and holes differently, such as a tilt, non-linear terms, or coupling to other bands.
A non-zero Fermi energy is not sufficient to exhibit the CME in an otherwise $CP$-symmetric Weyl material because a transition produces an electron and a hole that are related to each other by $CP$; therefore, this pair cannot contribute to a $CP$-odd coefficient. 
Since a Pauli blockade is uniform for all directions (see Fig.~\ref{NormalHME}) for an untilted Weyl cone, it does not affect the symmetry analysis.

In Ref.~\cite{yuta2018} a Pauli blockade on a tilted Weyl cone was suggested to realize the HME. In a tilted cone, one side has fast electrons and slow holes, and the other has slow electrons and fast holes. A finite Fermi level will blockade one side of the cone, allowing, for example, only transitions that create fast electrons and slow holes, as shown in Fig.~\ref{NormalHME}. This breaks the $CP$ symmetry connecting electrons and holes of opposite chiralities and allows the HME. 

In multifold fermions, because of the lack of tilt, there is still an approximate $CP$ symmetry, which can be seen from Eq~(\ref{eq:HMF}): since $k$ flips sign under $P$ and $S$ remains invariant, the coefficient $\chi v_0$ is odd under $P$. Since holes have opposite energy, momentum, and angular momentum as electrons, $\chi v_0$ also flips sign under $C$, and therefore is invariant under $CP$.
Thus, following the analysis of Weyl fermions, one would naively expect the HME to be absent for chiral multifold fermions with approximate $CP$ symmetry.
However, we will now show by explicit calculation that this is not the case, as long as the Fermi level is not exactly at the band-crossing point.
Heuristically, the non-vanishing HME in the presence of $CP$ symmetry results because there is a unique type of Pauli blockade possible for multifold fermions that is not possible for spin-1/2 Weyl fermions, as shown in Fig.~\ref{blockade}.

\section{HME in Multifold Fermions}\label{secHME}
\label{sec:HMEMF}
\begin{figure}[ht]
\centering
\includegraphics[scale=0.36]{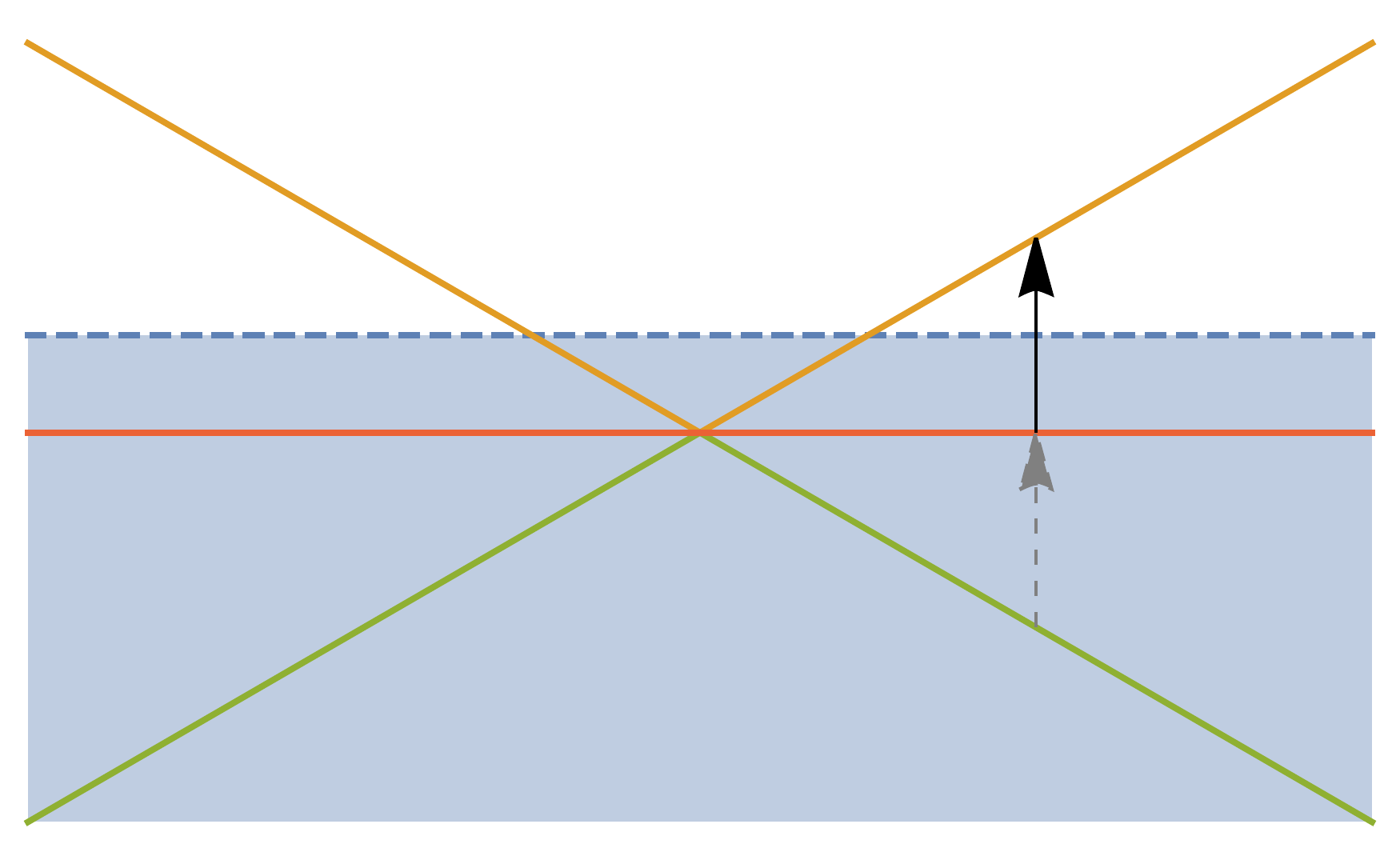}  \includegraphics[scale=0.36]{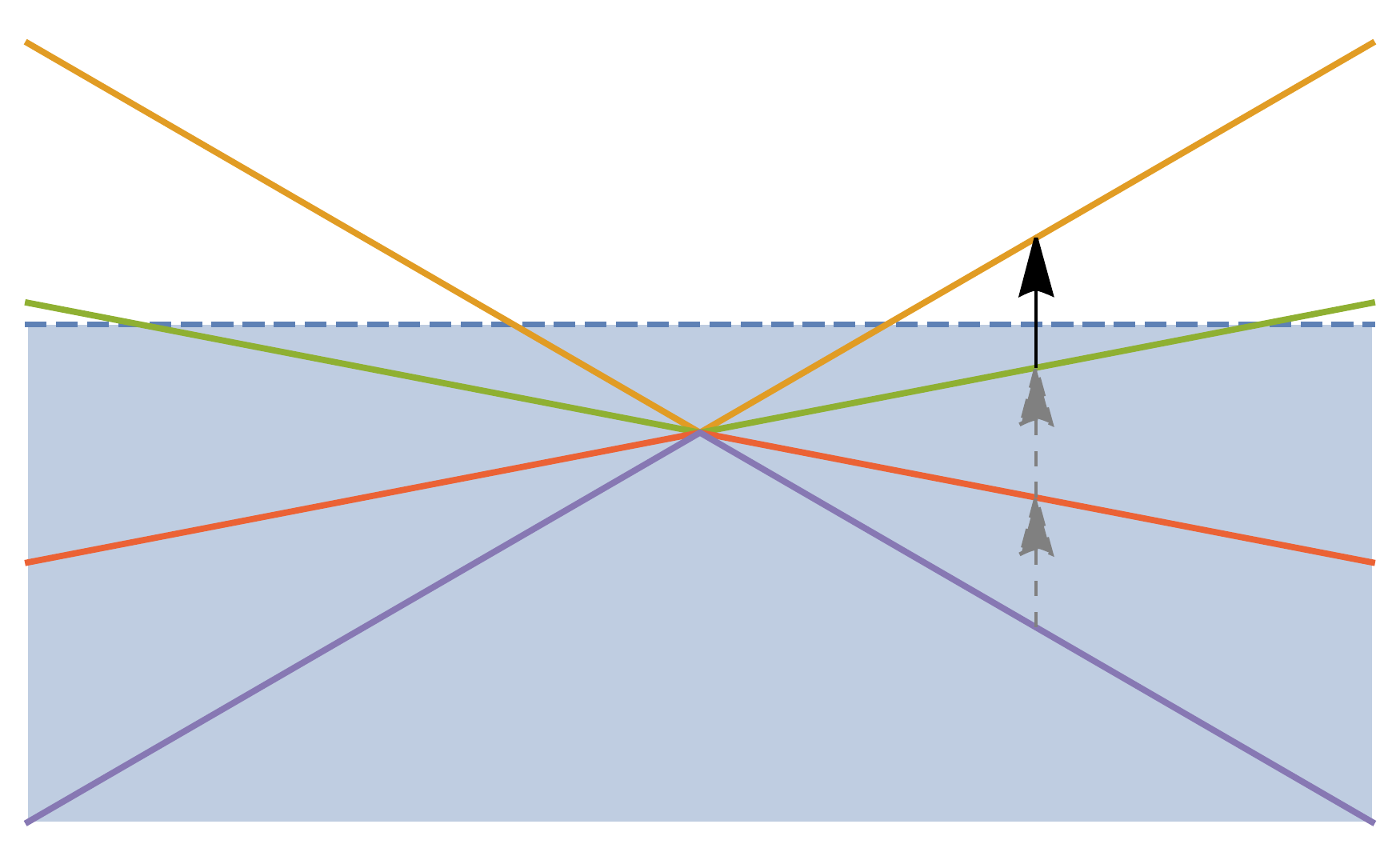}
\caption{Pauli blockade in a spin-1 (left) and spin-3/2 (right) multifold fermion. Solid black arrows indicate allowed transitions from occupied to empty bands, while dashed arrows indicate blockaded transitions. The shaded blue region indicates the Fermi sea. The Pauli blockaded multifold fermions exhibit the HME.}
\label{blockade}
\end{figure}

While multifold fermions described by Eqs~(\ref{eq:HMF}) and (\ref{eq:DHMF}) have $CP$ symmetry, \textit{transitions} between bands with different $|S_k|$ break $CP$ symmetry.
For example, as shown in Fig.~\ref{blockade}, if the Fermi level is such that transitions between bands with $S_k=m$ and $S_k = n \neq \pm m$ are allowed, while transitions to bands with $S_k =-m,-n$ are forbidden, then $CP$ symmetry is ``maximally'' broken compared to tilted simple Weyl cones.
Thus, as long as the Fermi level is not at the charge neutrality point of the multifold fermion, the HME will be present, even though the low-energy theory has $CP$ symmetry. 
This is very different than the situation for Weyl fermions described in the previous section, where the $CP$ symmetry of the Weyl cone must be explicitly broken to exhibit the HME, regardless of the Fermi level.

We now calculate the magnitude of the HME for a multifold fermion.
The effect of chiral Landau levels due to an external magnetic field on the velocity of fermions and the density of phase space can be modeled semiclassically by Chiral Kinetic Theory (CKT) \cite{xiao2005berry,Son2012Berry, stephanov2012chiral, 2013Son}, which prescribes:
\begin{align}\label{CKT}
\vec{v} &\to   \frac{\vec{v}+(\vec{v}\cdot \vec{\Omega}) e \vec{B}}{1+e\vec{\Omega}\cdot  \vec{B}}  \nonumber\\
d^3 k &\to (1+e\vec{\Omega}\cdot  \vec{B}) d^3 k
\end{align}
where the unperturbed velocity is $\vec{v} = \nabla_k\ E$, $\vec{\Omega}$ is the Berry curvature and $\vec{B}$ is the applied field.

If the system is in thermal equilibrium, there is no current. The photocurrent occurs because fermions are excited by photons, and have different velocities in their final states compared to initial states. The total DC current in response to a continuous wave is then given by:

\begin{align}\label{general}
\vec{j} = e\sum^{cones}\sum_{i<j}^{bands}\tau  \iint  & \frac{d^3 \vec{k}_i}{(2\pi)^3} (1+e\vec{\Omega}_i\cdot\vec{B})d^3 \vec{k}_j (1+e\vec{\Omega}_j\cdot  \vec{B})\nonumber\\\times &\  \delta(\vec{k}_i-\vec{k}_j)\delta(E_j - E_i - \hbar\omega) (f_i - f_j)\Gamma_{ij}\nonumber\\ \times &\ \left[\frac{\vec{v}_j + (\vec{v}_j\cdot \vec{\Omega}_j) e \vec{B}}{1+e\vec{\Omega}_j\cdot  \vec{B}} - \frac{\vec{v}_i+  (\vec{v}_i\cdot \vec{\Omega}_i)}{1+e\vec{\Omega}_i\cdot  \vec{B}} e \vec{B}\right] ,
\end{align}

where $\tau$ is the relaxation time, $f_{i,j}$ is the Fermi distribution function at energy $E_{i,j}$, and $\Gamma_{ij}$ is the transition rate from state $i$ to state $j$, given by Fermi's golden rule:
\begin{equation}
    \Gamma = 2\pi |\langle\psi_i|V_{+\omega}|\psi_j\rangle|^2
    \label{golden}
\end{equation}
where $\psi_{i,j}$ are the initial and final states, $V_{+\omega}$ is the perturbation induced by light.
For an ultrafast pulse shorter than the relaxation time, $\vec{j} = \tau (...)$ in Eq.~(\ref{general}) is replaced by $\frac{d\vec{j}}{dt} = (...)$.

In the absence of a magnetic field, circularly polarized light will result in a nonzero photocurrent because its electric field violates time-reversal; this is exactly the CPGE.
Since the electric field of linearly polarized light is time-reversal symmetric, in a material which is also time-reversal symmetric, this integral vanishes for linearly polarized light in the absence of magnetic field. By definition, the helical magnetic effect is defined as the photocurrent contingent on a magnetic field; thus, we now focus only on terms that depend on $\vec{B}$.

In a spherically symmetric multifold fermion, the energy, velocity, and Berry curvature of the band with $\chi S_k = n$  (Chern number $2n\chi$) is:
\begin{align}
E = nv_0k \nonumber\\
\vec{v} = n v_0 \hat{k}
\label{eq:EVO}\\
\vec{\Omega} = n\chi \frac{\hat{k}}{k^2}\nonumber
\end{align}
Eq.~(\ref{general}) then simplifies. The leading (linear order in $\vec{B}$) term is given by:

\begin{align}\label{special1}
\vec{j} = e\sum_{cones}\sum_{m<n}\tau \chi &\int \frac{d^3 \vec{k}}{(2\pi)^3}  \delta((n-m)v_0 k-\hbar\omega)\\& \left[\frac{n^2 v_0}{k^2} e\vec{B} - \frac{m^2v_0}{k^2} e\vec{B}\right](f_m - f_n)\Gamma_{mn},\nonumber
\end{align}
where $\Gamma$ is the unperturbed transition rate. The transition rate 
from lower to upper states for linearly polarized light with $\vec{A} = \vec{A}_{+\omega}\exp(-i\omega t) + \vec{A}_{-\omega}\exp(i\omega t)$ is
\begin{align}
\Gamma_{mn} =\ & 2\pi|\langle\psi_n|ev_0\vec{A}_{+\omega}\cdot\vec{S}|\psi_m\rangle|^2\nonumber\\ 
=\ & 2\pi e^2 v_0^2 A_{+\omega}^2 \sin^2 \theta |S^x_{nm}|^2
\label{eq:Gammamn}
\end{align}
where $\theta$ is the angle between the electric field and the crystal momentum and
\begin{equation}S^x_{nm} = \frac{1}{2}(\delta_{m,n+1}+\delta_{n,m+1})\sqrt{j(j+1)-mn},
\label{eq:Snm}
\end{equation}
which is obtained by the expressing the elements of the spin matrices in the first line of Eq.~(\ref{eq:Gammamn}) in the basis of $\psi_{m,n}$, i.e., the eigenstates of the Hamiltonian in Eq.~(\ref{eq:HMF}). The selection rules defined by $S^x_{nm}$ in Eq.~(\ref{eq:Snm}) only allow transitions with $n-m = \pm 1$. 


Thus, the photocurrent for each cone is
\begin{equation}\label{special2}
\vec{j} = \chi \frac{e^3I\tau}{6\pi\hbar^2\epsilon_0 c}\frac{2e\vec{B}v_0^2}{\hbar\omega^2} \sum_{m,n} (n^2 - m^2) |S^x_{nm}|^2 (f_m - f_n),
\end{equation}
where $I = 2\epsilon_0 c A^2_{+\omega}\omega^2$ is the intensity of the light, and the Fermi distribution function is $f_m = f(m\omega) = 1/[1+\exp((m\omega - \mu)/T)]$, because the $\delta$-function in Eq.~(\ref{special1}) enforces the transition at $k = \omega/v_0(n-m) = \omega/v_0$.

Since the current is summed over all bands, the HME will be non-zero if there are transitions between bands $m$ to $n$, but not between $-m$ to $-n$.
The factor $(n^2 - m^2)|S^x_{nm}|^2$ is $1/2$ for a spin-$1$ fermion and $3/2$ for a spin-$3/2$ fermion.
However, it is zero for a spin-1/2 Weyl cone because $|n| = |m| = 1/2$.
This explains why the HME vanishes for a symmetric, untilted spin-1/2 Weyl fermion.

Note that the HME does not require $\Gamma$ to have any special form. This means it will be non-zero for any non-zero $\Gamma$, including for linearly polarized or even unpolarized light. For a spherically symmetric linear multifold fermion, the photocurrent is always in the direction of the magnetic field, and is completely independent of the linear polarization of light. In the general case, it can have transverse terms and a polarization dependence.

\begin{figure}
    \centering
    \includegraphics[scale=0.6]{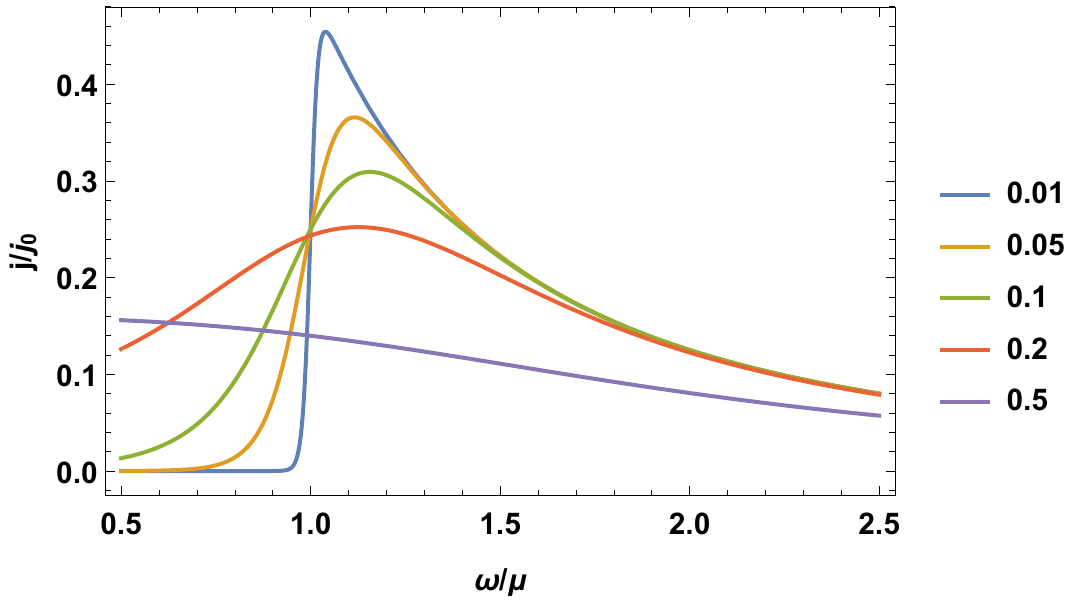} \includegraphics[scale=0.6]{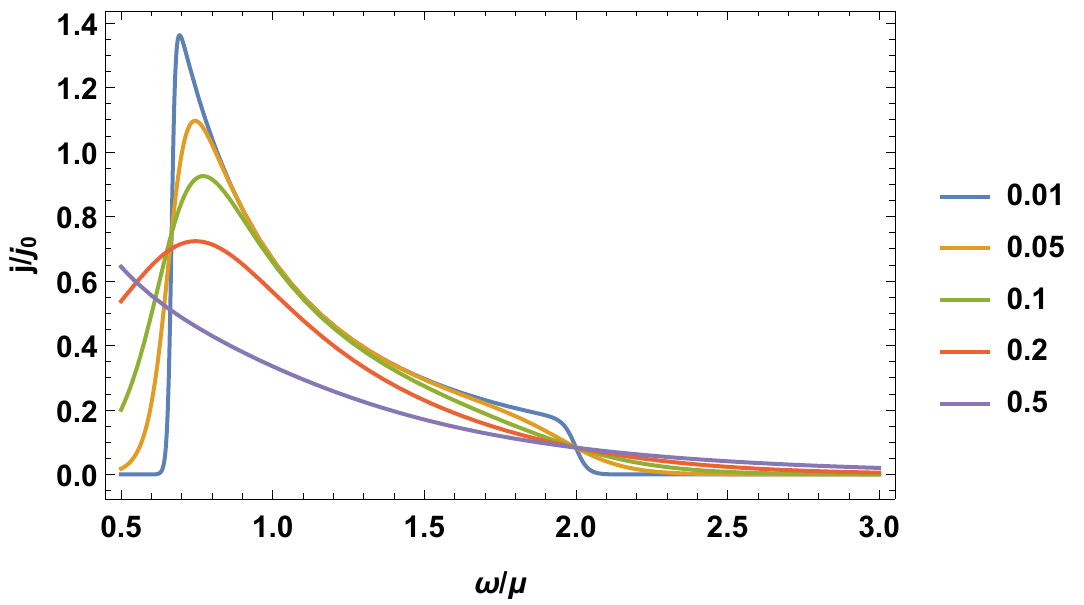}
    \caption{Normalized HME photocurrent vs normalized frequency for a spin-1 cone (left) and a spin-3/2 cone (right), for $T/\mu = 0.01,\allowbreak 0.05, 0.1, 0.2$, and $0.5$. The current is normalized in units of $j_0 = \frac{e^3I\tau}{3\pi\hbar^2\epsilon_0 c}\frac{eBv_F^2}{\hbar\mu^2}$ where $v_F$ is the velocity of the fastest band. The frequency is normalized in units of chemical potential $\mu$.
   At $T= 0$, the current is zero until a critical frequency where the upper band is unoccupied at the momentum required for the transition. In the spin-3/2 case, there is also a second critical frequency where the current drops to zero because both upper bands are unoccupied at the momentum required for the transition.}
    \label{linearHME}
\end{figure}

The normalized photocurrent vs normalized frequency are plotted in Figure \ref{linearHME} at different temperatures for both a spin-1 cone and a spin-3/2 cone.
For the spin-1 fermion, the HME is suppressed at low frequencies by a complete Pauli blocade. At frequencies above the chemical potential $\mu$, transitions from the middle band (which has no Berry curvature) to the upper band are allowed, and there is an HME current, which drops as the inverse square of frequency according to Eq.~(\ref{special2}).

For the spin-3/2 fermion, there are fast bands with velocity $3v_0/2$ and slow bands with velocity $v_0/2$. At frequencies less than $2\mu/3$, all transitions are blockaded, and there is no HME. Above this frequency, transitions are allowed because the slower band remains filled, while the faster band is empty.
Above $2\mu$, transitions from the slow to fast band are again suppressed because both bands are empty. 
Transitions from the lower slow band to the upper slow band are allowed at frequencies above $2\mu$, but they do not contribute to the HME as these bands have opposite Berry curvature and hence $n^2 = m^2$.
For frequencies between $2/3\mu$ and $2\mu$, the HME current decreases as the inverse square of frequency according to Eq.~(\ref{special2}).

At finite temperature, the frequency cutoffs described above become smoothed by the Fermi distribution function.

If the Fermi level is below the band degeneracy point, the photocurrent behaves similarly up to a minus sign because the charge carriers would be holes, not electrons.

The quantized CPGE can also be derived by integrating Eq~(\ref{general}) for circularly polarized light (i.e. $A_{+\omega} \sim \hat{x} + i \hat{y}$); the magnitude of the photocurrent is $\frac{e^3I\tau}{6\pi\hbar^2\epsilon_0 c}  |S^x_{nm}|^2$. The multifold HME has a relative factor of $\frac{2eBv^2}{\hbar\omega^2} (n^2 - m^2)$ vs the quantized CPGE. This means it is inversely proportional to the number of Landau levels involved in the photoexcited transitions.

\section{Non-linearity and Transitions from Multifold Fermions to Other Bands}\label{secNonLin}

In a crystal, multifold fermions do not have full spherical symmetry, only the symmetry of the little group at their crystal momentum.
Non-linear terms will generically be present and break spherical symmetry. Even in cubic crystals, spin-1, spin-3/2, and double spin-1 fermions without time-reversal symmetry can also have \textit{linear} terms that break the approximate spherical symmetry\cite{bradlyn2016beyond,FlickerMultifold}, such as:
\begin{equation}
    a (k_x S_x^3 + k_y S_y^3 + k_z  S_z^3)
\end{equation}
for spin-3/2 fermions;
\begin{equation}
    a \tau_2 (k_x \{S_y,S_z\} + k_y \{S_z,S_x\} + k_z  \{S_x,S_y\})
\end{equation}
for double spin-1 fermions; and
\begin{equation}
     a (k_x \{S_y,S_z\} + k_y \{S_z,S_x\} + k_z  \{S_x,S_y\})
\end{equation}
for spin-1 fermions not at TRIMs or in crystals with broken time-reversal.

In addition, transitions to other bands outside the multifold fermion are possible.
All these effects will contribute to the HME photocurrent, 
causing it to deviate from the idealized form in Eq.~(\ref{special2}).
In these cases, the HME photocurrent can be calculated from the general formula in Eq.~(\ref{general}). 

\begin{figure}
    \centering
    \includegraphics[scale=0.3]{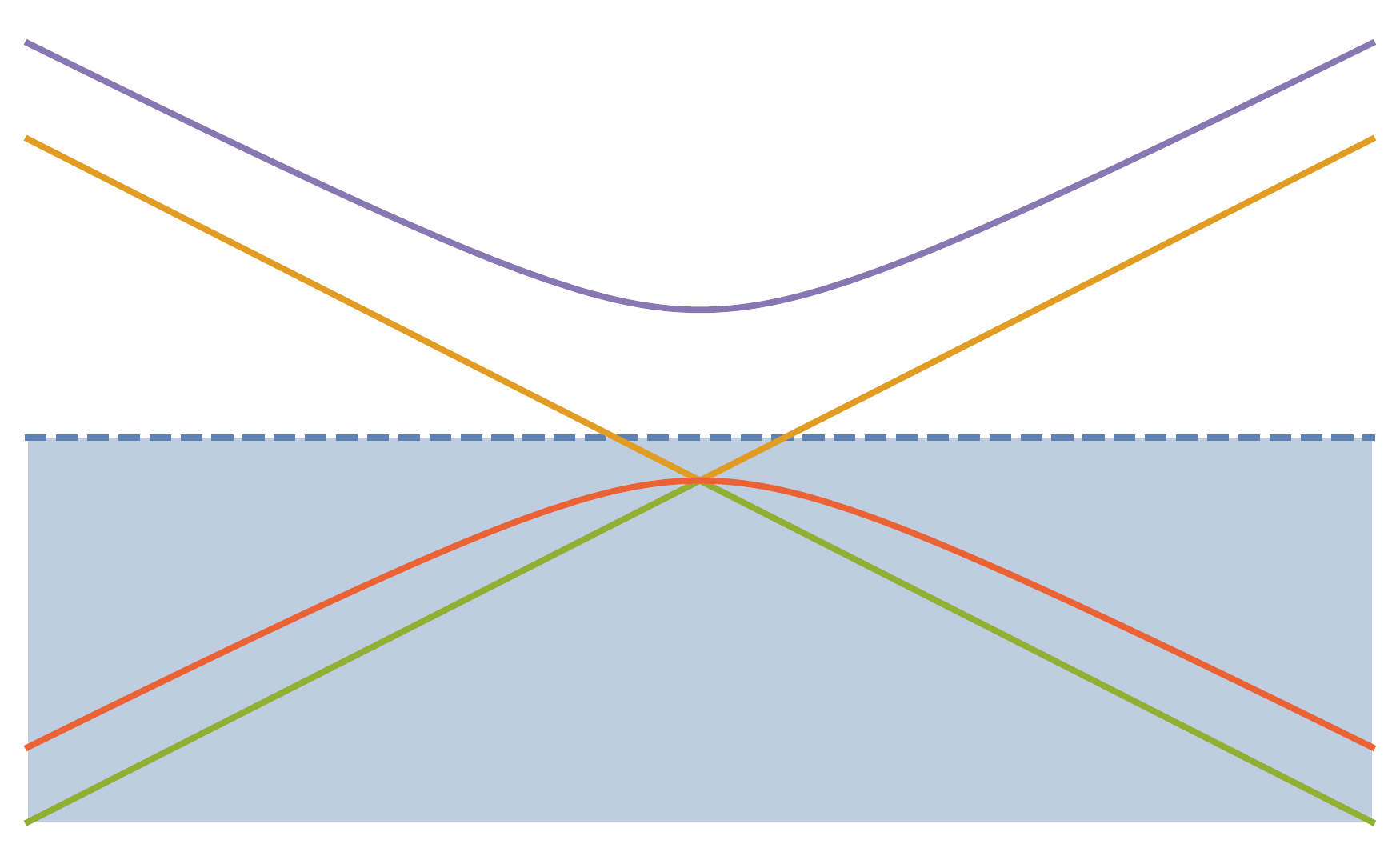}
    \includegraphics[scale=0.6]{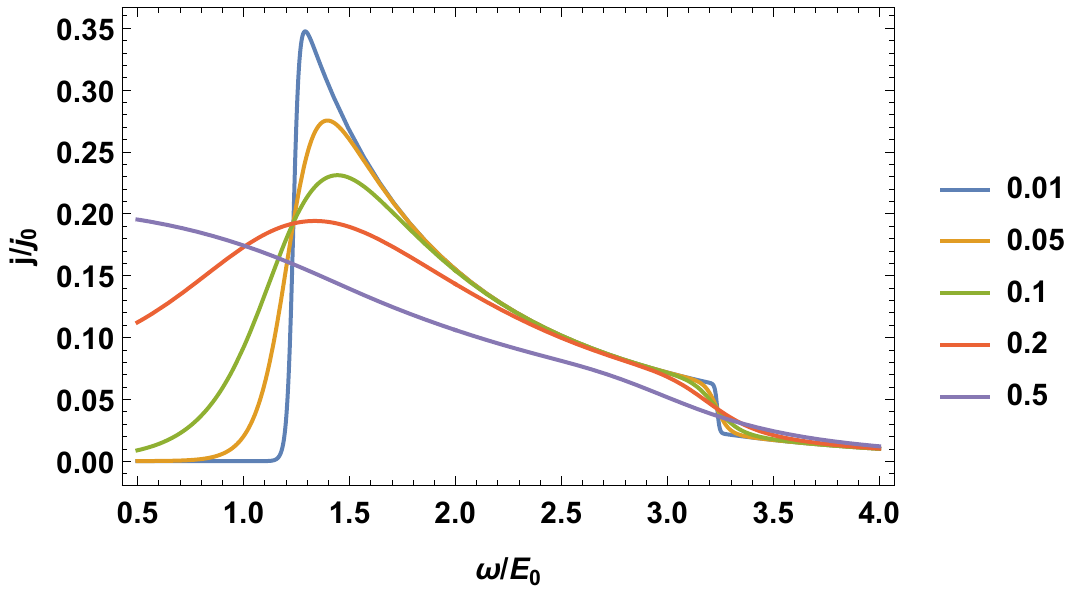}
    \caption{The dispersion relation (left) and normalized photocurrent vs normalized frequency (right) for the double spin-1/2 fermion split into a spin-1 fermion and trivial band, described by the Hamiltonian in Eq.~(\ref{bandgap}). 
    The photocurrent is normalized in units of $\frac{e^3I\tau}{3\pi\hbar^2\epsilon_0 c}\frac{eBv_F^2}{\hbar E_0^2}$ while the frequency is normalized in units of $E_0 = \Delta/4$. The chemical potential is taken to be $\mu = 0$. The polarization is along the magnetic field. The normalized temperature is $T/E_0 =  0.01, 0.05, 0.1, 0.2$, and $0.5$. At $T=0$, the current is zero for small frequencies. Above a critical frequency, the Pauli blockade is lifted only for transitions from trivial to chiral bands, and there is a current. Above a second critical frequency, the Pauli blockade is also lifted for transitions from chiral to trivial bands, and the current is \textit{approximately} cancelled, but is not exactly zero.}
    \label{splitCME}
\end{figure}

To illustrate the effects of non-linearity and transitions to trivial bands, we consider a double spin-1/2 fermion described by the Hamiltonian in Eq~(\ref{eq:DHMF}) plus a small perturbation of the form $\sum\tau_i\sigma_i$ that splits it into a spin-1 and a trivial fermion, as shown in Figure~\ref{splitCME}. Thus, this model is an example of a spin-1 fermion with both non-linear terms and transitions to another band.
This system can be described by a Hamiltonian of the form
\begin{equation}\label{bandgap}
H =  v_0 i \begin{pmatrix}0      & k_x  & k_y  & k_z  \\
-k_x & 0      & k_z  & -k_y \\
-k_y & -k_z & 0      & k_x  \\
-k_z & k_y  & -k_x & 0 \end{pmatrix}
+  \frac{\Delta}{4} \begin{pmatrix}3 & 0  & 0  & 0  \\
0 & -1 & 0  & 0  \\
0 & 0  & -1 & 0  \\
0 & 0  & 0  & -1\end{pmatrix}
\end{equation}

We plot the dispersion relation and HME current vs frequency for this system at different temperatures in Fig.~\ref{splitCME}. We consider the Fermi level to be between the Weyl node and the upper trivial band, i.e. between $-E_0$ and $3E_0$ where $E_0 = \Delta/4$. At low frequencies, all transitions are blockaded, and there is no photocurrent. Above $\omega = \mu  -E_0 + \sqrt{4E_0^2 + (\mu+E_0)^2}$, transitions from the lower trivial band to the upper chiral band are not blockaded, while those from the lower chiral band to the lower trivial band are blockaded, which marks the onset of the photocurrent. The system is similar to an isolated and symmetric spin-1 fermion (Fig.~\ref{linearHME}) and there is a large photocurrent. Above $\omega = \mu  +E_0 + \sqrt{4E_0^2 + (\mu+E_0)^2}$, transitions from the chiral bands to the upper trvial band are also allowed, which contribute a term with opposite sign to the photocurrent, which corresponds to the sharp drop in photocurrent. However, the cancellation is not exact because of non-linearity. At higher frequencies, the photocurrent is strongly suppressed because the system resembles a untilted double spin-1/2 cone. The photocurrent also has a very weak dependence on the polarization of light due to the non-linearity.

\section{Photocurrents and Landau Levels}\label{secLandau}

\begin{figure}
    \centering
    \includegraphics[scale=0.11]{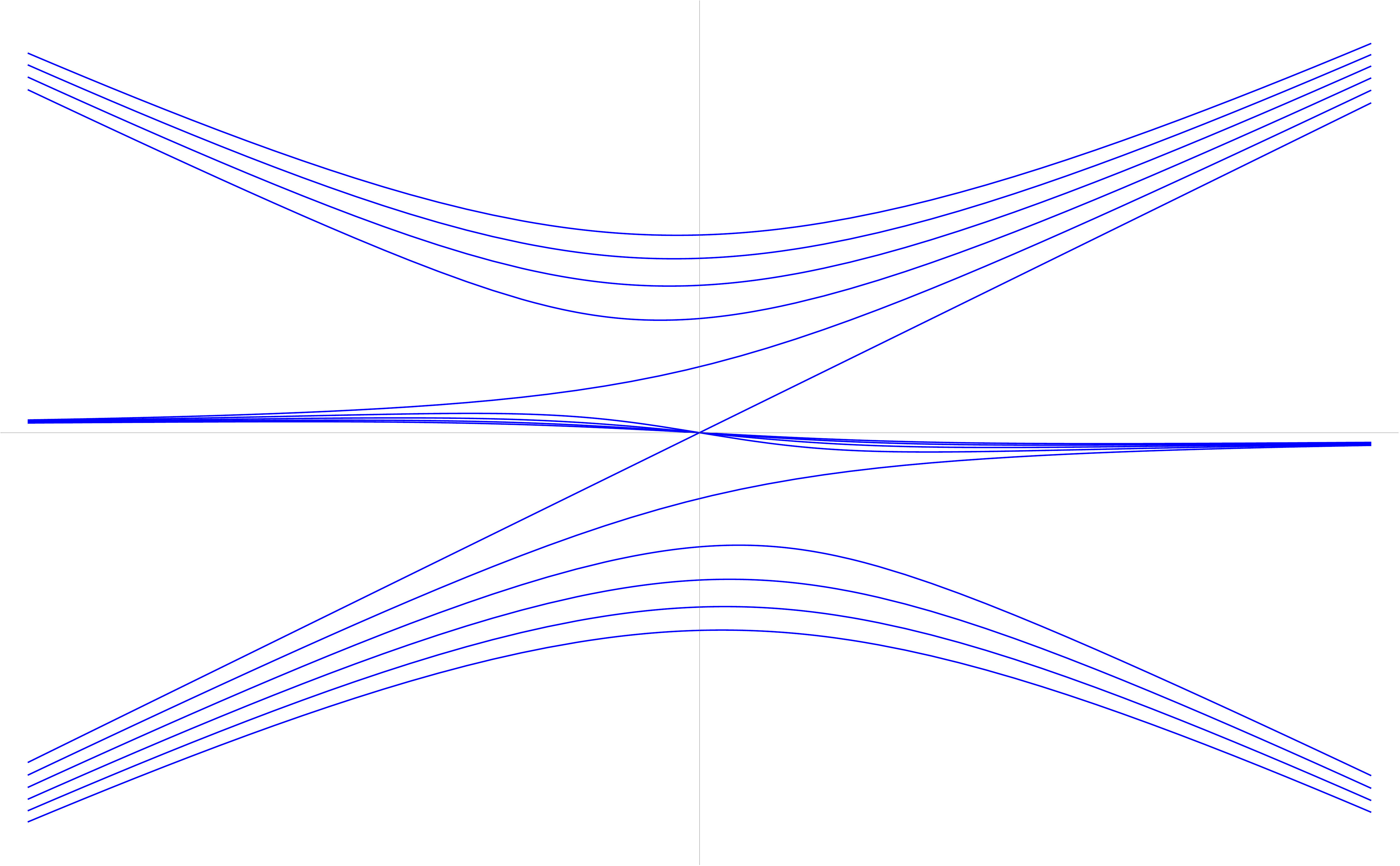}
    \includegraphics[scale=0.099]{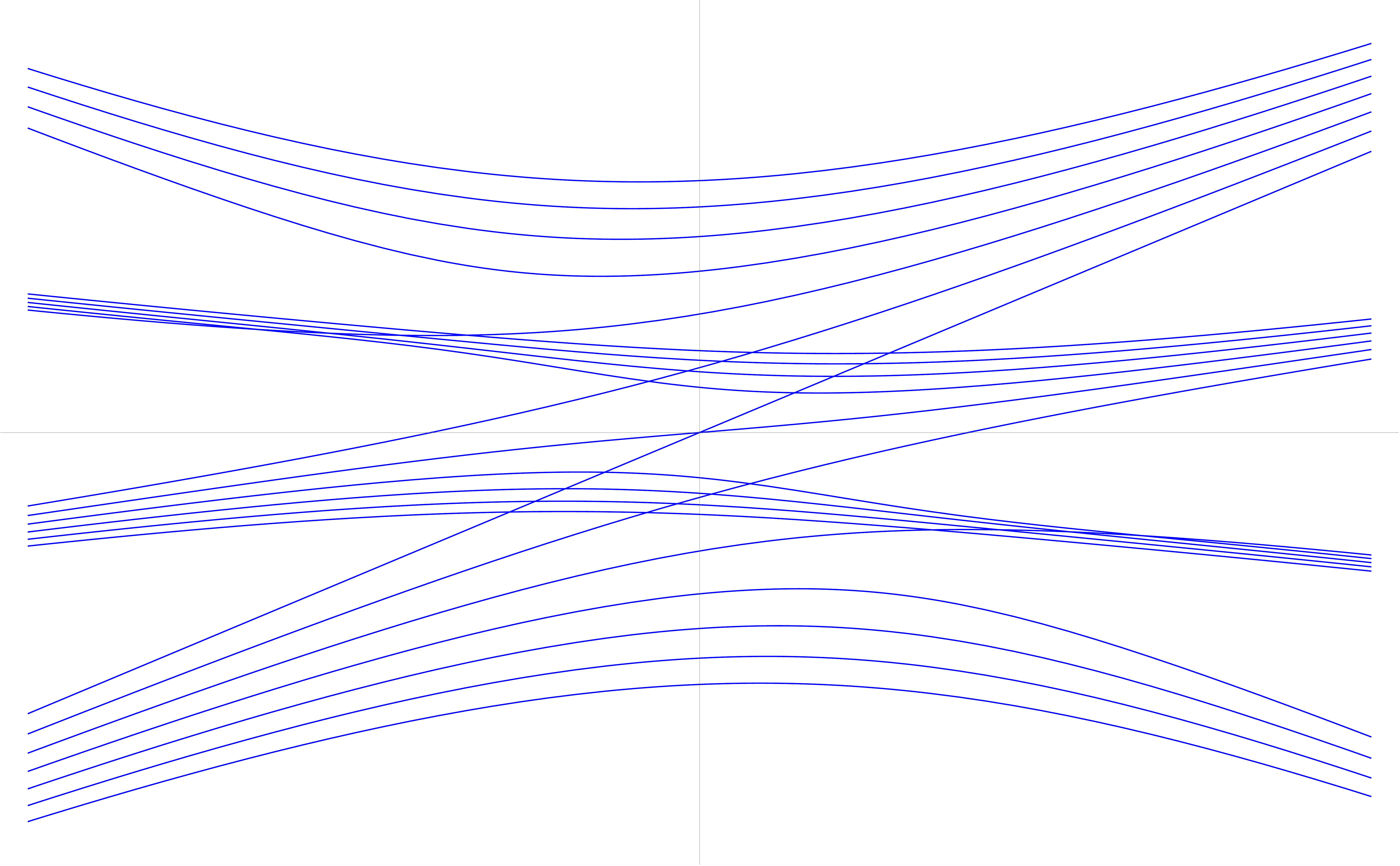}
    \caption{The chiral Landau levels and first five achiral Landau levels for a spin-1 fermion (left) and a spin-3/2 fermion (right).}
    \label{LL}
\end{figure}

The factor of $(n^2 - m^2) \frac{2eBv_0^2}{\omega^2}$ that appears in Eq.~(\ref{special2}) can also be understood from the Landau level spectrum of a multifold fermion.
In a magnetic field, chiral fermions exhibit chiral Landau levels, which propagate only in one direction.
Each band contributes a number of chiral Landau levels equal to its Chern number.
The Landau levels of a spin-1 fermion and a spin-3/2 fermion are illustrated in Fig.~\ref{LL}. 
The chiral anomaly can be interpreted as a consequence of these unpaired chiral Landau levels.

The effects of the unpaired Landau levels in the semiclassical calculation are captured by the deformation of phase space in Eq.~(\ref{CKT}). When the temperature or inverse scattering time is larger than the Landau splitting, we can ignore the quantum oscillations and focus only on the deformation of phase space \cite{2015gustavo}.
Here we show that the same scaling can be obtained for a spherically symmetric multifold fermion by counting the density of states in each chiral Landau level.

When the fermions are in thermal equilibrium, the total current vanishes because the Chern numbers of left and right handed cones cancel. 
However, a net current is possible when we disturb the distribution function by shining light.

Consider a spherically symmetric linear multifold fermion in a magnetic field along the $\hat{z}$ direction.
A band with $S_k = n$ contributes $2n$ chiral modes, corresponding to its Chern number.
In the semiclassical limit, the velocity of each chiral mode is $n\chi v_0$ along the direction of the magnetic field, the same as the speed of the unperturbed band (see Eq.~(\ref{eq:EVO})). 
The density of states per area within each Landau level in the $x-y$ plane is $eB/2\pi$; each Landau level corresponds to an area of $2\pi eB$ in the $k_x-k_y$ plane. 
The fermions that participate in transitions excited by light of frequency $\omega$ are located on a sphere in momentum space of radius $k = \omega/v$.
The number of Landau levels involved in these transitions is proportional to the cross section $\pi k^2$ of this sphere, so that $n_{LL} = \pi k^2/2\pi eB = \omega^2/2eBv_0^2$. 
The total number of fermions participating in transitions is proportional to the surface area of the sphere $4\pi k^2$, while the number of fermions in each chiral Landau level on the sphere (which have $k_x, k_y \sim 0$, because at fixed $k^2$ they have maximal $|k_z|$ for their band, and therefore minimal $k_x, k_y$) is proportional to the area of the Landau level projected onto the sphere, which for $\vec{k}\sim k\hat{z}$ is the same as the area of the Landau level $2\pi eB$. 
The fraction of fermions belonging to each chiral Landau level is $2\pi eB/4\pi k^2 = eB/2k^2 = eBv_0^2/2\omega^2$.
Because of these unpaired chiral modes, the average velocity along the magnetic field of the fermions in that band participating in the transition is 
\begin{equation}
   \langle \vec{v}_n \rangle = 2n\times n\chi v_0 \times \frac{e\vec{B}v_0^2}{2\omega^2} = n^2\chi \frac{e\vec{B}v_0^2}{\omega^2} v_0.
\end{equation}
 If fermions are excited from a band with $S_k = m$ to one with $S_k = n$, the average change in velocity is 
 \begin{equation}
\langle\Delta\vec{v}\rangle = \chi (n^2 - m^2) \frac{e\vec{B}v_0^2}{\omega^2} v_0
 \end{equation}
 along the magnetic field, which explains the scaling with $n^2-m^2$ and $B$ in Eq.~(\ref{special2}).

We now compare the HME and the CPGE. In the CPGE, the magnitude of the change in velocity for each transition is $v_0$. The rate of transition for circularly polarized light, from Eq.~(\ref{golden}), is proportional to $(1+\cos\theta)^2$ where $\theta$ is the angle between the momentum of the fermion and the angular momentum of light; the change in velocity projected along the angular momentum of light is $v_0 \cos\theta$.
Therefore, the change in \textit{velocity}, averaged over the whole sphere is
\begin{equation}
    \langle\Delta\vec{v}\rangle  = \frac{\int (1+\cos\theta)^2 \cos\theta \ 2\pi \sin\theta d\theta}{\int (1+\cos\theta)^2  \ 2\pi \sin\theta d\theta}v_0 = \frac{1}{2}v_0
\end{equation}
Therefore, the ratio of average change in velocity per excitation in the HME vs the CPGE is $(n^2 - m^2) \frac{2eBv_0^2}{\omega^2}$, which is precisely the ratio between the HME and the CPGE in Eq.~(\ref{special2}).
The factor $\frac{2eBv_0^2}{\omega^2} = n_{LL}^{-1}$ is the inverse number of Landau levels involved in the transitions.
This scaling obtained by counting the states in each chiral Landau level agrees with the calculation using chiral kinetic theory in Sec.~\ref{secHME}.

\section{Material realizations: Cobalt Silicide and Rhodium Silicide}\label{secMats}

The HME photocurrent derived in Sec.~\ref{secHME} should be present in the known multifold fermion materials in the B20 family, such as RhSi, CoSi, and AlPt \cite{tang2017multiple,HasanRhSi,sanchez2019topological,schroter2019chiral,RhSiCurrent,xu2020optical}.
These materials are in space group $P2_13$ (SG 198) with chiral tetrahedral symmetry. 
Ignoring spin-orbit coupling (SOC),
they exhibit a double spin-1 fermion, separated from a double trivial fermion by a large energy, at $\Gamma$ and a quadruple spin-1/2 fermion at $R$.
Thus, the multifold fermions are maximally separated in momentum space and not related to each other by even an approximate symmetry. 
The HME photocurrent in this case will come predominantly from the double spin-1 fermion at $\Gamma$, since the spin-1/2 fermions at $R$ do not contribute to the HME because symmetry forbids them from having a tilt.
For small frequencies, the HME photocurrent from $\Gamma$ will be approximately twice the contribution from a single spin-1 fermion, derived in Sec.~\ref{sec:HMEMF}. At larger frequencies, the quadratic dispersion of the middle band (which is flat to linear order, as shown in Fig.~\ref{blockade}) will cause the photocurrent to deviate from its idealized value; nonetheless, it should follow the general trend of the photocurrent plotted in Fig.~\ref{linearHME} (upper), where the photocurrent is nearly zero until a finite onset frequency (necessary to overcome the Pauli blockade) and then decreases.

SOC splits the bands at $\Gamma$ into a spin-3/2 fermion and a Weyl fermion and splits the bands at $R$ into a double spin-1 fermion and two quadratic bands, essentially two copies of the model described in Eq.~(\ref{bandgap}). 
In general, SOC will cause the HME photocurrent to deviate from its idealized form in Eq.~(\ref{special2}), and the general formula in Eq.~(\ref{general}) must be applied.
For mid-infrared light, the frequency is larger than the spin-orbit coupling, but significantly smaller than the separation at $\Gamma$, and we can approximate the HME current by considering only the double spin-1 fermion at $\Gamma$, and ignoring terms that break spherical symmetry.

We now compare the magnitude of the HME photocurrent to that of the CPGE, which has already been observed in RhSi \cite{RhSiCurrent,ni2020linear} and CoSi \cite{CoSiCurrent}. 
As discussed above, the HME photocurrent will come predominantly from the double spin-1 fermion at $\Gamma$.
In a material with a double spin-$1$ cone with Fermi velocity $3\times 10^5\  \mathrm{m/s}$, with the lower and middle bands fully occupied, but the upper band fully unoccupied, excited by light of energy $100\ \mathrm{meV}$, in a magnetic field of $5\ \mathrm{T}$, along the surface of the crystal, there would be $\sim 16$ Landau levels at the excitation energy. Ignoring the effect of the angle of incidence, the photocurrent would be $\sim 0.06$ times the CPGE contribution of the same cone. 
However, depending on the energy of light and the chemical potential, the CPGE could cancel, because the contributions to the CPGE from $\Gamma$ and $R$ enter with opposite sign, while the HME photocurrent does not have this putative cancellation because the spin-1/2 fermions at $R$ do not contribute.

Returning to the importance of the incident angle, the CPGE is always parallel (or anti-parallel) to the angular momentum of light, so the \textit{observed} CPGE, i.e. the component along the surface of crystal, has a factor of the sine of the angle of refraction. If the angle of refraction is $10^\circ$, the CPGE will have a factor of $\sim 0.16$, and the HME current will be $\sim 0.4$ times the CPGE current. Note that the HME current would always be in the direction of the magnetic field and roughly independent of the linear polarization (it is perfectly independent if the bands are perfectly linear).

\section{Discussion}\label{secDisc}

We have demonstrated that the Helical Magnetic Effect can occur in multifold Weyl fermions, as long as the Fermi level is not exactly at the degeneracy point. Unlike simple Weyl fermions, in multifold fermions this effect occurs even in the idealized limit, i.e., it does not require tilt, deviations from linearity or breaking of spherical symmetry.
We derived the HME photocurrent for an ideal multifold fermion in Eq.~(\ref{special2}), which is plotted in Fig.~\ref{linearHME} for a spin-1 and spin-3/2 fermion.
To demonstrate the effect of perturbations beyond the ideal case, we also considered a double spin-1/2 fermion split into a spin-1 fermion and a trivial band; the resulting HME photocurrent in this system is shown in Fig.~\ref{splitCME}.

The HME is of the order of the CPGE divided by the number of Landau levels at the energy of the incident light. However, unlike the CPGE, the HME will be observable even at normal incidence and not be suppressed by the sine of the refracted angle.

We predict that the HME is observable in materials known to exhibit multifold fermions such as RhSi, CoSi, and AlPt.
We estimate its magnitude will be within an order of magnitude of the CPGE photocurrent depending on the chemical potential, magnetic field, and incident angle of light.

\chapter{Conclusion}

\section{Summary}
In this work, we investigated the magnetic and optical responses of chiral fermions in condensed matter. 

In Chapter \ref{chOscillations} we derived the quantum oscillations in the \textit{anomalous} part of the longitudinal magnetoconductivity conductivity in Dirac and Weyl materials, and showed that they have a phase of $\pi/2$ compared to the oscillations in the \textit{Ohmic} part of the conductivity.

In Chapter \ref{chCMP} we showed that the circularly polarized light can transfer chirality to fermions. Each photon transfers, on average, $\alpha/\pi \approx 1/400$ units of chirality per Dirac cone or pair of Weyl cones. An imbalance in chiral fermions and an external magnetic field produces a current; this could lead to the \textit{chiral magnetic photocurrent}: a photocurrent parallel to an external magnetic field caused by circularly polarized light.

In Chapter \ref{chTHz}, we theoretically examined ultrafast photocurrents observed in response to circularly polarized near-infrared radiation in TaAs. We showed that these photocurrents can be explained by transitions from chiral Weyl fermions to massive bands. The selection rules involving chirality and angular momentum still operate on the Weyl fermions. This, combined with the tilt and anisotropy of Weyl cones, and the fact that left and right handed fermions are tilted in different directions, results in a net average change in velocity and a net current.

In Chapter \ref{chTHME}, we showed how the Berry curvature of Weyl fermions could result in a magnetic photocurrent in response to linearly polarized light even in symmetric Weyl materials such as TaAs. The combination of Berry curvature and external magnetic field affects the velocity and phase space volume of fermions. This results in a net average velocity change due to excitation, when the cones are tilted. We showed that the total current summed over all cones would not vanish for a material with the same symmetries as TaAs, and for light polarized along its prefered axis, the current would be perpendicular to both the axis and the magnetic field. However, it would also have polarization-dependent terms.

In Chapter \ref{chStrain}, we showed that in a material subjected to a constant twist (i.e. non-uniform strain), and a time-dependent uniform compression, the twist simulates magnetic field, and the time-dependent compression simulates magnetic field. This could result in chirality pumping even in the absence of an external electromagnetic chirality source. If the whole setup is placed in an external magnetic field, it could produce a net current due to the chiral magnetic effect.

In Chapter \ref{chTunable}, we showed that an external perturbation could break the symmetry between left and right handed fermions. We derived the conditions for a magnetic field to break this symmetry, considering both spatial symmetries and time-reversal. We showed this explicitly for a model of a material with the $T_d$ point group in a magnetic field, where the unperturbed Hamiltonian has an achiral 4-fold degeneracy which splits due to the magnetic field. For a magnetic field along a low symmetry direction, the left and right handed fermions have different energies and velocities; for some parameters, there is even an unequal number of type I (and type II) left and right handed fermions.

In Chapter \ref{chMHME}, we investigated magnetic photocurrents in materials with multifold fermions. We showed that for spin-1 and spin-3/2 fermions there is a magnetic photocurrent due to linearly polarized light when the chemical potential is not exactly at the Weyl node. Unlike simple linear Weyl fermions, the current does not the require the cones to be tilted or anisotropic. For materials such as RhSi and CoSi, only fermions of one chirality would contribute to the current and the current would not cancel, and could be within an order of magntitude of the CPGE observed in these materials.

\section{Future Directions}

Dirac and Weyl materials have diffusion of chiral charge \cite{zhang2017room} and are predicted to have the chiral magnetic wave \cite{CMW}, an excitation that travels along an external magnetic field involving both chiral charge and total charge. It would be interesting to investigate these effects in Weyl materials whose left and right handed cones are oriented in different directions, and also in asymmetic Weyl materials.

The effects described in this work have been calculated exclusively for bulk states, ignoring the contributions of Fermi arcs. For a more complete analysis, the Fermi arc states should also be considered. They could introduce a direction dependence in certain effects that have been calculated to be isotropic here. 

Recently, the transmission of fermions at an interface between two Weyl materials has been studied \cite{GrushinTransmit}. A synthesis of twinned CoSi has been reported \cite{SzczechWires}. It would be interesting to explore transport due to multifold fermions at interfaces, and the effect of interface states, if any.
\include{sections/Outlook}

\bibliographystyle{unsrtnat}
\renewcommand{\baselinestretch}{1}
\normalsize

\clearpage
\newpage
\phantomsection%
\addcontentsline{toc}{chapter}{\numberline{}{Bibliography}}%

\typeout{}
\bibliography{thesisrefsnew}

\begin{thebibliography}{185}
\providecommand{\natexlab}[1]{#1}
\providecommand{\url}[1]{\texttt{#1}}
\expandafter\ifx\csname urlstyle\endcsname\relax
  \providecommand{\doi}[1]{doi: #1}\else
  \providecommand{\doi}{doi: \begingroup \urlstyle{rm}\Url}\fi

\bibitem[Kaushik and Kharzeev(2017)]{SahalOsc}
Sahal Kaushik and Dmitri~E. Kharzeev.
\newblock Quantum oscillations in the chiral magnetic conductivity.
\newblock \emph{Phys. Rev. B}, 95:\penalty0 235136, June 2017.

\bibitem[{Kaushik} et~al.(2019){Kaushik}, {Kharzeev}, and
  {Philip}]{kaushik2019chiral}
Sahal {Kaushik}, Dmitri~E. {Kharzeev}, and Evan~John {Philip}.
\newblock {Chiral magnetic photocurrent in Dirac and Weyl materials}.
\newblock \emph{Physical Review B}, 99\penalty0 (7):\penalty0 075150, February
  2019.

\bibitem[{Gao} et~al.(2020){Gao}, {Kaushik}, {Philip}, {Li}, {Qin}, {Liu},
  {Zhang}, {Su}, {Chen}, {Weng}, {Kharzeev}, {Liu}, and {Qi}]{gao2020chiral}
Y.~{Gao}, S.~{Kaushik}, E.~J. {Philip}, Z.~{Li}, Y.~{Qin}, Y.~P. {Liu}, W.~L.
  {Zhang}, Y.~L. {Su}, X.~{Chen}, H.~{Weng}, D.~E. {Kharzeev}, M.~K. {Liu}, and
  J.~{Qi}.
\newblock {Chiral terahertz wave emission from the Weyl semimetal TaAs}.
\newblock \emph{Nature Communications}, 11:\penalty0 720, February 2020.

\bibitem[{Kaushik} et~al.(2020){Kaushik}, {Kharzeev}, and {Philip}]{SahalTHME}
Sahal {Kaushik}, Dmitri~E. {Kharzeev}, and Evan~John {Philip}.
\newblock {Transverse chiral magnetic photocurrent induced by linearly
  polarized light in mirror-symmetric Weyl semimetals}.
\newblock \emph{Physical Review Research}, 2\penalty0 (4):\penalty0 042011,
  October 2020.

\bibitem[Gao et~al.(2020)Gao, Kaushik, Kharzeev, and Philip]{LanLanStrain}
Lan-Lan Gao, Sahal Kaushik, Dmitri~E. Kharzeev, and Evan~John Philip.
\newblock {Chiral kinetic theory of anomalous transport induced by torsion}.
\newblock \emph{arXiv preprint arXiv:2010.07123}, October 2020.

\bibitem[{Kaushik} et~al.(2021){Kaushik}, {Philip}, and {Cano}]{SahalAsymm}
Sahal {Kaushik}, Evan~John {Philip}, and Jennifer {Cano}.
\newblock {Tunable chiral symmetry breaking in symmetric Weyl materials}.
\newblock \emph{Physical Review B}, 103\penalty0 (8):\penalty0 085106, February
  2021.

\bibitem[Kaushik and Cano(2021)]{SahalMHME}
Sahal Kaushik and Jennifer Cano.
\newblock {Magnetic Photocurrents in Multifold Weyl Fermions}.
\newblock \emph{arXiv preprint arXiv:2107.05106}, July 2021.

\bibitem[Kelvin(1894)]{kelvin}
William Thomson~Baron Kelvin.
\newblock \emph{The molecular tactics of a crystal}.
\newblock Clarendon Press, 1894.

\bibitem[Lee and Yang(1956)]{PViolationTh}
T.~D. Lee and C.~N. Yang.
\newblock Question of parity conservation in weak interactions.
\newblock \emph{Phys. Rev.}, 104:\penalty0 254--258, October 1956.

\bibitem[Wu et~al.(1957)Wu, Ambler, Hayward, Hoppes, and
  Hudson]{PViolationExpt}
C.~S. Wu, E.~Ambler, R.~W. Hayward, D.~D. Hoppes, and R.~P. Hudson.
\newblock Experimental test of parity conservation in beta decay.
\newblock \emph{Phys. Rev.}, 105:\penalty0 1413--1415, February 1957.

\bibitem[Dirac(1928)]{DiracEqn}
Paul Adrien~Maurice Dirac.
\newblock The quantum theory of the electron.
\newblock \emph{Proceedings of the Royal Society of London. Series A,
  Containing Papers of a Mathematical and Physical Character}, 117\penalty0
  (778):\penalty0 610--624, 1928.

\bibitem[Weyl(1929)]{WeylEqn}
Hermann Weyl.
\newblock Gravitation and the electron.
\newblock \emph{Proceedings of the National Academy of Sciences}, 15\penalty0
  (4):\penalty0 323--334, 1929.

\bibitem[{Young} et~al.(2012){Young}, {Zaheer}, {Teo}, {Kane}, {Mele}, and
  {Rappe}]{Young12}
S.~M. {Young}, S.~{Zaheer}, J.~C.~Y. {Teo}, C.~L. {Kane}, E.~J. {Mele}, and
  A.~M. {Rappe}.
\newblock {Dirac Semimetal in Three Dimensions}.
\newblock \emph{Physical Review Letters}, 108\penalty0 (14):\penalty0 140405,
  April 2012.

\bibitem[{Wang} et~al.(2012){Wang}, {Sun}, {Chen}, {Franchini}, {Xu}, {Weng},
  {Dai}, and {Fang}]{Wang12}
Zhijun {Wang}, Yan {Sun}, Xing-Qiu {Chen}, Cesare {Franchini}, Gang {Xu},
  Hongming {Weng}, Xi~{Dai}, and Zhong {Fang}.
\newblock {Dirac semimetal and topological phase transitions in A$_{3}$Bi
  (A=Na, K, Rb)}.
\newblock \emph{Physical Review B}, 85\penalty0 (19):\penalty0 195320, May
  2012.

\bibitem[{Liu} et~al.(2014{\natexlab{a}}){Liu}, {Zhou}, {Zhang}, {Wang},
  {Weng}, {Prabhakaran}, {Mo}, {Shen}, {Fang}, {Dai}, {Hussain}, and
  {Chen}]{Liu14}
Z.~K. {Liu}, B.~{Zhou}, Y.~{Zhang}, Z.~J. {Wang}, H.~M. {Weng},
  D.~{Prabhakaran}, S.~K. {Mo}, Z.~X. {Shen}, Z.~{Fang}, X.~{Dai},
  Z.~{Hussain}, and Y.~L. {Chen}.
\newblock {Discovery of a Three-Dimensional Topological Dirac Semimetal,
  Na$_{3}$Bi}.
\newblock \emph{Science}, 343\penalty0 (6173):\penalty0 864--867, February
  2014{\natexlab{a}}.

\bibitem[{Liu} et~al.(2014{\natexlab{b}}){Liu}, {Jiang}, {Zhou}, {Wang},
  {Zhang}, {Weng}, {Prabhakaran}, {Mo}, {Peng}, {Dudin}, {Kim}, {Hoesch},
  {Fang}, {Dai}, {Shen}, {Feng}, {Hussain}, and {Chen}]{Liu14a}
Z.~K. {Liu}, J.~{Jiang}, B.~{Zhou}, Z.~J. {Wang}, Y.~{Zhang}, H.~M. {Weng},
  D.~{Prabhakaran}, S.~K. {Mo}, H.~{Peng}, P.~{Dudin}, T.~{Kim}, M.~{Hoesch},
  Z.~{Fang}, X.~{Dai}, Z.~X. {Shen}, D.~L. {Feng}, Z.~{Hussain}, and Y.~L.
  {Chen}.
\newblock {A stable three-dimensional topological Dirac semimetal
  Cd$_{3}$As$_{2}$}.
\newblock \emph{Nature Materials}, 13\penalty0 (7):\penalty0 677--681, July
  2014{\natexlab{b}}.

\bibitem[{Steinberg} et~al.(2014){Steinberg}, {Young}, {Zaheer}, {Kane},
  {Mele}, and {Rappe}]{Steinberg14}
Julia~A. {Steinberg}, Steve~M. {Young}, Saad {Zaheer}, C.~L. {Kane}, E.~J.
  {Mele}, and Andrew~M. {Rappe}.
\newblock {Bulk Dirac Points in Distorted Spinels}.
\newblock \emph{Physical Review Letters}, 112\penalty0 (3):\penalty0 036403,
  January 2014.

\bibitem[{Yang} and {Nagaosa}(2014)]{nagaosa2014}
Bohm-Jung {Yang} and Naoto {Nagaosa}.
\newblock {Classification of stable three-dimensional Dirac semimetals with
  nontrivial topology}.
\newblock \emph{Nature Communications}, 5:\penalty0 4898, September 2014.

\bibitem[{Bradlyn} et~al.(2016){Bradlyn}, {Cano}, {Wang}, {Vergniory},
  {Felser}, {Cava}, and {Bernevig}]{bradlyn2016beyond}
Barry {Bradlyn}, Jennifer {Cano}, Zhijun {Wang}, M.~G. {Vergniory},
  C.~{Felser}, R.~J. {Cava}, and B.~Andrei {Bernevig}.
\newblock {Beyond Dirac and Weyl fermions: Unconventional quasiparticles in
  conventional crystals}.
\newblock \emph{Science}, 353\penalty0 (6299):\penalty0 aaf5037, August 2016.

\bibitem[{Cano} et~al.(2019){Cano}, {Bradlyn}, and
  {Vergniory}]{cano2019multifold}
Jennifer {Cano}, Barry {Bradlyn}, and M.~G. {Vergniory}.
\newblock {Multifold nodal points in magnetic materials}.
\newblock \emph{APL Materials}, 7\penalty0 (10):\penalty0 101125, October 2019.

\bibitem[{Klemenz} et~al.(2020){Klemenz}, {Schoop}, and
  {Cano}]{klemenz2020systematic}
Sebastian {Klemenz}, Leslie {Schoop}, and Jennifer {Cano}.
\newblock {Systematic study of stacked square nets: From Dirac fermions to
  material realizations}.
\newblock \emph{Physical Review B}, 101\penalty0 (16):\penalty0 165121, April
  2020.

\bibitem[{Wieder} et~al.(2020){Wieder}, {Wang}, {Cano}, {Dai}, {Schoop},
  {Bradlyn}, and {Bernevig}]{wieder2020strong}
Benjamin~J. {Wieder}, Zhijun {Wang}, Jennifer {Cano}, Xi~{Dai}, Leslie~M.
  {Schoop}, Barry {Bradlyn}, and B.~Andrei {Bernevig}.
\newblock {Strong and fragile topological Dirac semimetals with higher-order
  Fermi arcs}.
\newblock \emph{Nature Communications}, 11:\penalty0 627, January 2020.

\bibitem[{Wan} et~al.(2011){Wan}, {Turner}, {Vishwanath}, and
  {Savrasov}]{Wan11}
Xiangang {Wan}, Ari~M. {Turner}, Ashvin {Vishwanath}, and Sergey~Y. {Savrasov}.
\newblock {Topological semimetal and Fermi-arc surface states in the electronic
  structure of pyrochlore iridates}.
\newblock \emph{Physical Review B}, 83\penalty0 (20):\penalty0 205101, May
  2011.

\bibitem[{Weng} et~al.(2015){Weng}, {Fang}, {Fang}, {Bernevig}, and
  {Dai}]{Weng15}
Hongming {Weng}, Chen {Fang}, Zhong {Fang}, B.~Andrei {Bernevig}, and Xi~{Dai}.
\newblock {Weyl Semimetal Phase in Noncentrosymmetric Transition-Metal
  Monophosphides}.
\newblock \emph{Physical Review X}, 5\penalty0 (1):\penalty0 011029, January
  2015.

\bibitem[{Huang} et~al.(2015{\natexlab{a}}){Huang}, {Xu}, {Belopolski}, {Lee},
  {Chang}, {Wang}, {Alidoust}, {Bian}, {Neupane}, {Zhang}, {Jia}, {Bansil},
  {Lin}, and {Hasan}]{Huang15}
Shin-Ming {Huang}, Su-Yang {Xu}, Ilya {Belopolski}, Chi-Cheng {Lee}, Guoqing
  {Chang}, Baokai {Wang}, Nasser {Alidoust}, Guang {Bian}, Madhab {Neupane},
  Chenglong {Zhang}, Shuang {Jia}, Arun {Bansil}, Hsin {Lin}, and M.~Zahid
  {Hasan}.
\newblock {A Weyl Fermion semimetal with surface Fermi arcs in the transition
  metal monopnictide TaAs class}.
\newblock \emph{Nature Communications}, 6:\penalty0 7373, June
  2015{\natexlab{a}}.

\bibitem[{Xu} et~al.(2015{\natexlab{a}}){Xu}, {Alidoust}, {Belopolski}, {Yuan},
  {Bian}, {Chang}, {Zheng}, {Strocov}, {Sanchez}, {Chang}, {Zhang}, {Mou},
  {Wu}, {Huang}, {Lee}, {Huang}, {Wang}, {Bansil}, {Jeng}, {Neupert},
  {Kaminski}, {Lin}, {Jia}, and {Zahid Hasan}]{xu2015discovery}
Su-Yang {Xu}, Nasser {Alidoust}, Ilya {Belopolski}, Zhujun {Yuan}, Guang
  {Bian}, Tay-Rong {Chang}, Hao {Zheng}, Vladimir~N. {Strocov}, Daniel~S.
  {Sanchez}, Guoqing {Chang}, Chenglong {Zhang}, Daixiang {Mou}, Yun {Wu},
  Lunan {Huang}, Chi-Cheng {Lee}, Shin-Ming {Huang}, Baokai {Wang}, Arun
  {Bansil}, Horng-Tay {Jeng}, Titus {Neupert}, Adam {Kaminski}, Hsin {Lin},
  Shuang {Jia}, and M.~{Zahid Hasan}.
\newblock {Discovery of a Weyl fermion state with Fermi arcs in niobium
  arsenide}.
\newblock \emph{Nature Physics}, 11\penalty0 (9):\penalty0 748--754, September
  2015{\natexlab{a}}.

\bibitem[{Lv} et~al.(2015{\natexlab{a}}){Lv}, {Xu}, {Weng}, {Ma}, {Richard},
  {Huang}, {Zhao}, {Chen}, {Matt}, {Bisti}, {Strocov}, {Mesot}, {Fang}, {Dai},
  {Qian}, {Shi}, and {Ding}]{lv2015Nat}
B.~Q. {Lv}, N.~{Xu}, H.~M. {Weng}, J.~Z. {Ma}, P.~{Richard}, X.~C. {Huang},
  L.~X. {Zhao}, G.~F. {Chen}, C.~E. {Matt}, F.~{Bisti}, V.~N. {Strocov},
  J.~{Mesot}, Z.~{Fang}, X.~{Dai}, T.~{Qian}, M.~{Shi}, and H.~{Ding}.
\newblock {Observation of Weyl nodes in TaAs}.
\newblock \emph{Nature Physics}, 11\penalty0 (9):\penalty0 724--727, September
  2015{\natexlab{a}}.

\bibitem[{Xu} et~al.(2015{\natexlab{b}}){Xu}, {Belopolski}, {Alidoust},
  {Neupane}, {Bian}, {Zhang}, {Sankar}, {Chang}, {Yuan}, {Lee}, {Huang},
  {Zheng}, {Ma}, {Sanchez}, {Wang}, {Bansil}, {Chou}, {Shibayev}, {Lin}, {Jia},
  and {Hasan}]{xu2015}
Su-Yang {Xu}, Ilya {Belopolski}, Nasser {Alidoust}, Madhab {Neupane}, Guang
  {Bian}, Chenglong {Zhang}, Raman {Sankar}, Guoqing {Chang}, Zhujun {Yuan},
  Chi-Cheng {Lee}, Shin-Ming {Huang}, Hao {Zheng}, Jie {Ma}, Daniel~S.
  {Sanchez}, BaoKai {Wang}, Arun {Bansil}, Fangcheng {Chou}, Pavel~P.
  {Shibayev}, Hsin {Lin}, Shuang {Jia}, and M.~Zahid {Hasan}.
\newblock {Discovery of a Weyl fermion semimetal and topological Fermi arcs}.
\newblock \emph{Science}, 349\penalty0 (6248):\penalty0 613--617, August
  2015{\natexlab{b}}.

\bibitem[{Lv} et~al.(2015{\natexlab{b}}){Lv}, {Weng}, {Fu}, {Wang}, {Miao},
  {Ma}, {Richard}, {Huang}, {Zhao}, {Chen}, {Fang}, {Dai}, {Qian}, and
  {Ding}]{lv2015PRX}
B.~Q. {Lv}, H.~M. {Weng}, B.~B. {Fu}, X.~P. {Wang}, H.~{Miao}, J.~{Ma},
  P.~{Richard}, X.~C. {Huang}, L.~X. {Zhao}, G.~F. {Chen}, Z.~{Fang}, X.~{Dai},
  T.~{Qian}, and H.~{Ding}.
\newblock {Experimental Discovery of Weyl Semimetal TaAs}.
\newblock \emph{Physical Review X}, 5\penalty0 (3):\penalty0 031013, July
  2015{\natexlab{b}}.

\bibitem[{Xiong} et~al.(2015){Xiong}, {Kushwaha}, {Liang}, {Krizan},
  {Hirschberger}, {Wang}, {Cava}, and {Ong}]{2015Xiong}
Jun {Xiong}, Satya~K. {Kushwaha}, Tian {Liang}, Jason~W. {Krizan}, Max
  {Hirschberger}, Wudi {Wang}, R.~J. {Cava}, and N.~P. {Ong}.
\newblock {Evidence for the chiral anomaly in the Dirac semimetal Na$_{3}$Bi}.
\newblock \emph{Science}, 350\penalty0 (6259):\penalty0 413--416, October 2015.

\bibitem[Wehling et~al.(2014)Wehling, Black-Schaffer, and
  Balatsky]{BalatskyRev}
T.O. Wehling, A.M. Black-Schaffer, and A.V. Balatsky.
\newblock Dirac materials.
\newblock \emph{Advances in Physics}, 63\penalty0 (1):\penalty0 1--76, 2014.

\bibitem[{Armitage} et~al.(2018){Armitage}, {Mele}, and
  {Vishwanath}]{armitage2018}
N.~P. {Armitage}, E.~J. {Mele}, and Ashvin {Vishwanath}.
\newblock {Weyl and Dirac semimetals in three-dimensional solids}.
\newblock \emph{Reviews of Modern Physics}, 90\penalty0 (1):\penalty0 015001,
  January 2018.

\bibitem[{Soluyanov} et~al.(2015){Soluyanov}, {Gresch}, {Wang}, {Wu}, {Troyer},
  {Dai}, and {Bernevig}]{soluyanov2015}
Alexey~A. {Soluyanov}, Dominik {Gresch}, Zhijun {Wang}, Quansheng {Wu},
  Matthias {Troyer}, Xi~{Dai}, and B.~Andrei {Bernevig}.
\newblock {Type-II Weyl semimetals}.
\newblock \emph{Nature}, 527\penalty0 (7579):\penalty0 495--498, November 2015.

\bibitem[{Nielsen} and {Ninomiya}(1981)]{1981Nielsen}
H.~B. {Nielsen} and M.~{Ninomiya}.
\newblock {A no-go theorem for regularizing chiral fermions}.
\newblock \emph{Physics Letters B}, 105\penalty0 (2-3):\penalty0 219--223,
  October 1981.

\bibitem[{Huang} et~al.(2016){Huang}, {Xu}, {Belopolski}, {Lee}, {Chang},
  {Chang}, {Wang}, {Alidoust}, {Bian}, {Neupane}, {Sanchez}, {Zheng}, {Jeng},
  {Bansil}, {Neupert}, {Lin}, and {Zahid Hasan}]{huang2016}
Shin-Ming {Huang}, Su-Yang {Xu}, Ilya {Belopolski}, Chi-Cheng {Lee}, Guoqing
  {Chang}, Tay-Rong {Chang}, BaoKai {Wang}, Nasser {Alidoust}, Guang {Bian},
  Madhab {Neupane}, Daniel {Sanchez}, Hao {Zheng}, Horng-Tay {Jeng}, Arun
  {Bansil}, Titus {Neupert}, Hsin {Lin}, and M.~{Zahid Hasan}.
\newblock {New type of Weyl semimetal with quadratic double Weyl fermions}.
\newblock \emph{Proceedings of the National Academy of Science}, 113\penalty0
  (5):\penalty0 1180--1185, February 2016.

\bibitem[{Zhang} et~al.(2020){Zhang}, {Takahashi}, {Fang}, and
  {Murakami}]{FourWeyl}
Tiantian {Zhang}, Ryo {Takahashi}, Chen {Fang}, and Shuichi {Murakami}.
\newblock {Twofold quadruple Weyl nodes in chiral cubic crystals}.
\newblock \emph{Physical Review B}, 102\penalty0 (12):\penalty0 125148,
  September 2020.

\bibitem[Adler(1969)]{Adler1969}
Stephen~L. Adler.
\newblock {Axial-Vector Vertex in Spinor Electrodynamics}.
\newblock \emph{Physical Review}, 177\penalty0 (5):\penalty0 2426--2438,
  January 1969.

\bibitem[Bell and Jackiw(1969)]{Bell1969}
J.~S. Bell and R.~Jackiw.
\newblock {A PCAC puzzle: $\pi$0→$\gamma$ $\gamma$ in the $\sigma$-model}.
\newblock \emph{Il Nuovo Cimento A}, 60\penalty0 (1):\penalty0 47--61, March
  1969.

\bibitem[Nielsen and Ninomiya(1983)]{nielsen1983adler}
Holger~Bech Nielsen and Masao Ninomiya.
\newblock The adler-bell-jackiw anomaly and weyl fermions in a crystal.
\newblock \emph{Physics Letters B}, 130\penalty0 (6):\penalty0 389--396, 1983.

\bibitem[{Fukushima} et~al.(2008){Fukushima}, {Kharzeev}, and {Warringa}]{CME}
Kenji {Fukushima}, Dmitri~E. {Kharzeev}, and Harmen~J. {Warringa}.
\newblock {Chiral magnetic effect}.
\newblock \emph{Physical Review D}, 78\penalty0 (7):\penalty0 074033, October
  2008.

\bibitem[Kharzeev(2014)]{Kharzeev:2013ffa}
Dmitri~E. Kharzeev.
\newblock {The Chiral Magnetic Effect and Anomaly-Induced Transport}.
\newblock \emph{Prog. Part. Nucl. Phys.}, 75:\penalty0 133--151, 2014.

\bibitem[Kharzeev et~al.(2013)Kharzeev, Landsteiner, Schmitt, and
  Yee]{Kharzeev:2012ph}
Dmitri~E. Kharzeev, Karl Landsteiner, Andreas Schmitt, and Ho-Ung Yee.
\newblock {'Strongly interacting matter in magnetic fields': an overview}.
\newblock \emph{Lect. Notes Phys.}, 871:\penalty0 1--11, 2013.

\bibitem[Zhong et~al.(2016)Zhong, Moore, and Souza]{zhong2016gyrotropic}
Shudan Zhong, Joel~E Moore, and Ivo Souza.
\newblock Gyrotropic magnetic effect and the magnetic moment on the fermi
  surface.
\newblock \emph{Physical review letters}, 116\penalty0 (7):\penalty0 077201,
  2016.

\bibitem[Ma and Pesin(2015)]{ma2015chiral}
Jing Ma and DA~Pesin.
\newblock Chiral magnetic effect and natural optical activity in metals with or
  without weyl points.
\newblock \emph{Physical Review B}, 92\penalty0 (23):\penalty0 235205, 2015.

\bibitem[{Cortijo} et~al.(2016){Cortijo}, {Kharzeev}, {Landsteiner}, and
  {Vozmediano}]{cortijo2016}
Alberto {Cortijo}, Dmitri {Kharzeev}, Karl {Landsteiner}, and Maria A.~H.
  {Vozmediano}.
\newblock {Strain-induced chiral magnetic effect in Weyl semimetals}.
\newblock \emph{Physical Review B}, 94\penalty0 (24):\penalty0 241405, December
  2016.

\bibitem[Song et~al.(2016)Song, Zhao, Fang, and Dai]{song2016detecting}
Zhida Song, Jimin Zhao, Zhong Fang, and Xi~Dai.
\newblock Detecting chiral magnetic effect by lattice dynamics.
\newblock \emph{arXiv preprint arXiv:1609.05442}, September 2016.

\bibitem[{Kharzeev} et~al.(2018{\natexlab{a}}){Kharzeev}, {Kikuchi}, and
  {Meyer}]{meyer2018}
Dmitri~E. {Kharzeev}, Yuta {Kikuchi}, and Ren{\'e} {Meyer}.
\newblock {Chiral magnetic effect without chirality source in asymmetric Weyl
  semimetals}.
\newblock \emph{European Physical Journal B}, 91\penalty0 (5):\penalty0 83, May
  2018{\natexlab{a}}.

\bibitem[{Son} and {Yamamoto}(2012)]{Son2012Berry}
Dam~Thanh {Son} and Naoki {Yamamoto}.
\newblock {Berry Curvature, Triangle Anomalies, and the Chiral Magnetic Effect
  in Fermi Liquids}.
\newblock \emph{Physical Review Letters}, 109\penalty0 (18):\penalty0 181602,
  November 2012.

\bibitem[{Son} and {Spivak}(2013)]{2013Son}
D.~T. {Son} and B.~Z. {Spivak}.
\newblock {Chiral anomaly and classical negative magnetoresistance of Weyl
  metals}.
\newblock \emph{Physical Review B}, 88\penalty0 (10):\penalty0 104412,
  September 2013.

\bibitem[Zyuzin and Burkov(2012)]{Zyuzin:2012tv}
A.~A. Zyuzin and A.~A. Burkov.
\newblock {Topological response in Weyl semimetals and the chiral anomaly}.
\newblock \emph{Phys. Rev.}, B86:\penalty0 115133, 2012.

\bibitem[Basar et~al.(2014)Basar, Kharzeev, and Yee]{Basar:2013iaa}
Gokce Basar, Dmitri~E. Kharzeev, and Ho-Ung Yee.
\newblock {Triangle anomaly in Weyl semimetals}.
\newblock \emph{Phys. Rev. B}, 89\penalty0 (3):\penalty0 035142, 2014.

\bibitem[Vazifeh and Franz(2013)]{vazifeh2013electromagnetic}
MM~Vazifeh and M~Franz.
\newblock Electromagnetic response of weyl semimetals.
\newblock \emph{Physical Review Letters}, 111\penalty0 (2):\penalty0 027201,
  2013.

\bibitem[Goswami and Tewari(2013)]{goswami2013axionic}
Pallab Goswami and Sumanta Tewari.
\newblock Axionic field theory of (3+ 1)-dimensional weyl semimetals.
\newblock \emph{Physical Review B}, 88\penalty0 (24):\penalty0 245107, 2013.

\bibitem[{Burkov}(2014)]{2014Burkov}
A.~A. {Burkov}.
\newblock {Chiral Anomaly and Diffusive Magnetotransport in Weyl Metals}.
\newblock \emph{Physical Review Letters}, 113\penalty0 (24):\penalty0 247203,
  December 2014.

\bibitem[{Aji}(2012)]{aji2012adler}
Vivek {Aji}.
\newblock {Adler-Bell-Jackiw anomaly in Weyl semimetals: Application to
  pyrochlore iridates}.
\newblock \emph{Physical Review B}, 85\penalty0 (24):\penalty0 241101, June
  2012.

\bibitem[{Li} et~al.(2016){Li}, {Kharzeev}, {Zhang}, {Huang}, {Pletikosi{\'c}},
  {Fedorov}, {Zhong}, {Schneeloch}, {Gu}, and {Valla}]{2016Li}
Qiang {Li}, Dmitri~E. {Kharzeev}, Cheng {Zhang}, Yuan {Huang},
  I.~{Pletikosi{\'c}}, A.~V. {Fedorov}, R.~D. {Zhong}, J.~A. {Schneeloch},
  G.~D. {Gu}, and T.~{Valla}.
\newblock {Chiral magnetic effect in ZrTe$_{5}$}.
\newblock \emph{Nature Physics}, 12\penalty0 (6):\penalty0 550--554, June 2016.

\bibitem[Kim et~al.(2013)Kim, Kim, Wang, Sasaki, Satoh, Ohnishi, Kitaura, Yang,
  and Li]{kim2013dirac}
Heon-Jung Kim, Ki-Seok Kim, J-F Wang, M~Sasaki, N~Satoh, A~Ohnishi, M~Kitaura,
  M~Yang, and L~Li.
\newblock Dirac versus weyl fermions in topological insulators:
  Adler-bell-jackiw anomaly in transport phenomena.
\newblock \emph{Physical review letters}, 111\penalty0 (24):\penalty0 246603,
  2013.

\bibitem[Li et~al.(2015)Li, Wang, Liu, Wang, Liao, and Yu]{li2015giant}
Cai-Zhen Li, Li-Xian Wang, Haiwen Liu, Jian Wang, Zhi-Min Liao, and Da-Peng Yu.
\newblock Giant negative magnetoresistance induced by the chiral anomaly in
  individual cd3as2 nanowires.
\newblock \emph{Nature communications}, 6, 2015.

\bibitem[{Huang} et~al.(2015{\natexlab{b}}){Huang}, {Zhao}, {Long}, {Wang},
  {Chen}, {Yang}, {Liang}, {Xue}, {Weng}, {Fang}, {Dai}, and {Chen}]{Huang2015}
Xiaochun {Huang}, Lingxiao {Zhao}, Yujia {Long}, Peipei {Wang}, Dong {Chen},
  Zhanhai {Yang}, Hui {Liang}, Mianqi {Xue}, Hongming {Weng}, Zhong {Fang},
  Xi~{Dai}, and Genfu {Chen}.
\newblock {Observation of the Chiral-Anomaly-Induced Negative Magnetoresistance
  in 3D Weyl Semimetal TaAs}.
\newblock \emph{Physical Review X}, 5\penalty0 (3):\penalty0 031023, July
  2015{\natexlab{b}}.

\bibitem[{Wang} et~al.(2016){Wang}, {Zheng}, {Shen}, {Lu}, {Fang}, {Sheng},
  {Zhou}, {Yang}, {Li}, {Feng}, and {Xu}]{Wang2016}
Zhen {Wang}, Yi~{Zheng}, Zhixuan {Shen}, Yunhao {Lu}, Hanyan {Fang}, Feng
  {Sheng}, Yi~{Zhou}, Xiaojun {Yang}, Yupeng {Li}, Chunmu {Feng}, and Zhu-An
  {Xu}.
\newblock {Helicity-protected ultrahigh mobility Weyl fermions in NbP}.
\newblock \emph{Physical Review B}, 93\penalty0 (12):\penalty0 121112, March
  2016.

\bibitem[{Zhang} et~al.(2016){Zhang}, {Xu}, {Belopolski}, {Yuan}, {Lin},
  {Tong}, {Bian}, {Alidoust}, {Lee}, {Huang}, {Chang}, {Chang}, {Hsu}, {Jeng},
  {Neupane}, {Sanchez}, {Zheng}, {Wang}, {Lin}, {Zhang}, {Lu}, {Shen},
  {Neupert}, {Zahid Hasan}, and {Jia}]{Zhang2016}
Cheng-Long {Zhang}, Su-Yang {Xu}, Ilya {Belopolski}, Zhujun {Yuan}, Ziquan
  {Lin}, Bingbing {Tong}, Guang {Bian}, Nasser {Alidoust}, Chi-Cheng {Lee},
  Shin-Ming {Huang}, Tay-Rong {Chang}, Guoqing {Chang}, Chuang-Han {Hsu},
  Horng-Tay {Jeng}, Madhab {Neupane}, Daniel~S. {Sanchez}, Hao {Zheng}, Junfeng
  {Wang}, Hsin {Lin}, Chi {Zhang}, Hai-Zhou {Lu}, Shun-Qing {Shen}, Titus
  {Neupert}, M.~{Zahid Hasan}, and Shuang {Jia}.
\newblock {Signatures of the Adler-Bell-Jackiw chiral anomaly in a Weyl fermion
  semimetal}.
\newblock \emph{Nature Communications}, 7:\penalty0 10735, February 2016.

\bibitem[Shekhar et~al.(2015)Shekhar, Arnold, Wu, Sun, Schmidt, Kumar, Grushin,
  Bardarson, Reis, Naumann, et~al.]{shekhar2015large}
Chandra Shekhar, Frank Arnold, Shu-Chun Wu, Yan Sun, Marcus Schmidt, Nitesh
  Kumar, Adolfo~G Grushin, Jens~H Bardarson, Ricardo Donizeth~dos Reis, Marcel
  Naumann, et~al.
\newblock Large and unsaturated negative magnetoresistance induced by the
  chiral anomaly in the weyl semimetal tap.
\newblock \emph{arXiv preprint arXiv:1506.06577}, June 2015.

\bibitem[Yang et~al.(2015)Yang, Liu, Wang, Zheng, and Xu]{yang2015chiral}
Xiaojun Yang, Yupeng Liu, Zhen Wang, Yi~Zheng, and Zhu-an Xu.
\newblock Chiral anomaly induced negative magnetoresistance in topological weyl
  semimetal nbas month=jun.
\newblock \emph{arXiv preprint arXiv:1506.03190}, 2015.

\bibitem[{Arnold} et~al.(2016){Arnold}, {Shekhar}, {Wu}, {Sun}, {Dos Reis},
  {Kumar}, {Naumann}, {Ajeesh}, {Schmidt}, {Grushin}, {Bardarson}, {Baenitz},
  {Sokolov}, {Borrmann}, {Nicklas}, {Felser}, {Hassinger}, and
  {Yan}]{Arnold2016}
Frank {Arnold}, Chandra {Shekhar}, Shu-Chun {Wu}, Yan {Sun}, Ricardo~Donizeth
  {Dos Reis}, Nitesh {Kumar}, Marcel {Naumann}, Mukkattu~O. {Ajeesh}, Marcus
  {Schmidt}, Adolfo~G. {Grushin}, Jens~H. {Bardarson}, Michael {Baenitz},
  Dmitry {Sokolov}, Horst {Borrmann}, Michael {Nicklas}, Claudia {Felser},
  Elena {Hassinger}, and Binghai {Yan}.
\newblock {Negative magnetoresistance without well-defined chirality in the
  Weyl semimetal TaP}.
\newblock \emph{Nature Communications}, 7:\penalty0 11615, May 2016.

\bibitem[Murakawa et~al.(2013)Murakawa, Bahramy, Tokunaga, Kohama, Bell,
  Kaneko, Nagaosa, Hwang, and Tokura]{murakawa2013detection}
H~Murakawa, MS~Bahramy, M~Tokunaga, Y~Kohama, C~Bell, Y~Kaneko, N~Nagaosa,
  HY~Hwang, and Y~Tokura.
\newblock Detection of berry's phase in a bulk rashba semiconductor.
\newblock \emph{Science}, 342\penalty0 (6165):\penalty0 1490--1493, 2013.

\bibitem[{Chan} et~al.(2017){Chan}, {Lindner}, {Refael}, and {Lee}]{lee2017}
Ching-Kit {Chan}, Netanel~H. {Lindner}, Gil {Refael}, and Patrick~A. {Lee}.
\newblock {Photocurrents in Weyl semimetals}.
\newblock \emph{Physical Review B}, 95\penalty0 (4):\penalty0 041104, January
  2017.

\bibitem[{Ma} et~al.(2017){Ma}, {Xu}, {Chan}, {Zhang}, {Chang}, {Lin}, {Xie},
  {Palacios}, {Lin}, {Jia}, {Lee}, {Jarillo-Herrero}, and {Gedik}]{MaTaAs}
Qiong {Ma}, Su-Yang {Xu}, Ching-Kit {Chan}, Cheng-Long {Zhang}, Guoqing
  {Chang}, Yuxuan {Lin}, Weiwei {Xie}, Tom{\'a}s {Palacios}, Hsin {Lin}, Shuang
  {Jia}, Patrick~A. {Lee}, Pablo {Jarillo-Herrero}, and Nuh {Gedik}.
\newblock {Direct optical detection of Weyl fermion chirality in a topological
  semimetal}.
\newblock \emph{Nature Physics}, 13\penalty0 (9):\penalty0 842--847, September
  2017.

\bibitem[{de Juan} et~al.(2017){de Juan}, {Grushin}, {Morimoto}, and
  {Moore}]{dejuan17}
Fernando {de Juan}, Adolfo~G. {Grushin}, Takahiro {Morimoto}, and Joel~E.
  {Moore}.
\newblock {Quantized circular photogalvanic effect in Weyl semimetals}.
\newblock \emph{Nature Communications}, 8:\penalty0 15995, July 2017.

\bibitem[{Flicker} et~al.(2018){Flicker}, {de Juan}, {Bradlyn}, {Morimoto},
  {Vergniory}, and {Grushin}]{FlickerMultifold}
Felix {Flicker}, Fernando {de Juan}, Barry {Bradlyn}, Takahiro {Morimoto},
  Maia~G. {Vergniory}, and Adolfo~G. {Grushin}.
\newblock {Chiral optical response of multifold fermions}.
\newblock \emph{Physical Review B}, 98\penalty0 (15):\penalty0 155145, October
  2018.

\bibitem[{Rees} et~al.(2020){Rees}, {Manna}, {Lu}, {Morimoto}, {Borrmann},
  {Felser}, {Moore}, {Torchinsky}, and {Orenstein}]{RhSiCurrent}
Dylan {Rees}, Kaustuv {Manna}, Baozhu {Lu}, Takahiro {Morimoto}, Horst
  {Borrmann}, Claudia {Felser}, J.~E. {Moore}, Darius~H. {Torchinsky}, and
  J.~{Orenstein}.
\newblock {Helicity-dependent photocurrents in the chiral Weyl semimetal RhSi}.
\newblock \emph{Science Advances}, 6\penalty0 (29):\penalty0 eaba0509, July
  2020.

\bibitem[Ni et~al.(2021)Ni, Wang, Zhang, Pozo, Xu, Han, Manna, Paglione,
  Felser, Grushin, et~al.]{CoSiCurrent}
Zhuoliang Ni, K~Wang, Y~Zhang, O~Pozo, B~Xu, X~Han, K~Manna, J~Paglione,
  C~Felser, Adolfo~G Grushin, et~al.
\newblock Giant topological longitudinal circular photo-galvanic effect in the
  chiral multifold semimetal cosi.
\newblock \emph{Nature communications}, 12\penalty0 (1):\penalty0 1--8, 2021.

\bibitem[{Xiao} et~al.(2005){Xiao}, {Shi}, and {Niu}]{xiao2005berry}
Di~{Xiao}, Junren {Shi}, and Qian {Niu}.
\newblock {Berry Phase Correction to Electron Density of States in Solids}.
\newblock \emph{Physical Review Letters}, 95\penalty0 (13):\penalty0 137204,
  September 2005.

\bibitem[{Stephanov} and {Yin}(2012)]{stephanov2012chiral}
M.~A. {Stephanov} and Y.~{Yin}.
\newblock {Chiral Kinetic Theory}.
\newblock \emph{Physical Review Letters}, 109\penalty0 (16):\penalty0 162001,
  October 2012.

\bibitem[Lifshits and Kosevich(1956)]{lifshits_kosevich}
IM~Lifshits and AM~Kosevich.
\newblock Theory of magnetic susceptibility in metals at low temperatures.
\newblock \emph{Sov. Phys. JETP}, 2:\penalty0 636--645, 1956.

\bibitem[Shoenberg(1984)]{shoenberg_2009}
David Shoenberg.
\newblock \emph{Magnetic oscillations in metals}.
\newblock Cambridge University Press, 1984.

\bibitem[Liang et~al.(2015)Liang, Gibson, Ali, Liu, Cava, and
  Ong]{liang2015ultrahigh}
Tian Liang, Quinn Gibson, Mazhar~N Ali, Minhao Liu, RJ~Cava, and NP~Ong.
\newblock Ultrahigh mobility and giant magnetoresistance in the dirac semimetal
  cd3as2.
\newblock \emph{Nature materials}, 14\penalty0 (3):\penalty0 280--284, 2015.

\bibitem[Hu et~al.(2016)Hu, Liu, Graf, Radmanesh, Adams, Chuang, Wang,
  Chiorescu, Wei, Spinu, et~al.]{hu2016pi}
Jin Hu, JY~Liu, David Graf, SMA Radmanesh, DJ~Adams, Alyssa Chuang, Yu~Wang,
  Irinel Chiorescu, Jiang Wei, Leonard Spinu, et~al.
\newblock $\pi$ berry phase and zeeman splitting of weyl semimetal tap.
\newblock \emph{Scientific reports}, 6:\penalty0 18674, 2016.

\bibitem[Luk'yanchuk and Kopelevich(2006)]{luk2006dirac}
Igor~A Luk'yanchuk and Yakov Kopelevich.
\newblock Dirac and normal fermions in graphite and graphene: Implications of
  the quantum hall effect.
\newblock \emph{Physical review letters}, 97\penalty0 (25):\penalty0 256801,
  2006.

\bibitem[Kharzeev(2016)]{kharzeev2016chiral}
Dmitri~E Kharzeev.
\newblock Chiral magnetic superconductivity.
\newblock \emph{arXiv preprint arXiv:1612.05677}, December 2016.

\bibitem[{Monteiro} et~al.(2015){Monteiro}, {Abanov}, and
  {Kharzeev}]{2015gustavo}
Gustavo~M. {Monteiro}, Alexander~G. {Abanov}, and Dmitri~E. {Kharzeev}.
\newblock {Magnetotransport in Dirac metals: Chiral magnetic effect and quantum
  oscillations}.
\newblock \emph{Physical Review B}, 92\penalty0 (16):\penalty0 165109, October
  2015.

\bibitem[Lipkin(1964)]{Lipkin1964}
Daniel~M. Lipkin.
\newblock {Existence of a New Conservation Law in Electromagnetic Theory}.
\newblock \emph{Journal of Mathematical Physics}, 5\penalty0 (5):\penalty0
  696--700, May 1964.

\bibitem[Tang and Cohen(2010)]{Tang2010}
Yiqiao Tang and Adam~E. Cohen.
\newblock {Optical Chirality and Its Interaction with Matter}.
\newblock \emph{Physical Review Letters}, 104\penalty0 (16):\penalty0 163901,
  April 2010.

\bibitem[Bliokh and Nori(2011)]{Bliokh2011}
Konstantin~Y. Bliokh and Franco Nori.
\newblock {Characterizing optical chirality}.
\newblock \emph{Physical Review A}, 83\penalty0 (2):\penalty0 021803, February
  2011.

\bibitem[Coles and Andrews(2012)]{Coles2012}
Matt~M. Coles and David~L. Andrews.
\newblock {Chirality and angular momentum in optical radiation}.
\newblock \emph{Physical Review A}, 85\penalty0 (6):\penalty0 063810, June
  2012.

\bibitem[Oka and Aoki(2009)]{Oka2009}
Takashi Oka and Hideo Aoki.
\newblock {Photovoltaic Hall effect in graphene}.
\newblock \emph{Physical Review B}, 79\penalty0 (8):\penalty0 081406, February
  2009.

\bibitem[Yudin et~al.(2015)Yudin, Eriksson, and Katsnelson]{Yudin2015}
Dmitry Yudin, Olle Eriksson, and Mikhail~I. Katsnelson.
\newblock {Dynamics of quasiparticles in graphene under intense circularly
  polarized light}.
\newblock \emph{Physical Review B}, 91\penalty0 (7):\penalty0 075419, February
  2015.

\bibitem[Taguchi et~al.(2016)Taguchi, Imaeda, Sato, and Tanaka]{Taguchi2016}
Katsuhisa Taguchi, Tatsushi Imaeda, Masatoshi Sato, and Yukio Tanaka.
\newblock {Photovoltaic chiral magnetic effect in Weyl semimetals}.
\newblock \emph{Physical Review B}, 93\penalty0 (20):\penalty0 201202, May
  2016.

\bibitem[Ebihara et~al.(2016)Ebihara, Fukushima, and Oka]{Ebihara2016}
Shu Ebihara, Kenji Fukushima, and Takashi Oka.
\newblock {Chiral pumping effect induced by rotating electric fields}.
\newblock \emph{Physical Review B}, 93\penalty0 (15):\penalty0 155107, April
  2016.

\bibitem[Chan et~al.(2016)Chan, Lee, Burch, Han, and Ran]{Chan2016}
Ching-Kit Chan, Patrick~A. Lee, Kenneth~S. Burch, Jung~Hoon Han, and Ying Ran.
\newblock {When Chiral Photons Meet Chiral Fermions: Photoinduced Anomalous
  Hall Effects in Weyl Semimetals}.
\newblock \emph{Physical Review Letters}, 116\penalty0 (2):\penalty0 026805,
  January 2016.

\bibitem[Chern and Simons(1974)]{Chern1974}
Shiing-Shen Chern and James Simons.
\newblock {Characteristic Forms and Geometric Invariants}.
\newblock \emph{The Annals of Mathematics}, 99\penalty0 (1):\penalty0 48,
  January 1974.

\bibitem[Afanasiev and Stepanovsky(1996)]{Afanasiev1996}
G.~N. Afanasiev and Yu.~P. Stepanovsky.
\newblock {The helicity of the free electromagnetic field and its physical
  meaning}.
\newblock \emph{Il Nuovo Cimento A}, 109\penalty0 (3):\penalty0 271--279, March
  1996.

\bibitem[Woltjer(1958)]{Woltjer1958}
L~Woltjer.
\newblock {A THEOREM ON FORCE-FREE MAGNETIC FIELDS.}
\newblock \emph{Proceedings of the National Academy of Sciences of the United
  States of America}, 44\penalty0 (6):\penalty0 489--91, June 1958.

\bibitem[Moffatt(1969)]{Moffatt1969}
H.~K. Moffatt.
\newblock {The degree of knottedness of tangled vortex lines}.
\newblock \emph{Journal of Fluid Mechanics}, 35\penalty0 (01):\penalty0 117,
  January 1969.

\bibitem[Arnold and Khesin(1998)]{Arnold1998}
Vladimir~I Arnold and Boris~A Khesin.
\newblock \emph{{Topological Methods in Hydrodynamics}}, volume 125 of
  \emph{Applied Mathematical Sciences}.
\newblock Springer New York, New York, NY, 1998.
\newblock ISBN 978-0-387-94947-5.

\bibitem[Taylor(1974)]{Taylor1974}
J.~B. Taylor.
\newblock {Relaxation of Toroidal Plasma and Generation of Reverse Magnetic
  Fields}.
\newblock \emph{Physical Review Letters}, 33\penalty0 (19):\penalty0
  1139--1141, November 1974.

\bibitem[Kharzeev and Warringa(2009)]{Kharzeev2009}
Dmitri~E. Kharzeev and Harmen~J. Warringa.
\newblock {Chiral magnetic conductivity}.
\newblock \emph{Physical Review D}, 80\penalty0 (3):\penalty0 034028, August
  2009.

\bibitem[Burkov(2015)]{burkov2015negative}
A.~A. Burkov.
\newblock Negative longitudinal magnetoresistance in dirac and weyl metals.
\newblock \emph{Phys. Rev. B}, 91:\penalty0 245157, June 2015.

\bibitem[{Kharzeev} et~al.(2018{\natexlab{b}}){Kharzeev}, {Kikuchi}, {Meyer},
  and {Tanizaki}]{yuta2018}
Dmitri~E. {Kharzeev}, Yuta {Kikuchi}, Ren{\'e} {Meyer}, and Yuya {Tanizaki}.
\newblock {Giant photocurrent in asymmetric Weyl semimetals from the helical
  magnetic effect}.
\newblock \emph{Physical Review B}, 98\penalty0 (1):\penalty0 014305, July
  2018{\natexlab{b}}.

\bibitem[Wilczek(1987)]{Wilczek1987}
Frank Wilczek.
\newblock {Two applications of axion electrodynamics}.
\newblock \emph{Physical Review Letters}, 58\penalty0 (18):\penalty0
  1799--1802, May 1987.

\bibitem[Carroll et~al.(1990)Carroll, Field, and Jackiw]{Carroll1990}
Sean~M. Carroll, George~B. Field, and Roman Jackiw.
\newblock {Limits on a Lorentz- and parity-violating modification of
  electrodynamics}.
\newblock \emph{Physical Review D}, 41\penalty0 (4):\penalty0 1231--1240,
  February 1990.

\bibitem[Sikivie(1983)]{Sikivie1983}
P.~Sikivie.
\newblock {Experimental Tests of the "Invisible" Axion}.
\newblock \emph{Physical Review Letters}, 51\penalty0 (16):\penalty0
  1415--1417, October 1983.

\bibitem[Kampfrath et~al.(2013)Kampfrath, Battiato, Maldonado, Eilers,
  N{\"o}tzold, M{\"a}hrlein, Zbarsky, Freimuth, Mokrousov, Bl{\"u}gel,
  et~al.]{Kampfrath_NNano_2013}
Tobias Kampfrath, Marco Battiato, Pablo Maldonado, G~Eilers, J~N{\"o}tzold,
  Sebastian M{\"a}hrlein, V~Zbarsky, Frank Freimuth, Yuriy Mokrousov, Stefan
  Bl{\"u}gel, et~al.
\newblock Terahertz spin current pulses controlled by magnetic
  heterostructures.
\newblock \emph{Nature nanotechnology}, 8\penalty0 (4):\penalty0 256--260,
  2013.

\bibitem[Liu et~al.(2015)Liu, Richard, Song, Zhao, Fang, Chen, and
  Ding]{Liu_PRB_2015}
HW~Liu, P~Richard, ZD~Song, LX~Zhao, Z~Fang, G-F Chen, and H~Ding.
\newblock Raman study of lattice dynamics in the weyl semimetal taas.
\newblock \emph{Physical Review B}, 92\penalty0 (6):\penalty0 064302, 2015.

\bibitem[Osterhoudt et~al.(2019)Osterhoudt, Diebel, Gray, Yang, Stanco, Huang,
  Shen, Ni, Moll, Ran, et~al.]{Osterhoudt_arXiv_2018}
Gavin~B Osterhoudt, Laura~K Diebel, Mason~J Gray, Xu~Yang, John Stanco,
  Xiangwei Huang, Bing Shen, Ni~Ni, Philip~JW Moll, Ying Ran, et~al.
\newblock Colossal mid-infrared bulk photovoltaic effect in a type-i weyl
  semimetal.
\newblock \emph{Nature materials}, 18\penalty0 (5):\penalty0 471--475, 2019.

\bibitem[Ganichev and Prettl(2003)]{Ganichev_JPhys_2003}
Sergey~D Ganichev and Wilhelm Prettl.
\newblock Spin photocurrents in quantum wells.
\newblock \emph{Journal of physics: Condensed matter}, 15\penalty0
  (20):\penalty0 R935, 2003.

\bibitem[McIver et~al.(2012)McIver, Hsieh, Steinberg, Jarillo-Herrero, and
  Gedik]{McIver_NNano_2011}
JW~McIver, David Hsieh, Hadar Steinberg, Pablo Jarillo-Herrero, and Nuh Gedik.
\newblock Control over topological insulator photocurrents with light
  polarization.
\newblock \emph{Nature nanotechnology}, 7\penalty0 (2):\penalty0 96--100, 2012.

\bibitem[Xu et~al.(2015)Xu, Dai, Zhao, Wang, Yang, Zhang, Liu, Xiao, Chen,
  Taylor, et~al.]{Xu_PRB_2015}
Bing Xu, YM~Dai, LX~Zhao, Kai Wang, Run Yang, Wei Zhang, JY~Liu, Hong Xiao,
  GF~Chen, AJ~Taylor, et~al.
\newblock Optical signatures of weyl points in taas.
\newblock \emph{arXiv preprint arXiv:1510.00470}, October 2015.

\bibitem[Braun et~al.(2016)Braun, Mussler, Hruban, Konczykowski, Schumann,
  Wolf, M{\"u}nzenberg, Perfetti, and Kampfrath]{Braun_NC_2016}
Lukas Braun, Gregor Mussler, Andrzej Hruban, Marcin Konczykowski, Thomas
  Schumann, Martin Wolf, Markus M{\"u}nzenberg, Luca Perfetti, and Tobias
  Kampfrath.
\newblock Ultrafast photocurrents at the surface of the three-dimensional
  topological insulator bi 2 se 3.
\newblock \emph{Nature communications}, 7\penalty0 (1):\penalty0 1--9, 2016.

\bibitem[Sipe and Shkrebtii(2000)]{Sipe_PRB_2000}
JE~Sipe and AI~Shkrebtii.
\newblock Second-order optical response in semiconductors.
\newblock \emph{Physical Review B}, 61\penalty0 (8):\penalty0 5337, 2000.

\bibitem[Nastos and Sipe(2006)]{Nastos_PRB_2006}
F~Nastos and JE~Sipe.
\newblock Optical rectification and shift currents in gaas and gap response:
  Below and above the band gap.
\newblock \emph{Physical Review B}, 74\penalty0 (3):\penalty0 035201, 2006.

\bibitem[Buckeridge et~al.(2016)Buckeridge, Jevdokimovs, Catlow, and
  Sokol]{Buckeridge_PRB_2016}
J~Buckeridge, D~Jevdokimovs, CRA Catlow, and AA~Sokol.
\newblock Bulk electronic, elastic, structural, and dielectric properties of
  the weyl semimetal taas.
\newblock \emph{Physical Review B}, 93\penalty0 (12):\penalty0 125205, 2016.

\bibitem[Stern(1966)]{Stern_PR_1966}
Frank Stern.
\newblock Effect of band tails on stimulated emission of light in
  semiconductors.
\newblock \emph{Phys. Rev.}, 148:\penalty0 186--194, August 1966.

\bibitem[Van~Mieghem(1992)]{Mieghem_RMP_1992}
Piet Van~Mieghem.
\newblock Theory of band tails in heavily doped semiconductors.
\newblock \emph{Rev. Mod. Phys.}, 64:\penalty0 755--793, July 1992.

\bibitem[Zhang et~al.(2018{\natexlab{a}})Zhang, Ishizuka, van~den Brink,
  Felser, Yan, and Nagaosa]{Zhang_PRB_2018}
Yang Zhang, Hiroaki Ishizuka, Jeroen van~den Brink, Claudia Felser, Binghai
  Yan, and Naoto Nagaosa.
\newblock Photogalvanic effect in weyl semimetals from first principles.
\newblock \emph{Physical Review B}, 97\penalty0 (24):\penalty0 241118,
  2018{\natexlab{a}}.

\bibitem[Shalygin et~al.(2016)Shalygin, Moldavskaya, Danilov, Farbshtein, and
  Golub]{Shalygin_PRB_2016}
VA~Shalygin, MD~Moldavskaya, SN~Danilov, II~Farbshtein, and LE~Golub.
\newblock Circular photon drag effect in bulk tellurium.
\newblock \emph{Physical Review B}, 93\penalty0 (4):\penalty0 045207, 2016.

\bibitem[Dekorsy et~al.(1996)Dekorsy, Auer, Bakker, Roskos, and
  Kurz]{Dekorsy_PRB_1996}
Thomas Dekorsy, Holger Auer, Huib~J Bakker, Hartmut~G Roskos, and Heinrich
  Kurz.
\newblock Thz electromagnetic emission by coherent infrared-active phonons.
\newblock \emph{Physical Review B}, 53\penalty0 (7):\penalty0 4005, 1996.

\bibitem[Zhang et~al.(2017)Zhang, Narayan, Lu, Zhang, Zhang, Ni, Yuan, Liu,
  Park, Zhang, et~al.]{Zhang_NC_2017}
Cheng Zhang, Awadhesh Narayan, Shiheng Lu, Jinglei Zhang, Huiqin Zhang,
  Zhuoliang Ni, Xiang Yuan, Yanwen Liu, Ju-Hyun Park, Enze Zhang, et~al.
\newblock Evolution of weyl orbit and quantum hall effect in dirac semimetal cd
  3 as 2.
\newblock \emph{Nature communications}, 8\penalty0 (1):\penalty0 1--8, 2017.

\bibitem[{Bel'kov} et~al.(2005){Bel'kov}, {Ganichev}, {Ivchenko}, {Tarasenko},
  {Weber}, {Giglberger}, {Olteanu}, {Tranitz}, {Danilov}, {Schneider},
  {Wegscheider}, {Weiss}, and {Prettl}]{2005belkov}
V.~V. {Bel'kov}, S.~D. {Ganichev}, E.~L. {Ivchenko}, S.~A. {Tarasenko},
  W.~{Weber}, S.~{Giglberger}, M.~{Olteanu}, H.~P. {Tranitz}, S.~N. {Danilov},
  Petra {Schneider}, W.~{Wegscheider}, D.~{Weiss}, and W.~{Prettl}.
\newblock {Magneto-gyrotropic photogalvanic effects in semiconductor quantum
  wells}.
\newblock \emph{Journal of Physics Condensed Matter}, 17\penalty0
  (21):\penalty0 3405--3428, June 2005.

\bibitem[{Golub} and {Ivchenko}(2018)]{2018golub}
L.~E. {Golub} and E.~L. {Ivchenko}.
\newblock {Circular and magnetoinduced photocurrents in Weyl semimetals}.
\newblock \emph{Physical Review B}, 98\penalty0 (7):\penalty0 075305, August
  2018.

\bibitem[Silva et~al.(2020)Silva, Ferreira, Schreck, and
  Urrutia]{silva2020magneticconductivity}
Pedro D.~S. Silva, Manoel~M. Ferreira, Marco Schreck, and Luis~F. Urrutia.
\newblock Magnetic-conductivity effects on electromagnetic propagation in
  dispersive matter.
\newblock \emph{Phys. Rev. D}, 102:\penalty0 076001, October 2020.

\bibitem[{Son} and {Yamamoto}(2013)]{son2013kinetic}
Dam~Thanh {Son} and Naoki {Yamamoto}.
\newblock {Kinetic theory with Berry curvature from quantum field theories}.
\newblock \emph{Physical Review D}, 87\penalty0 (8):\penalty0 085016, April
  2013.

\bibitem[Guinea et~al.(2010)Guinea, Katsnelson, and Geim]{guinea2010energy}
Francisco Guinea, MI~Katsnelson, and AK~Geim.
\newblock Energy gaps and a zero-field quantum hall effect in graphene by
  strain engineering.
\newblock \emph{Nature Physics}, 6\penalty0 (1):\penalty0 30--33, 2010.

\bibitem[Vozmediano et~al.(2010)Vozmediano, Katsnelson, and
  Guinea]{vozmediano2010gauge}
Maria~AH Vozmediano, MI~Katsnelson, and Francisco Guinea.
\newblock Gauge fields in graphene.
\newblock \emph{Physics Reports}, 496\penalty0 (4-5):\penalty0 109--148, 2010.

\bibitem[Cortijo et~al.(2015)Cortijo, Ferreir{\'o}s, Landsteiner, and
  Vozmediano]{cortijo2015elastic}
Alberto Cortijo, Yago Ferreir{\'o}s, Karl Landsteiner, and Mar{\'\i}a~AH
  Vozmediano.
\newblock Elastic gauge fields in weyl semimetals.
\newblock \emph{Physical review letters}, 115\penalty0 (17):\penalty0 177202,
  2015.

\bibitem[Levy et~al.(2010)Levy, Burke, Meaker, Panlasigui, Zettl, Guinea, Neto,
  and Crommie]{levy2010strain}
N~Levy, SA~Burke, KL~Meaker, M~Panlasigui, A~Zettl, F~Guinea, AH~Castro Neto,
  and Michael~F Crommie.
\newblock Strain-induced pseudo--magnetic fields greater than 300 tesla in
  graphene nanobubbles.
\newblock \emph{Science}, 329\penalty0 (5991):\penalty0 544--547, 2010.

\bibitem[Zhou et~al.(2013)Zhou, Jiang, Niu, and Shi]{Zhou:2012ix}
Jianhui Zhou, Hua Jiang, Qian Niu, and Junren Shi.
\newblock {Topological Invariants of Metals and Related Physical Effects}.
\newblock \emph{Chin. Phys. Lett.}, 30:\penalty0 027101, 2013.

\bibitem[Huang et~al.(2019)Huang, Li, Zhou, and Zhang]{PhysRevB.99.155152}
Ze-Min Huang, Longyue Li, Jianhui Zhou, and Hong-Hao Zhang.
\newblock Torsional response and liouville anomaly in weyl semimetals with
  dislocations.
\newblock \emph{Phys. Rev. B}, 99:\penalty0 155152, April 2019.

\bibitem[Nissinen(2020)]{PhysRevLett.124.117002}
Jaakko Nissinen.
\newblock Emergent spacetime and gravitational nieh-yan anomaly in chiral
  $p+ip$ weyl superfluids and superconductors.
\newblock \emph{Phys. Rev. Lett.}, 124:\penalty0 117002, March 2020.

\bibitem[Nissinen and Volovik(2020)]{PhysRevResearch.2.033269}
J.~Nissinen and G.~E. Volovik.
\newblock Thermal nieh-yan anomaly in weyl superfluids.
\newblock \emph{Phys. Rev. Research}, 2:\penalty0 033269, August 2020.

\bibitem[Laurila and Nissinen(2020)]{laurila2020torsional}
Sara Laurila and Jaakko Nissinen.
\newblock Torsional landau levels and geometric anomalies in condensed matter
  weyl systems.
\newblock \emph{arXiv preprint arXiv:2007.10682}, July 2020.

\bibitem[Nieh and Yan(1982)]{nieh1982quantized}
HT~Nieh and ML~Yan.
\newblock Quantized dirac field in curved riemann-cartan background. i.
  symmetry properties, green's function.
\newblock \emph{Annals of Physics}, 138\penalty0 (2):\penalty0 237--259, 1982.

\bibitem[Kharzeev(2010)]{Kharzeev:2009fn}
Dmitri~E. Kharzeev.
\newblock {Topologically induced local P and CP violation in QCD x QED}.
\newblock \emph{Annals Phys.}, 325:\penalty0 205--218, 2010.

\bibitem[Pikulin et~al.(2016)Pikulin, Chen, and Franz]{Pikulin_2016}
D.I. Pikulin, Anffany Chen, and M.~Franz.
\newblock Chiral anomaly from strain-induced gauge fields in dirac and weyl
  semimetals.
\newblock \emph{Physical Review X}, 6\penalty0 (4), October 2016.

\bibitem[Song and Dai(2019)]{song2019hear}
Zhida Song and Xi~Dai.
\newblock Hear the sound of weyl fermions.
\newblock \emph{Physical Review X}, 9\penalty0 (2):\penalty0 021053, 2019.

\bibitem[Chernodub and Vozmediano(2019)]{Chernodub:2019lhw}
M.N. Chernodub and Mar\'\i{}a~A.H. Vozmediano.
\newblock {Chiral sound waves in strained Weyl semimetals}.
\newblock \emph{Phys. Rev. Res.}, 1\penalty0 (3):\penalty0 032040, 2019.

\bibitem[Sukhachov and Rostami(2020)]{PhysRevLett.124.126602}
P.~O. Sukhachov and H.~Rostami.
\newblock Acoustogalvanic effect in dirac and weyl semimetals.
\newblock \emph{Phys. Rev. Lett.}, 124:\penalty0 126602, March 2020.

\bibitem[Xiao et~al.(2010)Xiao, Chang, and Niu]{xiao2010berry}
Di~Xiao, Ming-Che Chang, and Qian Niu.
\newblock Berry phase effects on electronic properties.
\newblock \emph{Reviews of modern physics}, 82\penalty0 (3):\penalty0 1959,
  2010.

\bibitem[Chen(2013)]{Chen2013Berry}
Jiunn~Wei Chen.
\newblock Berry curvature and 4-dimensional monopole in relativistic chiral
  kinetic equation.
\newblock \emph{Phys.rev.lett}, 110, 2013.

\bibitem[Dwivedi and Stone(2013)]{Dwivedi:2013dea}
Vatsal Dwivedi and Michael Stone.
\newblock {Classical chiral kinetic theory and anomalies in even space-time
  dimensions}.
\newblock \emph{J. Phys. A}, 47:\penalty0 025401, 2013.

\bibitem[Basar et~al.(2013)Basar, Kharzeev, and Zahed]{Basar:2013qia}
Gokce Basar, Dmitri~E. Kharzeev, and Ismail Zahed.
\newblock {Chiral and Gravitational Anomalies on Fermi Surfaces}.
\newblock \emph{Phys. Rev. Lett.}, 111:\penalty0 161601, 2013.

\bibitem[Manuel and Torres-Rincon(2014)]{Manuel:2013zaa}
Cristina Manuel and Juan~M. Torres-Rincon.
\newblock {Kinetic theory of chiral relativistic plasmas and energy density of
  their gauge collective excitations}.
\newblock \emph{Phys. Rev. D}, 89\penalty0 (9):\penalty0 096002, 2014.

\bibitem[Chen et~al.(2014)Chen, Son, Stephanov, Yee, and Yin]{Chen:2014cla}
Jing-Yuan Chen, Dam~T. Son, Mikhail~A. Stephanov, Ho-Ung Yee, and Yi~Yin.
\newblock {Lorentz Invariance in Chiral Kinetic Theory}.
\newblock \emph{Phys. Rev. Lett.}, 113\penalty0 (18):\penalty0 182302, 2014.

\bibitem[Gorbar et~al.(2017)Gorbar, Miransky, Shovkovy, and
  Sukhachov]{Gorbar:2016ygi}
E.V. Gorbar, V.A. Miransky, I.A. Shovkovy, and P.O. Sukhachov.
\newblock {Consistent Chiral Kinetic Theory in Weyl Materials: Chiral Magnetic
  Plasmons}.
\newblock \emph{Phys. Rev. Lett.}, 118\penalty0 (12):\penalty0 127601, 2017.

\bibitem[Kharzeev et~al.(2017)Kharzeev, Stephanov, and Yee]{Kharzeev:2016sut}
Dmitri~E. Kharzeev, Mikhail~A. Stephanov, and Ho-Ung Yee.
\newblock {Anatomy of chiral magnetic effect in and out of equilibrium}.
\newblock \emph{Phys. Rev. D}, 95\penalty0 (5):\penalty0 051901, 2017.

\bibitem[Sundaram and Niu(1999)]{sundaram1999wave}
Ganesh Sundaram and Qian Niu.
\newblock Wave-packet dynamics in slowly perturbed crystals: Gradient
  corrections and berry-phase effects.
\newblock \emph{Physical Review B}, 59\penalty0 (23):\penalty0 14915, 1999.

\bibitem[Volovik(2003)]{volovik2003universe}
Grigory~E Volovik.
\newblock \emph{The universe in a helium droplet}, volume 117.
\newblock Oxford University Press on Demand, 2003.

\bibitem[Freimuth et~al.(2013)Freimuth, Bamler, Mokrousov, and
  Rosch]{freimuth2013phase}
Frank Freimuth, Robert Bamler, Yuriy Mokrousov, and Achim Rosch.
\newblock Phase-space berry phases in chiral magnets: Dzyaloshinskii-moriya
  interaction and the charge of skyrmions.
\newblock \emph{Physical Review B}, 88\penalty0 (21):\penalty0 214409, 2013.

\bibitem[McCormick et~al.(2017)McCormick, Kimchi, and
  Trivedi]{mccormick2017minimal}
Timothy~M McCormick, Itamar Kimchi, and Nandini Trivedi.
\newblock Minimal models for topological weyl semimetals.
\newblock \emph{Physical Review B}, 95\penalty0 (7):\penalty0 075133, 2017.

\bibitem[Hayata and Hidaka(2017)]{Hayata:2016wgy}
Tomoya Hayata and Yoshimasa Hidaka.
\newblock {Kinetic derivation of generalized phase space Chern-Simons theory}.
\newblock \emph{Phys. Rev. B}, 95\penalty0 (12):\penalty0 125137, 2017.

\bibitem[Grushin et~al.(2016)Grushin, Venderbos, Vishwanath, and
  Ilan]{grushin2016inhomogeneous}
Adolfo~G Grushin, J{\"o}rn~WF Venderbos, Ashvin Vishwanath, and Roni Ilan.
\newblock Inhomogeneous weyl and dirac semimetals: Transport in axial magnetic
  fields and fermi arc surface states from pseudo-landau levels.
\newblock \emph{Physical Review X}, 6\penalty0 (4):\penalty0 041046, 2016.

\bibitem[{Chang} et~al.(2017){Chang}, {Xu}, {Wieder}, {Sanchez}, {Huang},
  {Belopolski}, {Chang}, {Zhang}, {Bansil}, {Lin}, and {Hasan}]{HasanRhSi}
Guoqing {Chang}, Su-Yang {Xu}, Benjamin~J. {Wieder}, Daniel~S. {Sanchez},
  Shin-Ming {Huang}, Ilya {Belopolski}, Tay-Rong {Chang}, Songtian {Zhang},
  Arun {Bansil}, Hsin {Lin}, and M.~Zahid {Hasan}.
\newblock {Unconventional Chiral Fermions and Large Topological Fermi Arcs in
  RhSi}.
\newblock \emph{Physical Review Letters}, 119\penalty0 (20):\penalty0 206401,
  November 2017.

\bibitem[{Rao} et~al.(2019){Rao}, {Li}, {Zhang}, {Tian}, {Li}, {Fu}, {Tang},
  {Wang}, {Li}, {Fan}, {Li}, {Huang}, {Liu}, {Long}, {Fang}, {Weng}, {Shi},
  {Lei}, {Sun}, {Qian}, and {Ding}]{rao2019}
Zhicheng {Rao}, Hang {Li}, Tiantian {Zhang}, Shangjie {Tian}, Chenghe {Li},
  Binbin {Fu}, Cenyao {Tang}, Le~{Wang}, Zhilin {Li}, Wenhui {Fan}, Jiajun
  {Li}, Yaobo {Huang}, Zhehong {Liu}, Youwen {Long}, Chen {Fang}, Hongming
  {Weng}, Youguo {Shi}, Hechang {Lei}, Yujie {Sun}, Tian {Qian}, and Hong
  {Ding}.
\newblock {Observation of unconventional chiral fermions with long Fermi arcs
  in CoSi}.
\newblock \emph{Nature}, 567\penalty0 (7749):\penalty0 496--499, March 2019.

\bibitem[{Ray} et~al.(2020){Ray}, {Sadhukhan}, {Richter}, {Facio}, and {van den
  Brink}]{ray2020}
Rajyavardhan {Ray}, Banasree {Sadhukhan}, Manuel {Richter}, Jorge~I. {Facio},
  and Jeroen {van den Brink}.
\newblock {Tunable chirality of noncentrosymmetric magnetic Weyl semimetals}.
\newblock \emph{arXiv preprint arXiv:2006.10602}, page arXiv:2006.10602, June
  2020.

\bibitem[Barron(1986)]{barron1986}
Laurence~D Barron.
\newblock True and false chirality and absolute asymmetric synthesis.
\newblock \emph{Journal of the American Chemical Society}, 108\penalty0
  (18):\penalty0 5539--5542, 1986.

\bibitem[{Yang} et~al.(2015){Yang}, {Liu}, {Sun}, {Peng}, {Yang}, {Zhang},
  {Zhou}, {Zhang}, {Guo}, {Rahn}, {Prabhakaran}, {Hussain}, {Mo}, {Felser},
  {Yan}, and {Chen}]{yang2015}
L.~X. {Yang}, Z.~K. {Liu}, Y.~{Sun}, H.~{Peng}, H.~F. {Yang}, T.~{Zhang},
  B.~{Zhou}, Y.~{Zhang}, Y.~F. {Guo}, M.~{Rahn}, D.~{Prabhakaran},
  Z.~{Hussain}, S.~K. {Mo}, C.~{Felser}, B.~{Yan}, and Y.~L. {Chen}.
\newblock {Weyl semimetal phase in the non-centrosymmetric compound TaAs}.
\newblock \emph{Nature Physics}, 11\penalty0 (9):\penalty0 728--732, September
  2015.

\bibitem[{Li} et~al.(2017){Li}, {Wen}, {He}, {Zhang}, {Xia}, {Yu}, {Yang},
  {Zhu}, {Alshareef}, and {Zhang}]{li2017}
Peng {Li}, Yan {Wen}, Xin {He}, Qiang {Zhang}, Chuan {Xia}, Zhi-Ming {Yu},
  Shengyuan~A. {Yang}, Zhiyong {Zhu}, Husam~N. {Alshareef}, and Xi-Xiang
  {Zhang}.
\newblock {Evidence for topological type-II Weyl semimetal WTe$_{2}$}.
\newblock \emph{Nature Communications}, 8:\penalty0 2150, December 2017.

\bibitem[Zhang et~al.(2018{\natexlab{b}})Zhang, Yang, and Wang]{zhang2018}
Minping Zhang, Zongxian Yang, and Guangtao Wang.
\newblock Coexistence of type-i and type-ii weyl points in the weyl-semimetal
  osc2.
\newblock \emph{The Journal of Physical Chemistry C}, 122\penalty0
  (6):\penalty0 3533--3538, 2018{\natexlab{b}}.

\bibitem[{Xu} et~al.(2016){Xu}, {Weng}, {Lv}, {Matt}, {Park}, {Bisti},
  {Strocov}, {Gawryluk}, {Pomjakushina}, {Conder}, {Plumb}, {Radovic},
  {Aut{\`e}s}, {Yazyev}, {Fang}, {Dai}, {Qian}, {Mesot}, {Ding}, and
  {Shi}]{xu2016}
N.~{Xu}, H.~M. {Weng}, B.~Q. {Lv}, C.~E. {Matt}, J.~{Park}, F.~{Bisti}, V.~N.
  {Strocov}, D.~{Gawryluk}, E.~{Pomjakushina}, K.~{Conder}, N.~C. {Plumb},
  M.~{Radovic}, G.~{Aut{\`e}s}, O.~V. {Yazyev}, Z.~{Fang}, X.~{Dai}, T.~{Qian},
  J.~{Mesot}, H.~{Ding}, and M.~{Shi}.
\newblock {Observation of Weyl nodes and Fermi arcs in tantalum phosphide}.
\newblock \emph{Nature Communications}, 7:\penalty0 11006, March 2016.

\bibitem[{Modic} et~al.(2019){Modic}, {Meng}, {Ronning}, {Bauer}, {Moll}, and
  {Ramshaw}]{modic2019}
K.~A. {Modic}, Tobias {Meng}, Filip {Ronning}, Eric~D. {Bauer}, Philip J.~W.
  {Moll}, and B.~J. {Ramshaw}.
\newblock {Thermodynamic Signatures of Weyl Fermions in NbP}.
\newblock \emph{Scientific Reports}, 9:\penalty0 2095, February 2019.

\bibitem[{Yuan} et~al.(2020){Yuan}, {Zhang}, {Zhang}, {Yan}, {Lyu}, {Zhang},
  {Li}, {Song}, {Zhao}, {Leng}, {Ozerov}, {Chen}, {Wang}, {Shi}, {Yan}, and
  {Xiu}]{yuan2020}
Xiang {Yuan}, Cheng {Zhang}, Yi~{Zhang}, Zhongbo {Yan}, Tairu {Lyu}, Mengyao
  {Zhang}, Zhilin {Li}, Chaoyu {Song}, Minhao {Zhao}, Pengliang {Leng},
  Mykhaylo {Ozerov}, Xiaolong {Chen}, Nanlin {Wang}, Yi~{Shi}, Hugen {Yan}, and
  Faxian {Xiu}.
\newblock {The discovery of dynamic chiral anomaly in a Weyl semimetal NbAs}.
\newblock \emph{Nature Communications}, 11:\penalty0 1259, March 2020.

\bibitem[{Sun} et~al.(2020){Sun}, {Song}, {Weng}, and
  {Dai}]{sun2020topological}
Song {Sun}, Zhida {Song}, Hongming {Weng}, and Xi~{Dai}.
\newblock {Topological metals induced by the Zeeman effect}.
\newblock \emph{Physical Review B}, 101\penalty0 (12):\penalty0 125118, March
  2020.

\bibitem[{Choi} et~al.(2020){Choi}, {Villanova}, and {Park}]{choi2020}
Yichul {Choi}, John~W. {Villanova}, and Kyungwha {Park}.
\newblock {Zeeman-splitting-induced topological nodal structure and anomalous
  Hall conductivity in ZrTe$_{5}$}.
\newblock \emph{Physical Review B}, 101\penalty0 (3):\penalty0 035105, January
  2020.

\bibitem[{Ruan} et~al.(2016){Ruan}, {Jian}, {Yao}, {Zhang}, {Zhang}, and
  {Xing}]{ruan2016}
Jiawei {Ruan}, Shao-Kai {Jian}, Hong {Yao}, Haijun {Zhang}, Shou-Cheng {Zhang},
  and Dingyu {Xing}.
\newblock {Symmetry-protected ideal Weyl semimetal in HgTe-class materials}.
\newblock \emph{Nature Communications}, 7:\penalty0 11136, April 2016.

\bibitem[{Shekhar} et~al.(2018){Shekhar}, {Kumar}, {Grinenko}, {Singh},
  {Sarkar}, {Luetkens}, {Wu}, {Zhang}, {Komarek}, {Kampert}, {Skourski},
  {Wosnitza}, {Schnelle}, {McCollam}, {Zeitler}, {K{\"u}bler}, {Yan}, {Klauss},
  {Parkin}, and {Felser}]{shekhar2018}
Chandra {Shekhar}, Nitesh {Kumar}, V.~{Grinenko}, Sanjay {Singh}, R.~{Sarkar},
  H.~{Luetkens}, Shu-Chun {Wu}, Yang {Zhang}, Alexander~C. {Komarek}, Erik
  {Kampert}, Yurii {Skourski}, Jochen {Wosnitza}, Walter {Schnelle}, Alix
  {McCollam}, Uli {Zeitler}, J{\"u}rgen {K{\"u}bler}, Binghai {Yan}, H.~H.
  {Klauss}, S.~S.~P. {Parkin}, and C.~{Felser}.
\newblock {Anomalous Hall effect in Weyl semimetal half-Heusler compounds RPtBi
  (R = Gd and Nd)}.
\newblock \emph{Proceedings of the National Academy of Science}, 115\penalty0
  (37):\penalty0 9140--9144, September 2018.

\bibitem[{Ghimire} et~al.(2019){Ghimire}, {Facio}, {You}, {Ye}, {Checkelsky},
  {Fang}, {Kaxiras}, {Richter}, and {van den Brink}]{ghimire2019}
Madhav~Prasad {Ghimire}, Jorge~I. {Facio}, Jhih-Shih {You}, Linda {Ye},
  Joseph~G. {Checkelsky}, Shiang {Fang}, Efthimios {Kaxiras}, Manuel {Richter},
  and Jeroen {van den Brink}.
\newblock {Creating Weyl nodes and controlling their energy by magnetization
  rotation}.
\newblock \emph{Physical Review Research}, 1\penalty0 (3):\penalty0 032044,
  December 2019.

\bibitem[{H{\"u}bener} et~al.(2017){H{\"u}bener}, {Sentef}, {de Giovannini},
  {Kemper}, and {Rubio}]{hubener2017}
Hannes {H{\"u}bener}, Michael~A. {Sentef}, Umberto {de Giovannini},
  Alexander~F. {Kemper}, and Angel {Rubio}.
\newblock {Creating stable Floquet-Weyl semimetals by laser-driving of 3D Dirac
  materials}.
\newblock \emph{Nature Communications}, 8:\penalty0 13940, January 2017.

\bibitem[{He} et~al.(2018){He}, {Di Sante}, {Li}, {Chen}, {Rondinelli}, and
  {Franchini}]{he2018}
Jiangang {He}, Domenico {Di Sante}, Ronghan {Li}, Xing-Qiu {Chen}, James~M.
  {Rondinelli}, and Cesare {Franchini}.
\newblock {Tunable metal-insulator transition, Rashba effect and Weyl Fermions
  in a relativistic charge-ordered ferroelectric oxide}.
\newblock \emph{Nature Communications}, 9:\penalty0 492, February 2018.

\bibitem[{Sharma} et~al.(2019){Sharma}, {Xiang}, {Shao}, {Zhang}, {Tsymbal},
  {Hamilton}, and {Seidel}]{sharma2019}
Pankaj {Sharma}, Fei-Xiang {Xiang}, Ding-Fu {Shao}, Dawei {Zhang}, Evgeny~Y.
  {Tsymbal}, Alex~R. {Hamilton}, and Jan {Seidel}.
\newblock {A room-temperature ferroelectric semimetal}.
\newblock \emph{Science Advances}, 5\penalty0 (7):\penalty0 eaax5080, July
  2019.

\bibitem[{O'Brien} et~al.(2017){O'Brien}, {Beenakker}, and
  {Adagideli}]{obrien2017}
T.~E. {O'Brien}, C.~W.~J. {Beenakker}, and I.~{Adagideli}.
\newblock {Superconductivity Provides Access to the Chiral Magnetic Effect of
  an Unpaired Weyl Cone}.
\newblock \emph{Physical Review Letters}, 118\penalty0 (20):\penalty0 207701,
  May 2017.

\bibitem[{Hirschberger} et~al.(2016){Hirschberger}, {Kushwaha}, {Wang},
  {Gibson}, {Liang}, {Belvin}, {Bernevig}, {Cava}, and {Ong}]{hirschberger2016}
Max {Hirschberger}, Satya {Kushwaha}, Zhijun {Wang}, Quinn {Gibson}, Sihang
  {Liang}, Carina~A. {Belvin}, B.~A. {Bernevig}, R.~J. {Cava}, and N.~P. {Ong}.
\newblock {The chiral anomaly and thermopower of Weyl fermions in the
  half-Heusler GdPtBi}.
\newblock \emph{Nature Materials}, 15\penalty0 (11):\penalty0 1161--1165,
  November 2016.

\bibitem[{Cano} et~al.(2017){Cano}, {Bradlyn}, {Wang}, {Hirschberger}, {Ong},
  and {Bernevig}]{cano2017}
Jennifer {Cano}, Barry {Bradlyn}, Zhijun {Wang}, Max {Hirschberger}, N.~P.
  {Ong}, and B.~A. {Bernevig}.
\newblock {Chiral anomaly factory: Creating Weyl fermions with a magnetic
  field}.
\newblock \emph{Physical Review B}, 95\penalty0 (16):\penalty0 161306, April
  2017.

\bibitem[{Qu} et~al.(2016){Qu}, {van Veen}, {de Vries}, {Beukman}, {Wimmer},
  {Yi}, {Kiselev}, {Nguyen}, {Sokolich}, {Manfra}, {Nichele}, {Marcus}, and
  {Kouwenhoven}]{qu2016}
Fanming {Qu}, Jasper {van Veen}, Folkert~K. {de Vries}, Arjan J.~A. {Beukman},
  Michael {Wimmer}, Wei {Yi}, Andrey~A. {Kiselev}, Binh-Minh {Nguyen}, Marko
  {Sokolich}, Michael~J. {Manfra}, Fabrizio {Nichele}, Charles~M. {Marcus}, and
  Leo~P. {Kouwenhoven}.
\newblock {Quantized Conductance and Large g-Factor Anisotropy in InSb Quantum
  Point Contacts}.
\newblock \emph{Nano Letters}, 16\penalty0 (12):\penalty0 7509--7513, December
  2016.

\bibitem[{Nilsson} et~al.(2009){Nilsson}, {Caroff}, {Thelander}, {Larsson},
  {Wagner}, {Wernersson}, {Samuelson}, and {Xu}]{nilsson2009}
Henrik~A. {Nilsson}, Philippe {Caroff}, Claes {Thelander}, Marcus {Larsson},
  Jakob~B. {Wagner}, Lars-Erik {Wernersson}, Lars {Samuelson}, and H.~Q. {Xu}.
\newblock {Giant, Level-DependentgFactors in InSb Nanowire Quantum Dots}.
\newblock \emph{Nano Letters}, 9\penalty0 (9):\penalty0 3151--3156, September
  2009.

\bibitem[{Vurgaftman} et~al.(2001){Vurgaftman}, {Meyer}, and
  {Ram-Mohan}]{vurgaftman2001}
I.~{Vurgaftman}, J.~R. {Meyer}, and L.~R. {Ram-Mohan}.
\newblock {Band parameters for III-V compound semiconductors and their alloys}.
\newblock \emph{Journal of Applied Physics}, 89\penalty0 (11):\penalty0
  5815--5875, June 2001.

\bibitem[{Tang} et~al.(2017){Tang}, {Zhou}, and {Zhang}]{tang2017multiple}
Peizhe {Tang}, Quan {Zhou}, and Shou-Cheng {Zhang}.
\newblock {Multiple Types of Topological Fermions in Transition Metal
  Silicides}.
\newblock \emph{Physical Review Letters}, 119\penalty0 (20):\penalty0 206402,
  November 2017.

\bibitem[{Sanchez} et~al.(2019){Sanchez}, {Belopolski}, {Cochran}, {Xu}, {Yin},
  {Chang}, {Xie}, {Manna}, {S{\"u}{\ss}}, {Huang}, {Alidoust}, {Multer},
  {Zhang}, {Shumiya}, {Wang}, {Wang}, {Chang}, {Felser}, {Xu}, {Jia}, {Lin},
  and {Hasan}]{sanchez2019topological}
Daniel~S. {Sanchez}, Ilya {Belopolski}, Tyler~A. {Cochran}, Xitong {Xu},
  Jia-Xin {Yin}, Guoqing {Chang}, Weiwei {Xie}, Kaustuv {Manna}, Vicky
  {S{\"u}{\ss}}, Cheng-Yi {Huang}, Nasser {Alidoust}, Daniel {Multer},
  Songtian~S. {Zhang}, Nana {Shumiya}, Xirui {Wang}, Guang-Qiang {Wang},
  Tay-Rong {Chang}, Claudia {Felser}, Su-Yang {Xu}, Shuang {Jia}, Hsin {Lin},
  and M.~Zahid {Hasan}.
\newblock {Topological chiral crystals with helicoid-arc quantum states}.
\newblock \emph{Nature}, 567\penalty0 (7749):\penalty0 500--505, March 2019.

\bibitem[{Schr{\"o}ter} et~al.(2019){Schr{\"o}ter}, {Pei}, {Vergniory}, {Sun},
  {Manna}, {de Juan}, {Krieger}, {S{\"u}ss}, {Schmidt}, {Dudin}, {Bradlyn},
  {Kim}, {Schmitt}, {Cacho}, {Felser}, {Strocov}, and
  {Chen}]{schroter2019chiral}
Niels B.~M. {Schr{\"o}ter}, Ding {Pei}, Maia~G. {Vergniory}, Yan {Sun}, Kaustuv
  {Manna}, Fernando {de Juan}, Jonas.~A. {Krieger}, Vicky {S{\"u}ss}, Marcus
  {Schmidt}, Pavel {Dudin}, Barry {Bradlyn}, Timur~K. {Kim}, Thorsten
  {Schmitt}, Cephise {Cacho}, Claudia {Felser}, Vladimir~N. {Strocov}, and
  Yulin {Chen}.
\newblock {Chiral topological semimetal with multifold band crossings and long
  Fermi arcs}.
\newblock \emph{Nature Physics}, 15\penalty0 (8):\penalty0 759--765, May 2019.

\bibitem[{Xu} et~al.(2020){Xu}, {Fang}, {S{\'a}nchez-Mart{\'\i}nez},
  {Venderbos}, {Ni}, {Qiu}, {Manna}, {Wang}, {Paglione}, {Bernhard}, {Felser},
  {Mele}, {Grushin}, {Rappe}, and {Wu}]{xu2020optical}
Bing {Xu}, Zhenyao {Fang}, Miguel-{\'A}ngel {S{\'a}nchez-Mart{\'\i}nez}, Jorn
  W.~F. {Venderbos}, Zhuoliang {Ni}, Tian {Qiu}, Kaustuv {Manna}, Kefeng
  {Wang}, Johnpierre {Paglione}, Christian {Bernhard}, Claudia {Felser},
  Eugene~J. {Mele}, Adolfo~G. {Grushin}, Andrew~M. {Rappe}, and Liang {Wu}.
\newblock {Optical signatures of multifold fermions in the chiral topological
  semimetal CoSi}.
\newblock \emph{Proceedings of the National Academy of Science}, 117\penalty0
  (44):\penalty0 27104--27110, November 2020.

\bibitem[{Sekh} and {Mandal}(2021)]{MandalDichroism}
Sajid {Sekh} and Ipsita {Mandal}.
\newblock {Circular Dichroism As A Probe For Topology In 3D Semimetals}.
\newblock \emph{arXiv preprint arXiv:2105.05272}, May 2021.

\bibitem[{Fang} et~al.(2012){Fang}, {Gilbert}, {Dai}, and
  {Bernevig}]{fang2012multi}
Chen {Fang}, Matthew~J. {Gilbert}, Xi~{Dai}, and B.~Andrei {Bernevig}.
\newblock {Multi-Weyl Topological Semimetals Stabilized by Point Group
  Symmetry}.
\newblock \emph{Physical Review Letters}, 108\penalty0 (26):\penalty0 266802,
  June 2012.

\bibitem[{Ni} et~al.(2020){Ni}, {Xu}, {S{\'a}nchez-Mart{\'\i}nez}, {Zhang},
  {Manna}, {Bernhard}, {Venderbos}, {de Juan}, {Felser}, {Grushin}, and
  {Wu}]{ni2020linear}
Zhuoliang {Ni}, B.~{Xu}, M.~{\'A}. {S{\'a}nchez-Mart{\'\i}nez}, Y.~{Zhang},
  K.~{Manna}, C.~{Bernhard}, J.~W.~F. {Venderbos}, F.~{de Juan}, C.~{Felser},
  A.~G. {Grushin}, and Liang {Wu}.
\newblock {Linear and nonlinear optical responses in the chiral multifold
  semimetal RhSi}.
\newblock \emph{npj Quantum Materials}, 5:\penalty0 96, January 2020.

\bibitem[{Zhang} et~al.(2017){Zhang}, {Zhang}, {Wang}, {Liu}, {Chen}, {Lu},
  {Liang}, {Cao}, {Yuan}, {Tang}, {Li}, {Zhou}, {Gu}, {Wu}, {Zou}, and
  {Xiu}]{zhang2017room}
Cheng {Zhang}, Enze {Zhang}, Weiyi {Wang}, Yanwen {Liu}, Zhi-Gang {Chen},
  Shiheng {Lu}, Sihang {Liang}, Junzhi {Cao}, Xiang {Yuan}, Lei {Tang}, Qian
  {Li}, Chao {Zhou}, Teng {Gu}, Yizheng {Wu}, Jin {Zou}, and Faxian {Xiu}.
\newblock {Room-temperature chiral charge pumping in Dirac semimetals}.
\newblock \emph{Nature Communications}, 8:\penalty0 13741, January 2017.

\bibitem[Kharzeev and Yee(2011)]{CMW}
Dmitri~E. Kharzeev and Ho-Ung Yee.
\newblock Chiral magnetic wave.
\newblock \emph{Physical Review D}, 83\penalty0 (8), April 2011.

\bibitem[Tchoumakov et~al.(2021)Tchoumakov, Bujnowski, Noky, Gooth, Grushin,
  and Cayssol]{GrushinTransmit}
Serguei Tchoumakov, Bogusz Bujnowski, Jonathan Noky, Johannes Gooth, Adolfo~G.
  Grushin, and Jérôme Cayssol.
\newblock Conservation of chirality at a junction between two weyl semimetals.
\newblock \emph{arXiv preprint arXiv:2106.02462}, June 2021.

\bibitem[Szczech and Jin(2010)]{SzczechWires}
Jeannine~R. Szczech and Song Jin.
\newblock Epitaxially-hyperbranched fesi nanowires exhibiting merohedral
  twinning.
\newblock \emph{J. Mater. Chem.}, 20:\penalty0 1375--1382, 2010.

\end{thebibliography}

\clearpage
\newpage

\appendix
\include{sections/appendixSdkr}

\end{document}